\documentclass[10pt]{article}
 
\usepackage{natbib}
\usepackage{hyperref}

\usepackage{amsfonts}
\usepackage{amsthm}
\usepackage{amscd}

\usepackage{graphicx}
\usepackage{color}
\usepackage{graphics}
\usepackage{amsmath}
\usepackage{amssymb}
\usepackage{amsthm}
 \usepackage{rotating}
\usepackage{epstopdf}
\usepackage{epsfig}
\usepackage{url}
 \usepackage{xr}

\usepackage{verbatim}
\usepackage{float}
\usepackage{algorithm}
\usepackage{algorithmic}
\usepackage{lscape}
\usepackage{array,arydshln}

\usepackage{xcolor}

\usepackage{enumerate}
\usepackage{enumitem}

\usepackage{ulem}  

\newcommand\ba {\mathbf a}
\newcommand\bb {\mathbf b}
\newcommand\bc {\mathbf c}
\newcommand\bd {\mathbf d}
\newcommand\be {\mathbf e}

\newcommand\bg {\mathbf g}
\newcommand\bh {\mathbf h}

\newcommand\bt {\mathbf t}

\newcommand\bx {\mathbf x}

\newcommand\bz {\mathbf z}

\newcommand\bA {\mathbf A}
\newcommand\bB {\mathbf B}
\newcommand\bC {\mathbf C}
\newcommand\bD {\mathbf D}

\newcommand\bF {\mathbf F}

\newcommand\bR {\mathbf R}

\newcommand\bV {\mathbf V}
\newcommand\bW {\mathbf W}
\newcommand\bX {\mathbf X}

\newcommand\bZ {\mathbf Z}

\newcommand\bbG {\mathbb{G}}

\newcommand\indica {\mathbb{I}}

\newcommand\wc {\widehat{{c}}}

\newcommand\wbc {\widehat{\bc}}

\newcommand\wefe {\widehat{f}}

\newcommand\wh {\widehat{{h}}}

\newcommand\wbh {\widehat{\bh}}
\newcommand\wm {\widehat{m}}

\newcommand\wbA {\widehat{\bA}}
\newcommand\wbB {\widehat{\bB}}

\newcommand\wbD {\widehat{\bD}}

\newcommand\wta {\widetilde{a}}

\newcommand\wtbb {\widetilde{\bb}}

\newcommand\wtg {\widetilde{g}}

\newcommand\wtbt {\widetilde{\bt}}

\newcommand\wtbz {\widetilde{\bz}}
\newcommand\wtbZ {\widetilde{\bZ}}

\newcommand\wtB {\widetilde{B}}

\newcommand\itA {{\mathcal{A}}}
\newcommand\itB {{\mathcal{B}}}
\newcommand\itC {{\mathcal{C}}}

\newcommand\itE {{\mathcal{E}}}
\newcommand\itF {{\mathcal{F}}}
\newcommand\itG {{\mathcal{G}}}
\newcommand\itH {{\mathcal{H}}}
\newcommand\itI {{\mathcal{I}}}

\newcommand\itN {{\mathcal{N}}}

\newcommand\itS {{\mathcal{S}}}

\newcommand\itV {{\mathcal{V}}}

\newcommand{\eps}{\varepsilon}

\newcommand\bbe {\mbox{\boldmath $\beta$}}

\newcommand\bbech {\mbox{\scriptsize${\bbe}$}}

\newcommand\bphi {\mbox{\boldmath $\phi$}}

\newcommand\bla {\mbox{\boldmath $\lambda$}}

\newcommand\blach {\mbox{\footnotesize$\bla$}}

\newcommand\btau {\mbox{\boldmath $\tau$}}
\newcommand\bthe {\mbox{\boldmath $\theta$}}

\newcommand\bthech {\mbox{\footnotesize$\bthe$}}

\newcommand\btauch {\mbox{\footnotesize$\btau$}}

\newcommand\bxi {\mbox{\boldmath $\xi$}}
\newcommand\bxich {\mbox{\footnotesize $\bxi$}}

\newcommand\bSi {\mbox{\boldmath $\Sigma$}}

\newcommand\wbeta {\widehat{\beta}}
\newcommand\wbbe {\widehat{\bbe}}

\newcommand\weta {\widehat{\eta}}
\newcommand\weps {\widehat{\epsilon}}

\newcommand\wphi {\widehat{\phi}}
\newcommand\wbphi {\widehat{\bphi}}
\newcommand\wbphich {\mbox{\footnotesize$\widehat{\bphi}$}}

\newcommand\wmu {\widehat{\mu}}

\newcommand\wsigma {\widehat{\sigma}}

\newcommand\wtheta {\widehat{\theta}}
\newcommand\wbthe {\widehat{\bthe}}
\newcommand\wbthech  {\mbox{\footnotesize$\wbthe$}}

\newcommand\wup {\widehat{\upsilon}}
\newcommand\wxi {\widehat{\xi}}
\newcommand\wbxi {\widehat{\bxi}}

\newcommand\wbxich  {\mbox{\footnotesize$\wbxi$}}

\newcommand\wbSi {\widehat{\bSi}}

\newcommand\wbtau {\widehat{\btau}}

\newcommand\wteta {\widetilde{\eta}}

\newcommand{\tuk}{\mbox{\scriptsize \sc t}}

\newcommand{\eme}{\mbox{\scriptsize \sc m}}

\newcommand{\ini}{\mbox{\footnotesize \sc ini}}

\def\real{\mathbb{R}}
\def\natu{\mathbb{N}}
\def\qu{\mathbb{Q}}

\newcommand{\esp}{\mathbb{E}}
\newcommand{\prob}{\mathbb{P}}

\newcommand{\var}{\mbox{\sc Var}}

\newcommand{\as}{\mbox{\footnotesize a.s.}}

\newcommand{\convpp}{ \buildrel{a.s.}\over\longrightarrow}

\newcommand{\convprob  }{ \buildrel{p}\over\longrightarrow}

\newcommand{\convdist}{ \buildrel{D}\over\longrightarrow}

\newcommand{\trasp}{^{\mbox{\footnotesize \sc t}}}
\newcommand{\trass}{{\mbox{\footnotesize \sc t}}}

\newcommand\bcero {{\bf{0}}}

\def\dst{\displaystyle}

\def\argmin{\mathop{\mbox{argmin}}}

\newcommand\noi{\noindent}
\def\dst{\displaystyle}

\def\square{\ifmmode\sqr\else{$\sqr$}\fi}
\def\sqr{\vcenter{
         \hrule height.1mm
         \hbox{\vrule width.1mm height2.2mm\kern2.18mm
\vrule width.1mm}
         \hrule height.1mm}}

\newcommand\ls {\mbox{\scriptsize\sc ls}}

\newcommand\trim {\mbox{\footnotesize  \sc tr}}

 \theoremstyle{plain}

\newtheorem{theorem}{Theorem}[section]
\newtheorem{lemma}[theorem]{Lemma}

\newtheorem{proposition}[theorem]{Proposition} 
\newtheorem{remark}{Remark}[section]

\begin{document}
  
\title{A robust spline approach in partially linear additive models}
\author{Graciela Boente$^1$ and  Alejandra Mercedes Mart\'{\i}nez$^2$\\ 
\small $^1$  CONICET and Universidad de Buenos Aires,  Argentina\\
\small $^2$   CONICET and Universidad Nacional de Luj\'an,   Argentina} 
  
\date{}
\maketitle
 
\begin{abstract}
Partially linear additive models generalize  linear ones since they model the relation between a response variable and covariates by assuming that some covariates  have a linear relation with the response but each of the others enter through  unknown univariate smooth functions. The harmful effect of outliers either in the residuals or in the covariates involved in the linear component has been described in the situation of partially linear models, that is, when only one nonparametric component is involved in the model. When dealing with additive components, the problem of providing reliable estimators when atypical data arise  is  of practical importance motivating the need of robust procedures. Hence, we propose a family of robust estimators for partially linear additive models by combining $B-$splines with robust linear regression estimators. We obtain consistency results, rates of convergence and asymptotic normality for the linear components, under mild assumptions. A Monte Carlo study is carried out to compare the performance of the robust proposal  with its classical counterpart under different models and contamination schemes. The numerical experiments show the advantage of the proposed methodology for finite samples. We also illustrate the usefulness of the proposed approach on a real data set.
\end{abstract} 

\vskip0.2in

\section{Introduction}{\label{intro}}

 Different approaches have been considered in the literature to deal with  the well-known \lq \lq curse of dimensionality\rq \rq ~of fully nonparametric regression models. Among others, we can mention additive models, single--index models, varying coefficient models and partial linear models. Specifically, partial linear models allow the response variables to depend linearly on some covariates, while the others are modelled in a fully non-parametric way. More precisely, in such models we deal with  observations $(Y_i,\bZ_i\trasp,\bX_i\trasp)\trasp$ independent and identically distributed  with the same distribution as $(Y,\bZ\trasp,\bX\trasp)\trasp$ where  $Y\in\real$, $\bZ\in\real^q$ and $\bX\in\real^p$. The relationship between the response  and covariates is given through
\begin{equation}\label{plm}
Y = m(\bZ\trasp,\bX\trasp)+\sigma\;\eps = \bbe\trasp\bZ+\eta(\bX)+\sigma\;\eps 
\end{equation}
where the error  $\eps$   is independent from $(\bZ\trasp,\bX\trasp)\trasp$,  $\bbe\in\real^q$  is the regression parameter and the unknown multivariate function $\eta:\real^p\to \real$ is assumed to be smooth. Furthermore, in the classical setting, it is usually assumed that $\esp( \eps) =0$ and $\var(\eps)=1$, so $\sigma>0$ stands for the unknown   scale parameter. 

The particular situation where  $X\in \real$ is considered in   H\"ardle \textsl{et al.} (2000) and  H\"ardle \textsl{et al.} (2004) who describe different procedures based on  kernels  or splines to estimate the unknown quantities. As  in linear regression models, these estimators are very sensitive to atypical observations since they   are   based on least squares principle. To deal with more reliable procedures when atypical data arise, when $p=1$,  Bianco and Boente (2004) considered robust estimators based on local $M-$estimators, while   He and Shi (1996) and He \textsl{et al.} (2002) studied $M-$estimators based on splines. 

It is worth mentioning that model \eqref{plm} requires multivariate smoothing when $p\ne 1$, so that  the \lq \lq curse of dimensionality\rq \rq ~is not overcome in particular when  $p\ge 4$. Partially linear additive regression models (\textsc{plam}) provide an   attempt to solve this problem, since the covariates related to the nonparametric component enter to the model through an additive structure. Hence, under a \textsc{plam}, the relationship between the response variable and the covariates is given through 
\begin{equation}\label{plam}
Y=\bbe\trasp\bZ+\eta(\bX)+\sigma\;\eps= \mu + \bbe\trasp\bZ+\sum_{j=1}^p \eta_j(X_{j})+\sigma \;\eps 
\end{equation}
where the univariate unknown functions $\eta_j$ ($1\leq j\leq p$), the coefficients $\mu\in\real$, the scale parameter $\sigma >0$ and $\bbe\in\real^q$ are the quantities to  estimate. Usually, the functions $\eta_j$ are assumed to be   continuous  with support on a compact interval $\itI_j$  which is also the support of the distribution of $X_j$. To guarantee identifiability, additional constraints on the functions $\eta_j$ are required.  Some of the most common conditions consist  in assuming that $ \int_{\itI_j} \eta_j(x)\, dx=0$, for $j=1,\dots,p$, or $ \esp \eta_j(X_j)=0$, for $j=1,\dots,p$. H\"ardle \textsl{et al.} (2004) describe a least squares kernel approach to estimate the components of the model performing sequential estimations in the direction of interest, which increases the computational cost. A different family of kernel based estimators that reduces   it was studied in  Manzan and Zerom (2005).
A different point of view was followed by Liu \textsl{et al.} (2011) who developed a variable selection procedure based on least squares regression and spline approximation. All these estimators are based on a least squares approach, so, as in partial linear models, a small proportion  of atypical data  may seriously  affect the estimations. A more resistant approach based on quantile regression and spline approximation was suggested in Guo \textsl{et al.} (2013) and extended to censored partially linear additive models by   Liu \textsl{et al.} (2017). Note that quantile estimators are related to an unbounded loss  function and for that reason, as in linear regression models, they may be affected  by high--leverage outliers.

To define estimators robust against high-leverage outliers, we use instead a bounded loss function and a preliminary residual scale estimator. More precisely, our proposal  combines $B-$splines to approximate the additive components with $MM-$regression estimators (Yohai, 1987). Thus, in our approach, we  allow the error distribution to have heavy tails and instead of requiring $\esp ( \eps) =0$ and $\var(\eps)=1$ as in the classical setting, we only require that the error $\eps$ has a symmetric distribution $F(\cdot)$ with scale parameter 1.
The rest of the paper is organized as follows. In Section \ref{sec:proposals},  we describe the robust proposal  considered, while theoretical results regarding   consistency and rates of convergence are presented in Section \ref{sec:consistency}. Furthermore, asymptotic normality results for the estimators of $\bbe$ are derive in Section \ref{sec:asymptoticnormality}. The results of a numerical study conducted to compare  the 
finite-sample properties of the classical least squares and the robust $MM-$estimator are summarized in Section  \ref{montecarlo}, while Section \ref{realdata} contains the analysis of a real data set. Some final comments are presented in Section \ref{comentarios}. All proofs are relegated to the Appendix.


\section{The robust estimators}\label{sec:proposals}
\subsection{Preliminaries}
As mentioned in the introduction, we will consider  independent and identically distributed observations $(Y_i,\bZ_i\trasp,\bX_i\trasp)\trasp$ with the same distribution as $(Y,\bZ\trasp,\bX\trasp)\trasp$ where  $Y\in\real$, $\bZ\in\real^q$ and $\bX\in\real^p$ and the relationship between the responses and the covariates is given through \eqref{plam}. To ensure identifiability of the additive components  $\eta_j:\itI_j\to \real $, we will impose the constraint  $ \int_{\itI_j} \eta_j(x)\, dx=0$, for $j=1,\dots,p$. The errors $\eps$ are assumed to be independent of the explanatory variables $(\bZ\trasp,\bX\trasp)\trasp$.

Taking into account that we intend to define robust estimators, we avoid moment conditions for the errors distribution and  allow it to have heavy tails by
just requiring that    $\eps$ has a symmetric distribution $F(\cdot)$ with scale parameter 1.

Our robust proposal   is based on $B-$splines, that is, we use a spline basis to approximate each additive function $\eta_j$ in \eqref{plam}.   To define the $B-$splines based estimators, fix the desired spline order $\ell_j$ and the number of knots $N_{n,j}$ used to estimate $\eta_j$.  Therefore, the  dimension of the  $B-$spline basis used in the approximation of $\eta_j$ has dimension $k_j=k_{n,j}=N_{n,j}+\ell_j$ and we will denote this basis  $\{\wtB_s^{(j)}: 1\leq s\leq k_{n,j}\}$.  
 
It is worth mentioning that a spline of  order $\ell$   is a polynomial of degree $\ell - 1$ within each   subinterval.  The  results derived in Section \ref{sec:consistency} show that, when using cubic splines, consistency is obtained when the additive components are   twice continuously differentiable.

The robust estimators to be defined are based on $MM-$regression estimators after approximating the components $\eta_j$ by  a spline. Under a linear regression model, $MM-$estimators were introduced in Yohai (1987). To attain robustness and efficiency, a two--step procedure is implemented. In the first step,  an initial robust regression estimator is used to estimate the  residual scale, this initial estimator may be inefficient but it should have a high breakdown point. In the second step, a regression $M-$estimator is computed using a bounded loss function and standardized residuals. The final estimator will retain the high breakdown of the initial one but its efficiency is improved by the use of an appropriate loss function.

To define our estimators, for any vector  
$\bla^{(j)}=(\lambda_1^{(j)}, \dots, \lambda_{k_j}^{(j)})\trasp\in \real^{k_j}$, we consider a spline $\sum_{s=1}^{k_j} \lambda_{s}^{(j)} \wtB_{s}^{(j)}(t)$ to approximate $\eta_j$, for $1\le j\le p$. It is worth noticing that even when considering the classical least squares estimators, the  minimization should be carried out taking into account the constraints $ \sum_{s=1}^{k_j} \lambda_{s}^{(j)} \int_{\itI_j} \wtB_{s}^{(j)}(x)\, dx=0$, for $j=1,\dots,p$. As in Guo \textsl{et al.} (2013), to deal with an unconstrained optimization problem,  we  center the basis functions, that is, we define $ B_s^{(j)}(x)=\wtB_s^{(j)}(x)- \int_{\itI_j}\wtB_s^{(j)}(x) dx$ and the centered approximation candidates for $\eta_j$ as $ \sum_{s=1}^{k_j} \lambda_{s}^{(j)}  B_{s}^{(j)}(t)$.

For given values $a\in \real$, $\bb\in \real^q$ and $\bla^{(j)} \in \real^{k_j}$, 
the classical least squares estimator is obtained minimizing $\sum_{i=1}^n \left\{ Y_i -a-\bb\trasp\bZ_i-\sum_{j=1}^p \sum_{s=1}^{k_j}\lambda_{s}^{(j)} B_s^{(j)}(X_{ij})\right\}^2$. However, the design matrix for this problem is ill conditioned even when $p=1$ due to the intercept. Effectively,  taking into account that $\sum_{s=1}^{k_j} \wtB_s^{(j)}(x)=1$, for all $x\in \itI_j$,  we easily obtain that $\sum_{s=1}^{k_j}  B_s^{(j)}(x)=0$. Thus, we may rewrite the approximation as
$$\sum_{s=1}^{k_j}\lambda_{s}^{(j)} B_s^{(j)}(x)=\sum_{s=1}^{k_j-1} \left(\lambda_{s}^{(j)} -\lambda_{k_j}^{(j)}\right) B_s^{(j)}(x)\,.$$
 For that reason,   we define $\bc^{(j)}=(c_1^{(j)}, \dots, c_{k_j-1}^{(j)})\trasp \in \real^{k_j-1}$ with $c_s^{(j)}=\lambda_{s}^{(j)}-\lambda_{k_j}^{(j)}$ and for   $1\le i\le n$  the residuals as
  \begin{equation} \label{eq:residuals}
r_i(a,{\bb},  \bc)=r_i(a,{\bb}, \bc^{(1)}, \dots, \bc^{(p)})\, =  \, Y_i -a-\bb\trasp\bZ_i-\sum_{j=1}^p \sum_{s=1}^{k_j-1} c_{s}^{(j)} B_s^{(j)}(X_{ij})=  \, Y_i -a-\bb\trasp\bZ_i-\bc\trasp\bV_i\, ,
\end{equation}
where $\bc=( \bc^{(1)\mbox{\footnotesize{\sc t}}},\dots,\bc^{(p)\mbox{\footnotesize{\sc t}}})\trasp\in \real^K$, with $K=\sum_{j=1}^p k_j -p$,  $\bV_i=(\bV^{(1)}(X_{i1})\trasp,\dots,\bV^{(p)}(X_{ip})\trasp)\trasp$ and $\bV^{(j)}(t)=(B_1^{(j)}(t),\dots,B_{k_j-1}^{(j)}(t))\trasp$ for $j=1,\dots,p$, which leads to a well conditioned design matrix.

\subsection{The robust $MM-$estimators}{\label{sec:MMestimador}}
In what follows the loss functions to be considered will be bounded $\rho-$functions as defined in Maronna \textsl{et al.}  (2019) (see  assumption \ref{ass:rho_bounded_derivable}\textbf{(a)}). A widely used family of bounded $\rho-$functions the Tukey's bisquare function defined as
$\rho_{\,\tuk,\,c}(t) =\min\left(1 - (1-(t/c)^2)^3, 1\right)$, where $c > 0$ is a tuning parameter 
that determines the robustness and efficiency properties of the associated estimators. 
 
To define the robust estimators, as in linear regression, we first  compute an $S-$estimator  and its associated residual scale. 
For that purpose, let $\rho_0$ be a bounded $\rho-$function and $s_n(a,\bb, \bc)$ be the $M-$scale estimator of the residuals given
as the solution to the following equation:
\begin{equation} \label{eq:s-est}
\frac{1}{n-q-K}\sum_{i=1}^n \rho_{0}\left(\frac{r_i(a, \bb,\bc)}{s_n(a,\bb, \bc)}\right) \, = \, b\, ,
\end{equation}
where $K=\sum_{j=1}^p k_j -p$  and to ensure consistency of the scale estimators $b =\esp( \rho_{0} (\eps) )$. 
As described in Maronna \textsl{et al.}  (2019), we use $1 / (n-q-K)$ instead of $1/n$ in \eqref{eq:s-est} 
to control the effect of a possibly large number of parameters  relative to the sample size. 
When $\rho_{0}$ is the Tukey's bisquare function the choice  $c_0 = 1.54764$ for the tuning constant and $b = 1/2$ ensures that the scale estimator  
 has 50\% breakdown point and is Fisher-consistent when the errors have a normal distribution.

The initial $S-$estimators are defined as the minimizers of   $s_n(a, \bb,\bc)$, that is, $\weta_{j,\ini}(x)= \sum_{s=1}^{k_j-1} \wc_{s,\ini}^{(j)}  B_s^{(j)}(x)   $ where $\wbc_{\ini}=( \wbc_{\ini}^{(1)\mbox{\footnotesize{\sc t}}},\dots,\wbc_{\ini}^{(p)\mbox{\footnotesize{\sc t}}})\trasp$ and
\begin{equation*} \label{eq:m-scale}
(\wmu_{\ini}, \wbbe_{\ini},\wbc_{\ini}) \ = \ \argmin_{a\in \real, \bb\in \real^q, \bc\in \real^K}  \, s_n(a,\bb, \bc) \, .
\end{equation*}
The   residual scale estimator equals 
\begin{equation} \label{eq:scale_est}
\wsigma  \, =  s_n(\wmu_{\ini}, \wbbe_{\ini}, \wbc_{\ini}) \, .
\end{equation}
To define the final $M-$estimator, consider a $\rho-$function  $\rho_{1}$ such that  $\rho_{1} \le \rho_{0}$ 
and $\sup_t\rho_1(t)=\sup_t\rho_0(t)$.  For instance, when  $\rho_{0}=\rho_{\,\tuk,\,c_0}$ and $\rho_{1}=\rho_{\,\tuk,\,c_1}$, this last condition is satisfied   when $c_1>c_0$. We compute an $M-$estimator with the residual scale estimator $\wsigma$ defined in \eqref{eq:scale_est} and 
the loss function  $\rho_{1}$, that is, 
\begin{equation}
(\wmu, \wbbe,\wbc)  \ = \  \argmin_{a\in \real, \bb\in \real^q, \bc\in \real^K} \sum_{i=1}^n \rho_{1} \left ( \frac{r_i(a, \bb,\bc) }{\wsigma} \right )\,.
\label{eq:estfinitos}
\end{equation}
The resulting estimators of the additive functions $\eta_j$ is  given by
\begin{eqnarray}
\weta_j(x)= \sum_{s=1}^{k_j-1} \wc_{s}^{(j)}  B_s^{(j)}(x)   \, ,
\label{eq:estimadoresgj}
\end{eqnarray}
where $\wbc=( \wbc^{(1)\mbox{\footnotesize{\sc t}}},\dots,\wbc^{(p)\mbox{\footnotesize{\sc t}}})\trasp$ and  $\wbc^{(j)}=(\wc_1^{(j)}, \dots, \wc_{k_j-1}^{(j)})\trasp$. 
The estimator of the multivariate regression function is then defined as $\wm(\bz ,\bx )= \wmu+\wbbe\trasp\bz+\sum_{j=1}^p \weta_j(x_j)$, for any  $\bz\in \real^q$ and $\bx=(x_1,\dots,x_p)\trasp$.

\subsection{Selection of $k_j$}
 
An important topic is the choice of the number of knots and their location for the space of $B-$splines. Knot selection is more important for the estimate of $\eta_j$ than for the estimate of $\bbe$. One approach is to use uniform knots,  which are usually sufficient when the function $\eta_j$ does not exhibit dramatic changes in its derivatives. On the other hand, non--uniform knots are desirable when the function has very different local behaviours in different regions. A  commonly used approach in this last situation is to consider as knots the quantiles of the observed explanatory variables $X_{ji}$, $1\le i\le n$, with uniform percentile ranks.

The number of elements of the basis which approximates each additive function  may be determined by a model selection criterion. However, it is well known that,  to ensure robustness properties of the final  estimator, a robust criterion is needed. A robust $BIC$ criterion  may be defined as in  He \textsl{et al.}  (2002)  as follows
 \begin{equation}\label{genBIC}
RBIC(\textbf{k})=\log\left(\wsigma^2\sum_{i=1}^n \rho_1\left(\frac{r_i}{\wsigma}\right)\right)+\frac{\log(n)}{2n}\,\sum_{j=1}^p k_j
\end{equation}
where $\textbf{k}=(k_1,\dots, k_p)\trasp$,   $r_i=Y_i-\wm(\bZ_i,\bX_i)$ are the residuals obtained using a basis of dimension $k_j$ to compute the estimator of $\eta_j$, $\rho_1$ is the same $\rho-$function used to compute the $M-$estimator and $\wsigma$ is the corresponding $S-$estimator. It is worth noting that when $\rho(t)=t^2$ and $p=1$, the proposed generalized criteria reduces to the criteria considered in He \textsl{et al.} (2002). 
Note that, when  the same number of elements of the basis is used for each additive component, the $RBIC$ criteria reduces to $$RBIC(k)=\log\left(\wsigma^2\sum_{i=1}^n \rho_1\left(\frac{r_i}{\wsigma}\right)\right)+p \,\frac{\log(n)}{2n} k .$$

\section{Consistency results}\label{sec:consistency}
In this section we will derive consistency results for the estimators defined in Section \ref{sec:proposals} under assumptions \ref{ass:densidad} to \ref{ass:proba} below. As in Gou \textsl{et al.} (2003),   without loss of generality,  we will assume $\itI_j=[0,1]$ for $j=1,\dots,p$. In assumption \ref{ass:rho_bounded_derivable} below the function $\rho$ will correspond to either $\rho_0$ or $\rho_1$ according to the result to be derived. From now on $\itC^r(0,1)$ will stand for the space of functions continuously differentiable  up to order $r$, $\|\cdot\|$ refers to the Euclidean norm in $\real^q$ and for any continuous function $v\,:\,\real\to\real$,  $\|v\|_{\infty}=\sup_t|v(t)|$. We will denote as $\itG$ the class of functions $\itG=\{g:[0,1]\to \real \mbox{ such that } \int_0^1 g(x)dx=0\}$ and for any $r\ge 1$, we define
$$\itH_{r}=\{\eta\in\itC^{r}[0,1]\,:\,\|\eta^{(\ell)}\|_{\infty}<\infty \,,\, 0\leq \ell\leq {r}\,\mbox{ and}\, \sup_{z_1\ne z_2}\frac{|\eta^{(r)}(z_1)-\eta^{(r)}(z_2)|}{|z_1-z_2|} <\infty\}$$ 

\begin{enumerate}[label=\textbf{C\arabic*}]
\item\label{ass:densidad}
 The random variable $\eps$ has density function $f_0(t)$ that is even, monotone non--decreasing in $|t|$, and strictly decreasing for $|t|$ in a neighbourhood of $0$.

\item\label{ass:rho_bounded_derivable}
\begin{itemize}
\item[\textbf{(a)}] The function $\rho:\real\to[0;+\infty)$ is a bounded continuous, even, non--decreasing in $[0,+\infty)$ and such that $\rho(0)=0$. Furthermore,  $\lim_{u\to +\infty}\rho(u)\neq 0$ and if $0\leq u<v$ with $\rho(v)<\sup_u \rho(u)$ then $\rho(u)<\rho(v)$. Without loss of generality, since $\rho$ is bounded, we assume that $\sup_u \rho(u)=1$.
\item[\textbf{(b)}] $\rho$ is continuously differentiable with bounded derivative $\psi$ such that $\zeta(s)=s\psi(s)$ is bounded.
\end{itemize}

\item\label{ass:etajCr} For $1\leq j\leq p$, the true function $\eta_{j}\in \itH_{r_j}$ where  $ r_j \ge 1$. Furthermore, the splines order used to estimate $\eta_{j}$ satisfy $\ell_j\ge r_j+2$. 

 \item\label{ass:kj} The basis dimension $k_j$ is assumed to be of order $O(n^{\nu_j})$ with  $0<\nu_j <1/(2r_j)$, with $r_j$ given in \ref{ass:etajCr}.  Moreover, the ratio of maximum and minimum spacings of knots is uniformly bounded.

 \item\label{ass:wsigma}  $\wsigma$ is a strong consistent estimator of $\sigma$. 
 
\item \label{ass:probacond} For almost any $\bx_0\in \real^p$,  $\prob(\bb \trasp\bZ=a|\bX=\bx_0)<1$, for any 
 $a\in \real$, $\bb\in\real^q$, $(\bb,a)\ne 0$.

\item\label{ass:proba} There exists $0<c<1$ such that $\prob\left(\bb \trasp\bZ+\sum_{j=1}^p g_j(X_{j})=a\right)<c$ for any $a\in \real$, $\bb\in\real^q$, $g_j\in \itG$,  $(a,\bb, g_1,\dots, g_p)\ne 0$. 

\end{enumerate}

\vskip0.1in
\begin{remark}{\label{rem:remark1}} Conditions \ref{ass:densidad} and \ref{ass:rho_bounded_derivable} are standard conditions for the errors and for the loss function, respectively. The first one is a condition assumed in the context of robustness  to ensure Fisher--consistency. In this sense,   \ref{ass:probacond} is also a requirement for Fisher--consistency and it is the conditional counterpart of the usual assumption in linear regression models to guarantee Fisher--consistency. Note that if, for almost any $\bx_0\in \real^p$, the distribution of $\bZ$ given $\bX=\bx_0$ has a density, then $\prob(\bb \trasp\bZ=a|\bX=\bx_0)=0$, for any 
 $a\in \real$, $\bb\in\real^q$, $(\bb,a)\ne 0$, implying that \ref{ass:probacond} and \ref{ass:proba} hold. Furthermore, it is worth mentioning that   \ref{ass:probacond} holds whenever   \ref{ass:proba}  is fulfilled with $c=0$. 

Condition \ref{ass:etajCr} regards the smoothness of the additive  nonparametric components and $r_j$ corresponds to the   smoothness degree of the $j-$additive true functions $\eta_{j}$. The regularity of the additive components stated in \ref{ass:etajCr} is  related to the order of the $B-$splines used to approximate them, meaning that if for instance cubic splines are used, our results will be valid for twice continuously differentiable functions.  As mentioned in He \textsl{et al.} (2002), if we think that $\eta_{j}$ is less smooth, quadratic splines can be considered. 

The condition about the knots spacing given in \ref{ass:kj}  is a standard one when using $B-$spline approximations. 

Strong consistency of the preliminary scale estimator is required in \ref{ass:wsigma} to allow for  scale estimators besides the one introduced in Section \ref{sec:MMestimador}. Proposition \ref{prop:prop1} below states that the $S-$scale defined through \eqref{eq:scale_est} is indeed  strongly consistent as required in \ref{ass:wsigma}.  

\end{remark}

The following lemma regards the  Fisher-consistency of the proposed estimators. Fisher consistency guarantees that we are estimating the target quantities and is a first step when deriving consistency results.  

\vskip0.1in

\begin{lemma}{\label{lemma:FC}}
Assume that \ref{ass:densidad} holds and let $\rho$ be a $\rho-$function satisfying \ref{ass:rho_bounded_derivable}\textbf{(a)}. Then, for any $\varsigma>0$, we have that 
\begin{enumerate}[label=(\roman*)]
\item $L(\mu,\bbe,\eta_{1},\dots,\eta_{p},\varsigma)\le L(a,\bb,g_1,\dots,g_p,\varsigma)$, for any $a\in \real$, $\bb\in\real^q$, $g_1\in\itG,\dots,g_p\in\itG$ where
\begin{equation}{\label{eq:funcionL}}
L(a,\bb,g_1,\dots,g_p,\varsigma)=\esp\rho\left(\frac{Y_1-a-\bb\trasp\bZ-\sum_{j=1}^p g_j(X_{j})}{\varsigma}\right)\,.
\end{equation}
\item If in addition \ref{ass:probacond} holds,   $(\mu,\bbe,\eta_{1},\dots,\eta_{p})$ is the unique minimizer of $L(a,\bb,g_1,\dots,g_p,\varsigma)$.
\end{enumerate}
\end{lemma}

\vskip0.1in

Proposition \ref{prop:prop1} derives strong consistency results for the residual scale estimator $\wsigma$ defined through \eqref{eq:scale_est} meaning that the scale estimators satisfy \ref{ass:wsigma} under mild conditions. To derive this result we define the population counterpart of $\wsigma$. More precisely, let $S(a,\bb,g_1,\dots,g_p)$ be the $M-$scale functional related to the residuals $r(a,\bb,g_1,\dots,g_p)=Y-a-\bb \trasp \bZ-\sum_{j=1}^p g_j(X_j)$, that is, given $a\in \real$, $\bb\in \real^q$ and $g_j\in \itG$, $1\le j\le p$,  $S(a,\bb,g_1,\dots,g_p)$ satisfies 
$$\esp\rho_{0}\left(\frac{r(a,\bb,g_1,\dots,g_p)}{S(a,\bb,g_1,\dots,g_p)}\right)=b\,.$$
For simplicity, we will assume that the scale estimators are calibrated so that 
$$\sigma=S(\mu,\bbe,\eta_{1},\dots,\eta_{p})=\argmin_{a\in \real, \bb\in \real^q,   g_1\in \itG, \dots, g_p\in \itG} S(a,\bb,g_1,\dots,g_p)$$ 
 meaning that $\esp\rho_{0}(\eps)=b$. 

\vskip0.1in
 \begin{proposition}{\label{prop:prop1}}
Assume that the  function $\rho_{0}$   satisfies \ref{ass:rho_bounded_derivable} and that \ref{ass:densidad}, \ref{ass:etajCr} and \ref{ass:kj} hold. Then, we have that $\wsigma\convpp\sigma$.
\end{proposition}

\vskip0.1in

Denote as $\bthe=(\mu,\bbe\trasp,\eta_1,\dots,\eta_p)\trasp$ and $\wbthe=(\wmu,\wbbe\trasp,\weta_1,\dots,\weta_p)\trasp$. To measure the closeness between the estimators and the parameters, given $\bthe_\ell=(a_\ell, \bb_\ell\trasp, g_{\ell,1},\dots, g_{\ell,p})\trasp\in\real\times\real^q\times \itC([0,1])\times\dots\times  \itC([0,1]) $, $\ell=1,2$, we consider the metric $\pi(\bthe_1,\bthe_2)=|a_1-a_2|+\|\bb_1-\bb_2\|+\sum_{j=1}^p \|g_{1,j}- g_{2,j}\|_{\infty}$ and we will use the following norm for the space $\itH_r$
$$\|\eta\|_{\itH_r}=\max_{1\leq j\leq r}\|\eta^{(j)}\|+\sup_{z_1\neq z_2, z_1,z_2\in (0,1)}\frac{|\eta^{(r)}(z_1)-\eta^{(r)}(z_2)|}{|z_1-z_2|}\,.$$

\begin{theorem}{\label{teo:consist}} 
Let $(Y_i,\bZ_i\trasp,\bX_i\trasp)\trasp$ be i.i.d. observations satisfying \eqref{plam}. Assume that \ref{ass:densidad} to \ref{ass:wsigma} hold and that for any $M>0$ and   $\delta>0$, $\inf_{\bt\in\itA_{\delta}}L(\bt,\sigma)>L(\bthe,\sigma)$, where 
$$\itA_{\delta}=\{\bt=(a,\bb\trasp,g_1,\dots,g_p)\trasp: a\in\real, \bb\in\real^q, g_j\in\itG\cap\itH_{r_j}, |a-\mu|+\|\bb-\bbe\|+\sum_{j=1}^p{\|g_j-\eta_j}\|_{\itH_{r_j}}\leq M,  \pi(\bthe,\bt)\ge \delta\}\,.$$
 Then, if in addition $\esp\|\bZ\|^2<\infty$, we have that $\pi(\wbthe,\bthe)\convpp 0$.
\end{theorem}

Proposition \ref{lemma:lema5} supplies sufficient conditions in order to ensure assumption $\inf_{\bt\in\itA_\delta}L(\bt,\sigma)>L(\bthe,\sigma)$ in Theorem \ref{teo:consist}.

\begin{proposition}{\label{lemma:lema5}}
Let $\rho_1$ be a function satisfying \ref{ass:rho_bounded_derivable} and such that $L(\bthe,\sigma)=b_{\rho_1}<1$, where the function $L$ is defined in \eqref{eq:funcionL} with $\rho=\rho_1$. Assume  that \ref{ass:densidad} and   \ref{ass:etajCr} to \ref{ass:probacond}   hold and that \ref{ass:proba} holds with $c<1-b_{\rho_1}$. 
Then, $\inf_{\bt\in\itA_{\delta}}L(\bt,\sigma)>L(\bthe,\sigma)$, entailing that   $\pi(\wbthe,\bthe)\convpp 0$  if  $\esp\|\bZ\|^2<\infty$.
\end{proposition}

\subsection{Rates of convergence}

In this section we show the rate of convergence of the estimators when the distance between two pairs $\bthe_1=(a_1,\bb_1\trasp,g_{1,1}\dots, g_{1,p})\trasp$ and $\bthe_2=(a_2, \bb_2\trasp,g_{2,1}\dots, g_{2,p})\trasp$ is measured through the mean square of the prediction differences, that is, through 
$$\pi^2_\prob(\bthe_1,\bthe_2)=\esp \left[\left(a_1-a_2+(\bb_1-\bb_2)\trasp \bZ+\sum_{j=1}^p (g_{1,j}-g_{2,j})(X_j)\right)^2\right]\,.$$
Furthermore, let $\itS_{j}$, $1\le j\le p$, denote the linear spaces
spanned by the centered $B-$splines bases of order $\ell_j$ and size $k_j$. We omit the dependence of the knots to avoid burden notation. Note that since $\sum_{s=1}^{k_j} B_s^{(j)}(x)=0$ for all $x$, the linear spaces have dimension $k_j-1$, so
\begin{equation}
\label{eq:itSj}
\itS_{j}=\left\{  \sum_{s=1}^{k_j-1} c_s \,   B_s^{(j)}(x) \, ,\, \bc \in \real^{k_j-1}\right\}\,,\quad 1\le j\le p \, .
\end{equation}
For that purpose, we will need the following additional assumption.
\begin{enumerate}[label=\textbf{C\arabic*}]
\setcounter{enumi}{7}
\item\label{ass:lowerbound} 
There exists a neighbourhood $\itV$ of $\sigma$ with closure $\overline{\itV}$ strictly included in $(0;+\infty)$, and constants $\epsilon_0$ and $C_0$ such that $L(\bt,\varsigma)-L(\bthe,\varsigma)\geq C_0\pi^2_\prob(\bt,\bthe)$ for any $\bt=(a,\bb\trasp,g_1,\dots,g_p)\trasp\in\real\times\real^q\times\itS_1\times\dots\times\itS_p$ such that $|a-\mu|+\|\bb-\bbe\| +\sum_{j=1}^p \|g_j-\eta_j \|_\infty < \epsilon_0$ and any $\varsigma\in\itV$.  
\end{enumerate}
\vskip0.1in

 The following Theorem provides converge rates in terms of the prediction distance $\pi_\prob $.
 
\vskip0.1in
\begin{theorem}{\label{teo:ratesnew}}
Let $\rho_1$ be a  function   satisfying  \ref{ass:rho_bounded_derivable}  and assume that $\psi_1$ is continuously differentiable with bounded derivative.  Assume   that \ref{ass:densidad}, \ref{ass:etajCr} to \ref{ass:probacond} and \ref{ass:lowerbound}  hold. Furthermore, assume that  $\esp \|\bZ\|^2<\infty$ and   \ref{ass:proba} holds  with $c<1-b_{\rho_1}$ and $b_{\rho_1}=L(\bthe,\sigma)<1$. Let  $0<\gamma_n$ be such that $\gamma_n= O( n^{\lambda})$  where $\lambda=\min_{1\le j\le p}(r_j\, \nu_j)$, $\gamma_n \, \sqrt{\log(\gamma_n)}= O(n^{(1-\nu)/2})$  with   $\nu= \max_{1\leq j\leq p}{\nu_j}$. Then,   we have that $\gamma_n \pi_\prob(\wbthe,\bthe)=O_{\prob}(1)$, where $\wbthe=(\wmu,\wbbe\trasp,\weta_1,\dots,\weta_p)\trasp$ is defined through  \eqref{eq:estfinitos} and \eqref{eq:estimadoresgj}. Hence, if $\nu_j=1/(1+2r_j)$ in \ref{ass:kj}, we have that we can choose $\gamma_n= n^{(1-\nu)/2 -\omega} $ for $\omega > 0$ arbitrarily small. Moreover, when the same smoothness degree $r$ is assumed for all additive components, i.e., $r_j= r$, for all $1\le j\le p$ and $\nu_j=1/(1+2r_j)=1/(1+2r)$, a convergence rate $n^{r/(1+2r)-\omega}$  arbitrarily close to the optimal one is obtained.    
\end{theorem}
 
\begin{remark}{\label{rem:remark2}} 
 Analogous arguments to those considered in Lemma S.2.3 in Boente \textsl{et al.} (2020) allow to show that if the matrix $\esp \wtbZ\, \wtbZ\trasp$ is non-singular, where $\wtbZ=(1,\bZ\trasp)\trasp\in\real^{q+1}$, and $\prob(\bZ=\esp(\bZ|\bX))<1$, then given $\bthe_1=(a_1,\bb_1\trasp,g_{1,1}\dots, g_{1,p})\trasp$ and $\bthe_2=(a_2, \bb_2\trasp,g_{2,1}\dots, g_{2,p})\trasp$ 
\begin{equation}
\pi_\prob^2 (\bthe_1 ,\bthe_2 )\ge C \left\{(a_1-a_2)^2+ \|\bb_1-\bb_2\|^2+ \esp\left[\sum_{j=1}^p\left(g_{1,j}(X_j)-g_{2,j}(X_j)\right)\right]^2\right\}\,.
\label{eq:cotapiprob}
\end{equation}
In this situation,  Theorem \ref{teo:ratesnew} leads to convergence rates for the parametric components, i.e., we have that $\gamma_n\left(|\wmu-\mu|+\|\wbbe-\bbe\|\right)=O_{\prob}(1)$.   

Note that, under  \eqref{eq:cotapiprob}, if $\gamma_n \pi_\prob(\wbthe,\bthe)=O_{\prob}(1)$, then $\gamma_n^2 \esp\left[\sum_{j=1}^p\left(\weta_{j}(X_j)-\eta_{j}(X_j)\right)\right]^2=O_{\prob}(1)$ which implies that
$\gamma_n^2 \var\left[\sum_{j=1}^p\left(\weta_{j}(X_j)-\eta_{j}(X_j)\right)\right]=O_{\prob}(1)$. Thus, if $X_j$ has a density $f_j$ bounded away from 0 and infinity on $\itI_j$, for $1\le j\le p$, Lemma 1 of Stone (1985)  ensures that   
\begin{equation}
\gamma_n^2 \var \left(\weta_{j}(X_j)-\eta_{j}(X_j)\right)=O_{\prob}(1)\,,
\label{eq:convvar}
\end{equation}
 which corresponds to a convergence rate in $L_2(P)$ when the estimators and regression function are centered with respect to their expected values. 

Assume in addition that  $\gamma_n= O( n^{\lambda})$  where $\lambda=\min_{1\le j\le p}(r_j\, \nu_j)$. As in the proof of Proposition \ref{prop:prop1}, let   $\widetilde{ \wteta}_j(x)$   be the centered spline such that $\int_{\itI_j} \widetilde{ \wteta}_j(x) dx=0$ and  $\|\widetilde{ \wteta}_j-\eta_j\|_{\infty}=O(n^{-r_j\,\nu_j})$. Define $\weta_j^{\star}=\weta_j- \esp  \weta_j(X_j)$, where the expectation is  taken with respect to $X_j$, conditioned on the sample, $\eta_j^{\star}=\eta_j- \esp  \eta_j(X_j)$  and $\wteta_j^{\star}=\widetilde{ \wteta}_j- \esp \widetilde{ \wteta}_j(X_j)$. Then, \eqref{eq:convvar} and the facts that $\|\widetilde{ \wteta}_j-\eta_j\|_{\infty}=O(n^{-r_j\,\nu_j})$ and $\gamma_n= O( n^{\lambda})$  imply that
$$\gamma_n^2 \esp\left[\weta_j^{\star} (X_j)-\wteta_j^{\star}(X_j)\right]^2=O_\prob(1)\,.$$
Taking into account that both $\weta_j^{\star}$ and $\wteta_j^{\star}$ are linear combinations of the $B-$spline basis $\{\wtB_s^{(j)}: 1\leq s\leq k_{j}\}$, from Lemma 7 of Stone (1986), we obtain that  for some positive constant $A>0$ independent of the sample size, 
$$\|\weta_j^{\star} -\wteta_j^{\star}\|_{\infty}^2\le A k_{j}\esp\left[\weta_j^{\star} (X_j)-\wteta_j^{\star}(X_j)\right]^2 $$ 
which entails that
 $ k_j^{-1/2} \gamma_n \|\weta_j^{\star} -\wteta_j^{\star} \|_{\infty} =O_\prob(1)$. 

Assume now that $\nu_j=1/(1+2r_j)$, so   $\lambda=(1-\nu)/2$, where $\nu= \max_{1\leq j\leq p}{\nu_j}$ and $\nu<1/2$. Choose $\gamma_n= n^{(1-\nu)/2 -\omega} $ for $0<\omega < (1-2\nu)/2$ arbitrarily small. Taking into account that $k_j=O(n^{\nu_j})$, we conclude that 
 $ n^{\alpha}  \|\weta_j^{\star} -\wteta_j^{\star} \|_{\infty} =O_\prob(1)$, with $\alpha=(1-2\nu)/2-\omega$, leading to $ n^{\alpha}  \|\weta_j^{\star} -\eta_j^{\star} \|_{\infty} =O_\prob(1)$, for $1\le j\le p$.  
 
It should be noticed that, if $X_j$ has a density $f_j$ bounded away from 0 and infinity on $\itI_j$, for $1\le j\le p$, and \eqref{eq:cotapiprob} holds, similar arguments to those considered in Theorem \ref{teo:ratesnew} combined with those    considered in Shen and Wong (1994) when analysing the Case 3 in page 596, may allow to derive that   $ \gamma_n^{\star}  \|\weta_j^{\star} -\eta_j^{\star} \|_{L_2(P)} =O_\prob(1)$, where $ \gamma_n^{\star}=n^{(1-\nu)/2}$, obtaining the optimal rate of convergence $n^{r/(1+2r)}$ if $r_j= r$, for all $1\le j\le p$ and $\nu_j=1/(1+2r_j)=1/(1+2r)$. However, in Theorem \ref{teo:ratesnew}, we have tried to avoid additional assumptions regarding the distribution of the covariates and for that reason a lower rate is obtained.
\end{remark}

\section{Asymptotic normality of the regression estimators}\label{sec:asymptoticnormality}

In this section, we attempt to derive the asymptotic distribution of the estimators for the regression parameter $\bbe$ under mild assumptions. 
For that purpose, define $\bh^{*}(\bX)=(h_1^*(\bX),\dots,h_{q}^*(\bX))\trasp$   as 
\begin{equation}\label{eq:11}
\bh^{*}(\bX)=\esp(\bZ|\bX) 
\end{equation} 
and $\bA=\esp[\bZ-\bh^*(\bX)][\bZ-\bh^*(\bX)]\trasp$.  Note that if $\bZ$ and $\bX$ are independent $\bh^{*}(\bX)=\esp(\bZ)$, so that $\bA$ is the covariance matrix of $\bZ$.

To obtain the asymptotic  distribution of $\wbbe$, we will need the following additional assumptions.

\begin{enumerate}[label=\textbf{N\arabic*}]
\item\label{ass:nonsingular} 
The matrix $\bA$ is non-singular.
\item \label{ass:sobreZ} For $1\le j\le p$, $\nu_j=1/(2r_j+1)$ with $r_j\ge 1$. Let   $\wbthe=(\wmu, \wbbe\trasp,,\weta_1,\dots,\weta_p)\trasp $ be the estimators defined through  \eqref{eq:estfinitos} and \eqref{eq:estimadoresgj}      and $\gamma_n=n^{\,(1-\nu)/2 -\omega}$ where   $\nu= \max_{1\leq j\leq p}{\nu_j}$ and   $0\le \omega< (1-\nu)/8$. One of the following conditions hold
\begin{enumerate}
\item[a)] $r_j>1$, for $1\le j\le p$,  $\esp \|\bZ\|^6<\infty$, $\pi (\wbthe,\bthe) \convprob 0$, $ \gamma_n  \pi_\prob(\wbthe,\bthe)=O_{\prob}( 1)$ and  $0\le \omega< (1-3\,\nu)/6$.
\item[b)] For some $j_0\in \{1, \cdots, p\}$, $r_{j_0}=1$,  $\esp \|\bZ\|^{10}<\infty$, $\pi (\wbthe,\bthe) \convprob 0$, $ \gamma_n  \pi_\prob(\wbthe,\bthe)=O_{\prob}( 1)$ and  $0\le \omega< 1/21$. 
\end{enumerate} 
 \item\label{ass:hstar}
For each $1\leq m\leq q$, the function $h_m^*(\bx)$ is an additive function in $\bx$, that is, it can be written as 
$$h_m^*(\bx)=\phi_m+\sum_{j=1}^p h_{mj}^*(x_j)\,,$$
where $h_{mj}^* \in \itH_{r_j}\cap \itG$, for $1\leq j\leq p$.
 
\end{enumerate}

\begin{remark}{\label{rem:remark-hstar}} 
Condition \ref{ass:nonsingular} prevents any element of $\bZ$ from being a.s. perfectly predictable by $\bX$ since, in this case, the model would be fully nonparametric. Moreover, it is a standard requirement in robust regression to obtain root$-$n estimators of the linear components. Assumption  \ref{ass:nonsingular} together with \ref{ass:hstar} entail  that  $\bZ$ should not be perfectly predictable by a linear combination of the components of $\bX$.    Note that the additive structure required in assumption \ref{ass:hstar} is satisfied if, for instance, $\bZ$ and $\bX$ are independent in which case $h_{mj}^*\equiv 0$ or if each covariate $Z_m$ of $\bZ$ depends only on one covariate $X_j$ of $\bX$.  The smoothness requirement in  assumption \ref{ass:hstar} was also a condition in assumption \textbf{(A8)} in Ma and Yang (2011). 
Finally, it should be noticed that the rates of convergence required in \ref{ass:sobreZ} may be obtained from Theorem \ref{teo:ratesnew}. 
\end{remark}

From now on, without loss of generality by eventually modifying $\mu$, we will assume that the parameter $\phi_m$ in \ref{ass:hstar} equals 0, so we have that $h_m^*(\bx)= \sum_{j=1}^p h_{mj}^*(x_j) $.

\begin{theorem}{\label{teo:asymptnormal}}
 Assume that  $\rho_1$   satisfies \ref{ass:rho_bounded_derivable} and that $\psi_1=\rho_1^{\prime}$ is twice continuously differentiable with bounded derivative and  that  \ref{ass:densidad}, \ref{ass:etajCr}  to \ref{ass:wsigma} and \ref{ass:nonsingular}  to \ref{ass:hstar} hold.  Then, 
 $\sqrt{n}(\wbbe-\bbe)\convdist N\left(0,\bSi\right)$,
where  $\bSi=\sigma^2 \esp\psi^2(\eps)\{\esp  \psi^\prime(\eps)\}^{-2}\bA^{-1}\,.$ 
\end{theorem}

It is worth mentioning that, as in linear regression, the asymptotic covariance matrix  $\bSi$   is related to the loss function only through  the expression  $\esp\psi^2(\eps)\{\esp \psi^\prime(\eps)\}^{-2}$. Thus,  under the partial linear additive model \eqref{plam}, the efficiency of the robust regression estimator    $\wbbe$ is the same as in location models.

\subsection{An estimator of $\bSi$}{\label{sec:estimoSigma}}
In any analysis, computing the standard errors of the considered estimators is an important task. Clearly, as in other settings, a possible estimator of $\bSi$ can be obtained taking its empirical counterpart and replacing the unknown quantities by appropriate estimators. More precisely, let $\wmu, \wbbe, \weta_j$ and $\wsigma$ the estimators defined in \eqref{eq:estfinitos}, \eqref{eq:estimadoresgj} and \eqref{eq:scale_est}, respectively. As in linear regression models the term $\upsilon=\esp\psi^2(\eps)\{\esp \psi^\prime(\eps)\}^{-2}$ can be easily estimated by
\begin{equation}
 \wup =  \frac 1n \sum_{i=1}^n \psi^2(\weps_i) \left\{\dst\frac 1n \sum_{i=1}^n \psi^\prime(\weps_i)\right\}^{-2}  \quad \mbox{where} \quad 
\weps_i=  \frac{Y_i- \wmu- \wbbe\trasp \bZ_i-\sum_{j=1}^p \weta_j(X_{ij})  }{\wsigma}\,,
\label{eq:wepsi}
\end{equation}
while an estimator of the matrix $\bA$ can be constructed as 
\begin{equation}
\wbA=\frac 1n \sum_{i=1}^n \left\{\bZ_i-\wbh^*(\bX_i)\right\}\left\{\bZ_i-\wbh^*(\bX_i)\right\}\trasp\,,
\label{eq:wbA1}
\end{equation}
for a proper estimator $\wbh^*(\bx)$ of $\bh^*(\bx)$, leading to the plug--in estimator of 
\begin{equation}\label{eq:matrizwSigma1}
\wbSi =\wsigma^2 \wup \;\wbA^{-1}\,.  
\end{equation}

Some facts need to be highlighted regarding the estimator $\wbA$ defined in \eqref{eq:wbA1}. Note that $\wbA$ is an average of the covariate  residuals $r_{i,\bZ}=\bZ_i-\wbh^*(\bX_i)$, so that large values of them may distort its value.  A similar behaviour arises in linear regression models and has been discussed in Section 5.6 in  Maronna \textsl{et al.}  (2019). In our setting, the problem is increased since   high leverage observations  may also affect the estimators $\wbh^*$ of $\bh^{*}$ if not chosen appropriately (see the discussion below), in which case, all values of $r_{i,\bZ}$ will be distorted. In particular, the covariate residuals  related to the outliers will be smaller than expected producing larger estimated asymptotic variances for each component of $\wbbe$. 

In order to control this effect one may combine the ideas in Yohai \textsl{et al.} (1991) with a more stable estimator of $\bh^*$. To be more precise, let $w(t)=\psi(t)/t$ if $t\ne 0$ and $w(0)=\psi^{\prime}(0)$ the weight function related to the score function $\psi$ and denote, for brevity,  $w_i=w(\weps_i)$, where $\weps_i$ are defined in \eqref{eq:wepsi}.
Then, given an estimator $\wbh^*$ of $\bh^*$, an estimator of $\bA$ may be constructed as
\begin{equation}
\wbA=\left( \sum_{i=1}^n w_i \right)^{-1} \sum_{i=1}^n w_i \left\{\bZ_i-\wbh^*(\bX_i)\right\}\left\{\bZ_i-\wbh^*(\bX_i)\right\}\trasp\,.
\label{eq:wbA}
\end{equation}
The independence between the covariates and the errors ensure that, under appropriate convergence conditions for $\wbh^*$,   $\wbA\convprob \bA$. Besides, if  $\wbh^*$ is a resistant estimator, an observation with high leverage will still have a large residual $\bZ_i-\wbh^*(\bX_i)$. The effect of a bad leverage point will be downweighted by the weights $w_i$ which may be 0 for large values of the residuals, if for instance the bisquare loss function is chosen, controlling in this way the damaging effect on the estimated asymptotic variances. In contrast, if the $i-$th observation is such that $\bZ_i$ is    a good leverage point, that is, one with a small residual $\weps_i$, the enlargement effect of $\bZ_i-\wbh^*(\bX_i)$ will  be beneficial  on  $\wbA$ reducing the asymptotic variances. 

 Following  Markatou and He (1994), another estimator of the asymptotic covariance matrix can be implemented  besides the one defined in \eqref{eq:matrizwSigma1} with $\wbA$ given in \eqref{eq:wbA}. Indeed, taking into account that, from the proof of Theorem \ref{teo:asymptnormal},   
 $ \bSi =\bB^{-1}\;\bD\;\bB^{-1\, \trass}$, where
$$ \bB= \, - \, \frac{1}{\sigma^2} \; \esp \left\{\psi^\prime ( \eps)(\bZ-\bh^*(\bX))(\bZ-\bh^*(\bX))\trasp\right\} \qquad \mbox{ and }\qquad 
\bD= \,\frac{1}{\sigma^2} \; \esp \left\{ \psi^2\left( \eps \right)(\bZ-\bh^*(\bX))(\bZ-\bh^*(\bX))\trasp\right\}\,,
$$
we may consider the estimator $\wbSi=\wbB^{-1}\wbD\wbB^{-1\, \trass}$,  where
\begin{align*}
\wbB   =    - \,  \frac{1}{n\;\wsigma^2} \;   \sum_{i=1}^n \psi^\prime\left( \weps_i\right)(\bZ_i-\wbh^*(\bX_i))(\bZ_i-\wbh^*(\bX_i))\trasp 
\; \mbox{and}\;
\wbD   =   \frac{1}{n\; \wsigma^2} \; \sum_{i=1}^n  \psi^2\left( \weps_i \right)(\bZ_i-\wbh^*(\bX_i))(\bZ_i-\wbh^*(\bX_i))\trasp\,.
\end{align*}
Note that, when considering the bisquare function, this estimator automatically down-weights the effect of bad leverage covariates, since in such case, both $\psi^\prime( \weps_i)$ and $\psi^2( \weps_i)$ will be  0 for large values of the residuals. 

The key point in the above discussion is   that the practitioner should be able to handle an appropriate estimator of $\bh^*$. 
 Taking into account   \ref{ass:hstar}, one may estimate $\bh^*(\bx)$ using additive $B-$splines, that is, for each $1\le m\le q$ and $1\le j\le p$, the elements of $\itS_{j}$ defined in \eqref{eq:itSj} may be used to provide an appropriate estimator. Hence, noting that 
$\bh^*(\bx)$ minimizes $\esp \|\bZ-\bh^*(\bX)\|^2$ over the  space of $q-$dimensional measurable functions and that $h_{mj}^* \in \itH_{r_j}\cap \itG$, the initial attempt is to consider the quantity
\begin{equation}
\label{eq:Upsilon}
 \Upsilon(\ba,\bxi)=\sum_{i=1}^n \|\bZ_i- \bh_{\ba,\bxich}^*(\bX_i)\|^2\,,
 \end{equation}
 where $\bxi=(\bxi_1\trasp,\dots, \bxi_q\trasp)\trasp$, $\bh_{\ba,\bxich}^*(\bx)=(h_{a_1,\bxich_1}^*(\bx), \dots, h_{a_q,\bxich_q}^*(\bx) )\trasp$ with  $\bxi_m=( \bxi^{(m,1)\;\trass}, \dots, \bxi^{(m,p)\;\trass})\trasp$, $\bxi^{(m,j)}=(\xi_{1}^{(m,j)}, \dots, \xi_{k_j-1}^{(m,j)})\trasp$   and $h_{a_m,\bxich_m}^*(\bx)=a_m+\sum_{j=1}^p \sum_{s=1}^{k_j-1} \xi_{s}^{(m,j)} B_s^{(j)}(x_j)$. An  estimator $\wbh^*(\bx)=(\wh^*_1(\bx), \dots,$ $ \wh_q^*(\bx))\trasp$ of $\bh^*(\bx)$ can be defined as $\wbh^*(\bx)= \bh_{\wbphich,\wbxich}^*(\bx)$, i.e.,   $\wh_m^*(\bx)=h_{\wphi_m,\wbxich_m}^*(\bx)=\wphi_m+\sum_{j=1}^p   \sum_{s=1}^{k_j-1} \wxi_{s}^{(m,j)} B_s^{(j)}(x_j)$ where the  vectors $ \wbphi $ and   $\wbxi$   minimize $ \Upsilon(\ba,\bxi)$ over $\ba=(a_1,\dots, a_q)\trasp$ and $\bxi=(\bxi_1\trasp, \dots, \bxi_q\trasp)\trasp$.  However,  even when this estimator is appropriate when no outliers arise in the covariates related to the linear component of the model, it will not be resistant when high leverage points are present. A possible solution to solve this problem is discussed below and uses also an $MM-$approach. 

A first attempt to solve the lack of robustness of the estimator that minimizes $ \Upsilon(\ba,\bxi)$ is to mimic the arguments considered in the construction of $\wbA$ and to define $\wbh^*$ minimizing a weighted version of $\Upsilon(\ba,\bxi)$ with weights $\{w_i\}_{i=1}^n$. However, even though this proposal will control the effect of bad leverage points providing consistent estimators of $\bh^*$, good leverage points will still influence the estimation producing small values of $ \bZ_i-\wbh^*(\bX_i)$ for these observations in detriment to the other observations that will see their covariate residual $r_{i,\bZ}$ increased. For that reason, in order to provide a proper estimator of $\bh^*$, we will further assume that a model relates $Z_m$ with the covariates $X_1,\dots, X_p$, see He \textsl{et al.} (2002)  for a related model. From now on, we assume that
$$ Z_{im} = \phi_m+\sum_{j=1}^p h_{mj}^*(X_{ij}) + \sigma_m\; u_{im}\;,$$
where $u_{im}\sim F_m(\cdot)$  are  independent from $\bX_i$ and independent from each other,  $\sigma_m>0$ is the   scale parameter and $F_m$ is symmetric around $0$ with scale 1. A procedure similar to that described in section \ref{sec:proposals} can be implemented as follows leading to uniform consistent estimators.  For that purpose, for $1\le m\le q$, define   
\begin{equation*}  
r_{i,m}(a, \bxi _m)=    Z_{i,m} -a- \sum_{j=1}^p \sum_{s=1}^{k_j-1} \xi_{s}^{(m, j)} B_s^{(j)}(X_{ij}) = Z_{i,m} - h^{*}_{a,\bxich_m}(\bX_i)\;,
\end{equation*}
where for the sake of simplicity and to avoid burden notation, we have assumed that the same bases are used for each component $Z_m$ of $\bZ$.

For each $1\le m\le q$, we consider a preliminary robust $S-$estimator $\wsigma_m$ computed with loss function $\rho_0$, that is,  we define $\wsigma_m  \, =  s_{n,m}(\widetilde{\phi}_{m},   \widetilde{\bxi }_{m}) $  where $s_{n,m}(\widetilde{\phi}_{m},   \widetilde{\bxi }_{m}) $ minimizes over $(a,\bxi_m)$  the solution $s_{n,m}(a,   {\bxi }_{m}) $ of 
$$\frac{1}{n-K}\sum_{i=1}^n \rho_{0}\left(\frac{r_{i,m}(a, \bxi _m)}{s_{n,m}(a, \bxi _m)}\right) \, = \, b \,.$$
Let  
 $\rho_{1}$  be such that $\rho_1\le \rho_0$. The $M-$estimator of $\bh^{*}$ is then obtained as $\wbh^*(\bx)= \bh_{\wbphich,\wbxich}^*(\bx)$,  where
\begin{equation*}
(\wbphi,  \wbxi)  \ = \  \argmin_{\ba\in \real^q,   \bxich=(\bxich_1\trasp, \dots, \bxich_q\trasp)\trasp \in \real^{q K}} \sum_{m=1}^q\sum_{i=1}^n \rho_{1} \left ( \frac{r_{i,m}(a_m, \bxi _m) }{\wsigma_m} \right )\,.
\end{equation*}
Note that, for each $m$, $(\wphi_m,  \wbxi_m) $ can be obtained minimizing the quantity $\sum_{i=1}^n \rho_{1} \left (  {r_{i,m}(a_m, \bxi _m) }/{\wsigma_m} \right )$.

\section{Monte Carlo Study}\label{montecarlo}

This section contains the results of a simulation study conducted to compare,  under different models and contamination schemes, the performance of the robust  estimators defined in Section \ref{sec:proposals} with that of their classical counterparts. All computations were carried out using an \texttt{R} implementation of our algorithm which is available at \url{https://github.com/alemermartinez/rplam}.  The classical estimator corresponds to  a linear regression least squares estimator after the $B-$splines approximation was performed for each additive component, while for the robust estimator, we considered an $MM-$estimator based on   Tukey's bisquare functions. For the initial $S-$estimators, we choose $c_0 = 1.54764$ and $b = 1/2$, while for the $M-$step the  tuning constant equals $c_1=4.685$.  The values for the constants $c_0$ and  $c_1$ are based on the performance of the $MM-$estimators in linear regression models. Indeed, the value $c_0 = 1.54764$ ensures Fisher--consistency of the scale estimator when the errors have a normal distribution.  Besides, $c_1=4.685$ corresponds to the tuning constant that guarantees, for normal errors, a 95\% efficiency for the robust estimators of  $\bbe$   (see Theorem \ref{teo:asymptnormal} and  Maronna \textsl{et al.}, 2019).  In all tables and figures,  the classical least squares estimator will be labelled as \textsc{ls}  and the robust $MM-$procedure proposed in this paper as \textsc{mm}.

In all scenarios, we performed $N=500$ replications, the sample size was $n=100$ and we  used cubic splines with equally spaced knots for all nonparametric components. For numerical simplicity, we also chose the same number of terms in the spline approximation for each additive component, that is, $k_j=k$ for all $j$.
As in He \textsl{et al.} (2002),   the common number of elements in the basis $k$ varies between $\max\{n^{1/5}/2;4\}$ and $8+2n^{1/5}$, which for the sample size considered leads to  $4\leq k\leq 13$. To select the basis dimension, we minimized the  $RBIC$ criteria defined in \eqref{genBIC}  over the set $\{4,\dots,13\}$ with the same $\rho$--function considered in the estimation step. Hence, for the robust estimator, the $\rho$--function in \eqref{genBIC} is  the Tukey's bisquare loss function with tuning constant $c_1=4.685$,  while for  the classical estimator,  $\rho(t)=t^2$. Note that for the classical procedure  no scale estimator   is needed.

The samples $\left\{(Y_i,\bZ_i\trasp,\bX_i\trasp)\trasp\right\}_{i=1}^n$ are generated with the same distribution as $ (Y,\bZ\trasp,\bX\trasp)\trasp$, $\bX =(X_1,X_2)\trasp\in\real^2$,  $\bZ=(Z_1,Z_2)\trasp\in\real^2$. In all cases, the response and the covariates are related through  the partially linear additive model 
$$Y=\mu+\beta\trasp\bZ+\eta_1(X_{1})+\eta_2(X_{2})+\sigma\,\eps\,,$$ 
with $\beta=(\beta_1,\beta_2)\trasp=(3,3)\trasp$, $ \mu=0$, $\sigma =0.2$,  $\eta_1(x_1)=2\sin(\pi x_1)- {4}/{\pi}$ and $ \eta_2(x_2)=e^{x_2}-(e-1)$, so that $\int_0^1 \eta_1(x)\,dx=\int_0^1 \eta_2(x)\,dx=0$. Six possible models combining different choices for  covariates distributions were studied. For all models $X_1$ and $X_2$ have marginal uniform distribution but different correlation between the covariates are allowed. In Models 4 to 6, $Z_1$ and 
$Z_2$ have a discrete distribution while in Models 1 to 3, they correspond to continuous random variables.  For clean samples, denoted from now on as $C_0$,  the errors distribution  is $\eps\sim N(0,1)$ and the considered models for the covariates distribution are
\begin{itemize}
\item \textsc{Model 1:} $Z_1, Z_2, X_1$ and $X_2$ are i.i.d. $\mathcal{U}(0,1)$.
 
\item \textsc{Model 2:}   $Z_1, Z_2, X_1 $ and $X_2$ have marginal $\mathcal{U}(0,1)$ distribution,   $X_1$ and $Z_1$ have correlation $0.7$, while $Z_2$ and $X_2$ independent  and independent  of $Z_1$ and $X_1$.

\item \textsc{Model 3:}   $X_1 $ and $X_2$ are i.i.d. $ \mathcal{U}(0,1)$ and we  defined  $Z_1=X_1 + X_2^2+u_1$, $Z_2= \{\exp(X_1 )-1\}/2+u_2$, where $u_1$ and $u_2$ i.i.d.  $u_j\sim N(0,(0.1)^2)$.  

 \item \textsc{Model 4:}   $X_1$ and $X_2$ are i.i.d. with distribution $ \mathcal{U}(0,1)$, $Z_1=W_1/3$ and $Z_2=W_2/5$ where $W_1\sim Bi(3,1/2)$ and $W_2\sim Bi(5,1/5)$ are independent of each other.   
 
\item \textsc{Model 5:}   $X_1$ and $X_2$ are i.i.d. $X_j\sim \mathcal{U}(0,1)$, $Z_1=W_1/10$ and $Z_2=W_2/10$ with $(W_1,W_2,W_3)\sim \mathcal{M}(10,1/4,1/2,1/4)$, that is, having a multinomial distribution, besides $X_1,X_2$ and $(W_1,W_2,W_3)$ are independent.
 
\item \textsc{Model 6:}   $X_1$ and $X_2$ are i.i.d. $X_j\sim \mathcal{U}(0,1)$, $Z_1=W_1/5$ with $W_1\sim Bi(5,1/4)$, $X_1, X_2$ and $W_1$ are independent. Moreover, $Z_2=(1/2)\left\{I_{(0;2/3)}(X_1)+W\right\}$ with $W\sim Bi(1, 1/2)$ independent of $X_1$. 
\end{itemize}
 Note that all considered models  satisfy the additivity required in assumption \ref{ass:hstar}. However, under Model 6, $h_2^*(\bx)=(1/2) I_{(0;2/3)}(x_1)+(1/4)$ is not a smooth function. 

\begin{figure}[ht!]
\begin{center}
\includegraphics[scale=0.45]{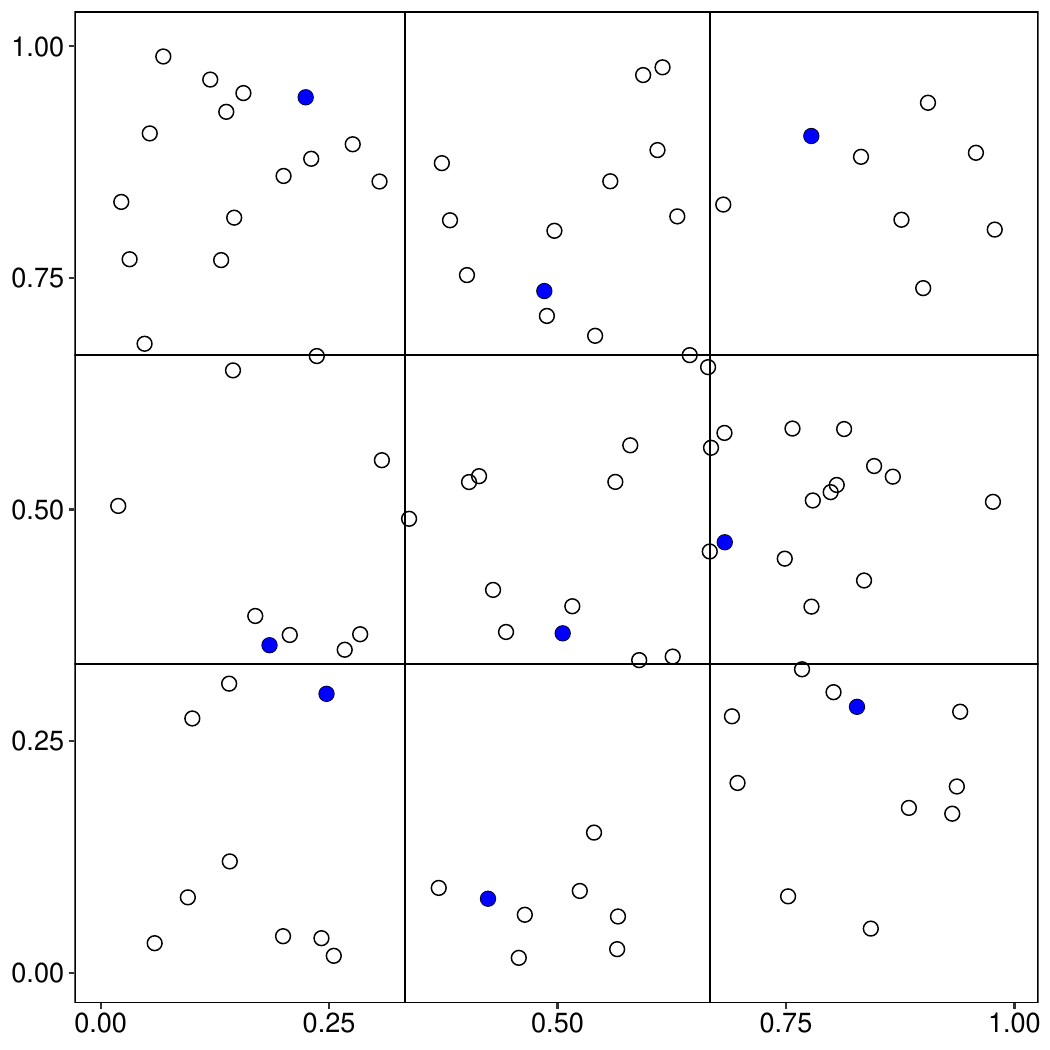}
\caption{\label{fig:Cont3}Scatter plot of the covariates $(X_{1i},X_{2i})\trasp$ corresponding to one of the samples contaminated according to scenario $C_3$. The solid circles indicate observations where $(Z_{1i},Z_{2i})\trasp$ is contaminated.}
\end{center}
\end{figure}

To study the effect of atypical data on  the estimators,   we considered  three contamination schemes which are described as
\begin{itemize}
\item $C_1$: $\eps_i\sim 0.9 N(0,1) + 0.1 N(0,100)$.
\item $C_2$: $u_i\sim 0.85 N(0,\sigma^2) + 0.15 N(15,0.1^2)$ where $u_i=\sigma\eps_i$.
\item $C_3$:  We divided the square $[0,1]\times[0,1]$   into 9 equally-sized quadrants as shown in  Figure \ref{fig:Cont3}. In this scenario the errors are not contaminated, that is, $\eps_i\sim N(0,1)$ but artificially 9 observations were modified in such a way that $(Z_{1i},Z_{2i})\trasp=(20,20)\trasp$ and its related pair $(X_{1i},X_{2i})\trasp$ of covariates belonged to a different quadrant in the square $[0,1]\times[0,1]$.  
\end{itemize}
Contaminations  $C_1$ and $C_2$ correspond to vertical outliers    and it is expected that they will affect 
mainly the  estimation of $\mu$ and eventually that of the additive components. In particular, $C_1$ corresponds to a $10\%$ of the errors with a larger variance and will affect mainly the mean square error and not the bias of the estimates. In contrast, scenario  $C_2$ corresponds to an asymmetric  \textit{gross error} model in which a $15\%$ of the errors are shifted in order to produce ``vertical outliers'' and will have more effect on the bias. In contrast, contamination $C_3$ aims to affect the regression parameter also through the high leverage points introduced.  Figure \ref{fig:Cont3} shows the 100 pairs of $\bX_i$ for one sample with $X_1$ and $X_2$ following a $\mathcal{U}(0,1)$ distribution. The solid blue circles correspond to the covariates $\bX_i$ where  $\bZ_i$ has been contaminated. When no generated data points $\bX_i$ were found in a quadrant, the observation $\bZ_i$ was not contaminated, so the total amount of contaminated observations was less than 9.

\begin{figure}[ht!]
\begin{center}
\newcolumntype{M}{>{\centering\arraybackslash}m{\dimexpr.01\linewidth-1\tabcolsep}}
   \newcolumntype{G}{>{\centering\arraybackslash}m{\dimexpr.33\linewidth-1\tabcolsep}}
\renewcommand{\arraystretch}{0.1}
\begin{tabular}{GG}
  \textsc{ls} &  \textsc{mm} \\ [-0.1in]
\includegraphics[scale=0.33]{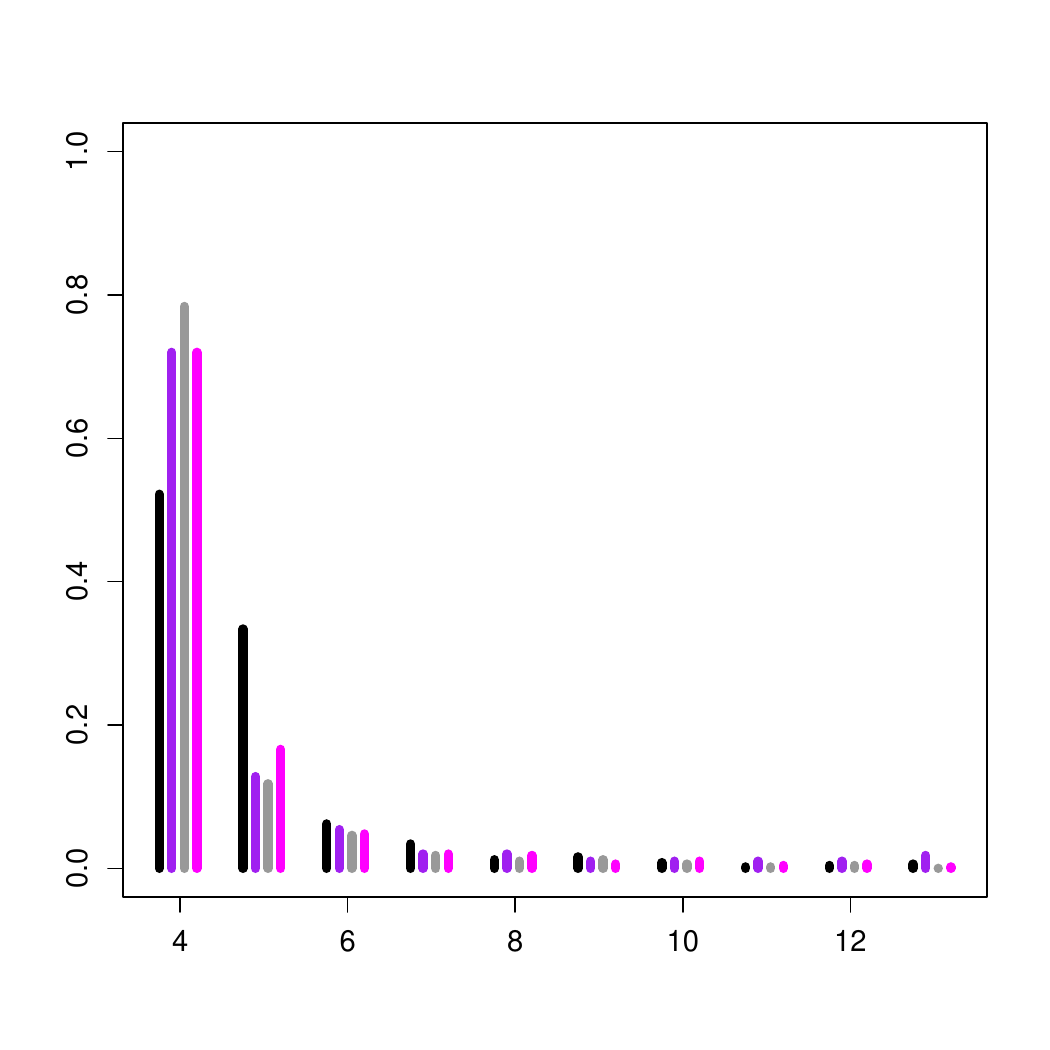}&
 \includegraphics[scale=0.33]{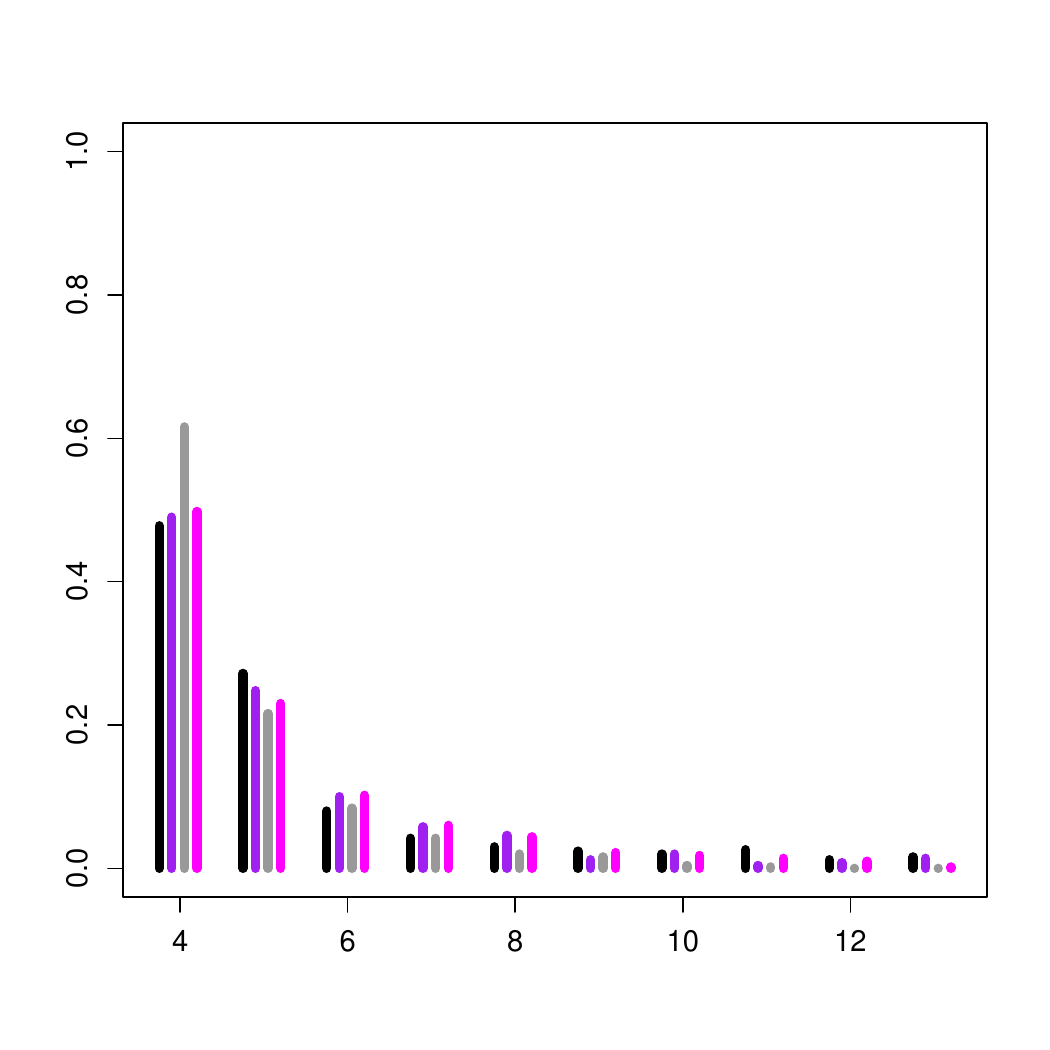}\\[-0.35in]
\end{tabular}
\caption{\small \label{fig:NBasis2} Plots of the proportion of the number of basis selected under Model 3. The  black bars correspond to  $C_0$, while the purple, grey and magenta  ones to contaminations $C_1$, $C_2$ and $C_3$, respectively.}
\end{center}
\end{figure}

\begin{figure}[ht!]
\begin{center}
\newcolumntype{M}{>{\centering\arraybackslash}m{\dimexpr.01\linewidth-1\tabcolsep}}
   \newcolumntype{G}{>{\centering\arraybackslash}m{\dimexpr.33\linewidth-1\tabcolsep}}
\renewcommand{\arraystretch}{0.1}
\begin{tabular}{GG}
  \textsc{ls} &   \textsc{mm} \\ [-0.1in]
\includegraphics[scale=0.33]{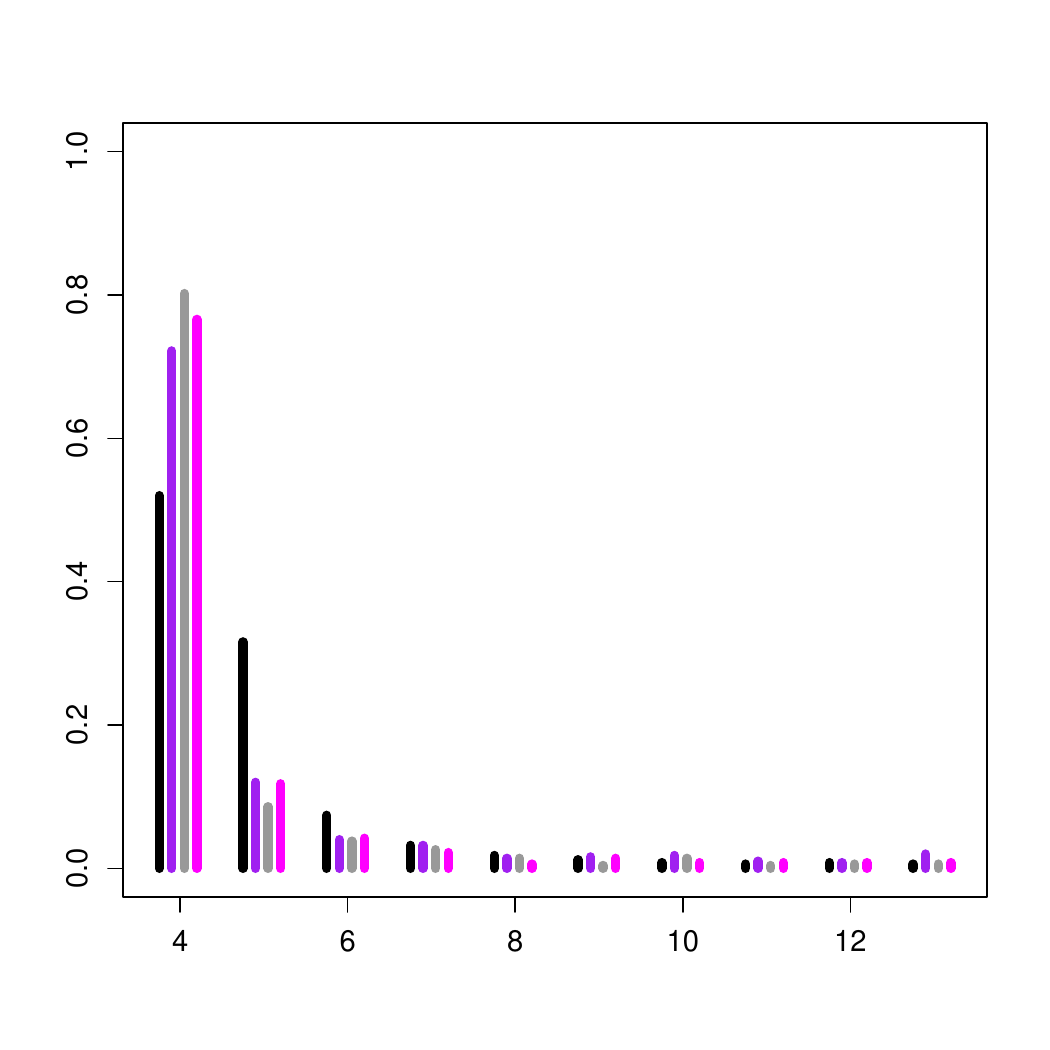}&
 \includegraphics[scale=0.33]{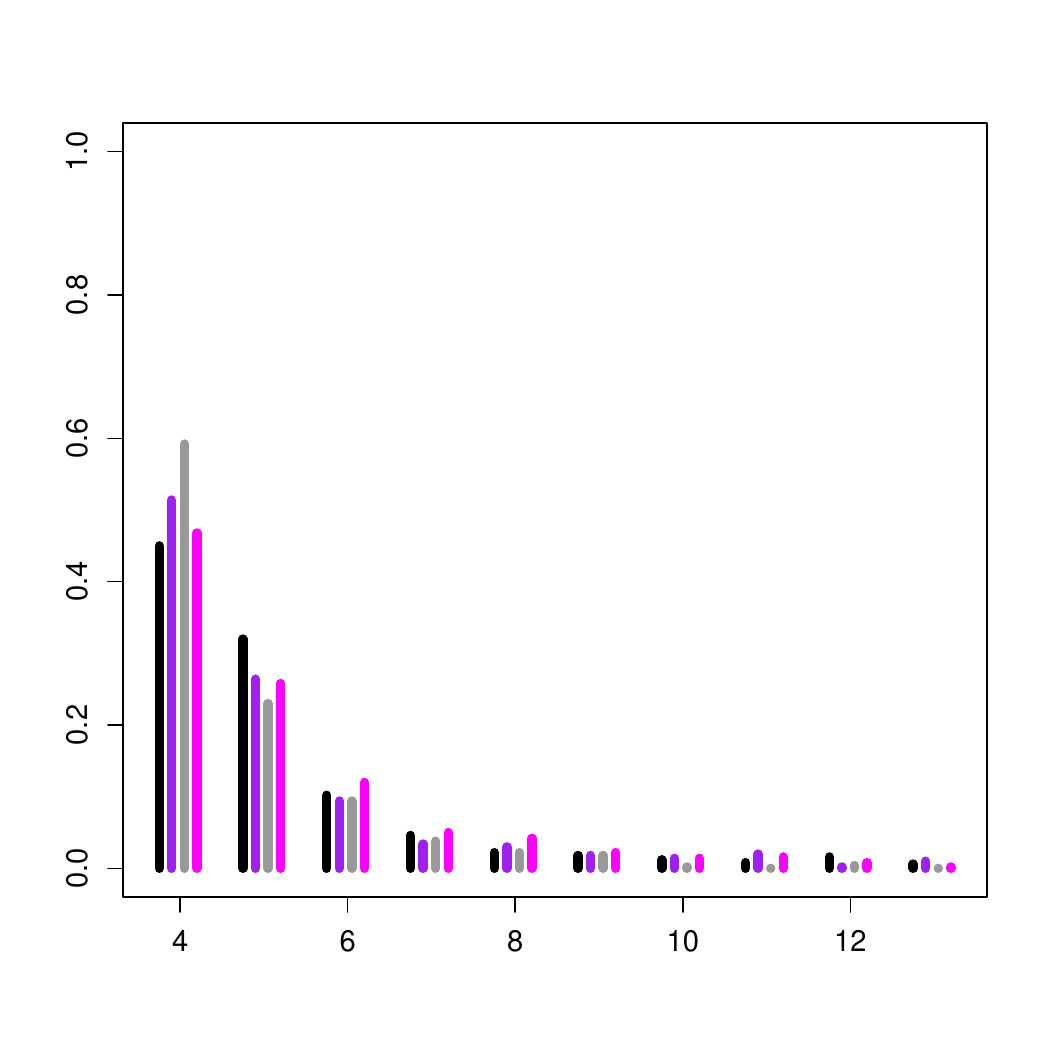}\\[-0.35in]
\end{tabular}
\caption{\small \label{fig:NBasis8} Plots of the proportion of the number of basis selected under Model 4. The  black bars correspond to  $C_0$, while the purple, grey and magenta  ones to contaminations $C_1$, $C_2$ and $C_3$, respectively.}
\end{center}
\end{figure}

To study how the outliers affect the selection of the basis dimension, Figures \ref{fig:NBasis2} and \ref{fig:NBasis8}  show  the plots of the proportion of times that the value $k$ is selected  by the $BIC$ criterion, under Models 3 and 4, respectively. Black bars correspond to the no contamination setting $C_0$, while  purple, grey and magenta ones to the contamination settings $C_1$, $C_2$ and $C_3$, respectively. The sensitivity to outliers of the classical $BIC$ criterion is reflected through the performance of the support of the selected basis dimension  which is more concentrated at 4 for contaminated data.   In contrast,  when using the robust procedure combined with the robust $BIC$ all bars have similar heights showing the stability of the selection method. Note that for the $MM-$estimators, only under $C_2$, dimension $k=4$ is selected more frequently than under $C_0$. It is also worth mentioning that the results reported in Figures \ref{fig:NBasis2}  and \ref{fig:NBasis8} illustrate that, for clean samples,
both the classical and the robust $BIC$ lead to similar choices for the basis dimension.
 
To evaluate the behaviour of the additive component estimators, we measured the performance of an estimator $\wefe$ of a general function $f:[0,1]\to \real$    approximating the integrated squared error (\textsc{ise}) over an equally spaced grid of points $\{t_\ell\}_{\ell=1}^M$, $0\le t_1<\dots< t_M\le 1$ with $M=1000$, that is, 
$$\mbox{\textsc{ise}}=\frac{1}{M}\sum_{s=1}^M \left(f(t_s)-\wefe(t_s)\right)^2\,.
$$ 
Taking into account that  a few large values of the \textsc{ise}   may have a huge impact on its mean  over replications and to prevent us for this distorted effect, instead of the mean integrated square error we considered two    measures less affected by extreme values: the median  of the \textsc{ise}, denoted  \textsc{medise} and the   mean of the  \textsc{ise} obtained after trimming the 5\% largest values labelled  \textsc{5\%-mise}. The obtained results for the \textsc{ise} are given in Table  \ref{tab:EstGsplines}.

 \begin{table}[ht!]
\begin{center}
\footnotesize
\renewcommand{\arraystretch}{1.2}
\setlength{\tabcolsep}{3pt}
 \begin{tabular}{|c|c|c|c|c||c|c||c|c||c|c||c|c||c|c|}
\cline{4-15}
\multicolumn{1}{c}{} & \multicolumn{1}{c}{} & \multicolumn{1}{c|}{} &\multicolumn{2}{c||}{Model 1} &\multicolumn{2}{c||}{Model 2} &\multicolumn{2}{c||}{Model 3}  &\multicolumn{2}{c||}{Model 4} & \multicolumn{2}{c||}{Model 5}& \multicolumn{2}{c|}{Model 6} \\\cline{4-15}
\multicolumn{1}{c}{} & \multicolumn{1}{c}{} & \multicolumn{1}{c|}{} & \textsc{ls} & \textsc{mm} & \textsc{ls} & \textsc{mm} & \textsc{ls} & \textsc{mm} & \textsc{ls} & \textsc{mm} & \textsc{ls} & \textsc{mm} & \textsc{ls} & \textsc{mm} \\ \hline
  & $C_0$ & \textsc{5\%-mise} & 0.002 &   0.003 &  0.003  & 0.003 &  0.007  & 0.008 & 0.002  & 0.003 & 0.002  & 0.003 &  0.003 & 0.003 \\ 
  & & \textsc{medise} & 0.002 &  0.002 &  0.002 &  0.003 &  0.005 &  0.006 &  0.002 &   0.002 &  0.002 &  0.002 &  0.003  & 0.003 \\   
 \cdashline{2-15}
& $C_1$ & \textsc{5\%-mise} & 0.019 &   0.003 & 0.023 &  0.004 & 0.081 &   0.009 & 0.019 &   0.003 & 0.019 &   0.003 & 0.022 &  0.004 \\ 
  $\eta_1$ & & \textsc{medise} & 0.014 &  0.003 & 0.018 &   0.003 & 0.056   & 0.007 & 0.014  & 0.003 & 0.014   & 0.003 & 0.016   & 0.003 \\  \cdashline{2-15}
 & $C_2$ & \textsc{5\%-mise} & 1.175 &   0.003 & 1.484 &   0.003 & 4.866 &  0.008 & 1.127   & 0.003 & 1.147  & 0.003 & 1.432  & 0.003 \\ 
 & & \textsc{medise} & 0.920  & 0.003 & 1.254   & 0.003 & 3.561  & 0.006 & 0.870   & 0.003 & 0.855 & 0.003 & 1.129  & 0.003 \\  \cdashline{2-15}
 & $C_3$ & \textsc{5\%-mise} & 0.063  & 0.003 & 0.660  & 0.004 & 2.553  & 0.009 & 0.038  & 0.003 & 0.009  & 0.003 & 0.281  & 0.003 \\ 
 & & \textsc{medise} & 0.054  & 0.003 & 0.664  & 0.003 & 2.564  & 0.007 & 0.031  & 0.003 & 0.008 & 0.003 & 0.280  & 0.003 \\  \hline
 & $C_0$ & \textsc{5\%-mise} & 0.002 & 0.002 & 0.002  & 0.002 & 0.005  & 0.006 & 0.002  & 0.002 & 0.002   & 0.002 & 0.002 & 0.002 \\ 
 & & \textsc{medise} & 0.001 &   0.002 & 0.002  & 0.002 & 0.003   & 0.004 & 0.001   & 0.002 & 0.001  & 0.002 & 0.002  & 0.002 \\  \cdashline{2-15}
 & $C_1$ & \textsc{5\%-mise} & 0.018  & 0.003 & 0.019   & 0.003 & 0.051 &  0.006 & 0.018   & 0.003 & 0.018 & 0.003 & 0.017  & 0.003 \\ 
  $\eta_2$ & & \textsc{medise} & 0.012  & 0.002 & 0.013  & 0.002 & 0.035   & 0.005 & 0.013  & 0.002 & 0.012   & 0.002 & 0.012 & 0.002 \\  \cdashline{2-15}
   & $C_2$ & \textsc{5\%-mise} & 1.282   & 0.002 & 1.256 & 0.002 & 3.470  & 0.006 & 1.244  & 0.002 & 1.231  & 0.002 & 1.271  & 0.002 \\ 
 & & \textsc{medise} & 1.008  & 0.002 & 0.949  & 0.002 & 2.713  & 0.004 & 1.001   & 0.002 & 0.983  & 0.002 & 1.000   & 0.002 \\  \cdashline{2-15}
  & $C_3$ & \textsc{5\%-mise} & 0.063  & 0.002 & 0.046  & 0.002 & 0.793   & 0.006 & 0.038  & 0.002 & 0.009  & 0.002 & 0.047  & 0.002 \\ 
 & & \textsc{medise} & 0.051   & 0.002 & 0.037   & 0.002 & 0.802 & 0.004 & 0.031   & 0.002 & 0.007 & 0.002 & 0.039  & 0.002 \\  \hline
 	\end{tabular}
\caption{\label{tab:EstGsplines}\footnotesize Summary measures for the   additive components estimates $\weta_1$ and $\weta_2$ based on the \textsc{ise}. The classical and robust procedures are labelled \textsc{ls} and \textsc{mm}, respectively.}
\end{center}
\end{table}

\begin{figure}[ht!]
\begin{center}
\renewcommand{\arraystretch}{1.2}
\begin{tabular}{cc}
\multicolumn{2}{c}{Model 1}\\  
$C_0$ &  $C_1$\\[-0.2in]
\includegraphics[scale=0.5]{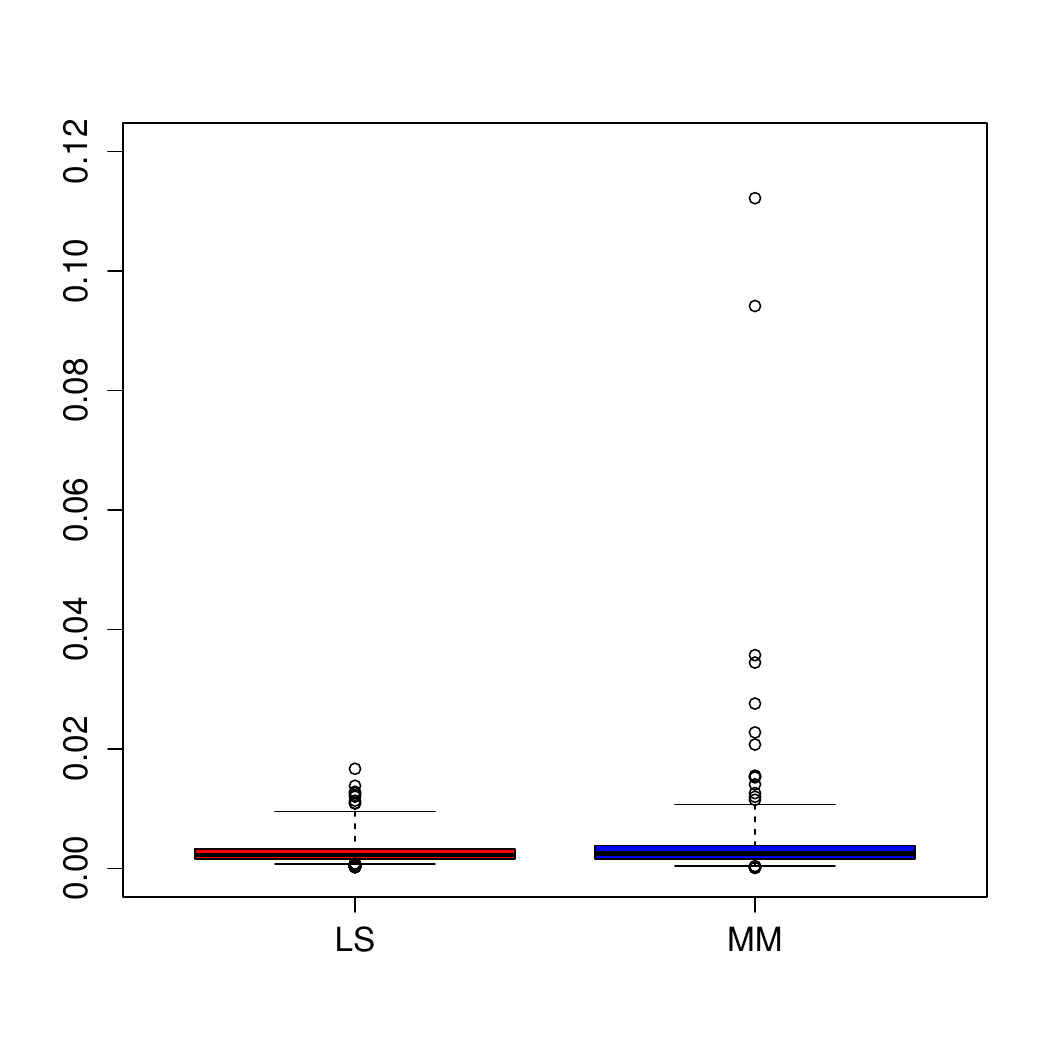} & 
\includegraphics[scale=0.5]{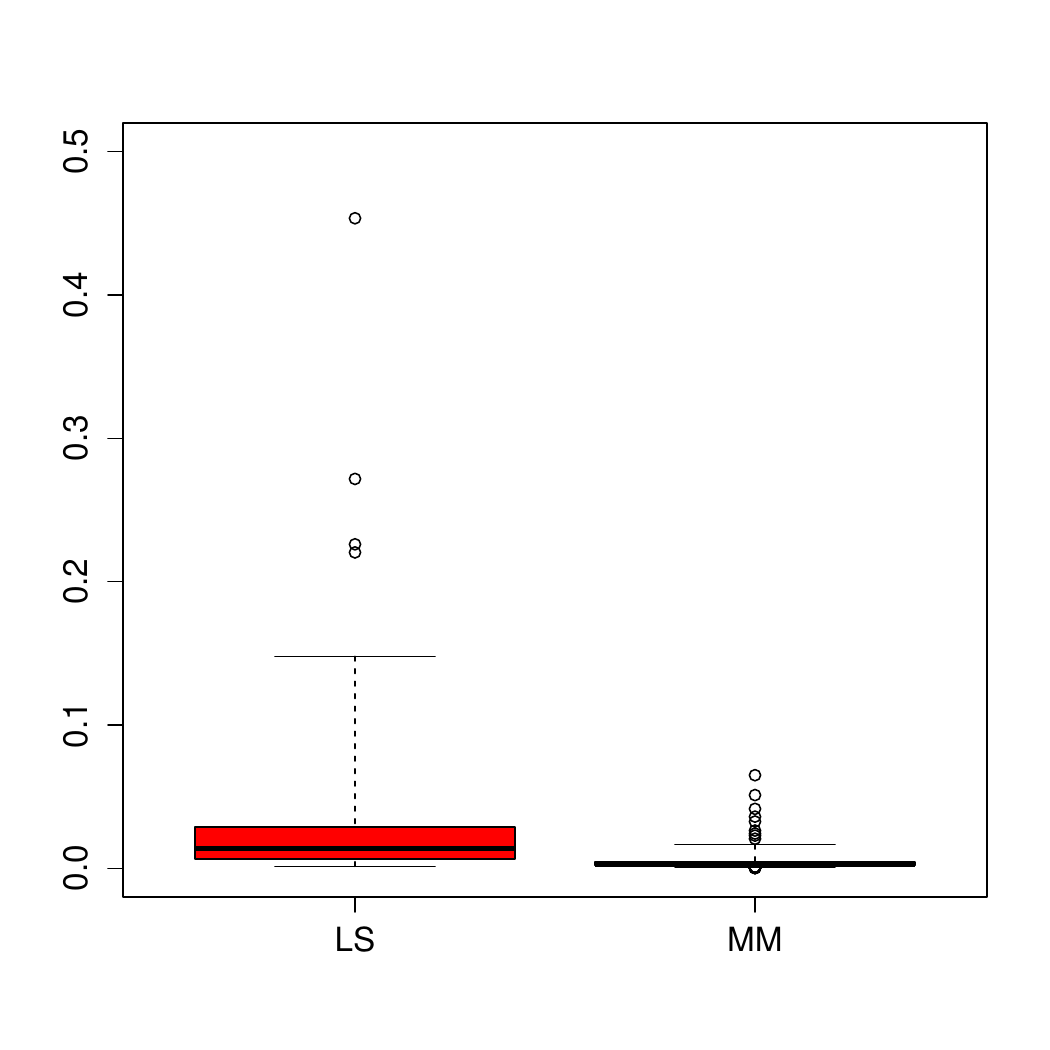} \\
$C_2$ &  $C_3$\\[-0.2in]
\includegraphics[scale=0.5]{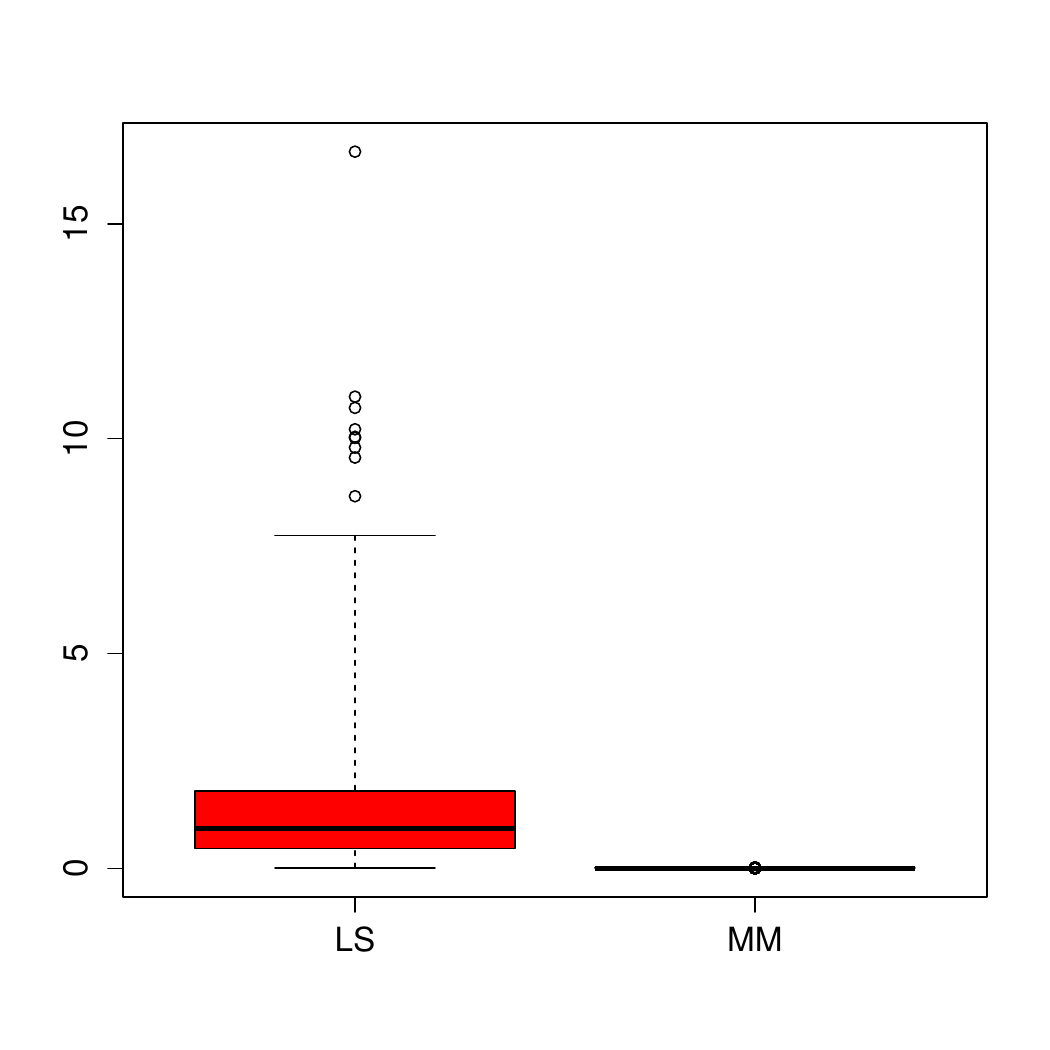} & 
\includegraphics[scale=0.5]{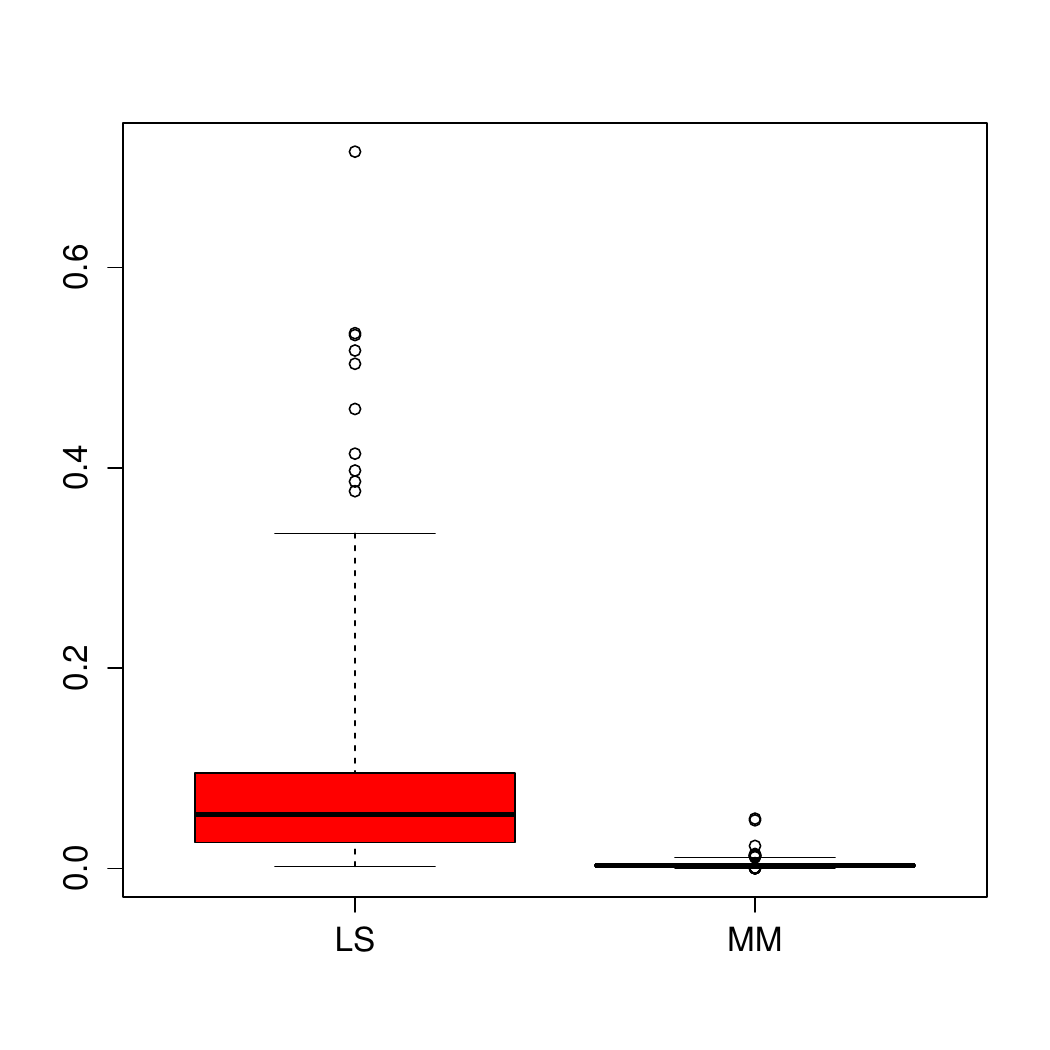}
\end{tabular}
\vskip-0.2in
\caption{\small\label{fig:ISE-M7-G1-ClasicoYRobusto}  Boxplots of the \textsc{ise} values for estimating the additive function $\eta_1$ for each contamination setting under Model 1, for the classical fit (in red) and for the robust fit (in blue).}

\end{center}
\end{figure}

\begin{figure}[ht!]
\begin{center}
\renewcommand{\arraystretch}{1.2}
\begin{tabular}{cc}
\multicolumn{2}{c}{Model 1}\\  
$C_0$ &  $C_1$\\[-0.1in]
\includegraphics[scale=0.5]{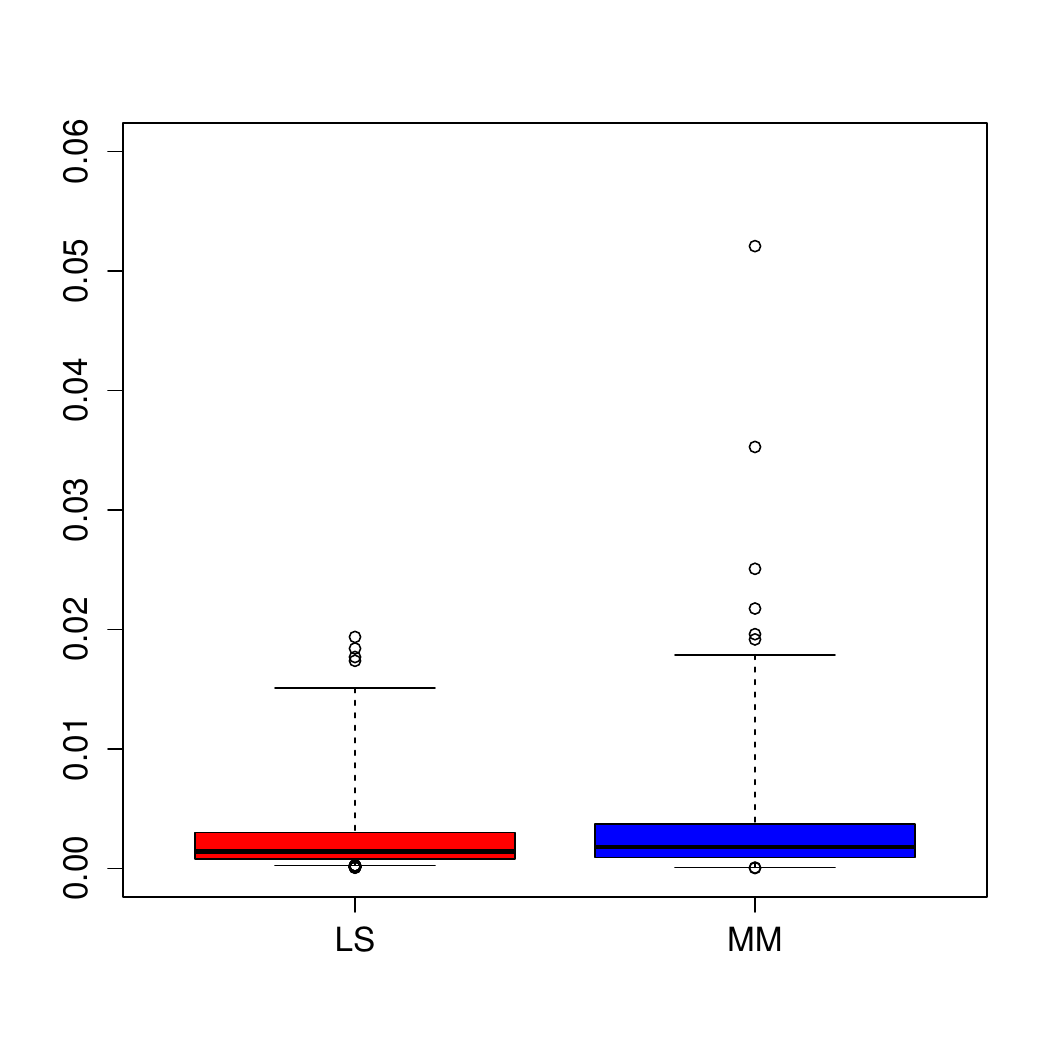} & 
\includegraphics[scale=0.5]{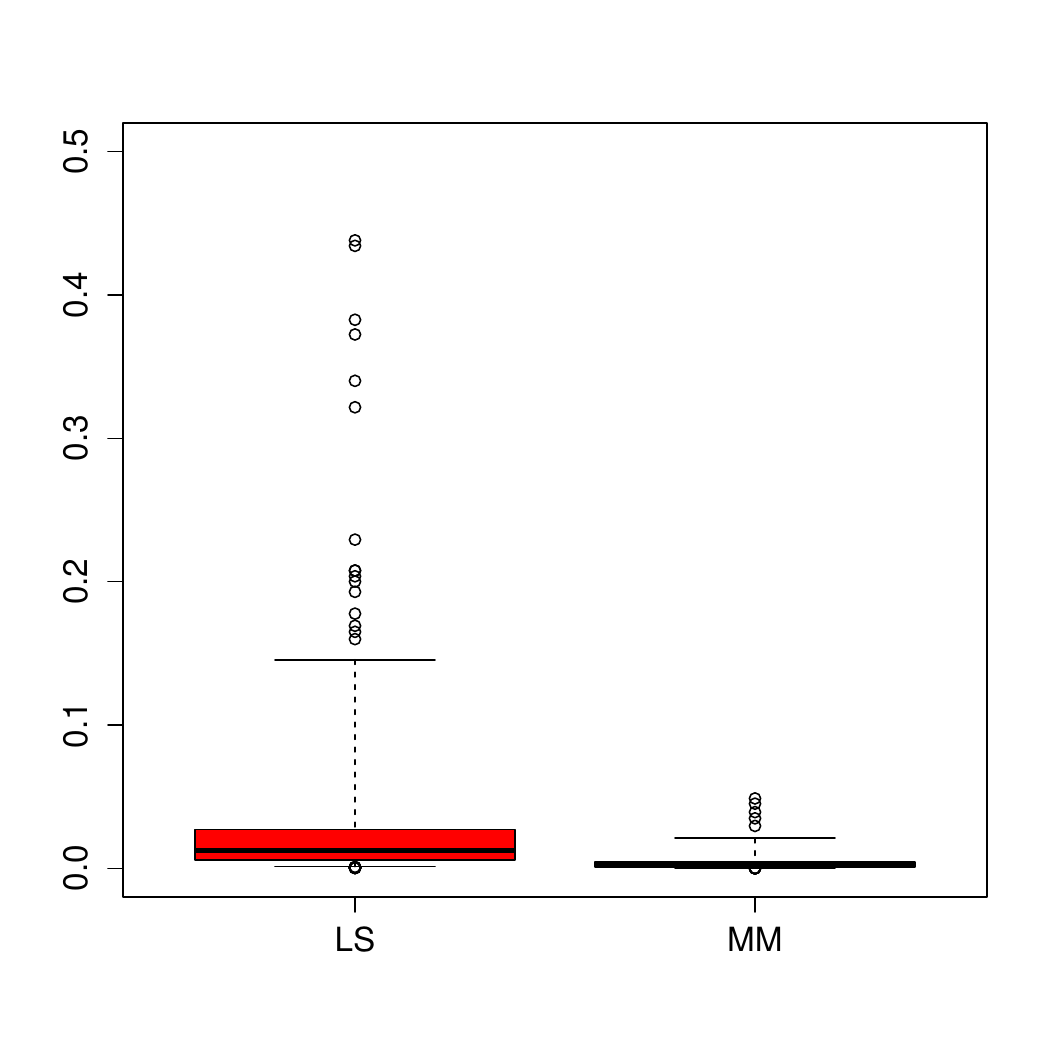} \\
$C_2$ &  $C_3$\\[-0.2in]
\includegraphics[scale=0.5]{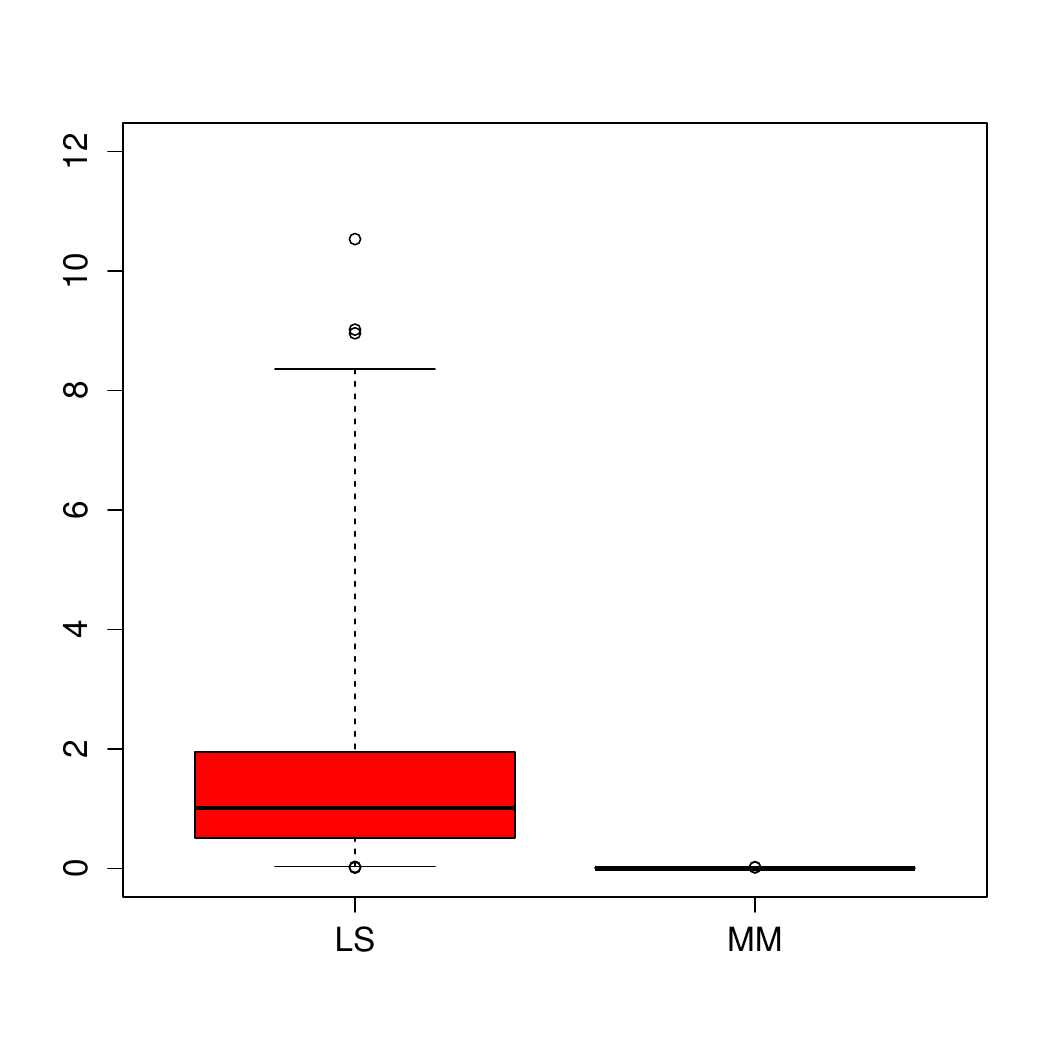} & 
\includegraphics[scale=0.5]{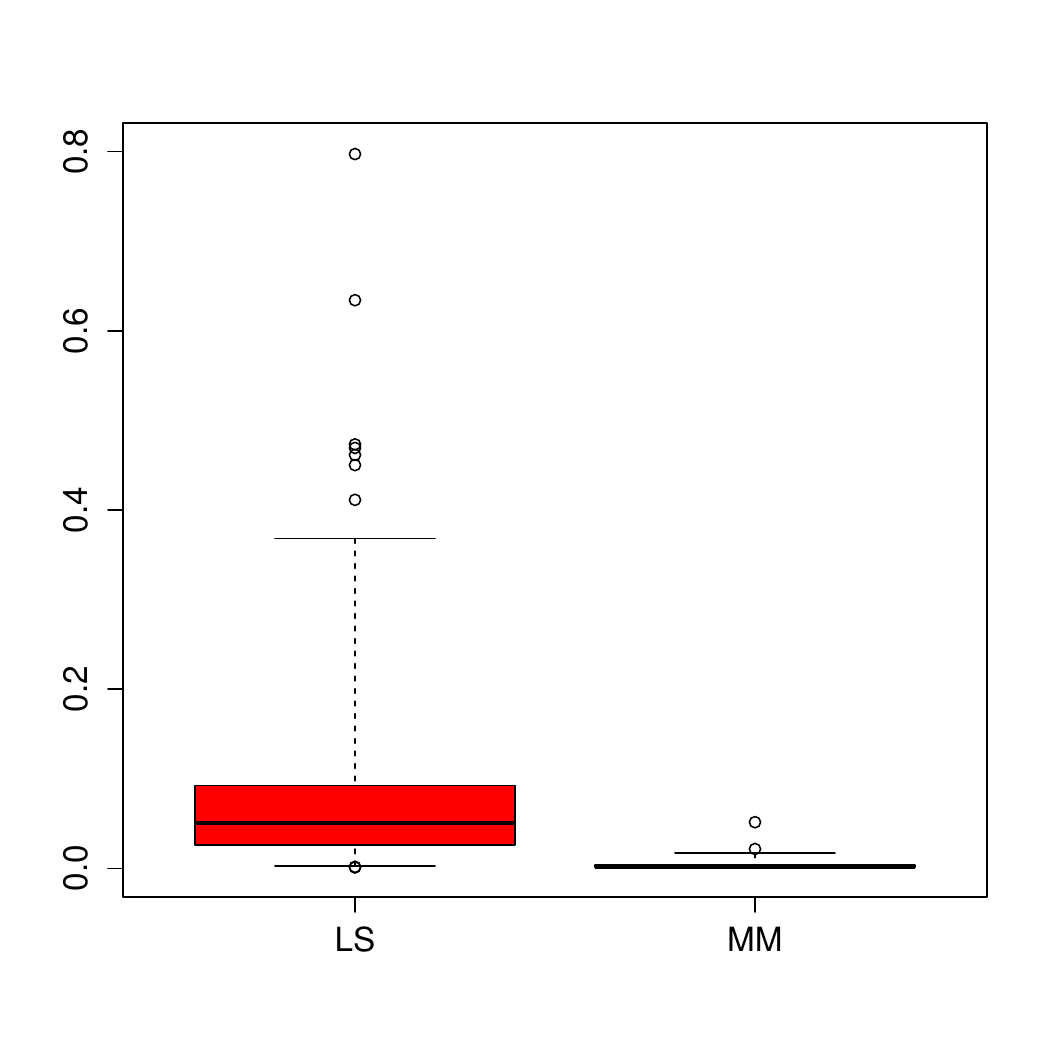}
\end{tabular}
\vskip-0.2in
\caption{\small  \label{fig:ISE-M7-G2-ClasicoYRobusto} Boxplots of the \textsc{ise} values for estimating the additive function $\eta_2$ for each contamination setting under Model 1, for the classical fit (in red) and for the robust fit (in blue).}

\end{center}
\end{figure}

\begin{figure}[ht!]
\begin{center}
\renewcommand{\arraystretch}{1.2}
\begin{tabular}{cc}
\multicolumn{2}{c}{Model 6}\\  
$C_0$ &  $C_1$\\[-0.2in]
\includegraphics[scale=0.5]{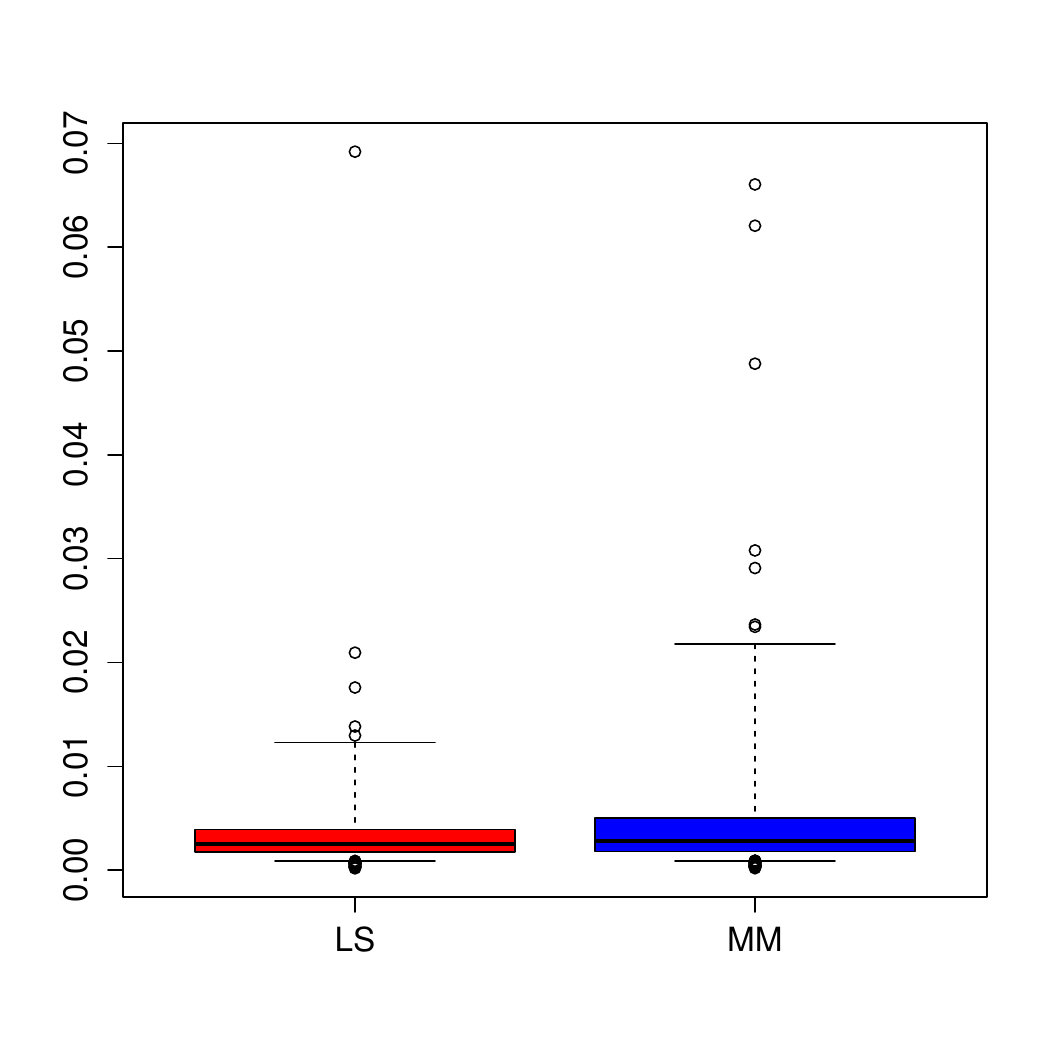} & 
\includegraphics[scale=0.5]{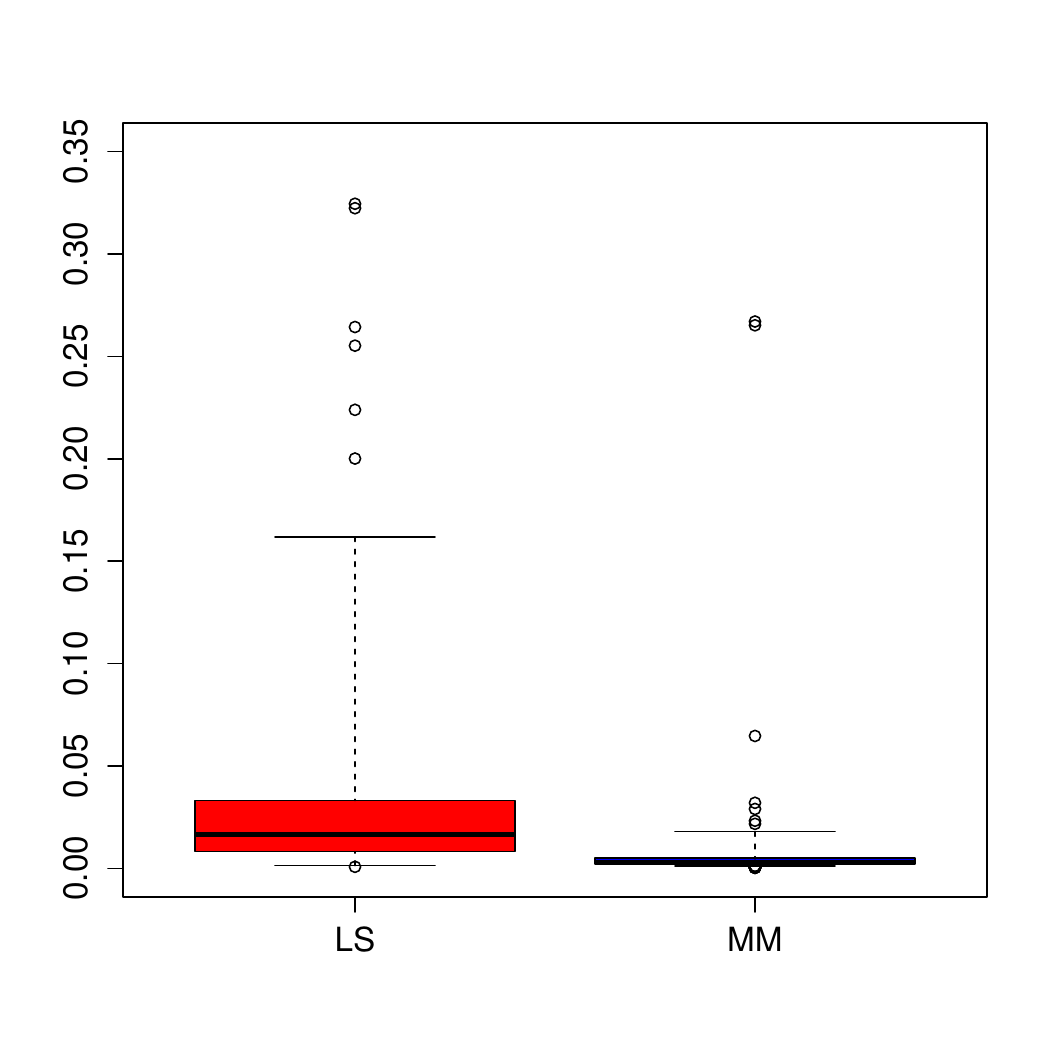} \\
$C_2$ &  $C_3$\\[-0.2in]
\includegraphics[scale=0.5]{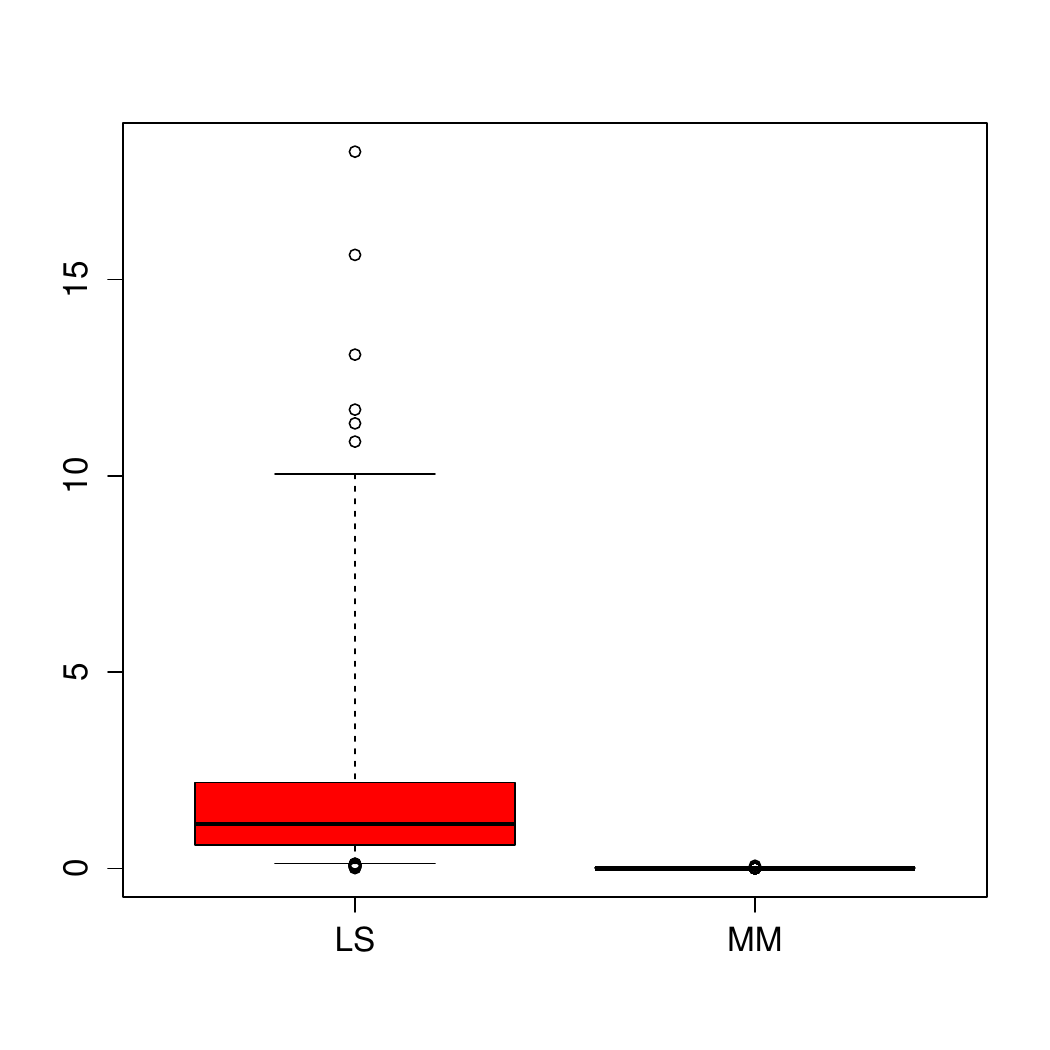} & 
\includegraphics[scale=0.5]{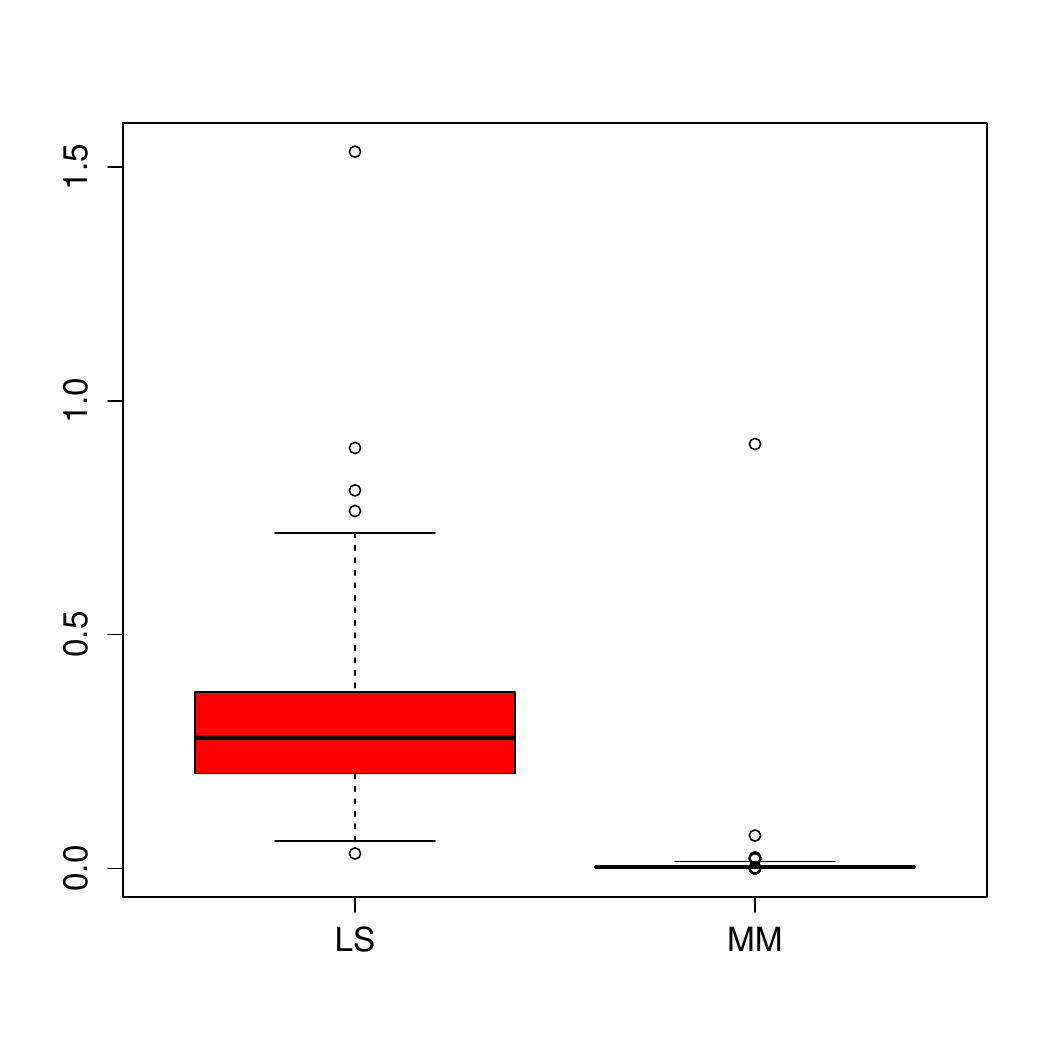}
\end{tabular}
\vskip-0.2in
\caption{\small \label{fig:ISE-M10-G1-ClasicoYRobusto}  Boxplots of the \textsc{ise} values for estimating the additive function $\eta_1$ for each contamination setting under Model 6, for the classical fit (in red) and for the robust fit (in blue).}

\end{center}
\end{figure}

\begin{figure}[ht!]
\begin{center}
\renewcommand{\arraystretch}{1.2}
\begin{tabular}{cc}
\multicolumn{2}{c}{Model 6}\\  
$C_0$ &  $C_1$\\[-0.2in]
\includegraphics[scale=0.5]{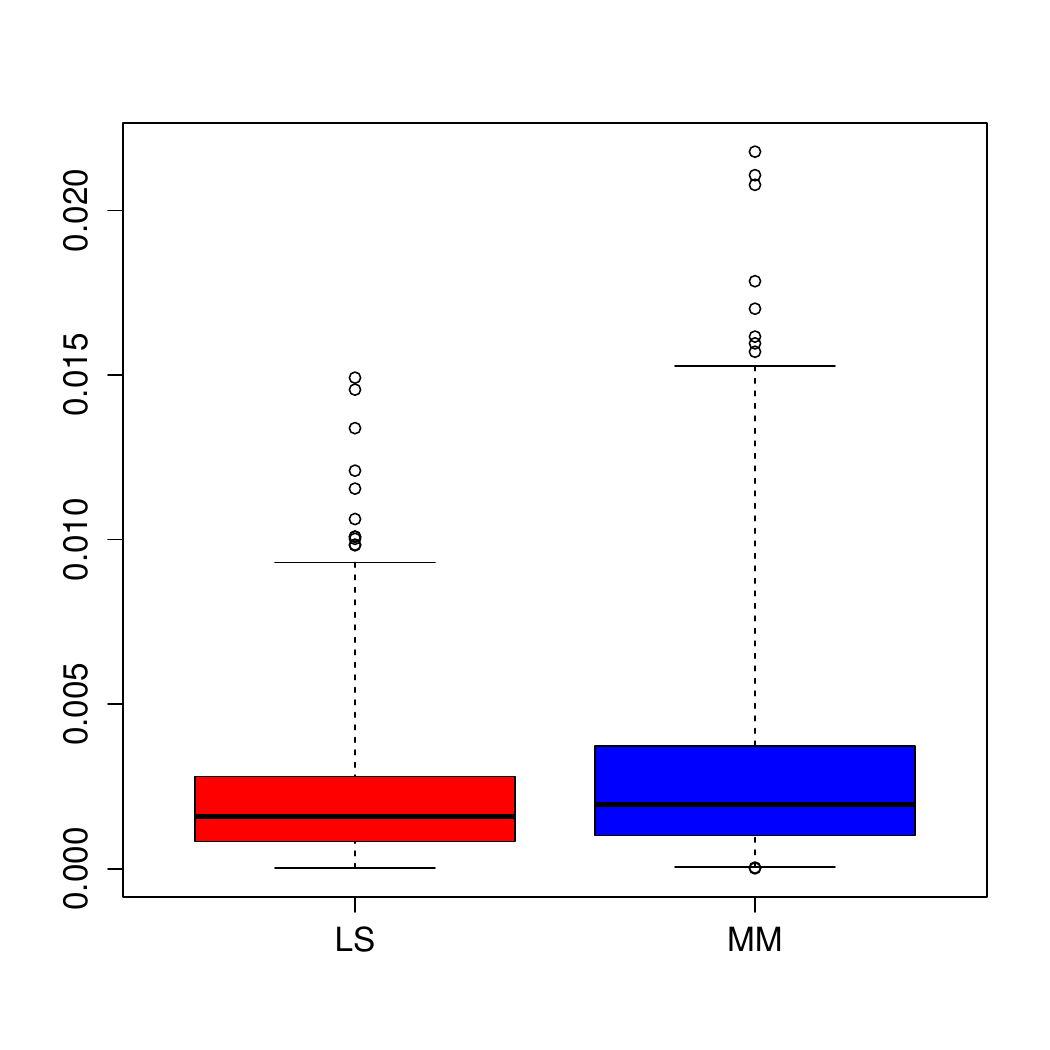} & 
\includegraphics[scale=0.5]{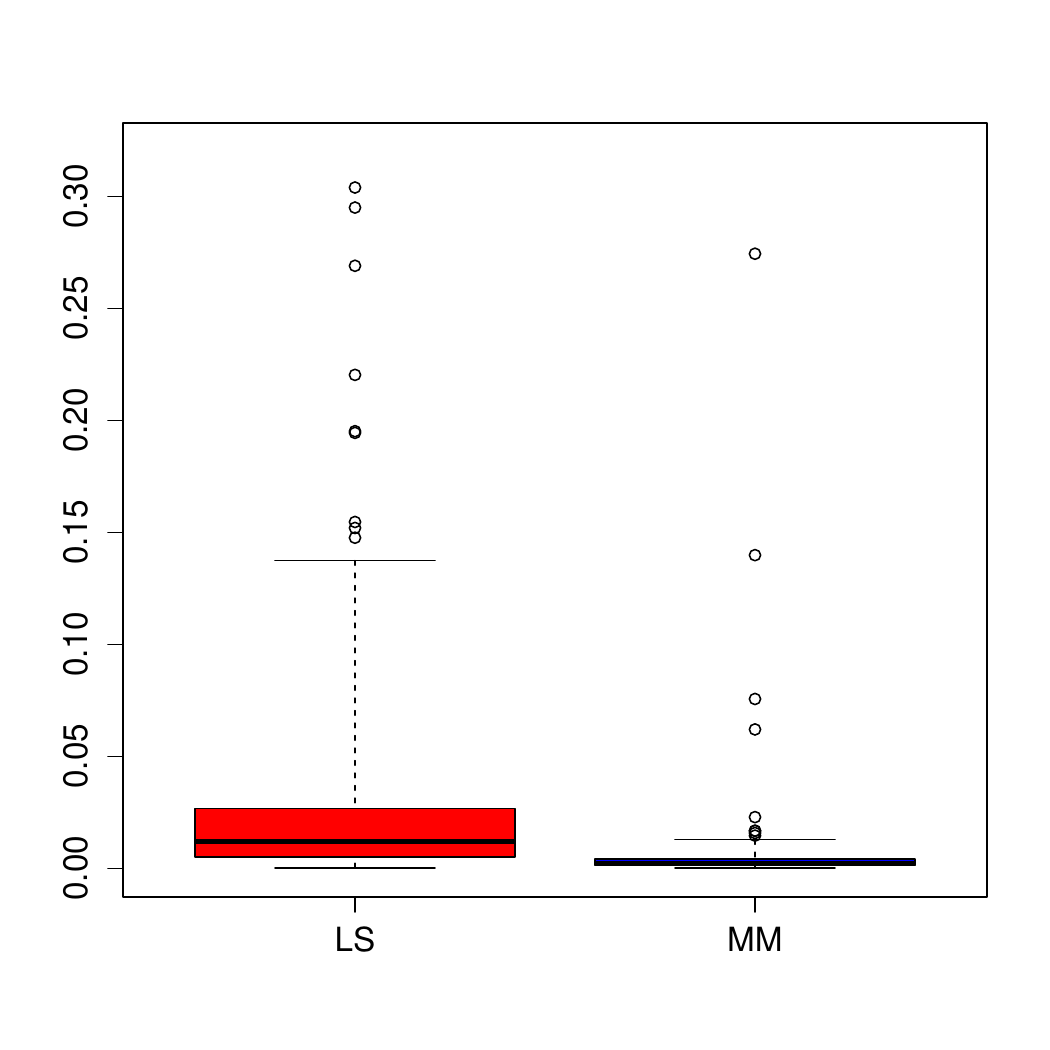} \\
$C_2$ &  $C_3$\\[-0.2in]
\includegraphics[scale=0.5]{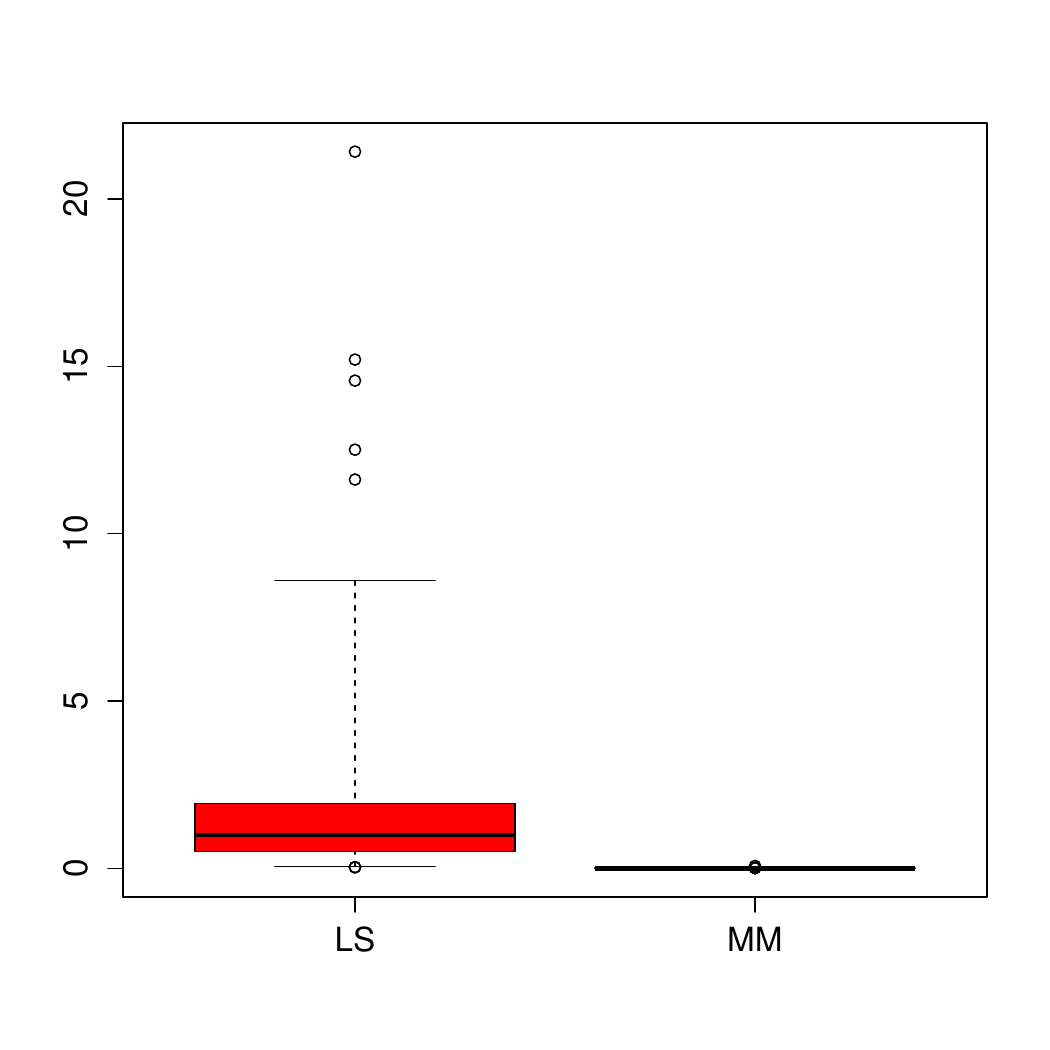} & 
\includegraphics[scale=0.5]{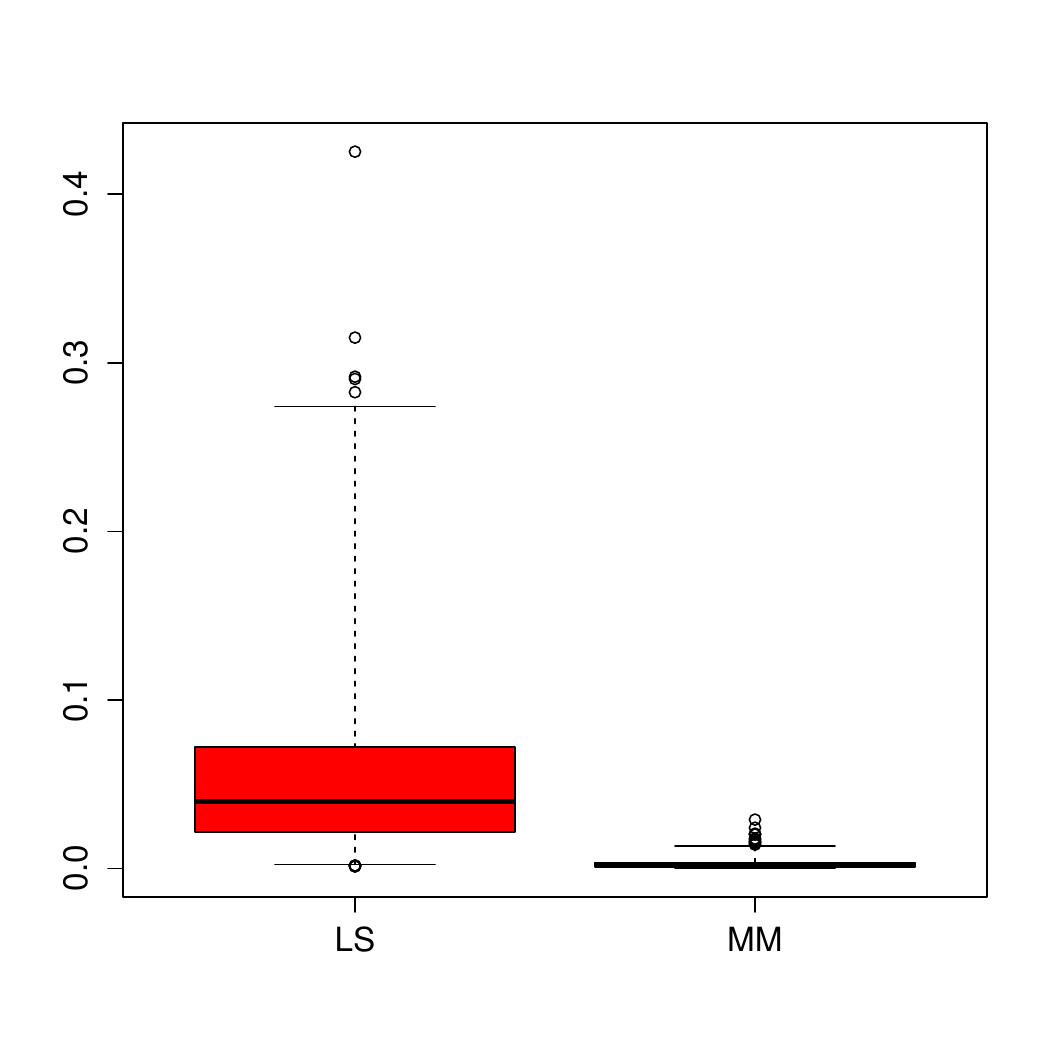}
\end{tabular}
\vskip-0.2in
\caption{\small \label{fig:ISE-M10-G2-ClasicoYRobusto}  Boxplots of the \textsc{ise} values for estimating the additive function $\eta_2$ for each contamination setting under Model 6, for the classical fit (in red) and for the robust fit (in blue).}

\end{center}
\end{figure}

Note that the \textsc{ise} is non-negative and expected to have a skewed distribution, for that reason, Figures \ref{fig:ISE-M7-G1-ClasicoYRobusto} and \ref{fig:ISE-M7-G2-ClasicoYRobusto}  present skewed-adjusted boxplots, as defined in Hubert  and Vandervieren (2008),   to display the obtained \textsc{ise} results  for the estimates of $\eta_1$ and $\eta_2$, respectively, under Model 1. The red and blue boxes correspond to the classical and robust procedures, respectively.
Similarly, Figures \ref{fig:ISE-M10-G1-ClasicoYRobusto} and \ref{fig:ISE-M10-G2-ClasicoYRobusto} contain the skewed-adjusted boxplots for the estimates of $\eta_1$ and $\eta_2$, respectively, under Model 6.

\begin{figure}[ht!]
\begin{center}
\newcolumntype{M}{>{\centering\arraybackslash}m{\dimexpr.05\linewidth-1\tabcolsep}}
   \newcolumntype{G}{>{\centering\arraybackslash}m{\dimexpr.5\linewidth-1\tabcolsep}}
\renewcommand{\arraystretch}{1.8}

\begin{tabular}{M GG}
& \multicolumn{2}{c}{Model 1}\\  
& \textsc{ls} &  \textsc{mm} \\ 
 $\eta_1$&
\includegraphics[scale=0.45]{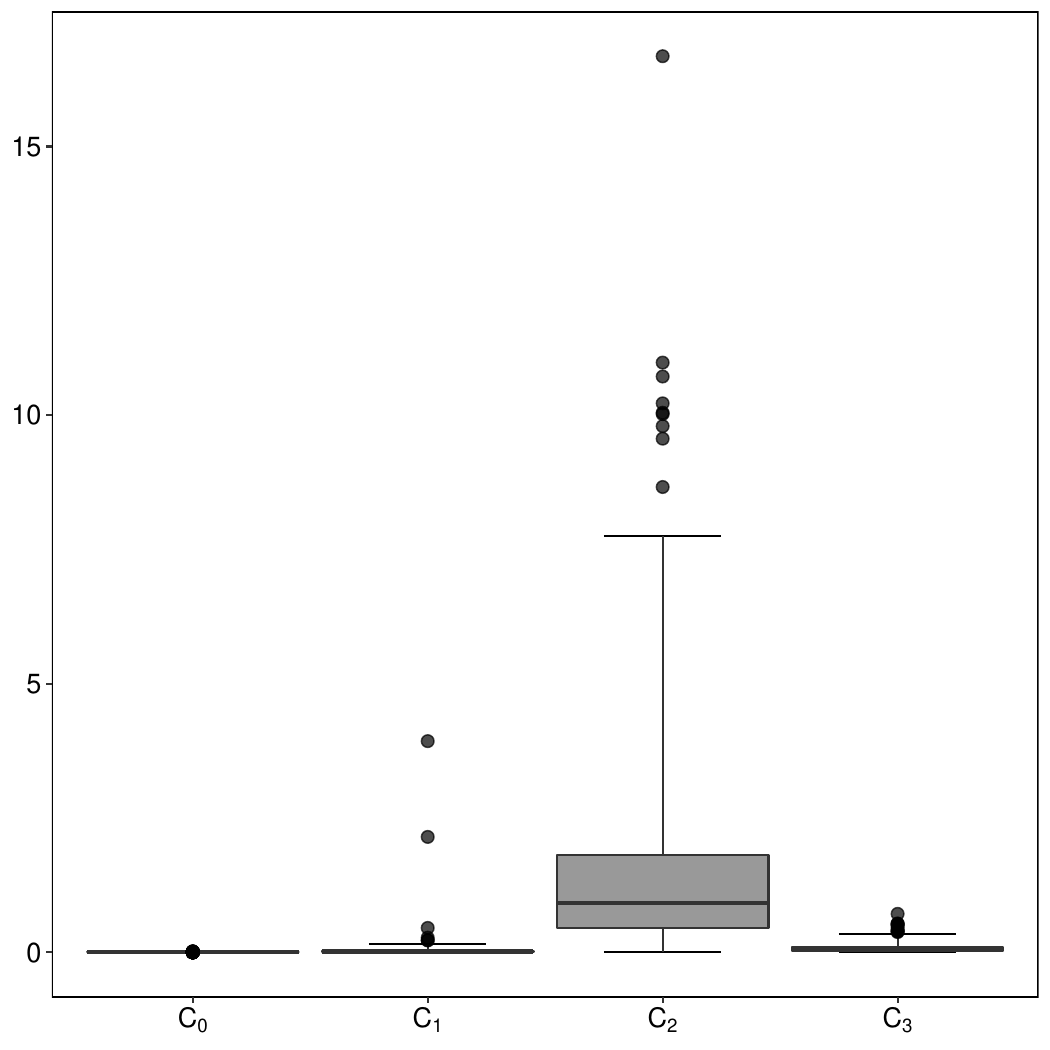}&
\includegraphics[scale=0.45]{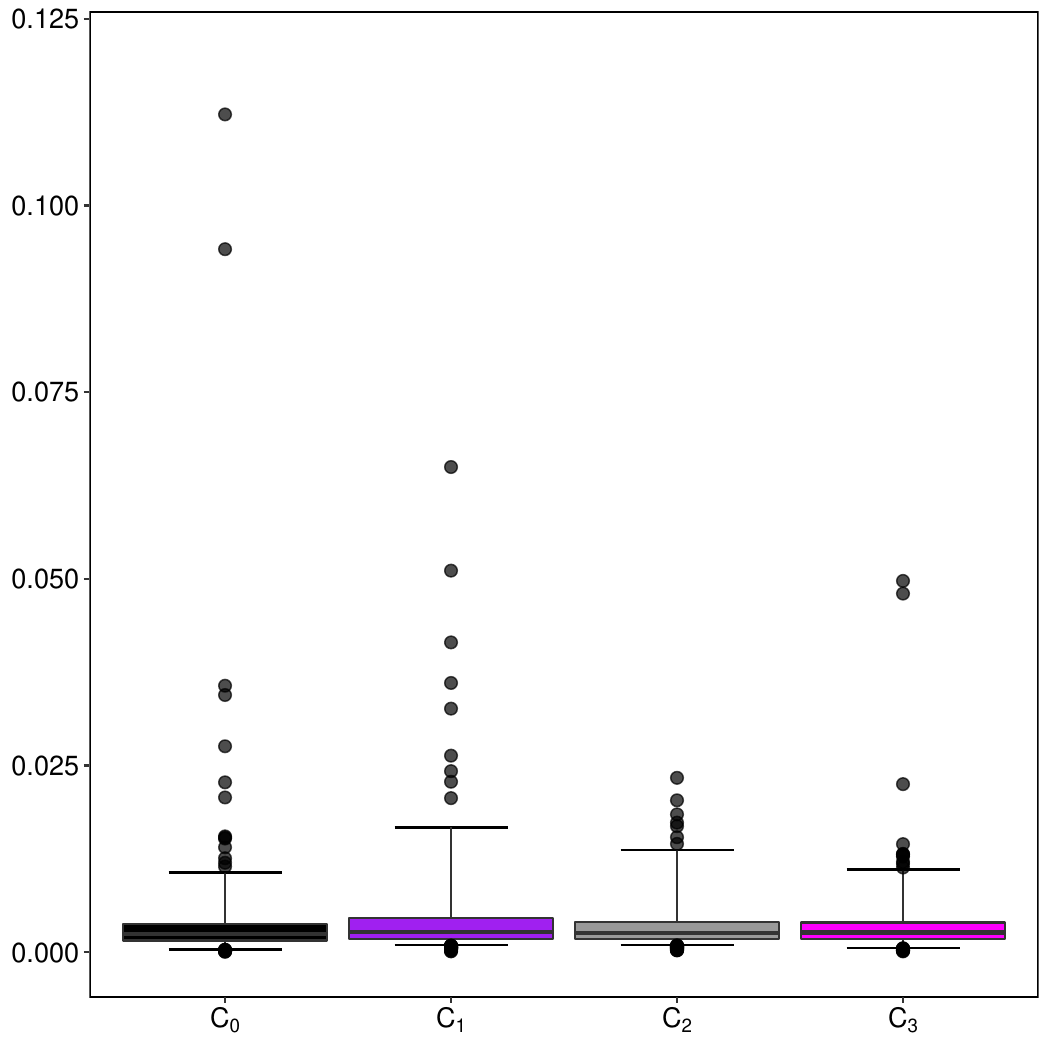}\\
 $\eta_2$ & 
\includegraphics[scale=0.45]{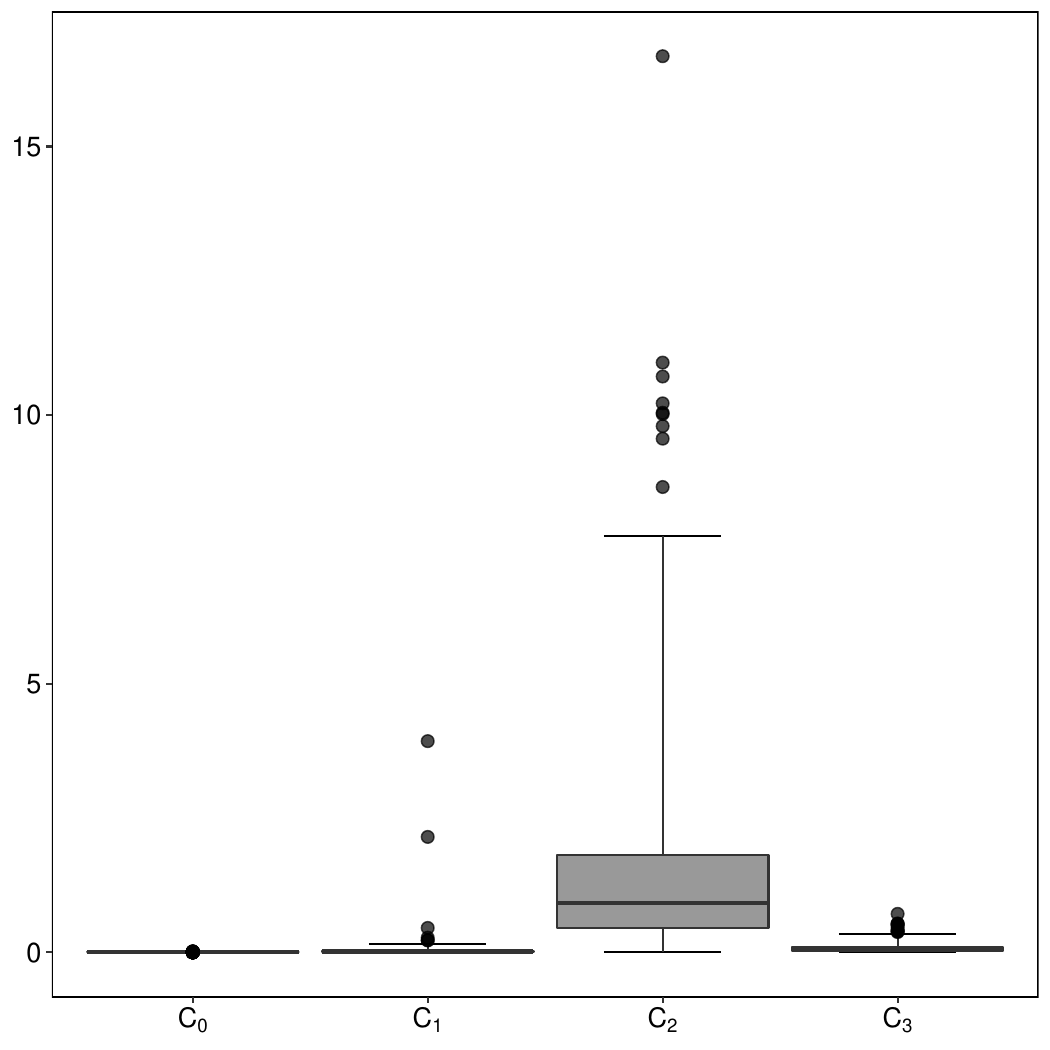}&
 \includegraphics[scale=0.45]{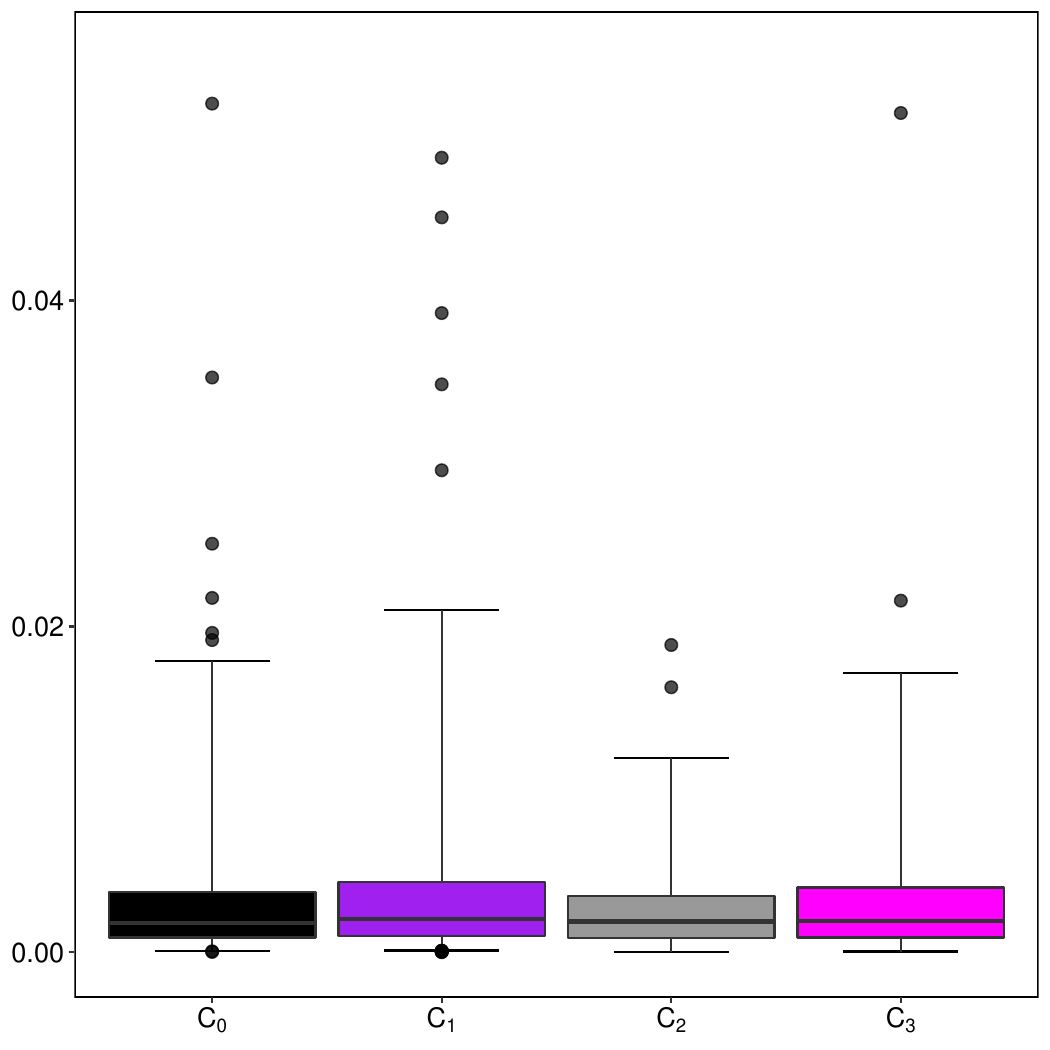}
 \end{tabular}
\caption{\small  \label{fig:ISE-M7-BP} Boxplots of the \textsc{ise} values obtained for the classical and robust estimations of the additive functions $\eta_1$ and $\eta_2$ for each contamination setting under Model 1.}

\end{center}
\end{figure}

\begin{figure}[ht!]
\begin{center}
\newcolumntype{M}{>{\centering\arraybackslash}m{\dimexpr.05\linewidth-1\tabcolsep}}
   \newcolumntype{G}{>{\centering\arraybackslash}m{\dimexpr.5\linewidth-1\tabcolsep}}
\renewcommand{\arraystretch}{1.8}
\begin{tabular}{M GG}
& \multicolumn{2}{c}{Model 6}\\  
& \textsc{ls} &  \textsc{mm} \\ 
 $\eta_1$&
\includegraphics[scale=0.45]{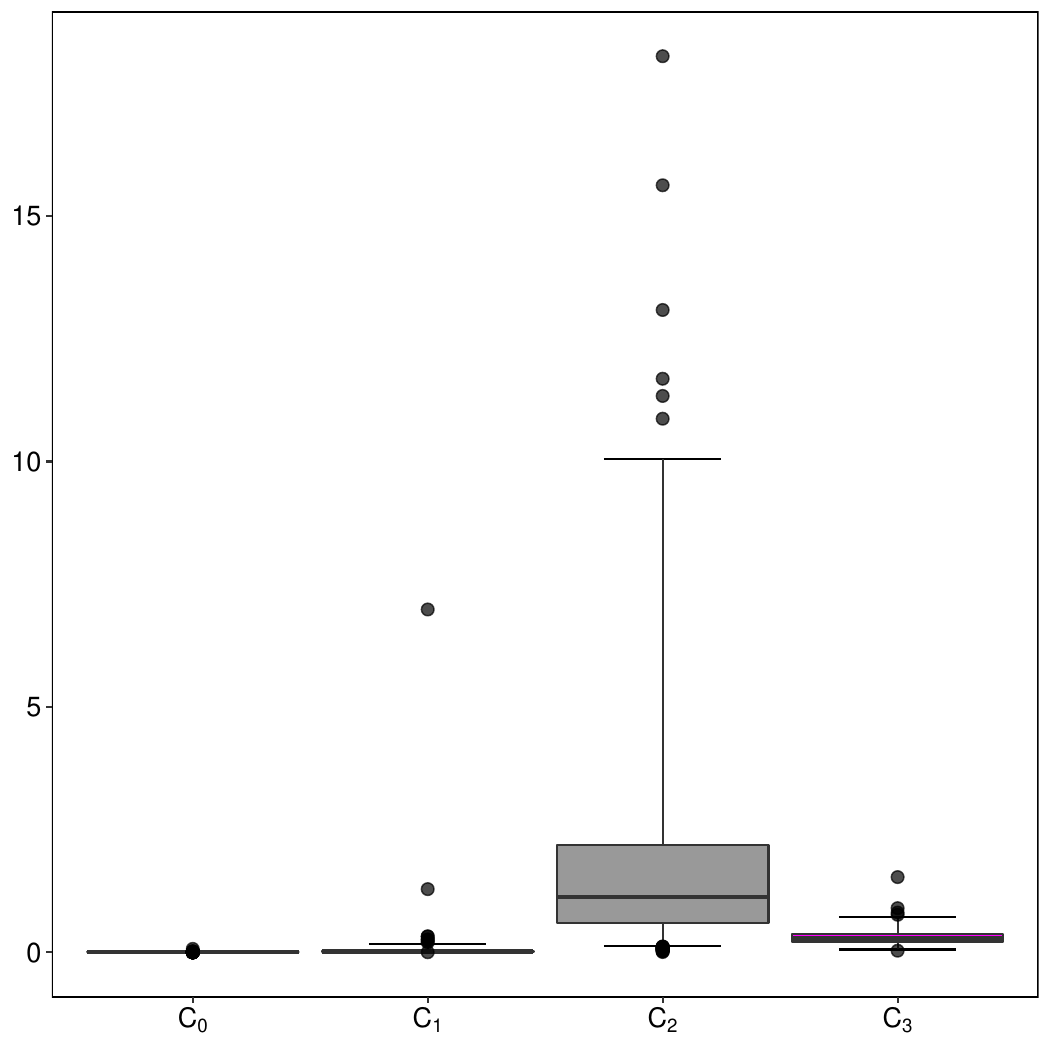}&
\includegraphics[scale=0.45]{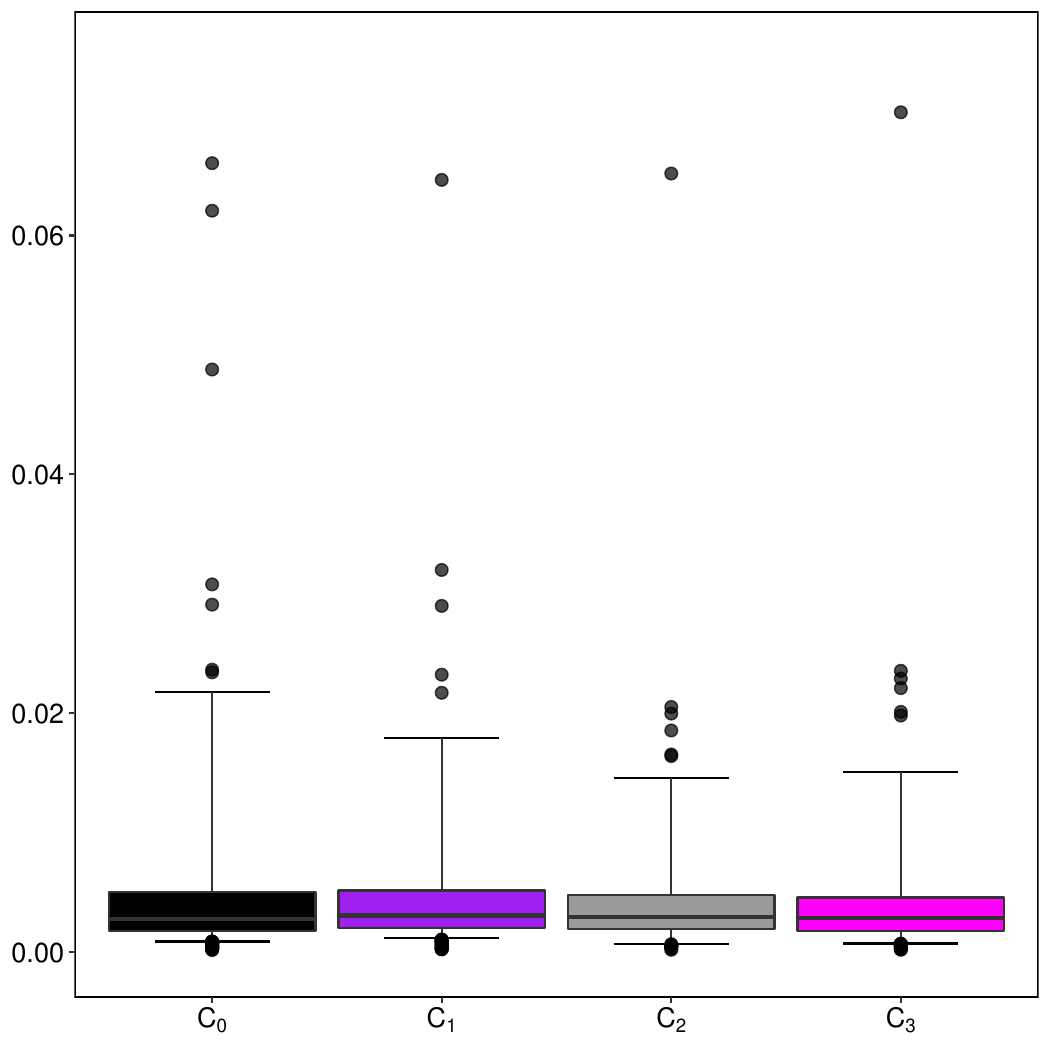}\\
 $\eta_2$&  
\includegraphics[scale=0.45]{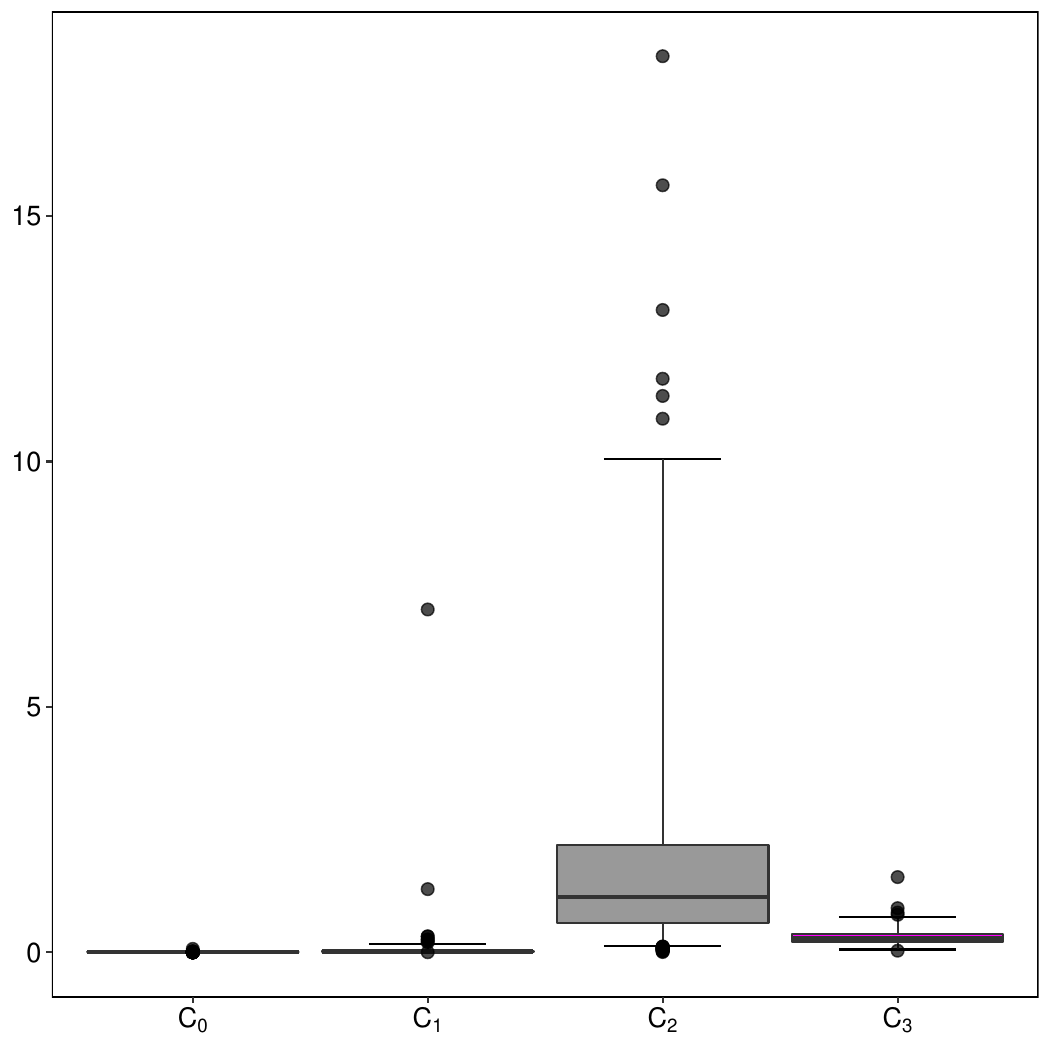} &
\includegraphics[scale=0.45]{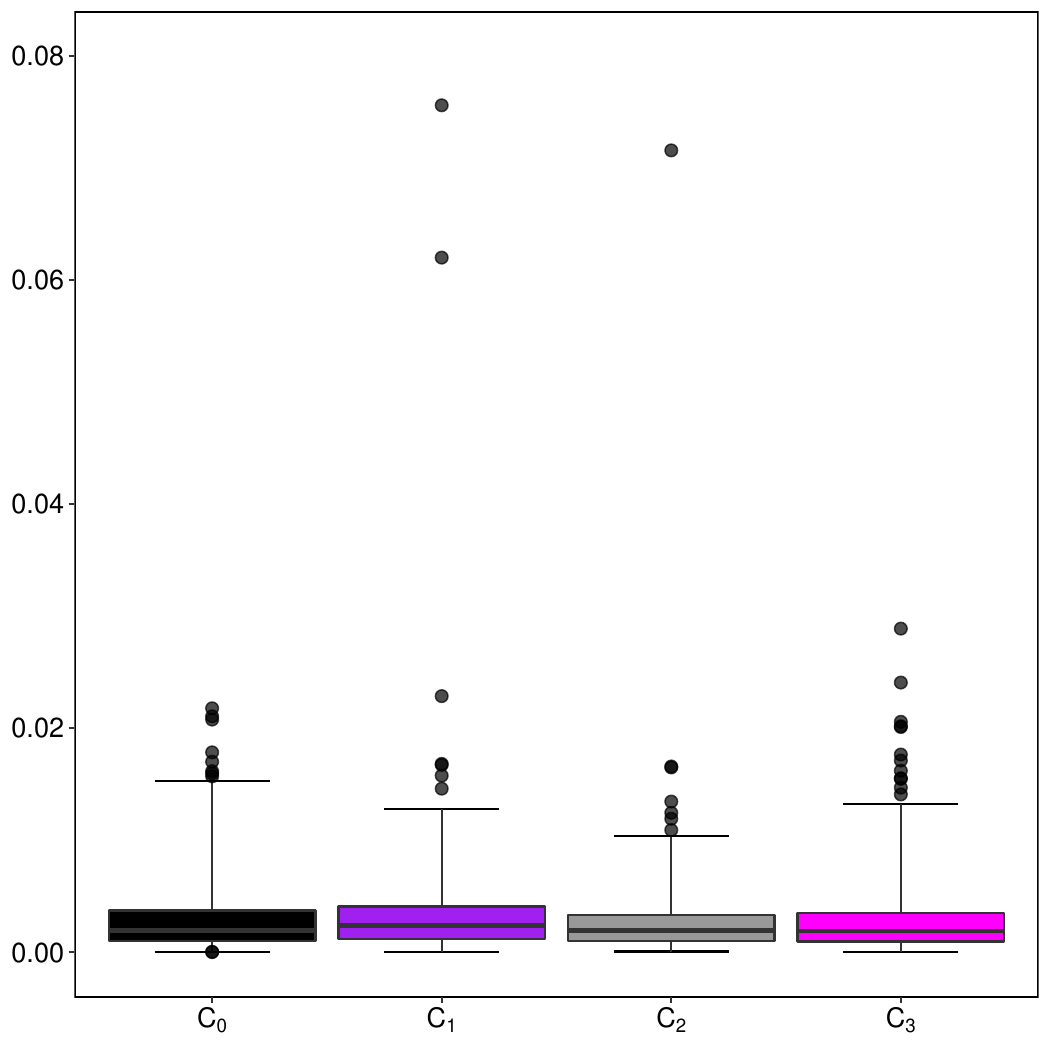}
\end{tabular}
\caption{\small \label{fig:ISE-M10-BP}  Boxplots of the \textsc{ise} values obtained for the classical and robust estimations of the additive functions $\eta_1$ and $\eta_2$ for each contamination setting under Model 6.}

\end{center}
\end{figure}

As seen in Table   \ref{tab:EstGsplines}, for clean samples, both classical and robust procedures present a similar behaviour. This fact is also reflected in the boxplots of the \textsc{ise} presented in Figures \ref{fig:ISE-M7-G1-ClasicoYRobusto} to \ref{fig:ISE-M10-G2-ClasicoYRobusto}. Note that, due to the efficiency loss, the  \textsc{ise}   for the robust method are slightly larger and show more wider boxes than when considering  the classical procedure. In contrast,   under the contamination schemes considered, the obtained values of the \textsc{ise} for the classical estimates are considerably enlarged which respect to those obtained for the robust procedure. Furthermore, the boxplots under $C_1$ to $C_3$ for the robust method are shifted towards $0$ with respect to those of the classical one. To see this effect, Figures \ref{fig:ISE-M7-BP} and \ref{fig:ISE-M10-BP} shows the same adjusted boxplots as parallel ones, for Models 1 and 6, respectively. The colours black, purple, grey and magenta identify the boxplots for the contamination settings $C_0$, $C_1$, $C_2$ and $C_3$, respectively. The stability of the robust procedure is reflected in the right panel of Figures \ref{fig:ISE-M7-BP} and \ref{fig:ISE-M10-BP}, since all the boxplots  are very similar. On the contrary, when observing the left panel, we  appreciate the sensitivity of the classical fits, in particular under $C_2$ where the \textsc{ise} values are extremely large leading to boxplots which are much higher than the other ones.
One of the reason of observing large values for the  $\mbox{\textsc{ise}}$ is that it may be heavily influenced by numerical errors at or near the boundaries of the grid, that is why,  following  He and Shi (1998), we consider a trimmed version  of the  \textsc{ise} computed without the $q$ first and last points on the grid, that is,
 \begin{eqnarray*}
  \mbox{\textsc{ise}}_{\trim}   & = &  \frac 1{M-2q} \sum_{s=q+1}^{M-q}  \left(f(t_s)-\wefe(t_s)\right)^2\, ,
 \end{eqnarray*}
 We chose $q=[M\times 0.05]$ which uses the central 90\% interior points in the grid.    Table \ref{tab:EstGsplinesTrim} reports  the  trimmed mean and  the median  of the   $ \mbox{\textsc{ise}}_{\trim} $  denoted   \textsc{mise}$_{\trim}$ and \textsc{medise}$_{\trim}$, respectively.

 \begin{table}[ht!]
\begin{center}
\footnotesize
\renewcommand{\arraystretch}{1.2}
\setlength{\tabcolsep}{3pt}
 \begin{tabular}{|c|c|c|c|c||c|c||c|c||c|c||c|c||c|c|}
\cline{4-15}
\multicolumn{1}{c}{} & \multicolumn{1}{c}{} & \multicolumn{1}{c|}{} &\multicolumn{2}{c||}{Model 1} &\multicolumn{2}{c||}{Model 2} &\multicolumn{2}{c||}{Model 3}  &\multicolumn{2}{c||}{Model 4} & \multicolumn{2}{c||}{Model 5}& \multicolumn{2}{c|}{Model 6} \\\cline{4-15}
\multicolumn{1}{c}{} & \multicolumn{1}{c}{} & \multicolumn{1}{c|}{} & \textsc{ls} & \textsc{mm} & \textsc{ls} & \textsc{mm} & \textsc{ls} & \textsc{mm} & \textsc{ls} & \textsc{mm} & \textsc{ls} & \textsc{mm} & \textsc{ls} & \textsc{mm} \\ \hline
		  & $C_0$ & \textsc{mise}$_{\trim}$ & 0.002 & 0.007 & 0.002 & 0.003 & 0.007 & 0.008 & 0.002 & 0.002 & 0.002 & 0.006 & 0.002 & 0.003 \\ 
  & & \textsc{medise}$_{\trim}$ & 0.002 & 0.002 & 0.002 & 0.002 & 0.004 & 0.004 & 0.002 & 0.002 & 0.002 & 0.002 & 0.002 & 0.002 \\  \cdashline{2-15}
  & $C_1$ & \textsc{mise}$_{\trim}$ & 0.019 & 0.003 & 0.023 & 0.003 & 0.082 & 0.009 & 0.019 & 0.003 & 0.018 & 0.003 & 0.022 & 0.003 \\  
  $\eta_1$ & & \textsc{medise}$_{\trim}$ & 0.010 & 0.002 & 0.013 & 0.002 & 0.043 & 0.005 & 0.010 & 0.002 & 0.010 & 0.002 & 0.013 & 0.002 \\  \cdashline{2-15}
  & $C_2$ & \textsc{mise}$_{\trim}$ & 1.104 & 0.002 & 1.335 & 0.003 & 4.845 & 0.008 & 1.052 & 0.002 & 1.063 & 0.002 & 1.351 & 0.003 \\ 
  & & \textsc{medise} & 0.612 & 0.002 & 0.847 & 0.002 & 2.761 & 0.005 & 0.583 & 0.002 & 0.580 & 0.002 & 0.819 & 0.002 \\  \cdashline{2-15}
  & $C_3$ & \textsc{mise}$_{\trim}$ & 0.057 & 0.002 & 0.505 & 0.003 & 2.068 & 0.009 & 0.035 & 0.002 & 0.008 & 0.002 & 0.269 & 0.003 \\ 
  & & \textsc{medise}$_{\trim}$ & 0.037 & 0.002 & 0.489 & 0.002 & 2.058 & 0.005 & 0.021 & 0.002 & 0.005 & 0.002 & 0.252 & 0.002 \\  \hline
  & $C_0$ & \textsc{mise}$_{\trim}$ & 0.002 & 0.011 & 0.002 & 0.002 & 0.005 & 0.006 & 0.002 & 0.002 & 0.002 & 0.002 & 0.002 & 0.002 \\ 
  & & \textsc{medise} & 0.001 & 0.001 & 0.001 & 0.002 & 0.003 & 0.001 & 0.001 & 0.001 & 0.001 & 0.001 & 0.001 & 0.001 \\  \cdashline{2-15}
  & $C_1$ & \textsc{mise}$_{\trim}$ & 0.018 & 0.002 & 0.018 & 0.003 & 0.050 & 0.006 & 0.018 & 0.004 & 0.017 & 0.002 & 0.017 & 0.003 \\ 
  $\eta_2$ & & \textsc{medise}$_{\trim}$ & 0.009 & 0.002 & 0.010 & 0.001 & 0.026 & 0.004 & 0.010 & 0.002 & 0.009 & 0.001 & 0.008 & 0.002 \\  \cdashline{2-15}
  & $C_2$ & \textsc{mise}$_{\trim}$ & 1.182 & 0.041 & 1.219 & 0.002 & 3.357 & 0.005 & 1.148 & 0.002 & 1.134 & 0.007 & 1.206 & 0.002 \\  
  & & \textsc{medise}$_{\trim}$ & 0.707 & 0.001 & 0.679 & 0.001 & 2.075 & 0.003 & 0.679 & 0.001 & 0.704 & 0.001 & 0.708 & 0.001 \\  \cdashline{2-15}
  & $C_3$ & \textsc{mise}$_{\trim}$ & 0.058 & 0.068 & 0.043 & 0.002 & 0.651 & 0.006 & 0.036 & 0.002 & 0.008 & 0.002 & 0.042 & 0.002 \\ 
  & & \textsc{medise}$_{\trim}$ & 0.034 & 0.001 & 0.026 & 0.001 & 0.645 & 0.004 & 0.021 & 0.001 & 0.005 & 0.001 & 0.030 & 0.001 \\  \hline
	\end{tabular}
\caption{\label{tab:EstGsplinesTrim}\footnotesize Summary measures for the   additive components estimates $\weta_1$ and $\weta_2$ based on the  $\mbox{\textsc{ise}}_{\trim}$. The classical and robust procedures are labelled \textsc{ls} and \textsc{mm}, respectively. }
\end{center}
\end{table}

Under $C_0$, the trimmed \textsc{mise} and \textsc{medise} of the classical estimators of the additive components and regression function are equal or very similar to those  of the robust ones. 
Tables \ref{tab:EstGsplines}  and \ref{tab:EstGsplinesTrim}  and also Figures \ref{fig:ISE-M7-G1-ClasicoYRobusto}  to \ref{fig:ISE-M10-BP}     illustrate the damage caused to the classical estimators by  contamination $C_2$.  The \textsc{ise} and the trimmed  \textsc{ise}   of the least squares estimators of $\eta_1$ and $\eta_2$,  across all models, are consistently  higher than those of the robust estimators. In particular, the \textsc{5\%-mise} and the  \textsc{mise}$_{\trim}$ of the classical procedure are more than  400  times those obtained with  the robust method in all models.  The other two contaminations also affect the classical estimators of the additive components but in a smaller degree since the ratio between the summary measure obtained for the classical and robust estimates is close to 4. In particular,   vertical outliers have a low impact on the least squares estimators of $\eta_1$ and $\eta_2$ only increasing their variability but not affecting their bias as it will be shown below in the functional boxplots of the estimated curves.  In contrast to the described behaviour of the least squares method, our robust proposal provides more reliable estimates    of the additive components which are almost unaffected by the different types of outliers, see Figures \ref{fig:ISE-M7-BP} and \ref{fig:ISE-M10-BP}.

\begin{figure}[H]
 \begin{center}
 \newcolumntype{M}{>{\centering\arraybackslash}m{\dimexpr.1\linewidth-1\tabcolsep}}
   \newcolumntype{G}{>{\centering\arraybackslash}m{\dimexpr.4\linewidth-1\tabcolsep}}
\renewcommand{\arraystretch}{0.1}
\begin{tabular}{M GG}
& \textsc{ls} &  \textsc{mm}\\[-0.3in]
$C_{0}$ &  
\includegraphics[scale=0.35]{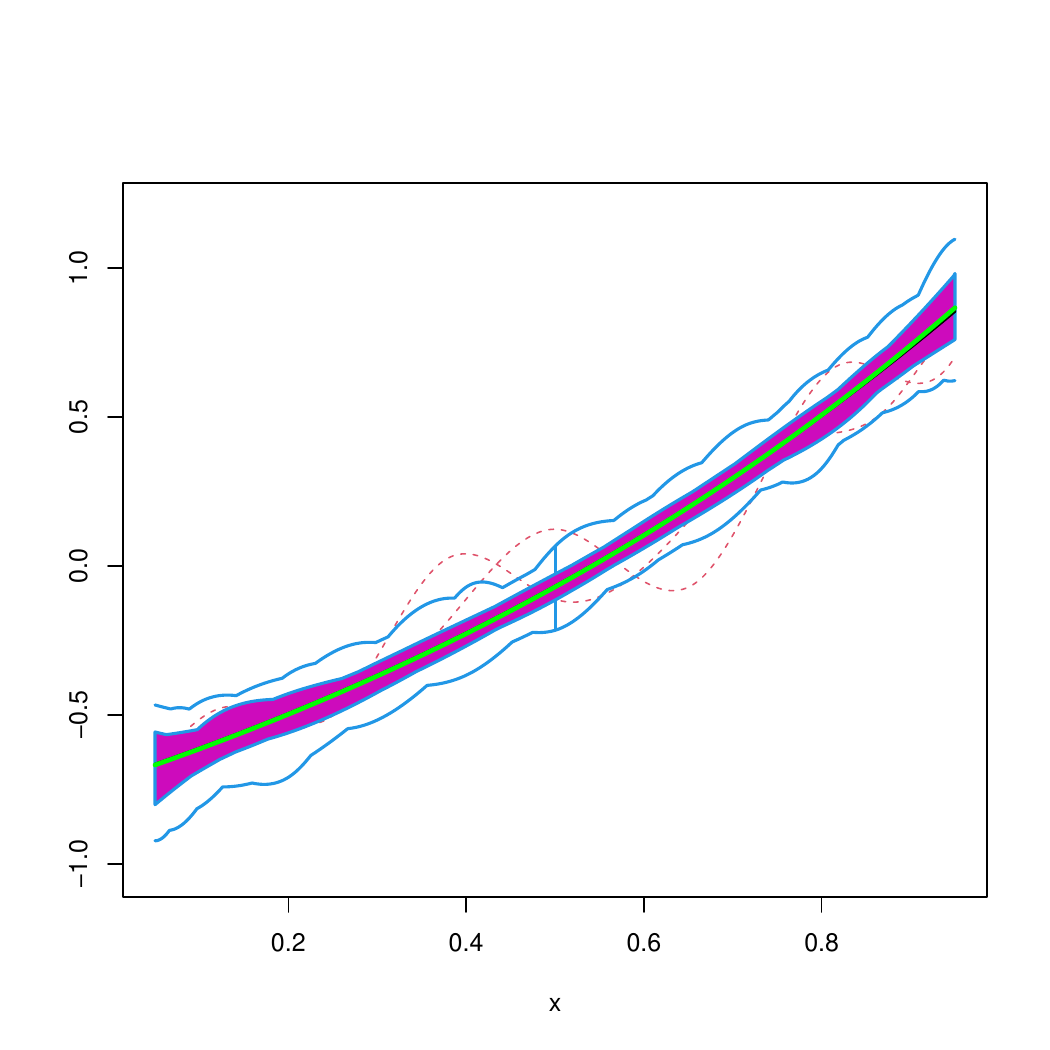} &
\includegraphics[scale=0.35]{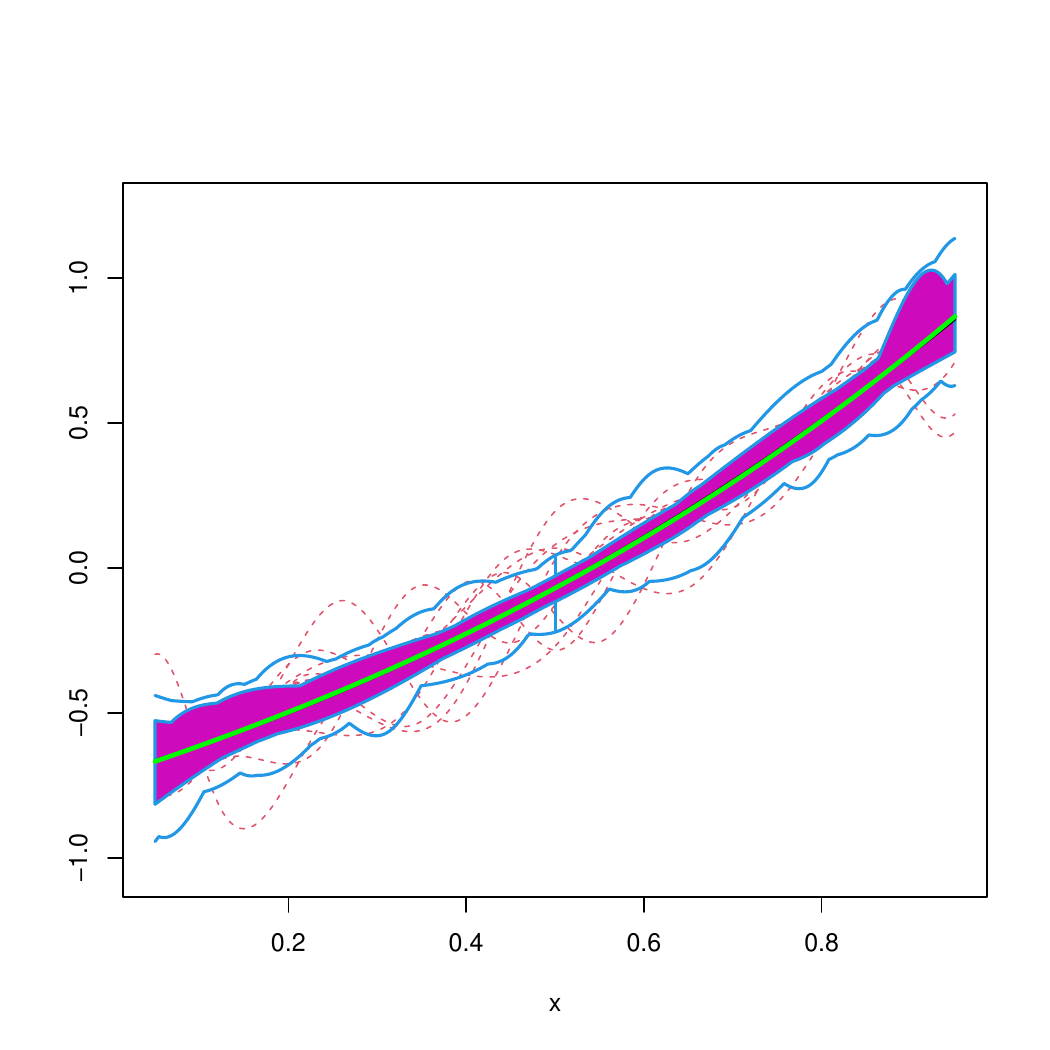}
 \\[-0.35in]
$C_1$& 
\includegraphics[scale=0.35]{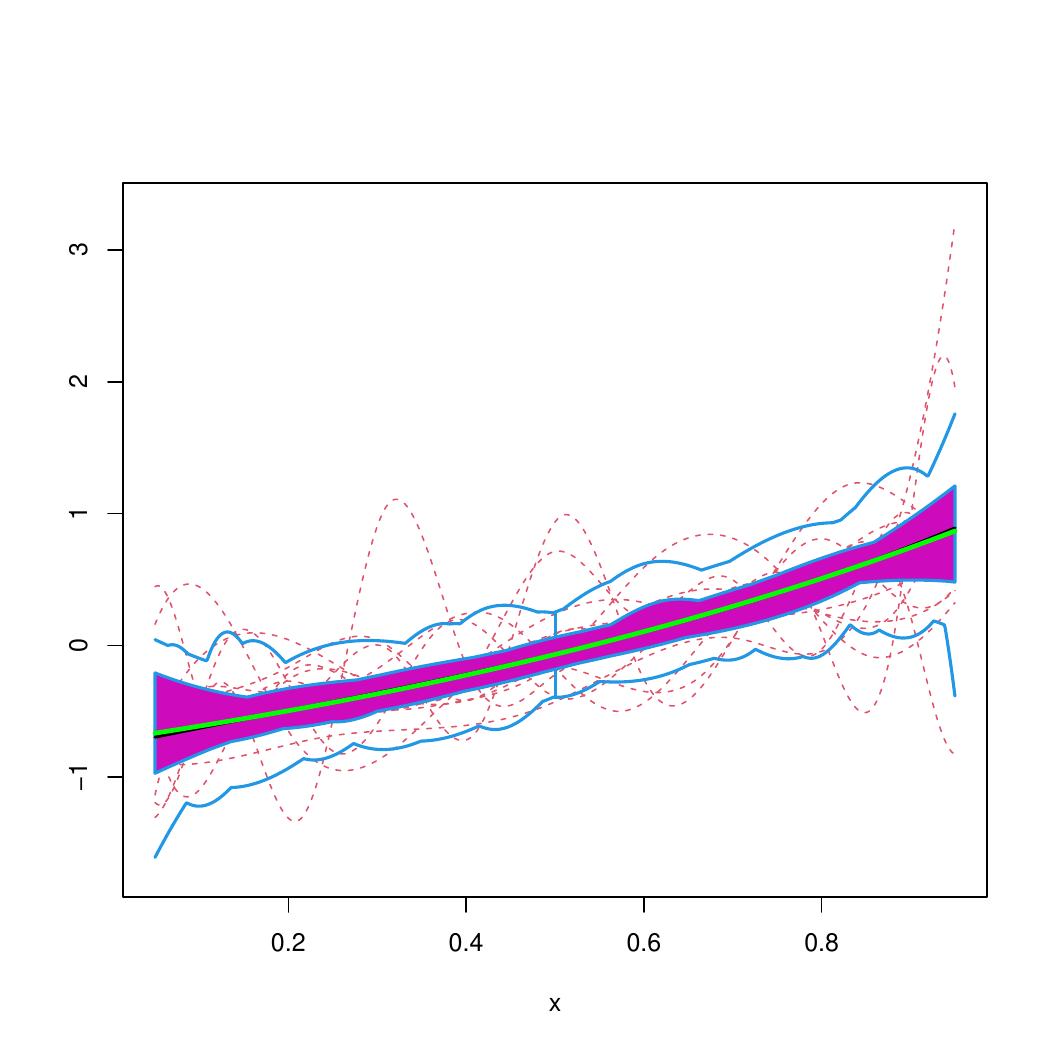} &
 \includegraphics[scale=0.35]{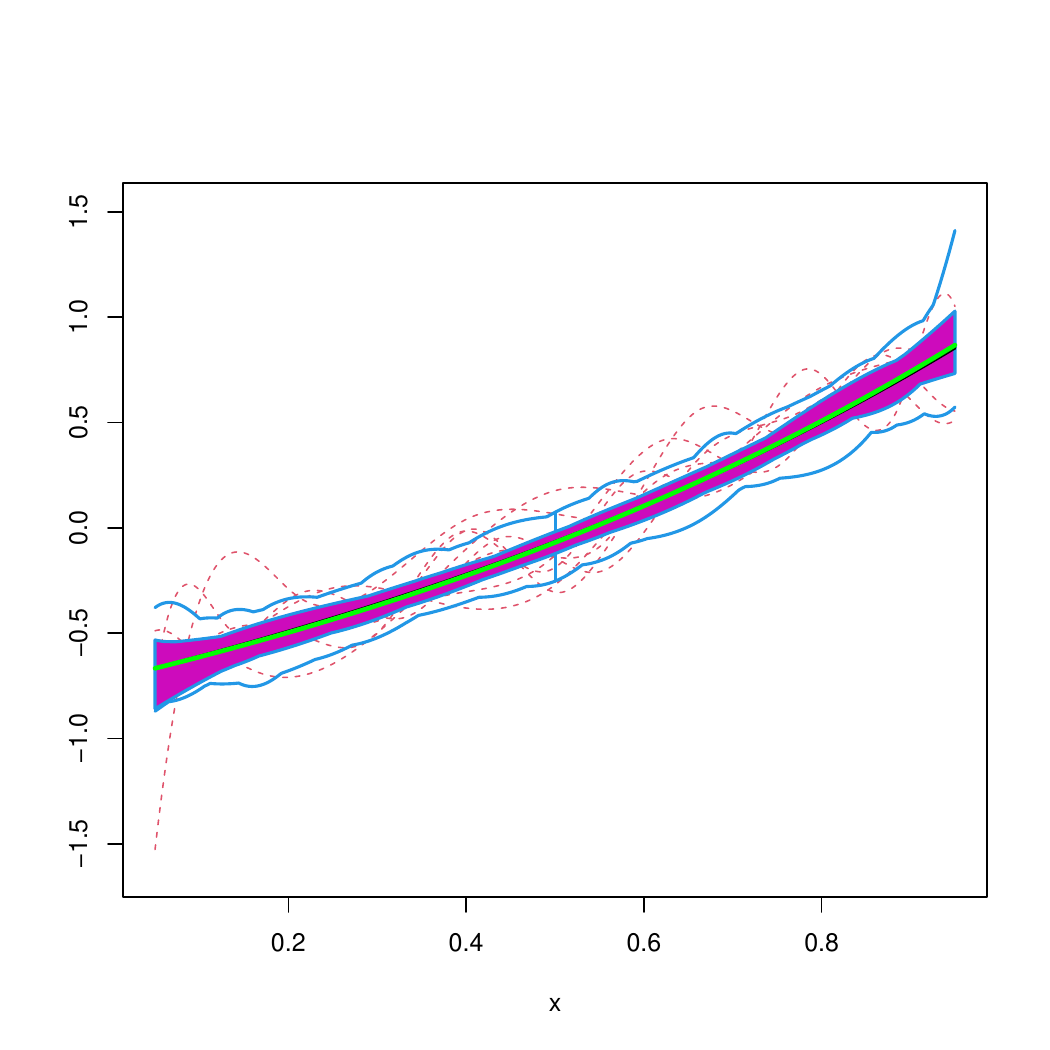}\\[-0.35in]
$C_2$&
\includegraphics[scale=0.35]{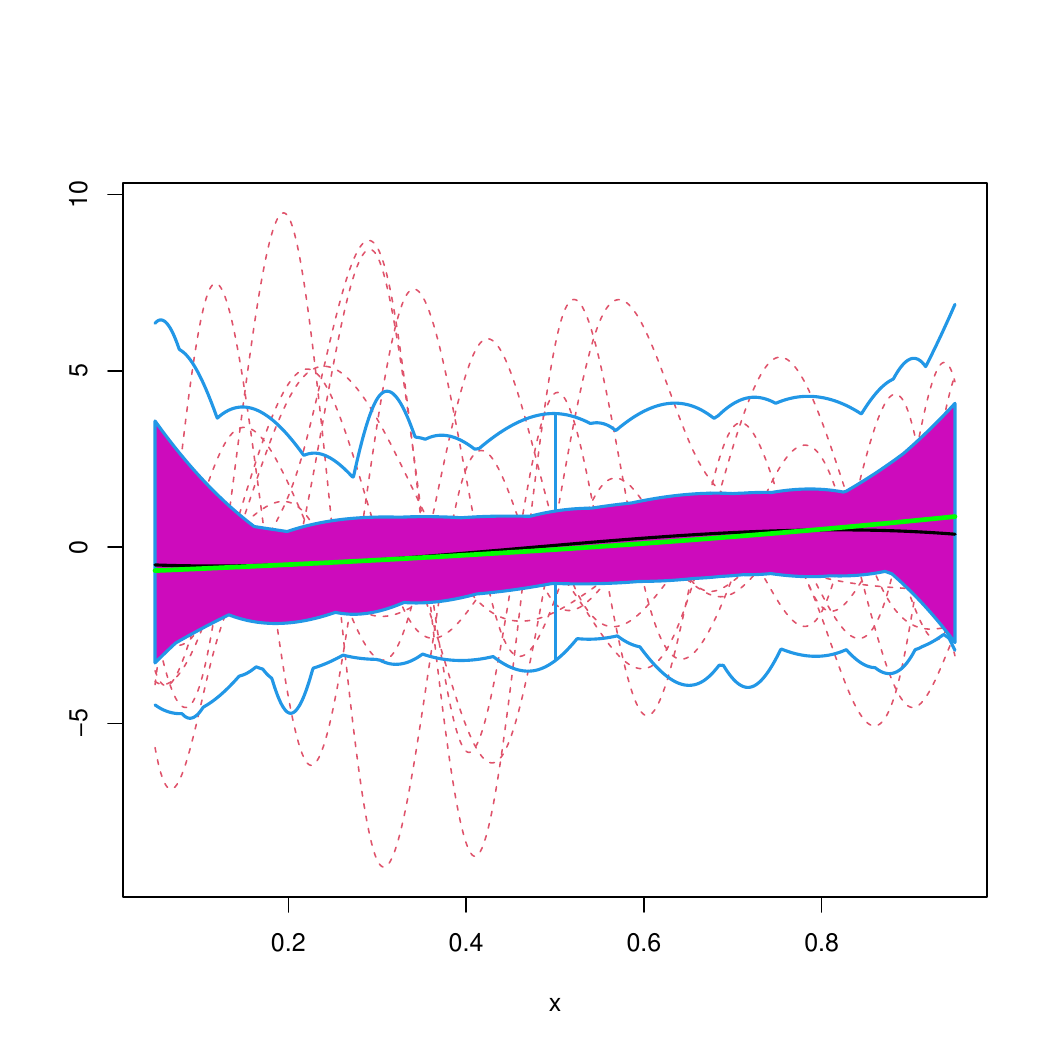} &
 \includegraphics[scale=0.35]{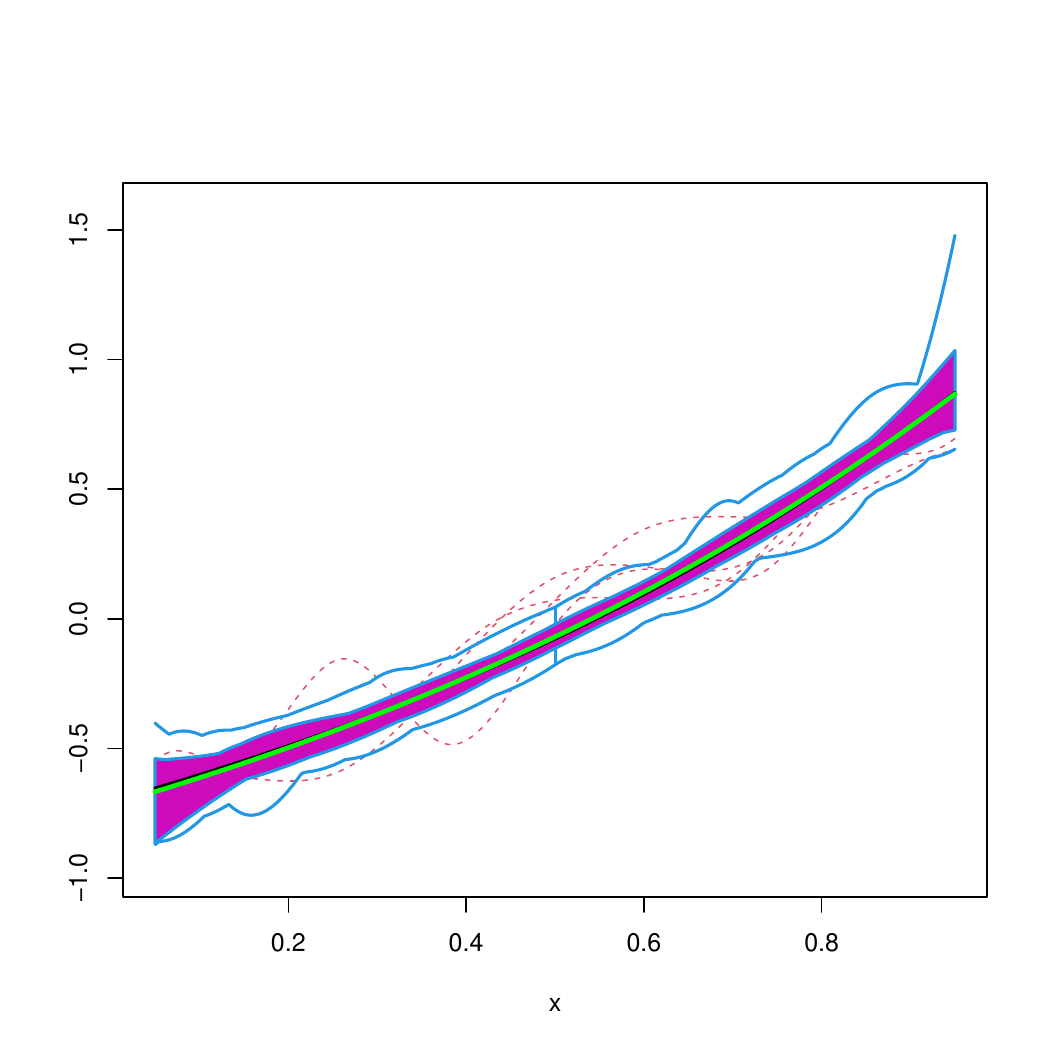}\\[-0.35in]
$C_3$ &
\includegraphics[scale=0.35]{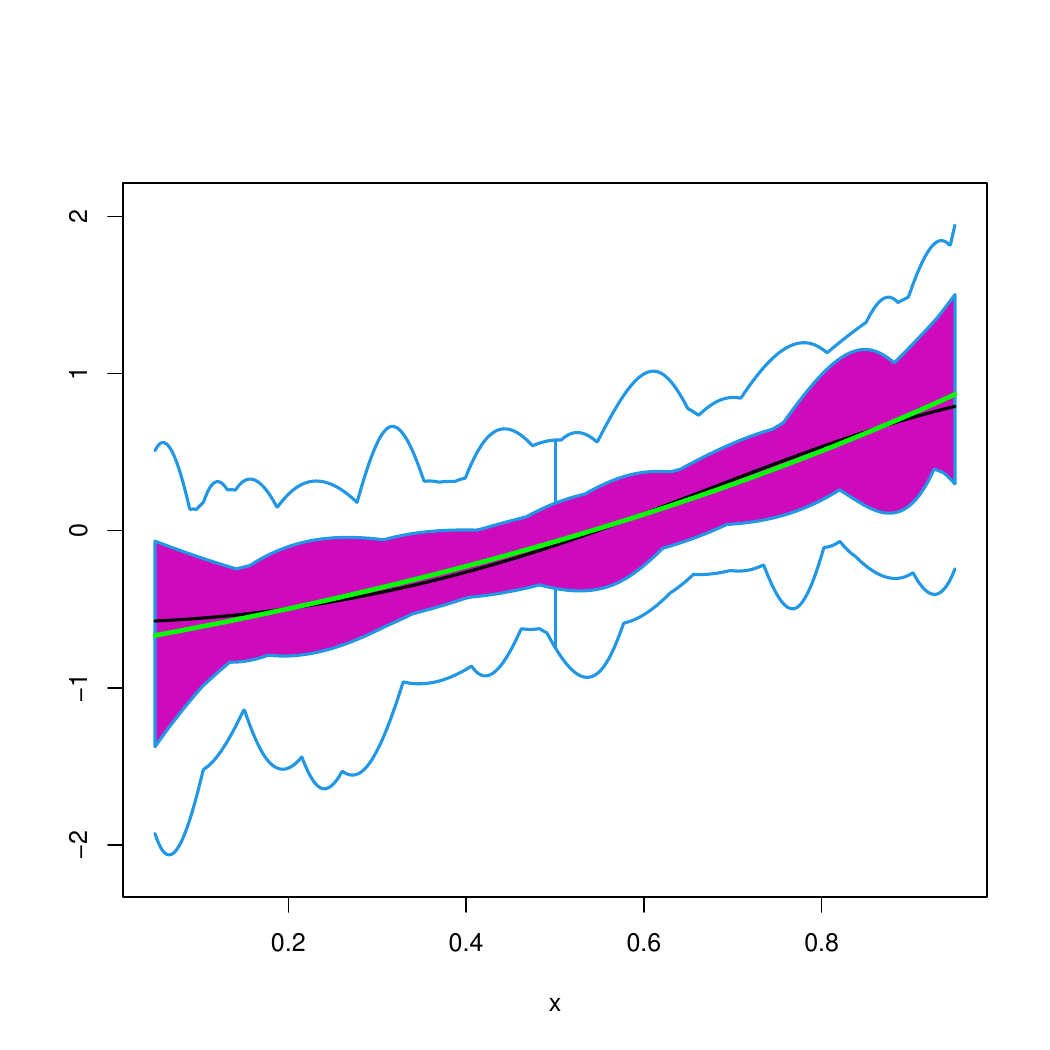} &
 \includegraphics[scale=0.35]{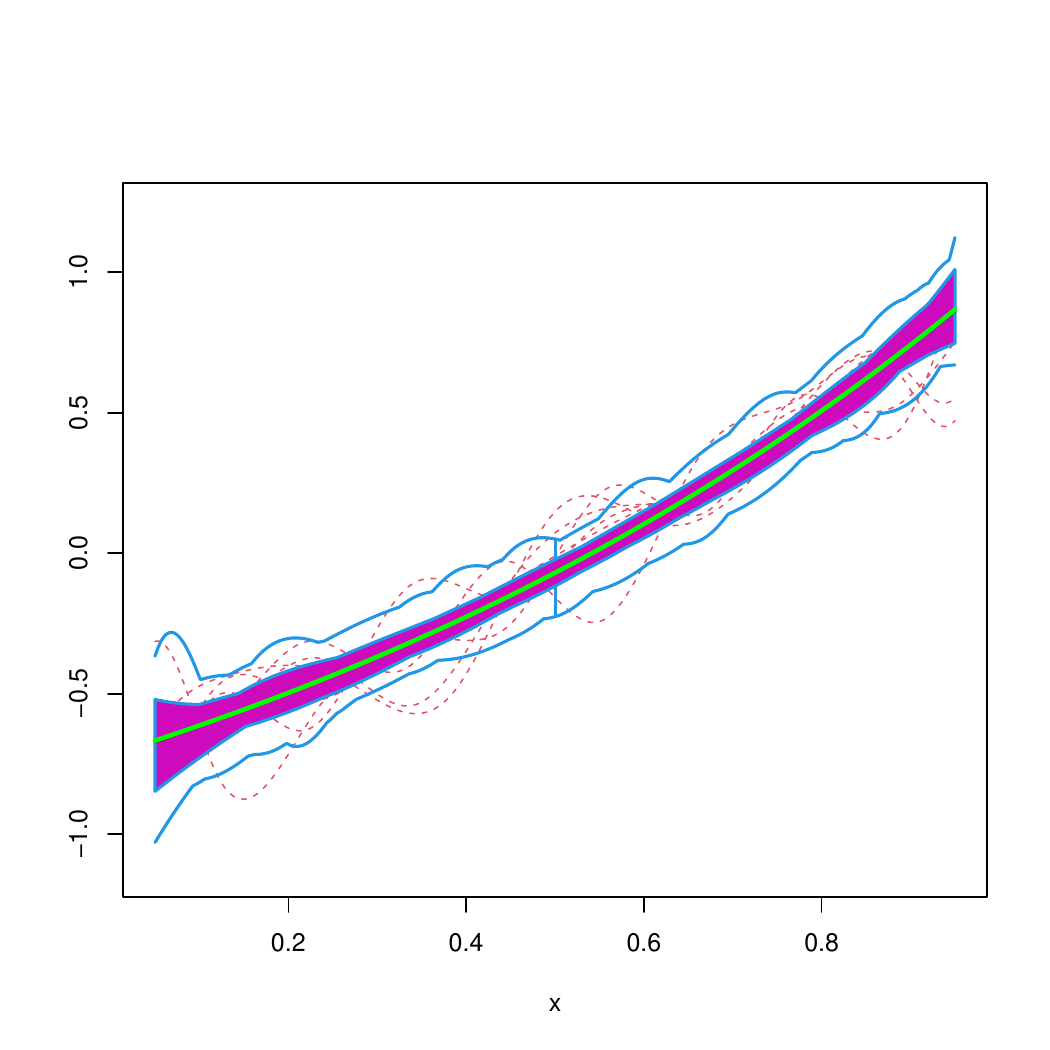}
 \end{tabular}
\caption{\small \label{fig:fboxplot-g2-m6}  Functional boxplots of the estimated additive functions $\eta_2$ obtained by the classical and  robust fit under Model 6. The solid green line corresponds to  the true function $\eta_2$.} 

\end{center}
\end{figure}

To illustrate  the performance of the estimated curves  $\weta_1$ and $\weta_2$, Figure \ref{fig:fboxplot-g2-m6} displays their functional boxplots when using the  classical and robust procedures, under Model 6.  As it is well-known when using splines, for both classical and robust fits, estimating problems may arise near the boundaries, for that reason,  we show here the different estimates $\weta_1$ and $\weta_2$ evaluated on  a grid of 100 equispaced points within the interval $[0.05, 0.95]$. Functional boxplots, introduced by  Sun and Genton (2011),  are    useful to visualize a collection of curves. The area in purple represents the 50\% inner band of curves, the dotted red lines correspond to outlying curves and the blue lines to the whiskers while the black line indicates the  deepest  function. The true functions is shown in solid green line in all plots.   As it is expected, for clean samples, box and whiskers for the robust fits are slightly larger than those obtained by the classical approach. However, for contamination settings $C_1$ to $C_3$ the general structure of the additive functions have been better captured by the robust fit noted on narrower envelops  In particular, when estimating $\eta_1$, contamination   $C_3$ distorts the classical estimators since the true function $\eta_1$ is not contained in the region containing the 50\% central curves.

\begin{figure}[H]
 \begin{center}
 \newcolumntype{M}{>{\centering\arraybackslash}m{\dimexpr.1\linewidth-1\tabcolsep}}
   \newcolumntype{G}{>{\centering\arraybackslash}m{\dimexpr.4\linewidth-1\tabcolsep}}
\renewcommand{\arraystretch}{0.1}
\begin{tabular}{M GG}
& \textsc{ls} &  \textsc{mm} \\[-0.3in]
$C_{0}$ &  
\includegraphics[scale=0.35]{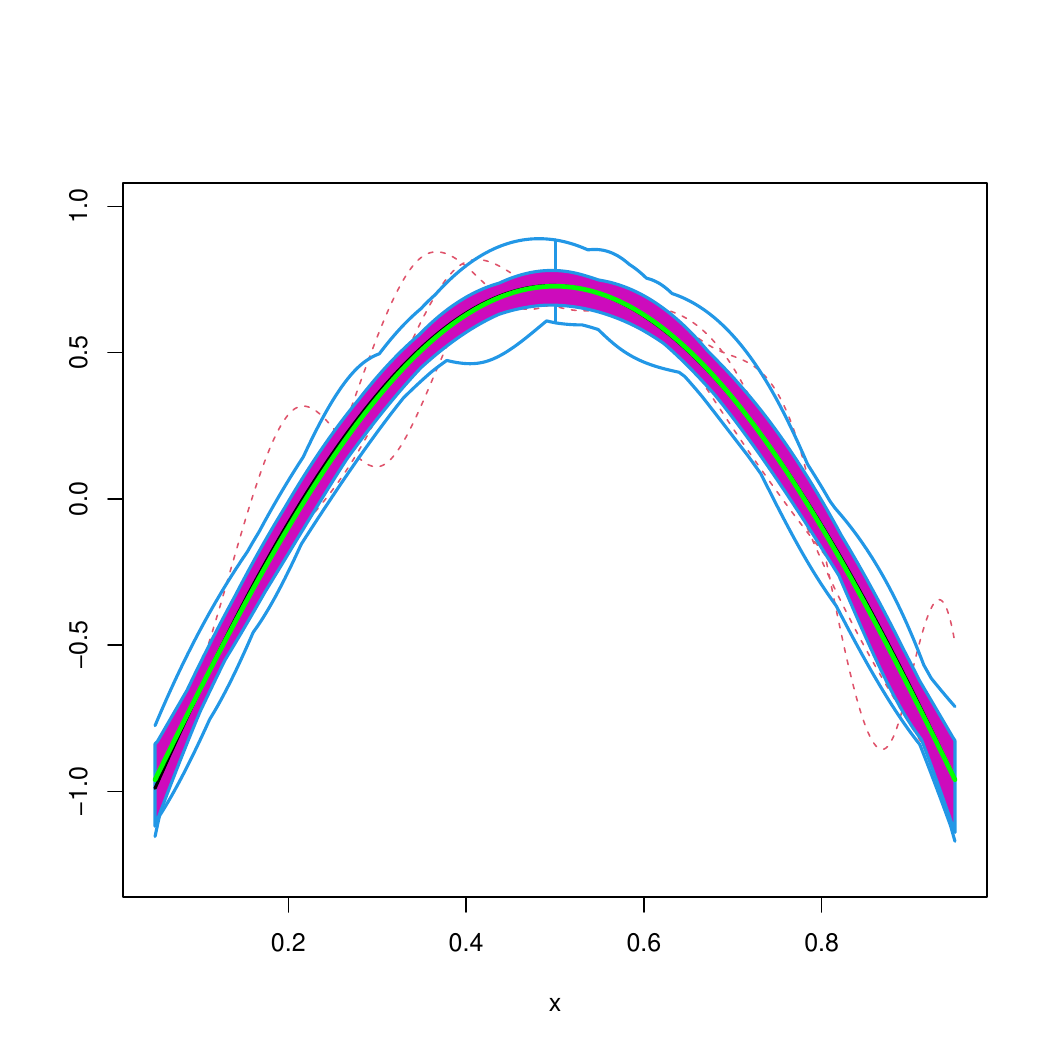}&
 \includegraphics[scale=0.35]{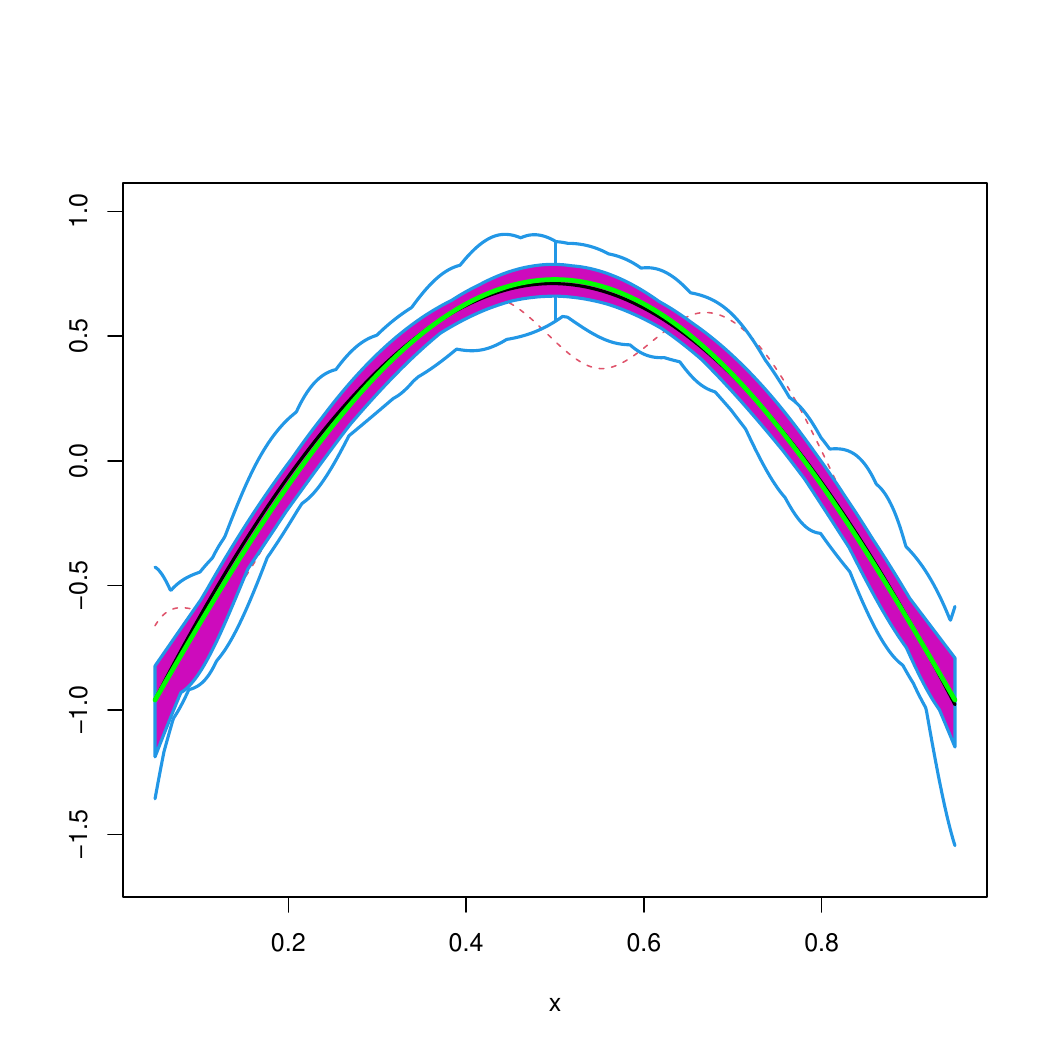}\\[-0.35in]
$C_1$&
\includegraphics[scale=0.35]{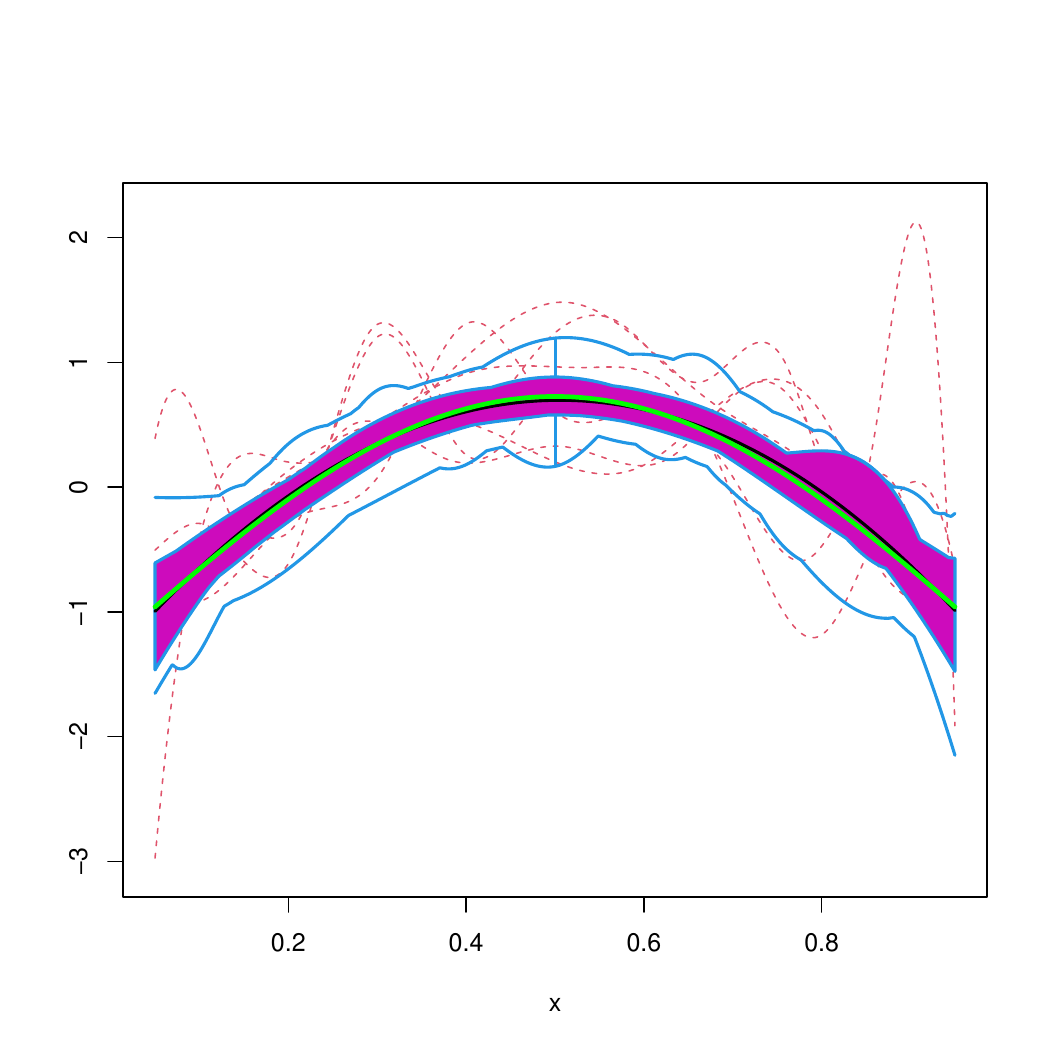} &
 \includegraphics[scale=0.35]{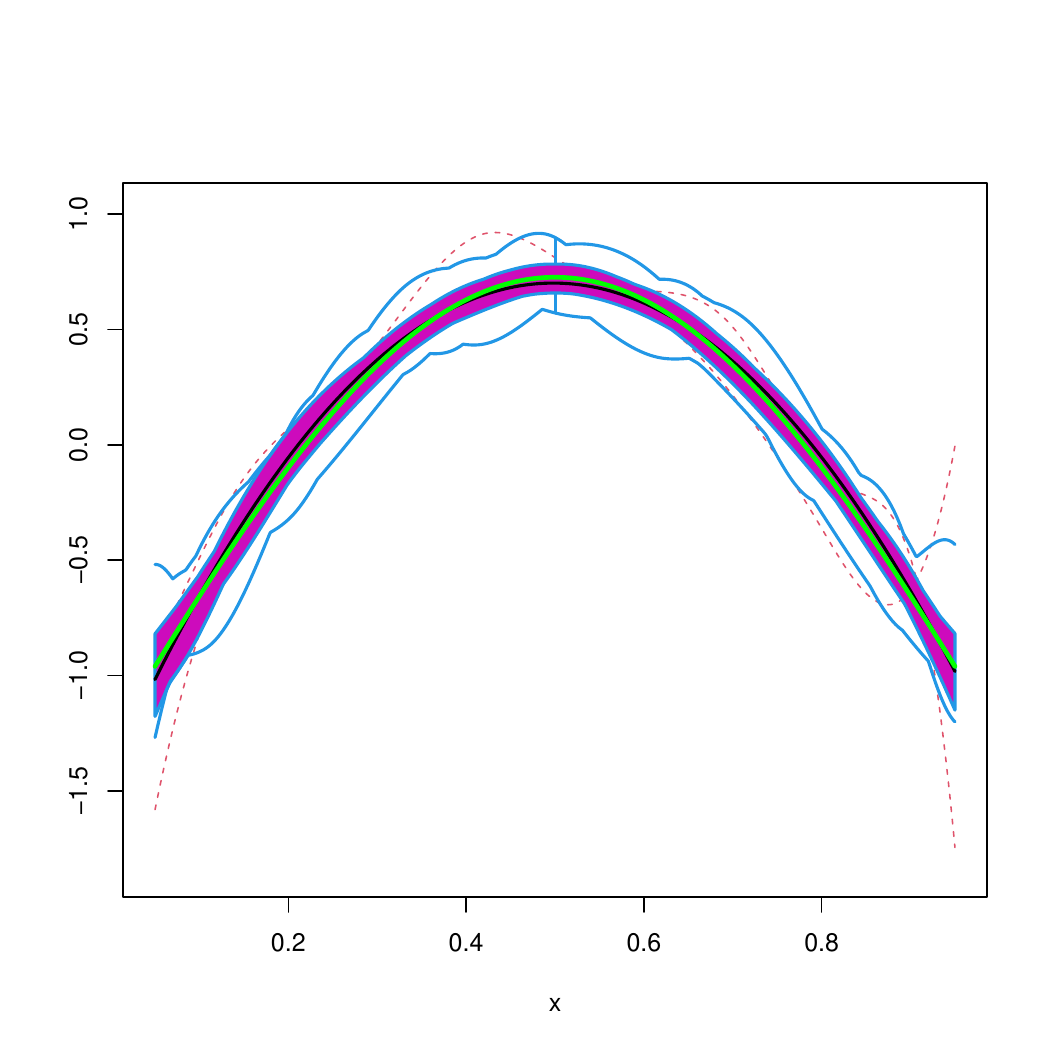}\\[-0.35in]
$C_2$&
\includegraphics[scale=0.35]{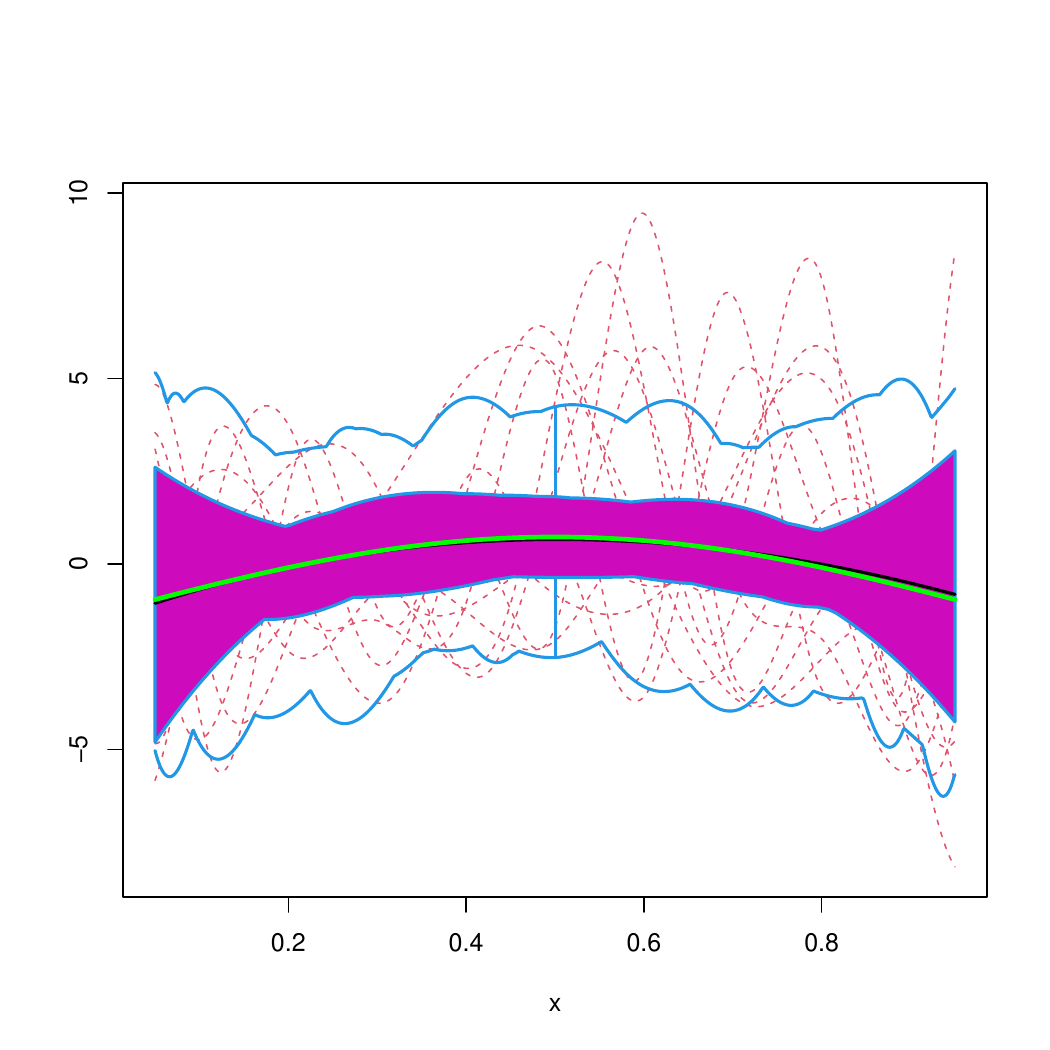} &
 \includegraphics[scale=0.35]{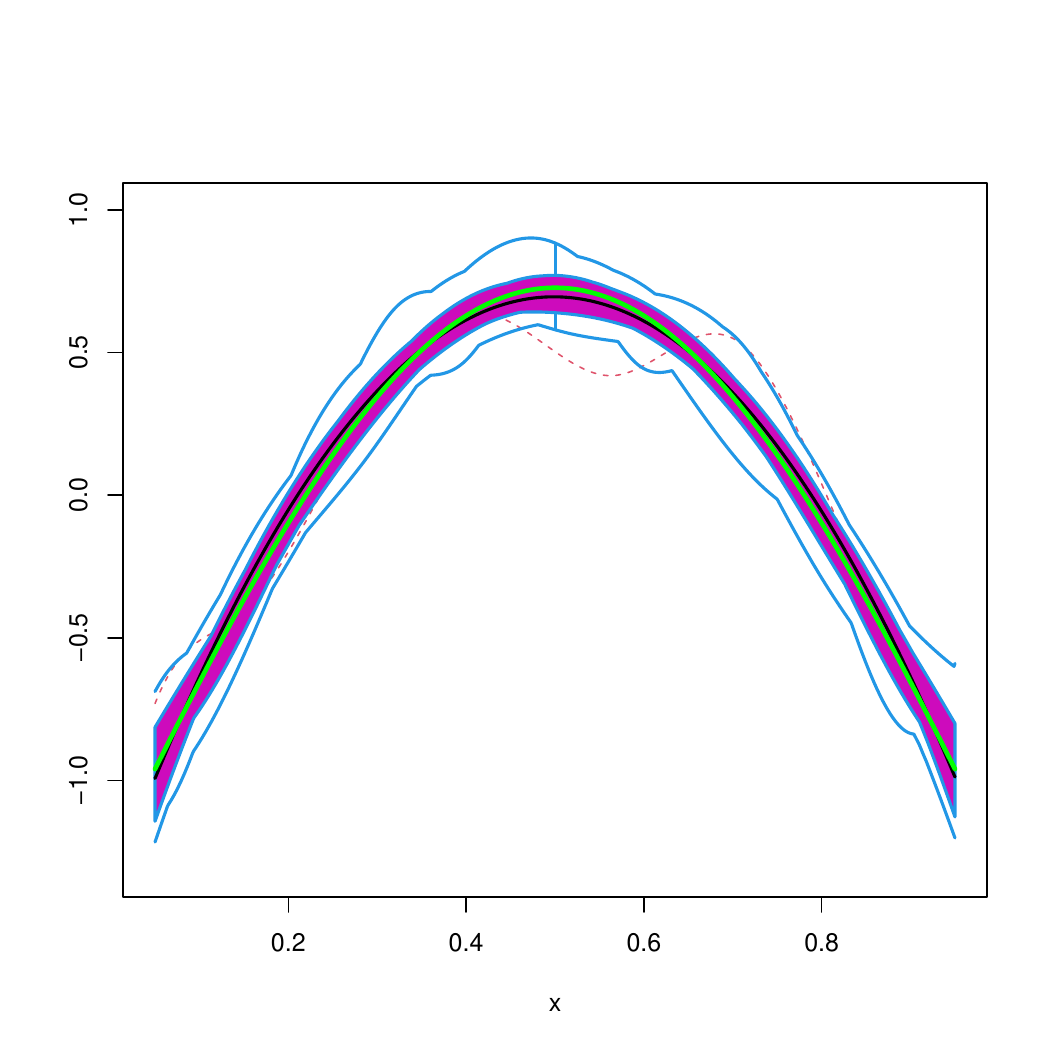}\\[-0.35in]
$C_3$&
\includegraphics[scale=0.35]{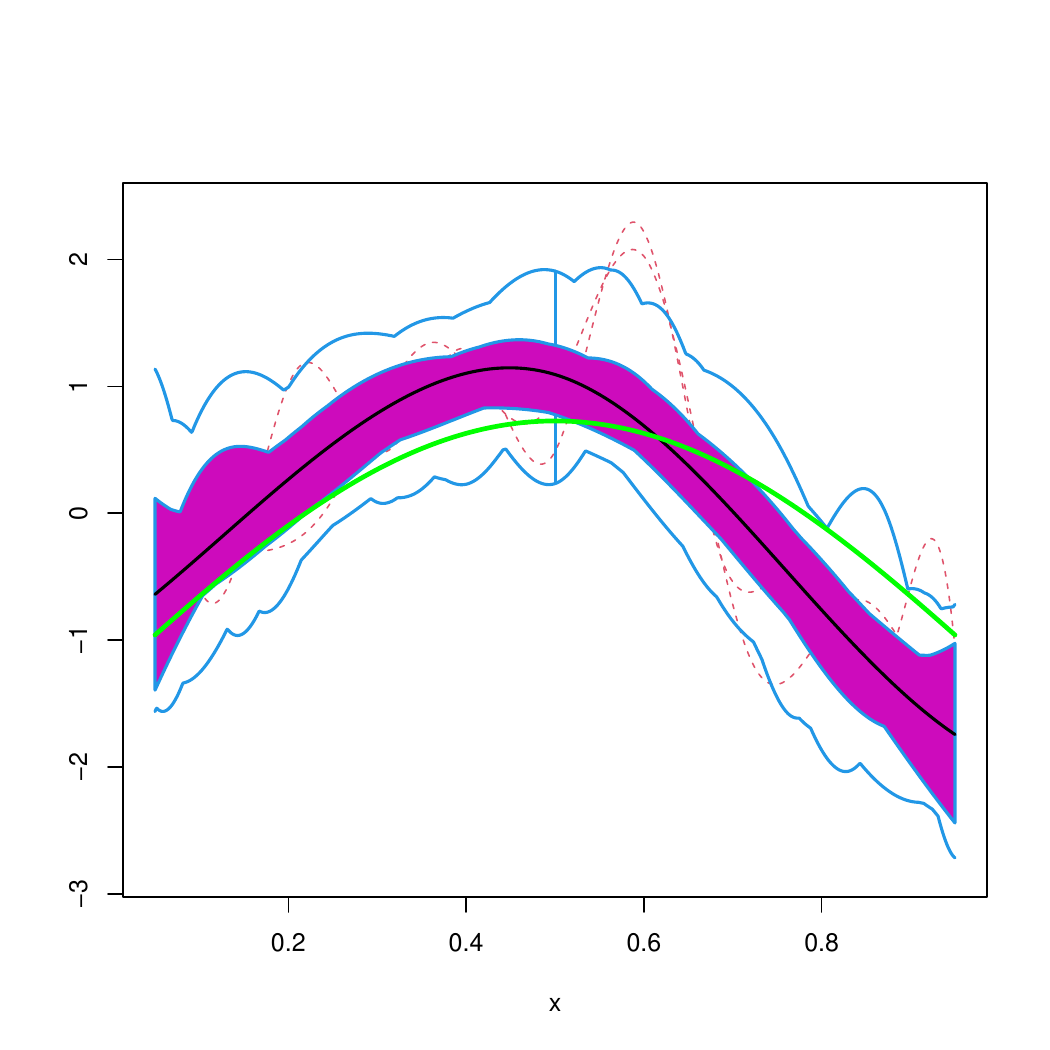} &
 \includegraphics[scale=0.35]{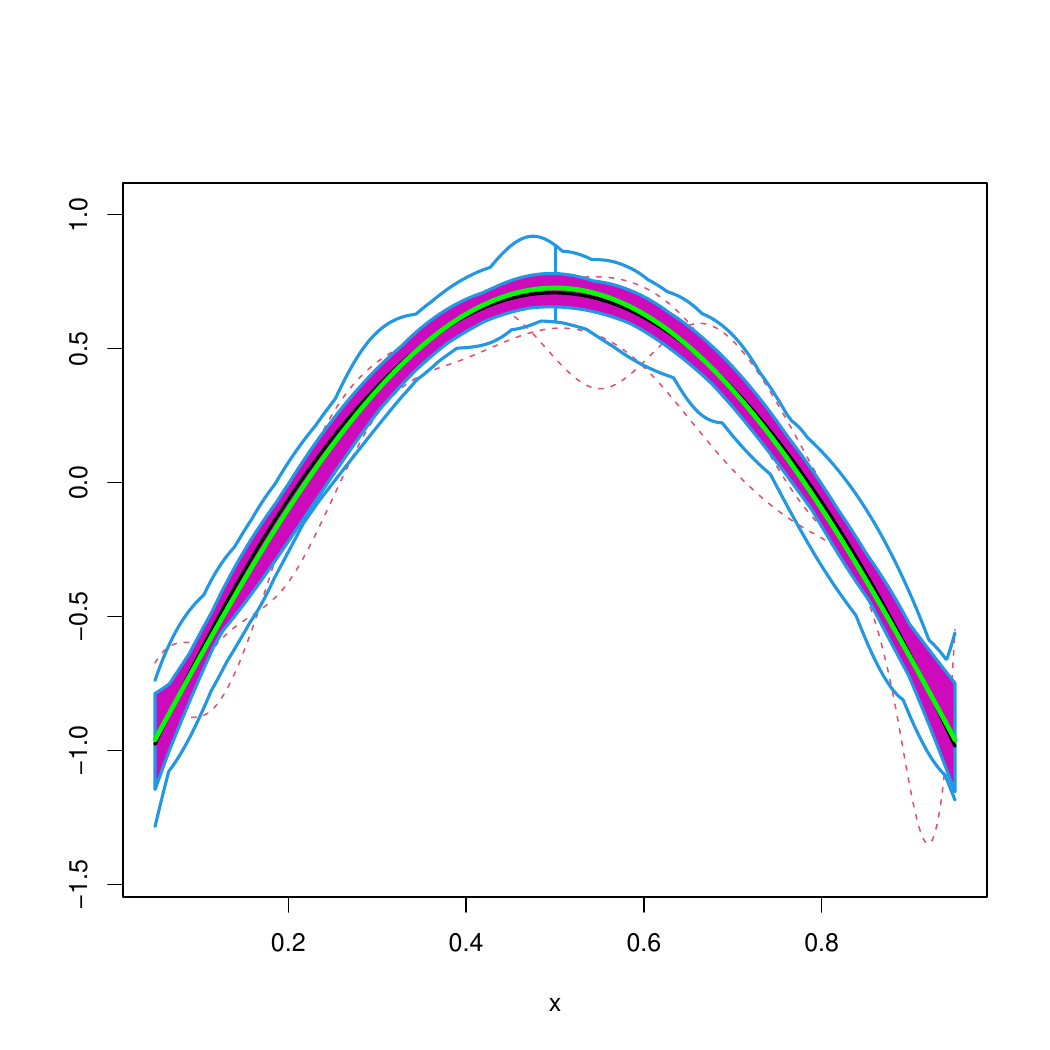}
 \end{tabular}
\caption{\small \label{fig:fboxplot-g1-m6}  Functional boxplots of the estimated additive functions $\eta_1$ obtained by the classical and robust fit under Model 6. The solid green line corresponds to  the true function $\eta_1$.} 

\end{center}
\end{figure}

For the estimators of the regression parameter $\bbe$, we considered as summary measures the bias, standard deviation and the mean square error of each component, denoted \textsc{bias}, \textsc{sd} and \textsc{mse}, respectively. Tables \ref{tab:beta1splines} and \ref{tab:beta2splines} report  the results obtained for the estimators of $\beta_1$ and $\beta_2$, respectively.

\begin{table} [ht!]
\begin{center}
\footnotesize
\setlength{\tabcolsep}{3pt}
 \begin{tabular}{|c|c|c|c||c|c||c|c||c|c||c|c||c|c|}
\cline{3-14}
\multicolumn{1}{c}{} & \multicolumn{1}{c|}{} &\multicolumn{2}{c||}{Model 1} &\multicolumn{2}{c||}{Model 2} &\multicolumn{2}{c||}{Model 3} &\multicolumn{2}{c||}{Model 4} & \multicolumn{2}{c||}{Model 5} & \multicolumn{2}{c|}{Model 6} \\\cline{3-14}
\multicolumn{1}{c}{} & \multicolumn{1}{c|}{} & \textsc{ls} & \textsc{mm} & \textsc{ls}  & \textsc{mm} & \textsc{ls}  & \textsc{mm} & \textsc{ls}  & \textsc{mm} & \textsc{ls}  & \textsc{mm} & \textsc{ls}  & \textsc{mm}\\ \hline
 & \textsc{bias} & 0.000 & -0.002 & 0.000 & 0.002 & -0.018 & -0.016 & 0.000 & -0.001 & -0.007 & -0.002 & 0.002 & 0.001 \\  
 $C_0$  
 & \textsc{sd} & 0.075 & 0.079 & 0.098 & 0.106 & 0.201 & 0.214 & 0.074 & 0.078 & 0.199 & 0.209 & 0.109 & 0.113 \\
 & \textsc{mse} & 0.006 & 0.006 & 0.010 & 0.011 & 0.040 & 0.046 & 0.005 & 0.006 & 0.040 & 0.044 & 0.012 & 0.013 \\ 
  \hline
  
  & \textsc{bias} & -0.010 & -0.004 & -0.006 & 0.003 & -0.010 & -0.020 & -0.011 & -0.004 & -0.060 & -0.003 & -0.013 & 0.001 \\ 
 $C_1$ 
  & \textsc{sd} & 0.245 & 0.086 & 0.334 & 0.113 & 0.679 & 0.226 & 0.251 & 0.087 & 0.623 & 0.226 & 0.386 & 0.123 \\
  & \textsc{mse} & 0.060 & 0.007 & 0.111 & 0.013 & 0.461 & 0.051 & 0.063 & 0.008 & 0.391 & 0.051 & 0.149 & 0.015 \\
 \hline
    & \textsc{bias} & -0.087 & -0.002 & 0.068 & 0.006 & 0.009 & -0.021 & -0.080 & -0.001 & -0.269 & 0.003 & 0.049 & 0.001 \\ 
 $C_2$ 
  & \textsc{sd} & 1.879 & 0.085 & 2.811 & 0.111 & 5.618 & 0.223 & 1.869 & 0.084 & 5.113 & 0.223 & 2.951 & 0.118 \\ 
  & \textsc{mse} & 3.530 & 0.007 & 7.888 & 0.012 & 31.495 & 0.050 & 3.491 & 0.007 & 26.168 & 0.050 & 8.693 & 0.014 \\ 
  \hline
  & \textsc{bias} & -2.989 & -0.002 & -3.963 & 0.004 & -2.975 & -0.010 & -1.629 & -0.002 & -3.287 & -0.005 & -3.884 & 0.001 \\   
$C_3$ 
  & \textsc{sd} & 0.231 & 0.082 & 0.272 & 0.111 & 0.317 & 0.224 & 0.287 & 0.080 & 0.184 & 0.218 & 0.270 & 0.119 \\ 
  & \textsc{mse} & 8.986 & 0.007 & 15.776 & 0.012 & 8.948 & 0.050 & 2.737 & 0.006 & 10.839 & 0.048 & 15.155 & 0.014 \\ 
  \hline
\end{tabular}
\caption{\label{tab:beta1splines}\footnotesize Bias, standard deviations and mean square errors for the estimates  of the first coordinate $\beta_1$ of $\bbe$. The classical and robust procedures are labelled \textsc{ls} and \textsc{mm}, respectively.}
\end{center}
\end{table}

 \begin{table}[ht!]
\begin{center}
\footnotesize
 \setlength{\tabcolsep}{3pt}
 \begin{tabular}{|c|c|c|c||c|c||c|c||c|c||c|c||c|c|}
\cline{3-14}
\multicolumn{1}{c}{} & \multicolumn{1}{c|}{} &\multicolumn{2}{c||}{Model 1} &\multicolumn{2}{c||}{Model 2} &\multicolumn{2}{c||}{Model 3} &\multicolumn{2}{c||}{Model 4} & \multicolumn{2}{c||}{Model 5} & \multicolumn{2}{c|}{Model 6} \\\cline{3-14}
\multicolumn{1}{c}{} & \multicolumn{1}{c|}{} & \textsc{ls} & \textsc{mm} & \textsc{ls}  & \textsc{mm} & \textsc{ls}  & \textsc{mm} & \textsc{ls}  & \textsc{mm} & \textsc{ls}  & \textsc{mm} & \textsc{ls}  & \textsc{mm} \\ \hline
& \textsc{bias} & 0.002 & 0.001 & 0.002 & 0.001 & 0.002 & 0.003 & 0.002 & 0.001 & -0.006 & -0.002 & 0.011 & 0.007 \\ 
  $C_0$
& \textsc{sd} &  0.076 & 0.081 & 0.075 & 0.078 & 0.210 & 0.215 & 0.123 & 0.128 & 0.172 & 0.179 & 0.087 & 0.091 \\ 
& \textsc{mse} & 0.006 & 0.007 & 0.006 & 0.006 & 0.044 & 0.046 & 0.015 & 0.016 & 0.030 & 0.032 & 0.008 & 0.008 \\ 
  \hline 
  & \textsc{bias} & -0.017 & -0.001 & -0.002 & 0.002 & 0.004 & -0.002 & -0.039 & -0.005 & -0.029 & -0.004 & 0.024 & 0.012 \\ 
  $C_1$
  & \textsc{sd} &  0.239 & 0.088 & 0.233 & 0.086 & 0.709 & 0.236 & 0.391 & 0.143 & 0.560 & 0.196 & 0.267 & 0.101 \\ 
  &\textsc{mse} & 0.057 & 0.008 & 0.054 & 0.007 & 0.502 & 0.056 & 0.154 & 0.020 & 0.314 & 0.038 & 0.072 & 0.010 \\ 
 \hline
  & \textsc{bias} & -0.051 & -0.001 & -0.035 & 0.001 & -0.218 & 0.003 & -0.166 & -0.002 & -0.274 & 0.004 & -0.006 & 0.012 \\ 
  $C_2$  
  & \textsc{sd} & 2.054 & 0.085 & 2.185 & 0.082 & 5.959 & 0.227 & 3.174 & 0.137 & 4.536 & 0.195 & 2.238 & 0.097 \\ 
  & \textsc{mse} & 4.214 & 0.007 & 4.766 & 0.007 & 35.491 & 0.052 & 10.080 & 0.019 & 20.605 & 0.038 & 5.001 & 0.010 \\ 
\hline
  & \textsc{bias} & -3.005 & 0.001 & -2.033 & 0.003 & -3.024 & 0.002 & -4.346 & 0.001 & -2.709 & 0.001 & -2.097 & 0.010 \\ 
  $C_3$  
  & \textsc{sd} & 0.231 & 0.081 & 0.272 & 0.081 & 0.309 & 0.225 & 0.283 & 0.131 & 0.187 & 0.184 & 0.277 & 0.094 \\ 
  & \textsc{mse} & 9.082 & 0.006 & 4.208 & 0.007 & 9.241 & 0.050 & 18.970 & 0.017 & 7.371 & 0.034 & 4.472 & 0.009 \\ 
   \hline
\end{tabular}
\caption{\label{tab:beta2splines}\footnotesize Bias, standard deviations and mean square errors for the estimates  of the second coordinate $\beta_2$ of $\bbe$. The classical and robust procedures are labelled \textsc{ls} and \textsc{mm}, respectively.}
\end{center}
\end{table}

 As expected, for clean data sets, the classical and robust estimators of the regression parameter behave similarly (see Tables \ref{tab:beta1splines} and \ref{tab:beta2splines}). When estimating the regression coefficient $\bbe$, the
less efficient robust estimator naturally results in higher standard deviations, even though the mean square error is equal to that of the classical counterpart due to a bias reduction. 

\begin{figure}[ht!]
\begin{center}
\renewcommand{\arraystretch}{1.6}
\begin{tabular}{cc}
\textsc{ls}&  \textsc{mm}\\ 
\includegraphics[scale=0.5]{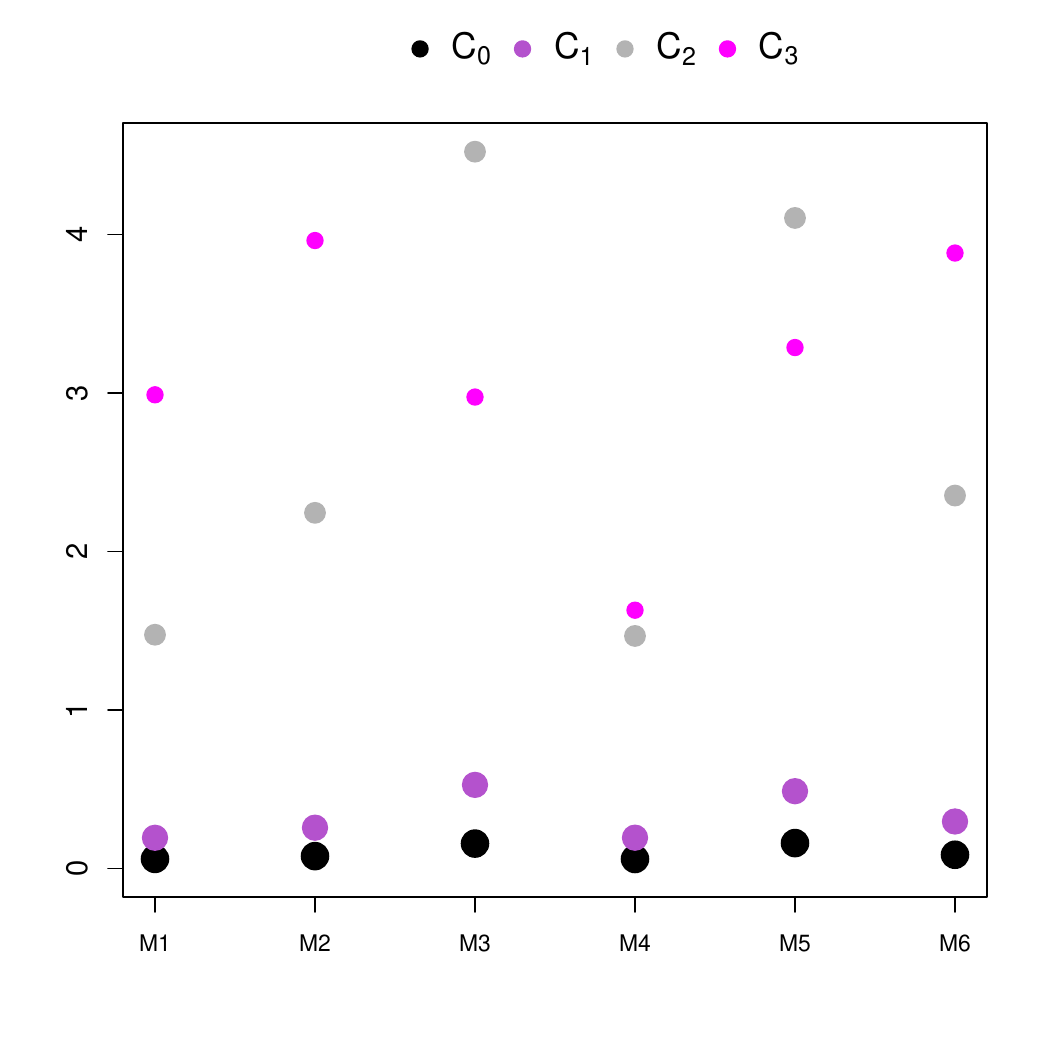} &
 \includegraphics[scale=0.5]{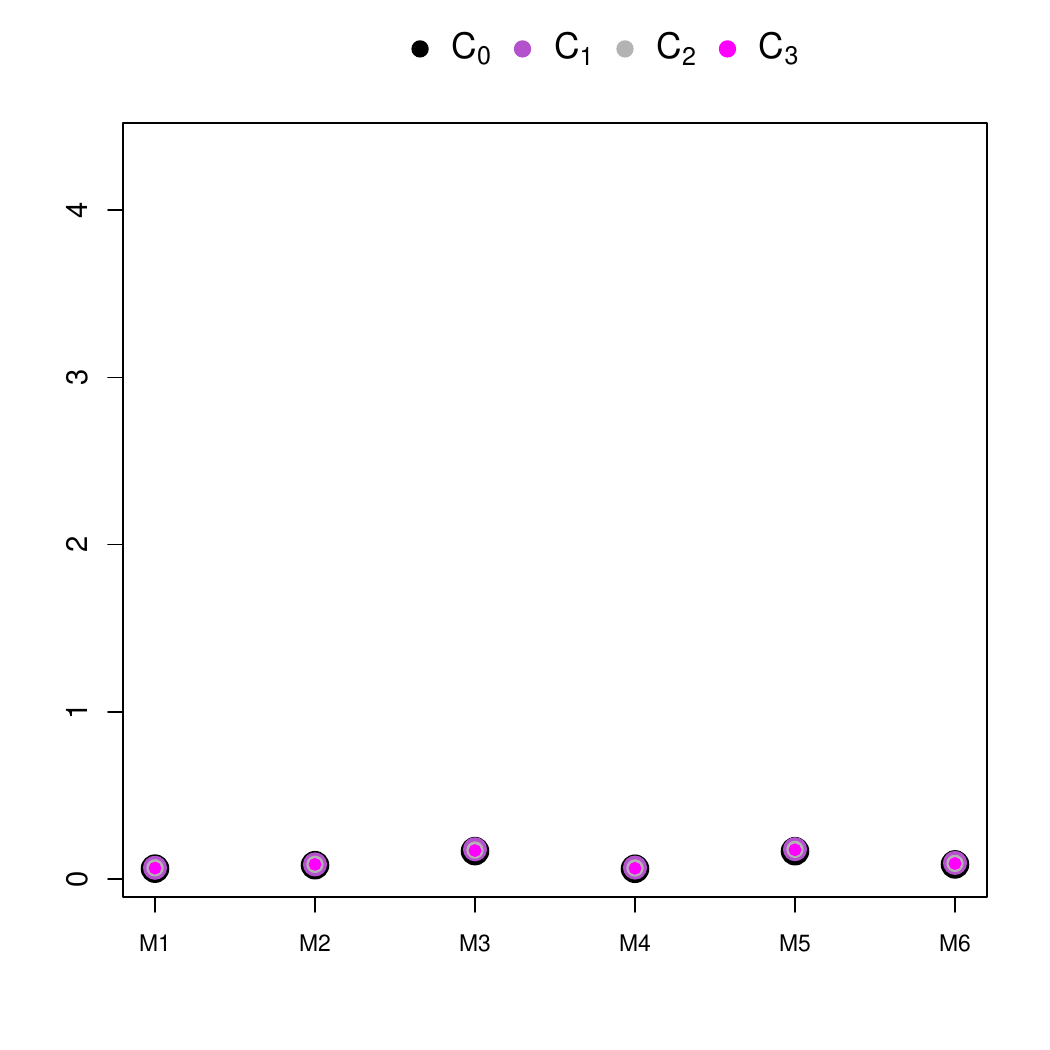}
\end{tabular}
\vskip-0.2in
\caption{\small \label{fig:Bias-Comp-b1} Plots of the  mean over replications of $|\wbeta_1-\beta_1|$ for each contamination setting (identified through colours  in the legend) and for the different models (denoted   \textsc{M1} to \textsc{M6} in the horizontal axis). The left panel corresponds to the classical procedure, while the right one to the  robust method.}

\end{center}
\end{figure}

\begin{figure}[ht!]
\begin{center}
\begin{tabular}{cc}
\textsc{ls}&  \textsc{mm}\\ 
\includegraphics[scale=0.5]{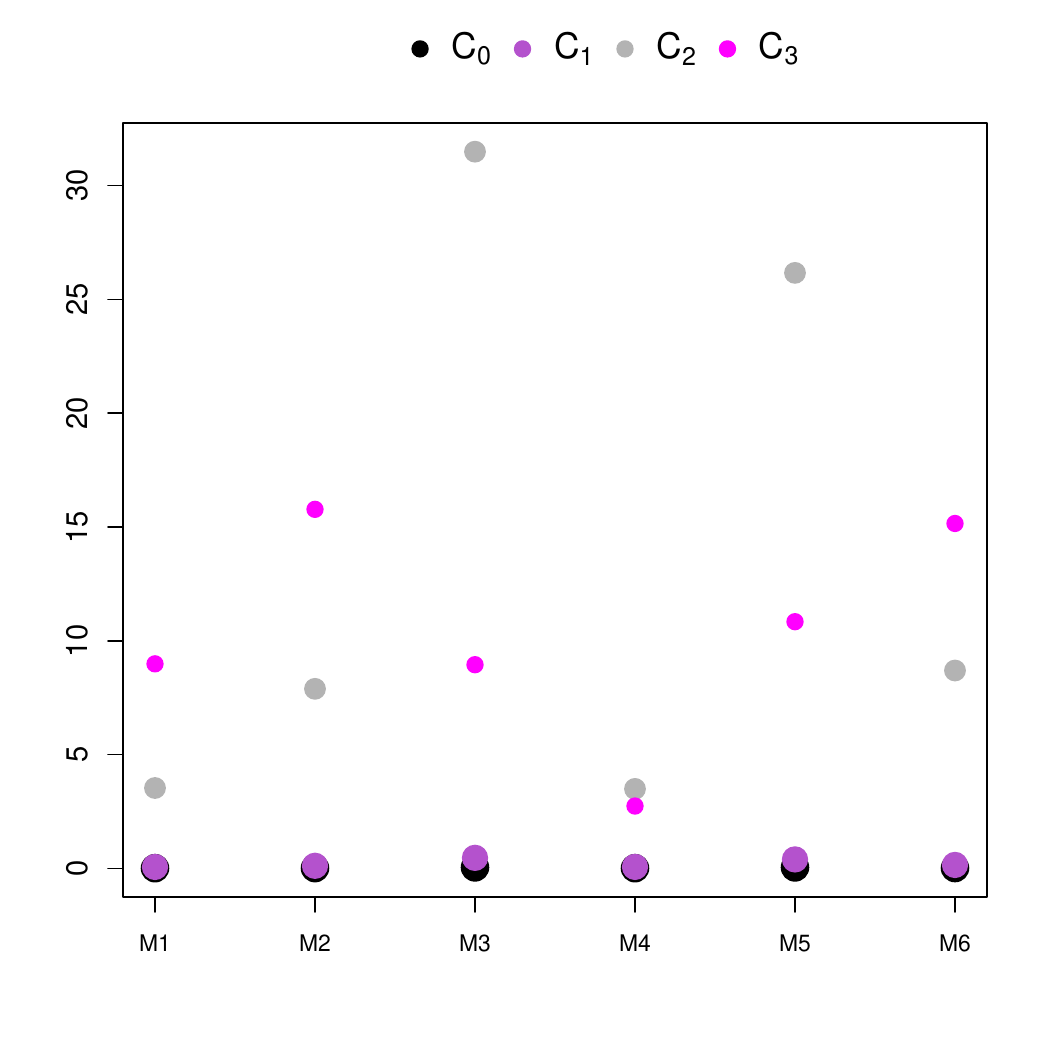} &
 \includegraphics[scale=0.5]{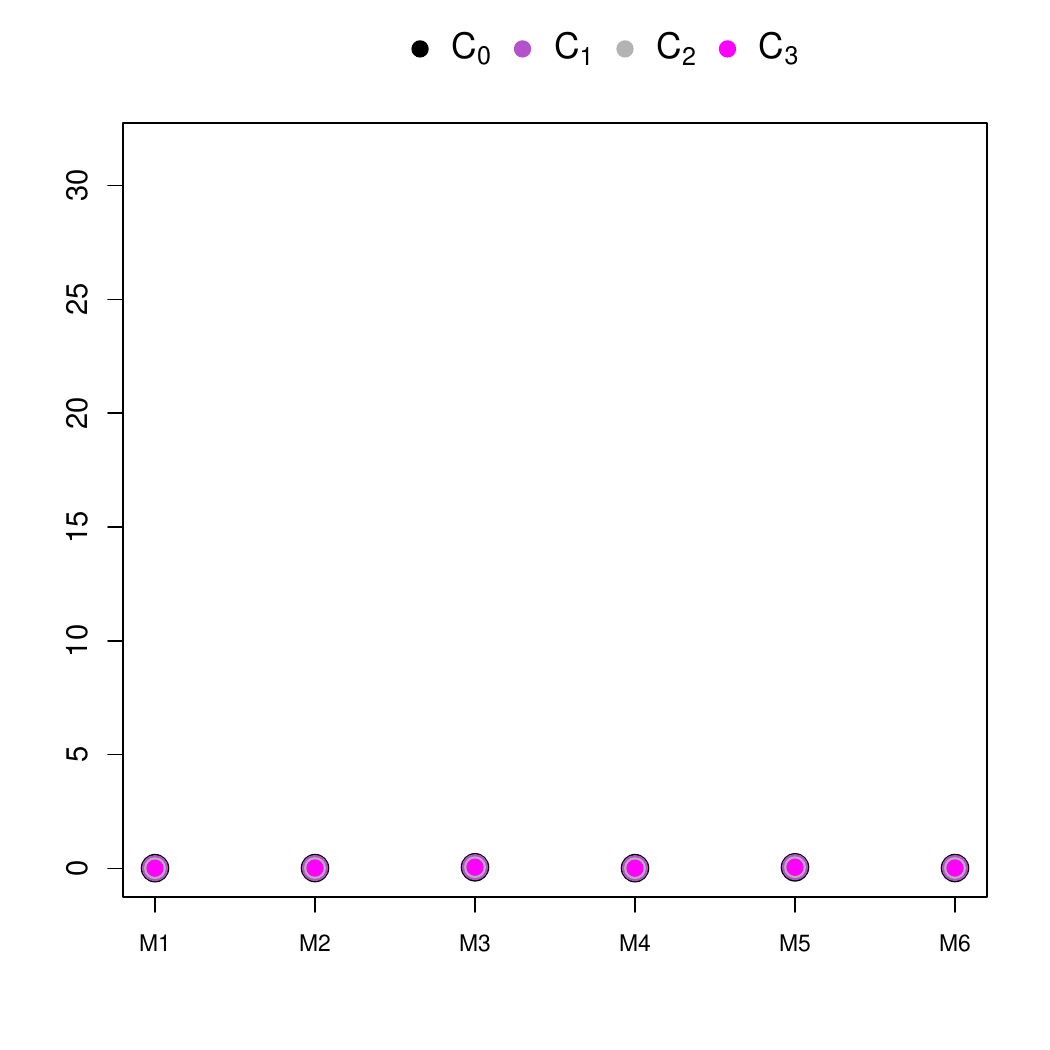}
\end{tabular}
\vskip-0.2in

\caption{\small \label{fig:MSE-Comp-b1} Plots of the \textsc{mse} error obtained when estimating $\beta_1$ for each contamination setting (identified through colours in the legend) and for the different models (denoted as \textsc{M1} to  \textsc{M6} in the horizontal axis). The left panel corresponds to the classical procedure, while the right one to the  robust method.}

\end{center}
\end{figure}

Regarding the performance of the estimates of $\bbe$ under  contamination, scenario  $C_{1}$ mainly affects the variability of the classical regression estimator (see Tables \ref{tab:beta1splines} and  \ref{tab:beta2splines}). This performance is related to the fact that the errors are still centered for this scheme  but with  a large dispersion.  
Finally, the high-leverage outliers introduced under $C_{3}$  have a damaging effect on the classical estimators of $\bbe$ that become completely uninformative, see Tables \ref{tab:beta1splines} and \ref{tab:beta2splines} and Figures \ref{fig:Bias-Comp-b1} and \ref{fig:MSE-Comp-b1}, since the absolute bias is enlarged more than 800 times affecting the mean square error. 
In contrast, the proposed robust estimators show a very stable  behaviour across all contamination and model settings. The  bias and mean square error of the estimators of $\bbe$, reported in  Tables \ref{tab:beta1splines} and \ref{tab:beta2splines}, show that the robust procedure is highly resistant against the contamination scenarios considered here, even when bad high leverage points are introduced in the sample. 

\begin{figure}[ht!]
\begin{center}
\renewcommand{\arraystretch}{1.6}
\begin{tabular}{cc}
$C_0$ &  $C_1$\\[-0.3in]
\includegraphics[scale=0.5]{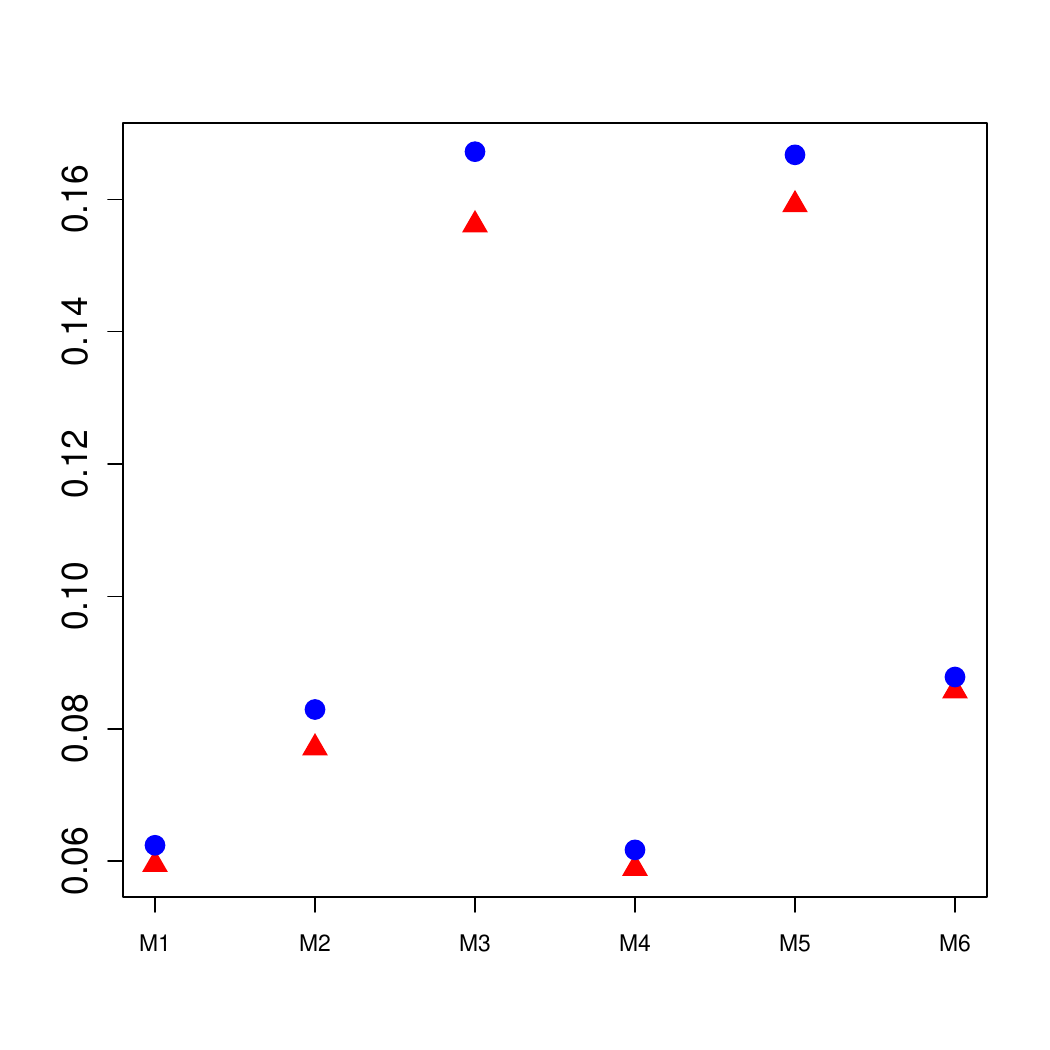} &
 \includegraphics[scale=0.5]{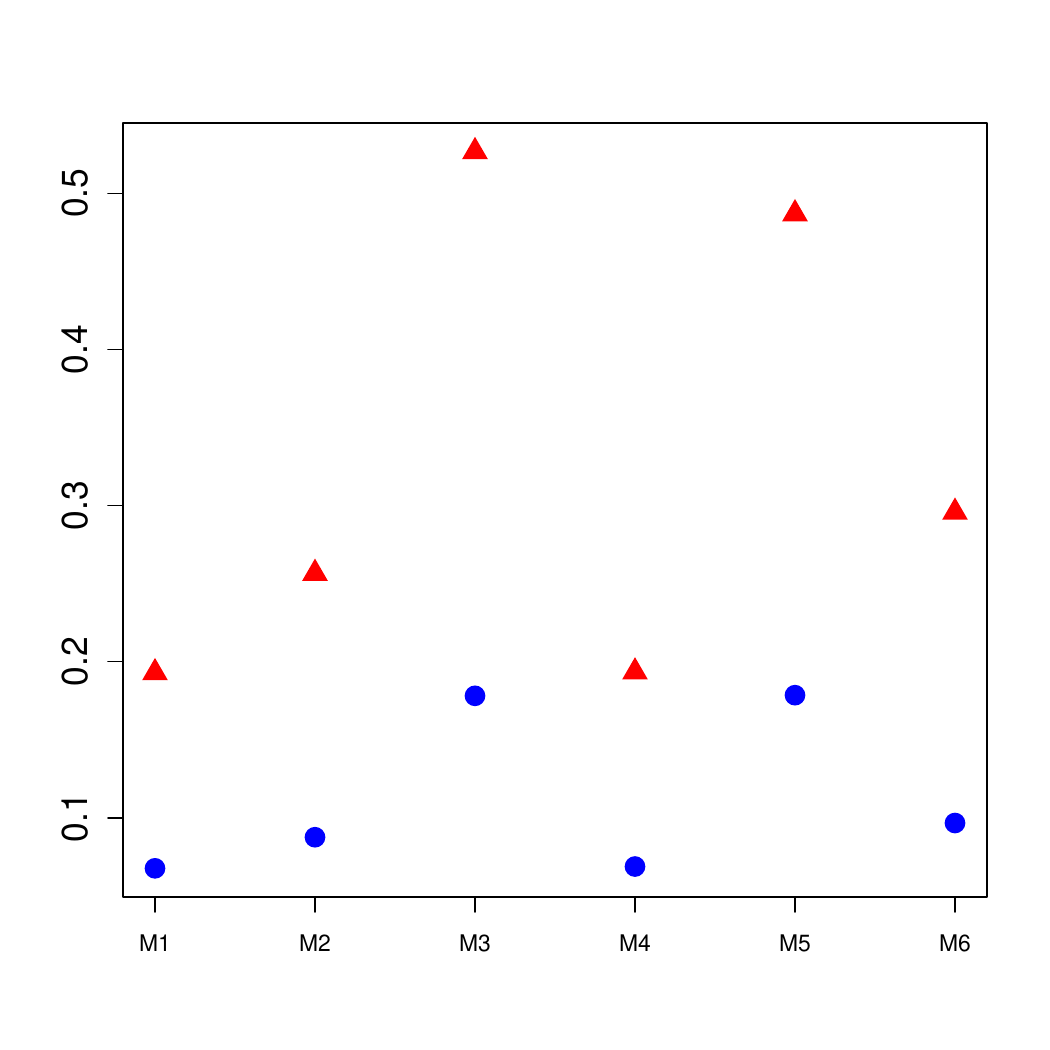}\\
 $C_2$ & $C_3$\\[-0.3in]
\includegraphics[scale=0.5]{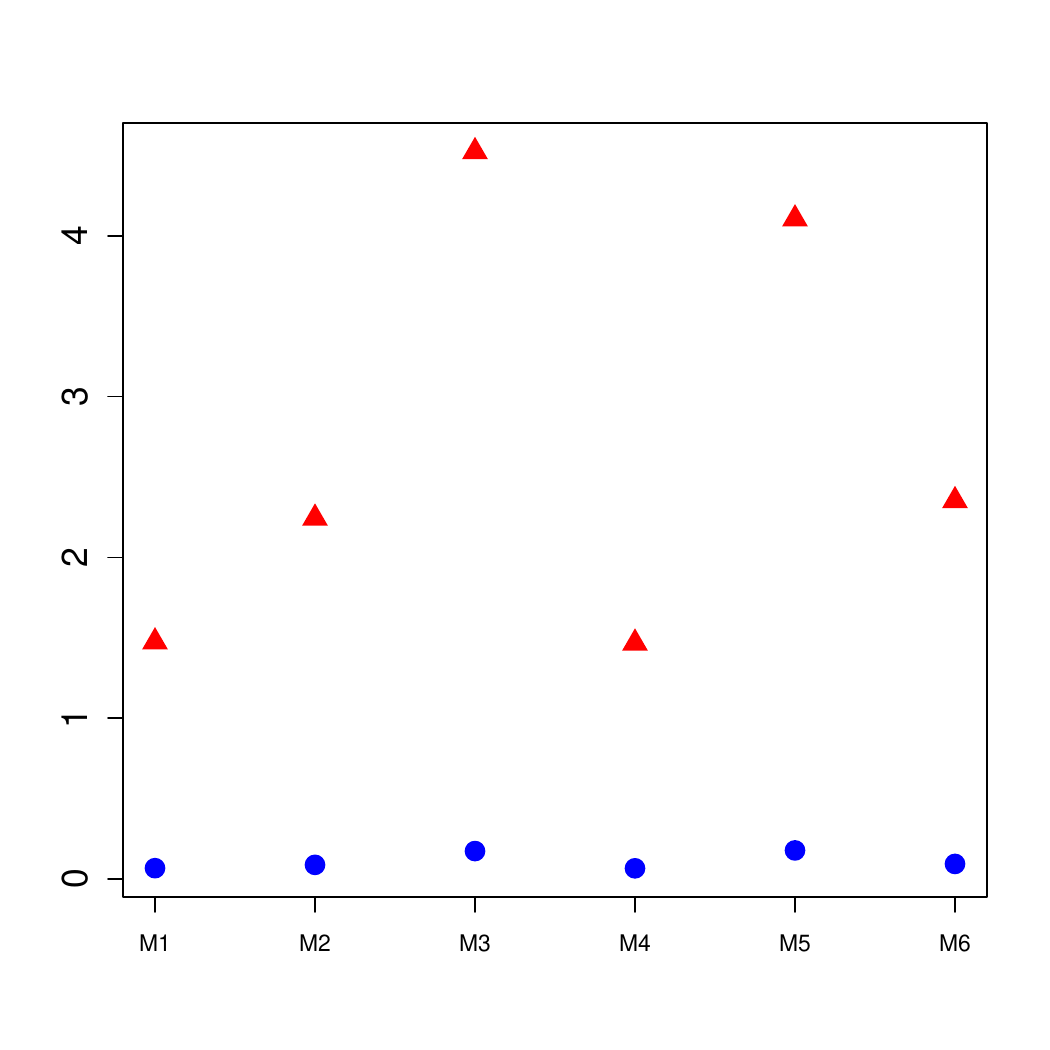} &
\includegraphics[scale=0.5]{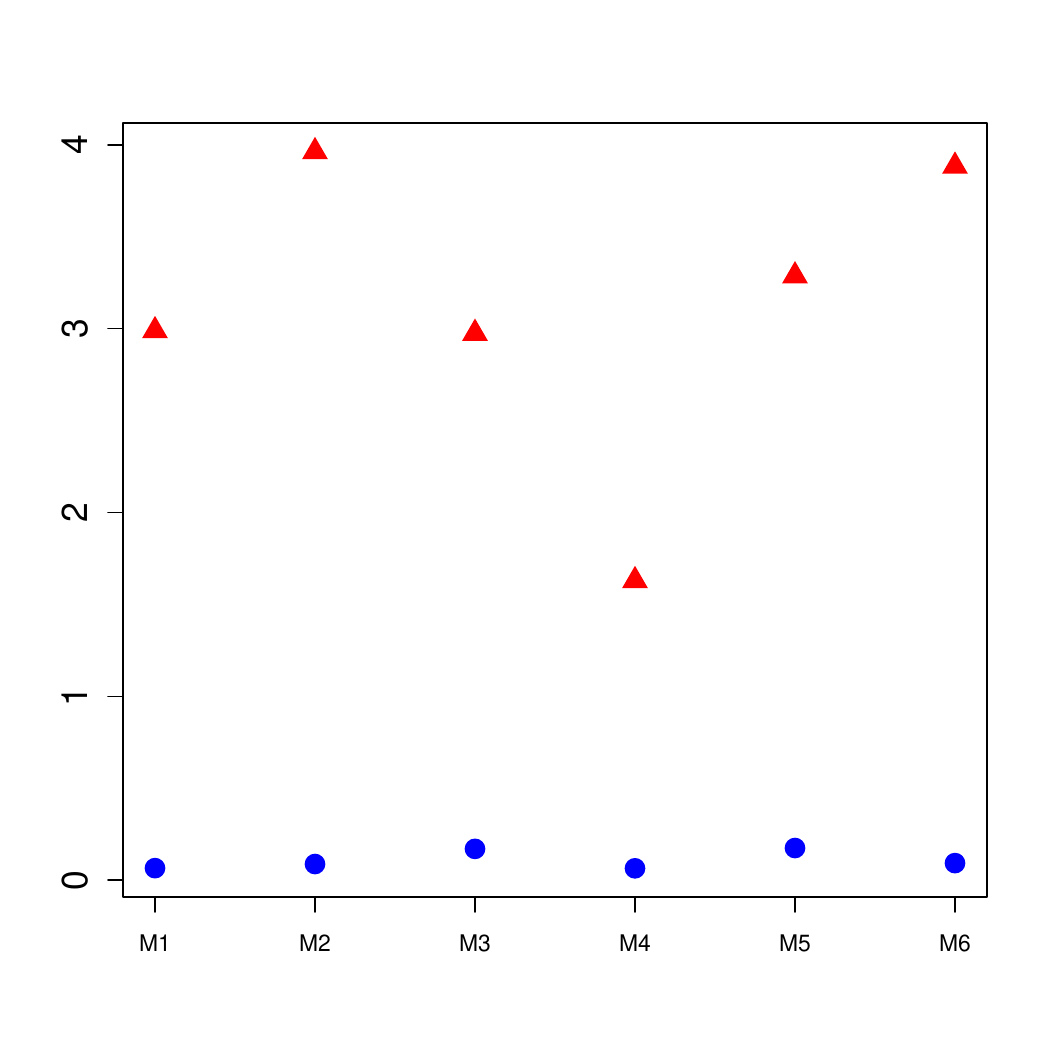}
\end{tabular}
\caption{\small \label{fig:Bias-b1}  Plots of the mean over replications of $|\wbeta_1-\beta_1|$,   for each contamination setting. The red filled triangles correspond to the classical estimator, while the blue circles to the robust ones.}

\end{center}
\end{figure}

\begin{figure}[ht!]
\begin{center}
\renewcommand{\arraystretch}{1.6}
\begin{tabular}{cc}
$C_0$ &  $C_1$\\[-0.3in]
\includegraphics[scale=0.5]{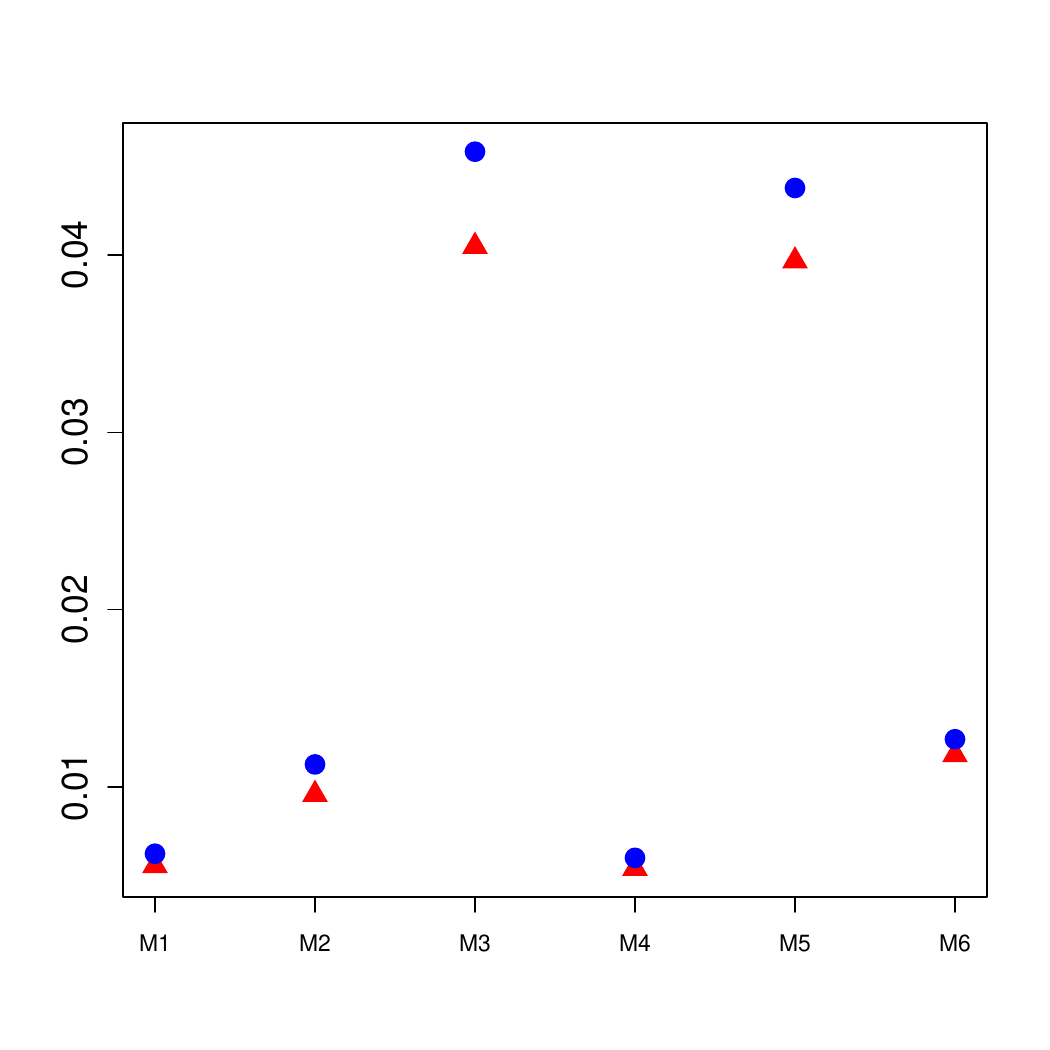}&
 \includegraphics[scale=0.5]{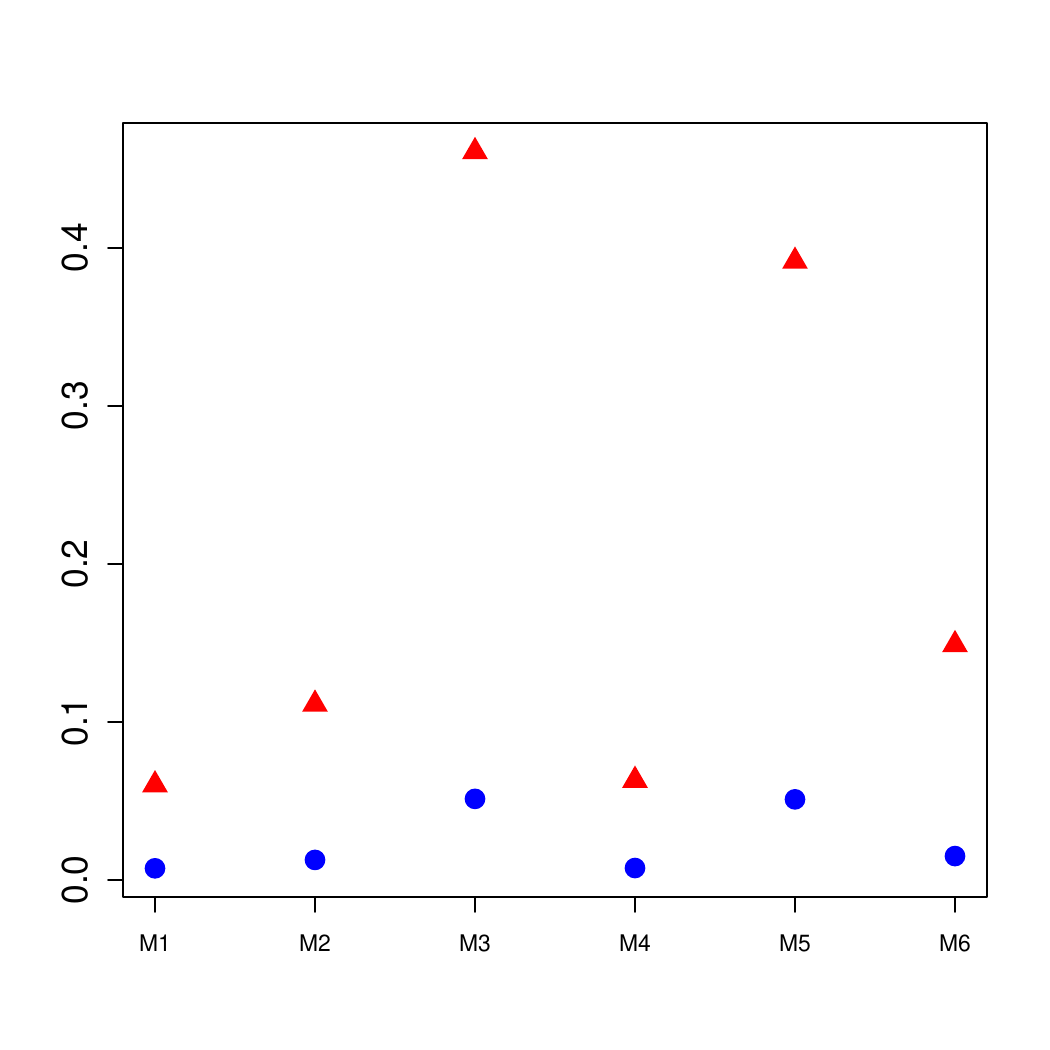}\\ 
$C_2$ &  $C_3$\\[-0.3in]
\includegraphics[scale=0.5]{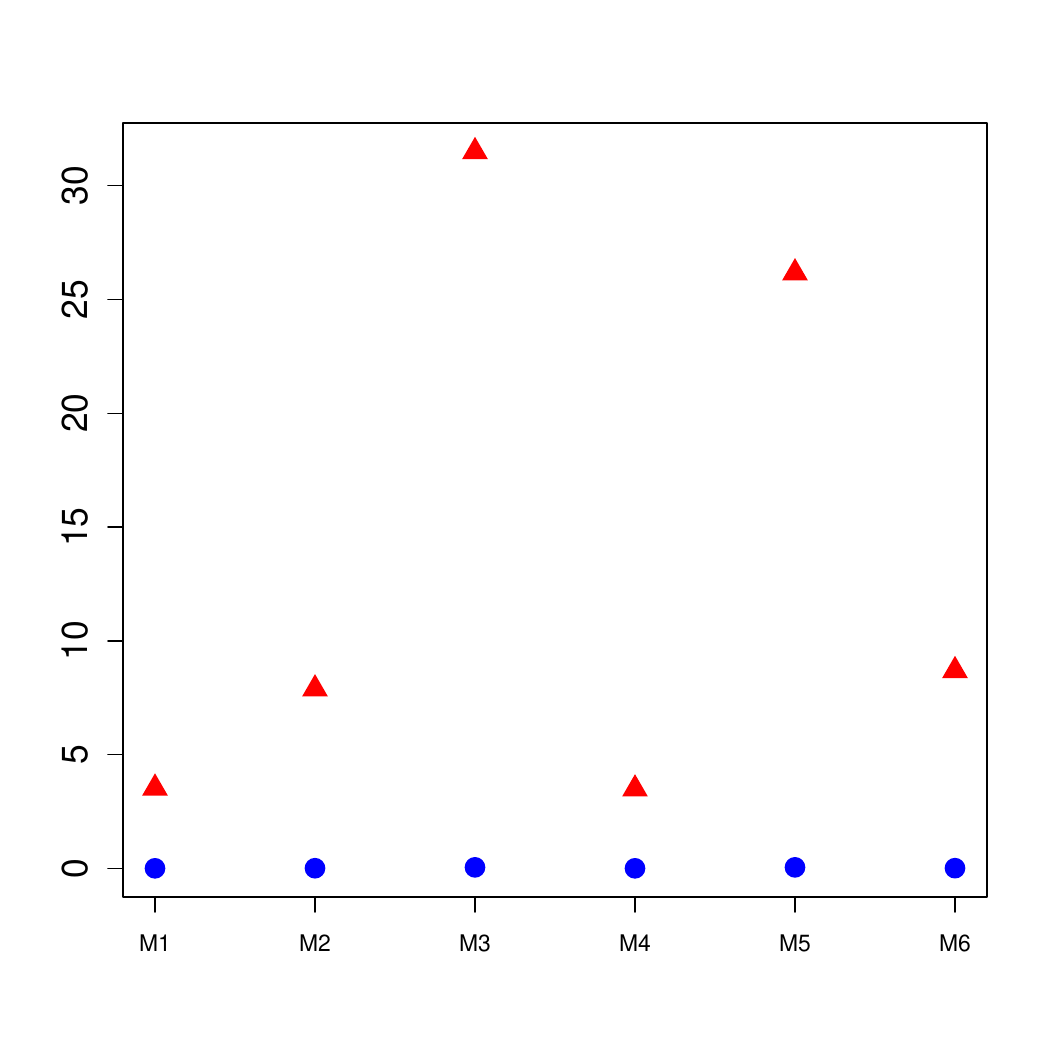} &
\includegraphics[scale=0.5]{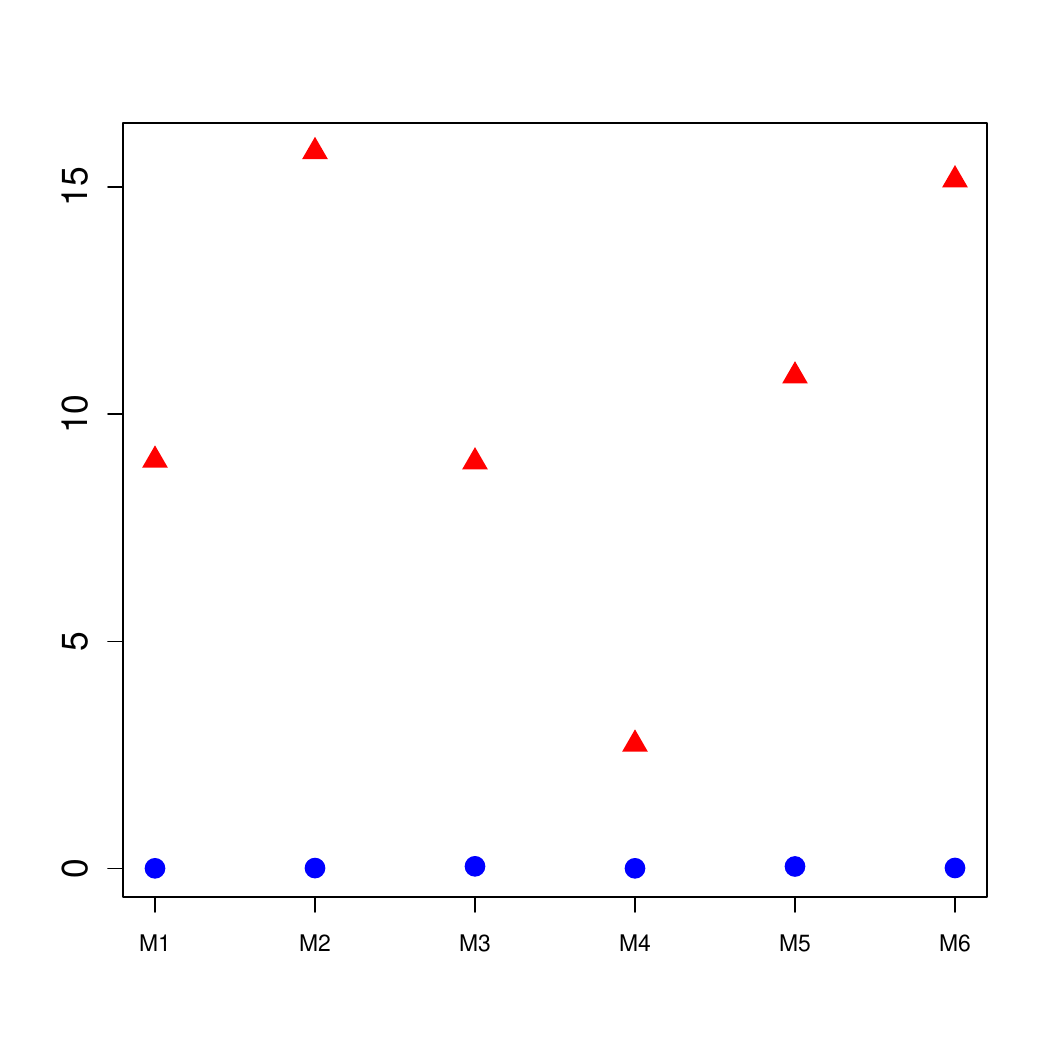}
\end{tabular}

\caption{\small \label{fig:MSE-b1} Plots of the \textsc{mse} error obtained when estimating $\beta_1$ for each contamination setting. The red filled triangles correspond to the classical estimator, while the blue circles to the robust ones.}

\end{center}
\end{figure}

With the goal of analysing the outliers impact on the bias of the estimators of $\beta_1$,  Figure \ref{fig:Bias-b1} shows the mean over replications of  the absolute bias $|\wbeta_1-\beta_1|$, under  the six models considered. Again,  red filled triangles and  blue filled circles correspond to the classical and robust estimators, respectively. The harmful effect of contamination schemes $C_2$ and $C_3$  on the classical procedure may be appreciate on the plots. To facilitate the comparison across contaminations,  Figure \ref{fig:Bias-Comp-b1} reports the mean over   replications of the absolute bias $|\wbeta_1-\beta_1|$ but with all contamination settings in the same plot. The different sizes of the filled circles allow to appreciate that for the robust estimators the results are almost the same. As shown in Figures \ref{fig:Bias-Comp-b1} and  \ref{fig:Bias-b1}, the impact of outliers on the bias of the classical method is particularly high under scheme $C_3$ which corresponds to high leverage points. In contrast, the robust procedure is stable providing reliable estimations. This effect is also  visualized in Figure \ref{fig:MSE-b1} that reports  the \textsc{mse} values obtained when estimating $\beta_1$. As above, the red filled triangles correspond to the classical estimators, while the blue filled circles to the robust ones. Regarding the \textsc{mse}, as expected, for clean samples  the classical and robust approaches lead to similar results, although they are slightly higher for the robust fit due to the lack of efficiency. On the other hand, for all models and contamination settings, the classical procedure leads to values of the \textsc{mse} that are much larger than those obtained by the robust one. The dramatic effect of contaminations on the classical estimators is more striking  in the left panel of Figure \ref{fig:MSE-Comp-b1} which present with different colours the contamination schemes and the six models in the horizontal axis. This Figure allows to  appreciate that, while for the robust approach the obtained values are  close to each other, for the classical one the \textsc{mse} values are much higher than those obtained for clean samples, specially when considering the contamination schemes $C_2$ and $C_3$.

\begin{figure}[ht!]
\begin{center}
\renewcommand{\arraystretch}{1.6}
\begin{tabular}{cc}
$C_0$ &  $C_1$\\[-0.3in]
\includegraphics[scale=0.5]{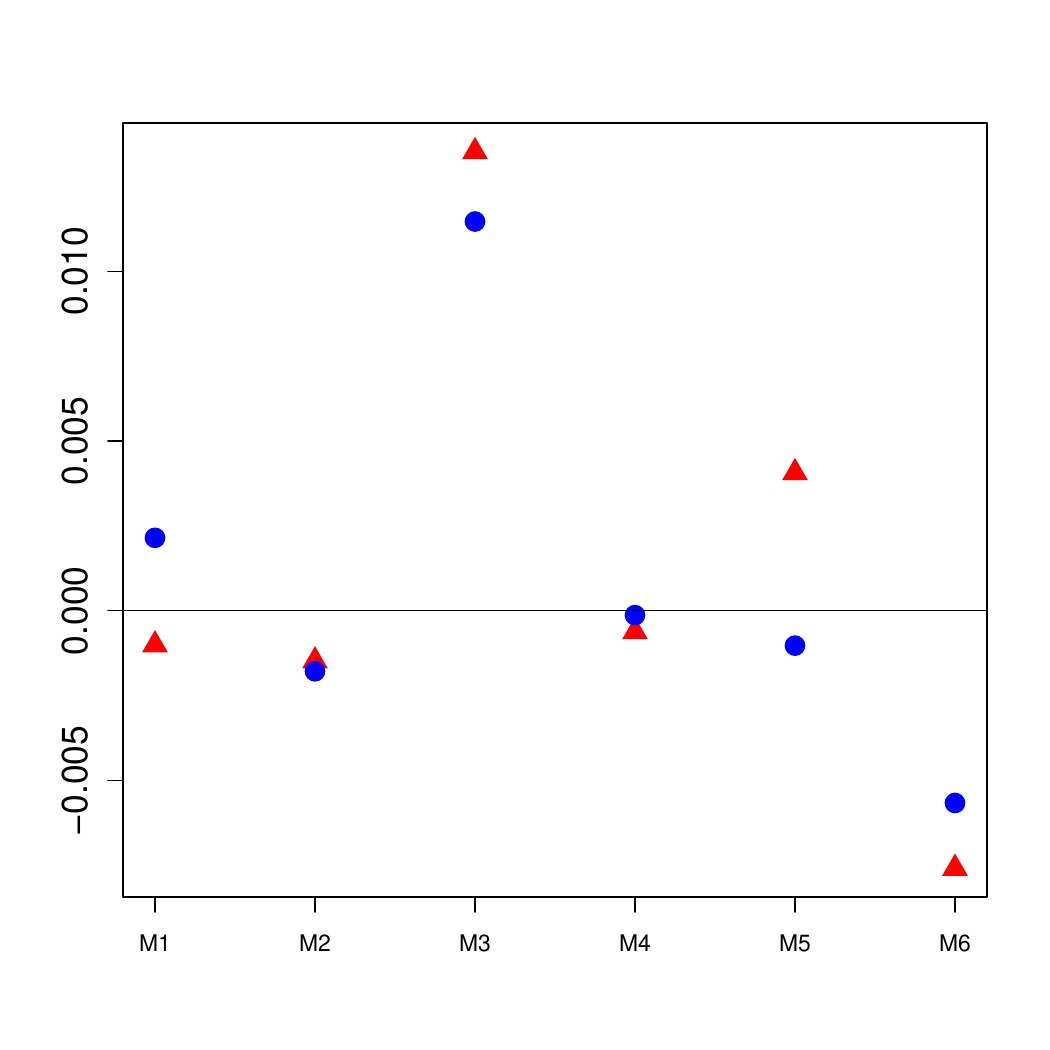} &
 \includegraphics[scale=0.5]{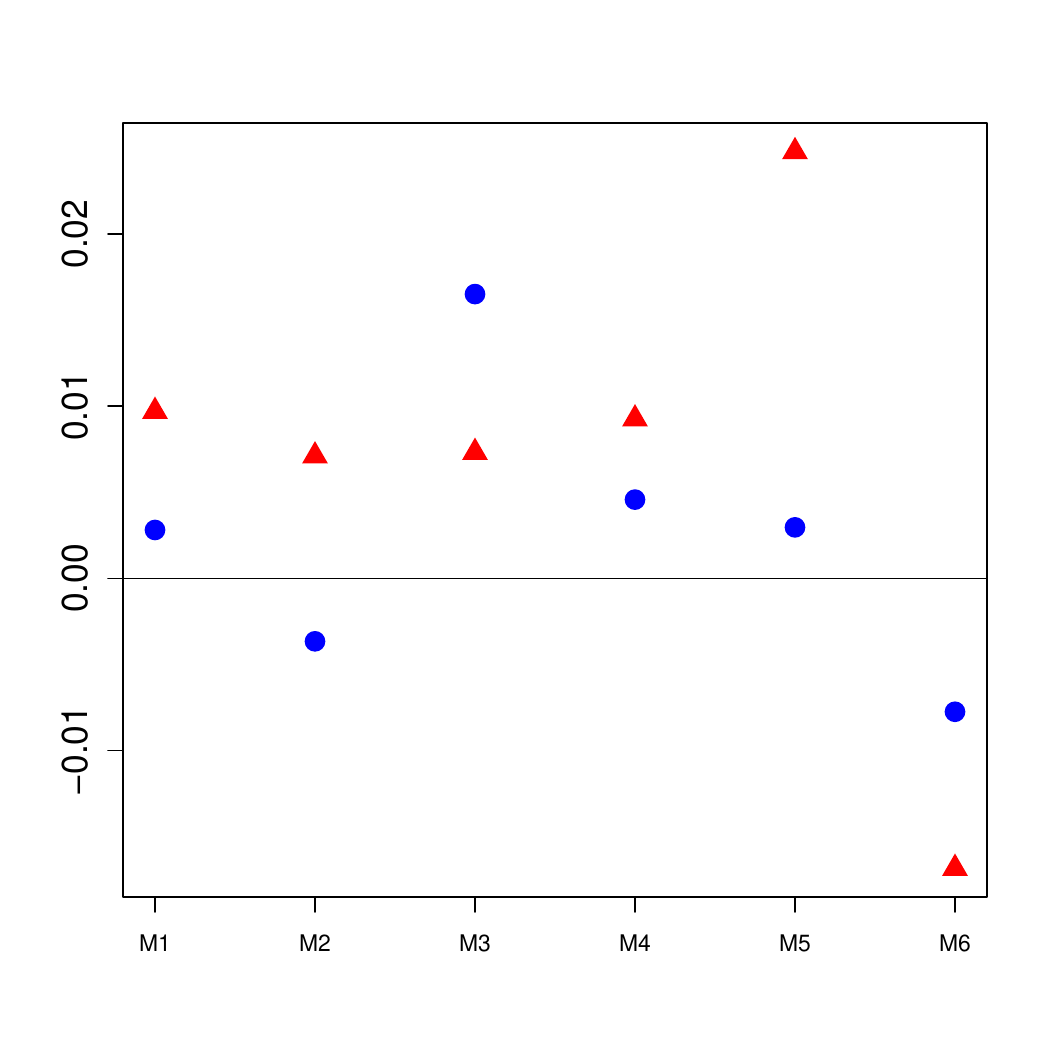}\\
 $C_2$ &  $C_3$\\[-0.3in]
\includegraphics[scale=0.5]{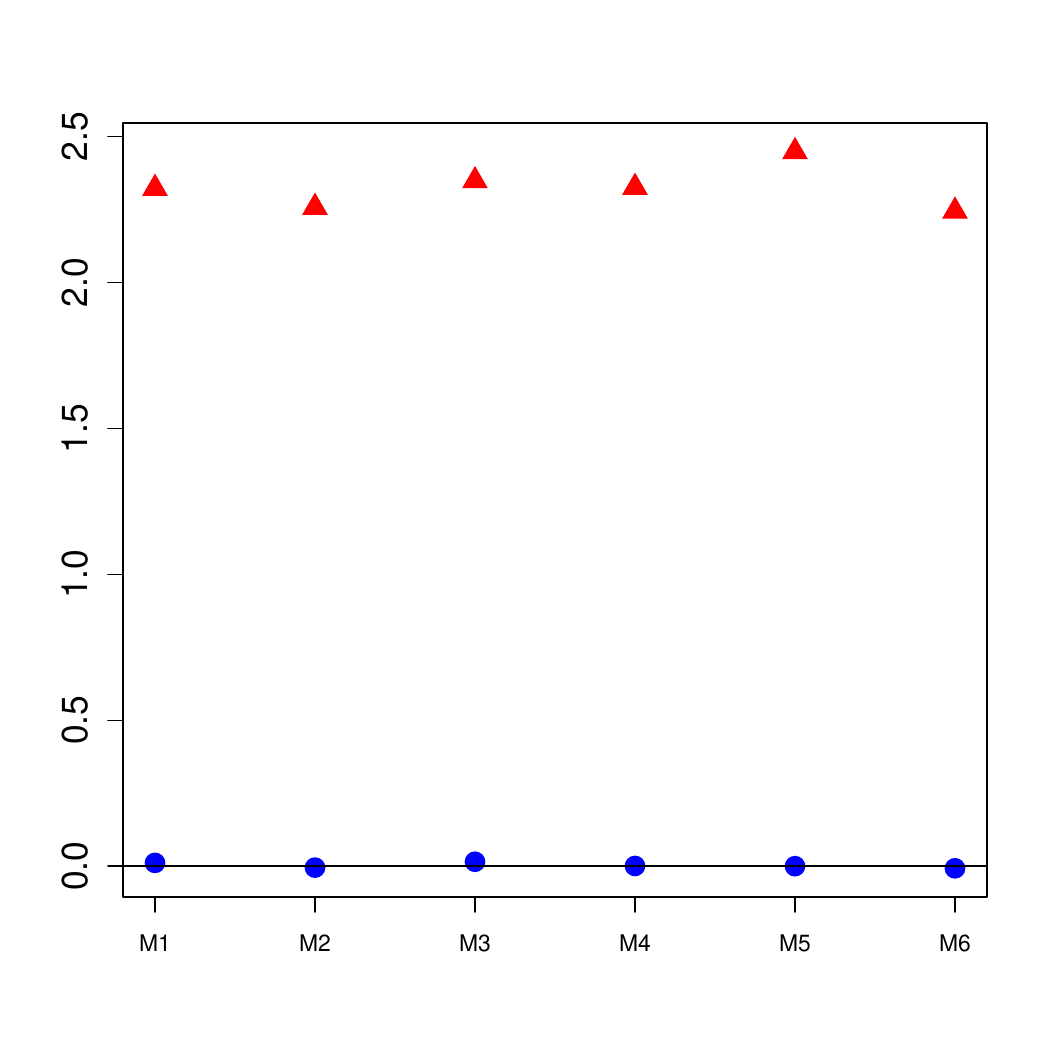} &
\includegraphics[scale=0.5]{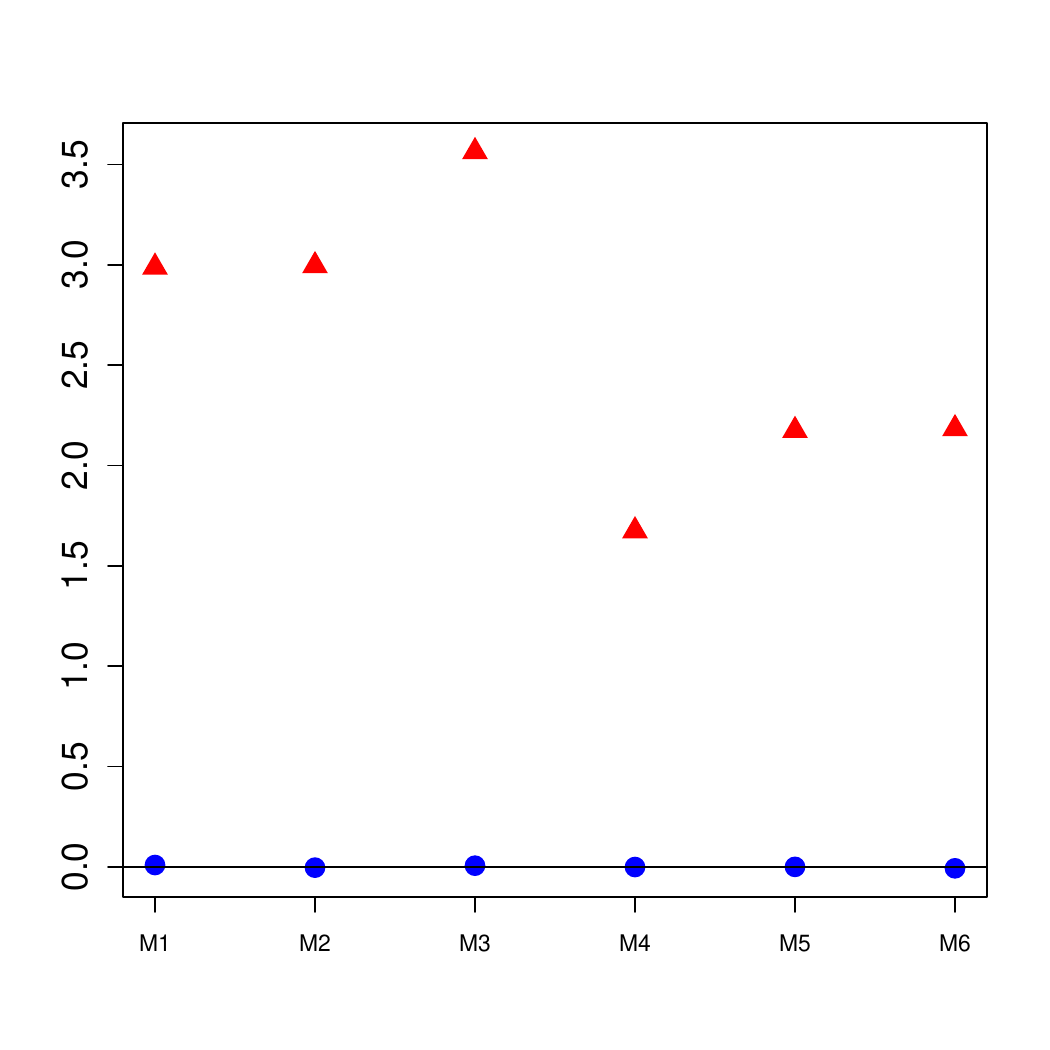}
\end{tabular}

\caption{\small \label{fig:Mu}  Plots of the mean over replications of $\wmu$, for each contamination setting. The red filled triangles correspond to the classical estimator, while the blue circles to the robust ones.}
\end{center} 
\end{figure}

With respect to the estimation of the constant $\mu$, Table \ref{tab:mu} reports the mean and standard deviation  over replications of the  classical and robust estimates, for all models and contamination settings. Besides, Figure \ref{fig:Mu} shows the mean over replications of estimation of the coefficient $\mu$ for all the models and contamination settings considered. Red triangles and blue circles correspond to the classical and robust estimates, respectively. The true value of the parameter ($\mu=0$) is shown using an horizontal  gray line. Under $C_0$, the mean  of the classical and robust estimates are similar, however, due to the loss of efficiency of the robust proposal, the   standard deviation  of the classical estimator is smaller than that of the robust one. Contamination scheme $C_1$ do not affect the bias of the classical estimates but it enlarges its variability, while the robust proposal remains stable. The only exception is Model 3, where the robust procedure presents a larger bias, see also Figure \ref{fig:Mu}. Under schemes $C_2$ and $C_3$, the mean of the classical estimators is distorted under  the six models (see Figure \ref{fig:Mu}). Besides, the standard deviations of the classical estimators are also larger than those of the robust ones under these two contamination settings, specially under  $C_2$, except for Model 1 and $C_3$ where the variability is smaller for the classical procedure but with a huge bias making the estimates unreliable.

\begin{table} [ht!]
\begin{center}
\footnotesize
\setlength{\tabcolsep}{3pt}
 \begin{tabular}{|c|c|c|c||c|c||c|c||c|c||c|c||c|c|}
\cline{3-14}
\multicolumn{1}{c}{} & \multicolumn{1}{c|}{} &\multicolumn{2}{c||}{Model 1} &\multicolumn{2}{c||}{Model 2} &\multicolumn{2}{c||}{Model 3} &\multicolumn{2}{c||}{Model 4} & \multicolumn{2}{c||}{Model 5} & \multicolumn{2}{c|}{Model 6} \\\cline{3-14}
\multicolumn{1}{c}{} & \multicolumn{1}{c|}{} & \textsc{ls} & \textsc{mm} & \textsc{ls}  & \textsc{mm} & \textsc{ls}  & \textsc{mm} & \textsc{ls}  & \textsc{mm} & \textsc{ls}  & \textsc{mm} & \textsc{ls}  & \textsc{mm}\\ \hline
& \textsc{mean} & -0.001 & 0.002 & -0.001 & -0.002 & 0.014 & 0.011 & -0.001 & -0.000 & 0.004 & -0.001 & -0.008 & -0.006 \\ 
  $C_0$ & \textsc{sd} & 0.058 & 0.107 & 0.064 & 0.070 & 0.183 & 0.196 & 0.049 & 0.052 & 0.123 & 0.130 & 0.059 & 0.061 \\ 
   \hline
  & \textsc{mean} & 0.010 & 0.003 & 0.007 & -0.004 & 0.007 & 0.017 & 0.009 & 0.005 & 0.025 & 0.003 & -0.017 & -0.008 \\ 
  $C_1$ & \textsc{sd} & 0.188 & 0.069 & 0.211 & 0.076 & 0.641 & 0.207 & 0.166 & 0.064 & 0.398 & 0.139 & 0.198 & 0.070 \\ 
  \hline
  & \textsc{mean} & 2.322 & 0.012 & 2.258 & -0.005 & 2.348 & 0.015 & 2.326 & 0.001 & 2.448 & 0.000 & 2.244 & -0.007 \\ 
  $C_2$ & \textsc{sd} & 1.586 & 0.185 & 1.928 & 0.072 & 5.161 & 0.202 & 1.326 & 0.056 & 3.263 & 0.155 & 1.530 & 0.066 \\ 
   \hline
  & \textsc{mean} & 2.988 & 0.010 & 2.995 & -0.003 & 3.564 & 0.006 & 1.675 & -0.000 & 2.174 & 0.001 & 2.183 & -0.007 \\ 
  $C_3$ & \textsc{sd} & 0.143 & 0.231 & 0.123 & 0.074 & 0.166 & 0.207 & 0.152 & 0.053 & 0.072 & 0.133 & 0.160 & 0.064 \\ 
     \hline
   \end{tabular}
\caption{\label{tab:mu}\footnotesize Mean and standard deviations for the estimates of $\mu$. The classical and robust procedures are labelled \textsc{ls} and \textsc{mm}, respectively.}
\end{center}
\end{table}

\section{Real data example}{\label{realdata}}

In this section, we analyse the  \texttt{airquality}  data set available in  \texttt{R} which considers different variables to evaluate daily air quality   in the New York region between May and September, 1973 (see Chambers \textsl{et al.}, 1983). The goal of our analysis is to model the Ozone concentration (\lq\lq $O_3$\rq\rq, measured in ppb) using as explanatory variables: the month, the temperature, the wind speed and the solar radiation  labelled \lq\lq Month\rq\rq, \lq\lq Temp\rq\rq, \lq\lq Wind\rq\rq,  and \lq\lq Solar.R\rq\rq, respectively. Regarding these covariates the month is a categorical variable with categories going from 5 to 9, the other variables may be considered as continuous. Temperature is measured in  Fahrenheit degrees,  wind speed   in mph and solar radiation in Langleys in the frequency band 4000-7700.  We considered  $n=111$ observations corresponding to cases that do not contain   missing values neither in the response variable nor in the covariates.

As   mentioned in Boente \textsl{et al.} (2017),  Dengyi and Kawagochi (1986) and Lacour \textsl{et al.} (2006) report a positive correlation between the ozone concentration and temperature in the Antarctica during Spring and also, in France during the 2003 heat wave. Besides, Cleveland (1985) finds that the relationship between ozone concentration and wind speed is nonlinear, with higher wind speeds associated to lower  concentrations of ozone.  Atypical data have been previously detected in this data set. Effectively,   Boente \textsl{et al.} (2017) used a robust fit to an additive model using as covariates \lq\lq Temp\rq\rq, \lq\lq Wind\rq\rq,  and \lq\lq Solar.R\rq\rq, while Bianco and Spano (2019)  consider  a robust fit for an exponential growth model that aims to explain the ozone daily behaviour in terms of wind speed.  For that reason, it is important to identify possible atypical observations and evaluate the   behaviour of the robust and classical estimators when all the covariates described above are included in the model. We then consider the following partially linear additive model
$$O_3 = \mu + \bbe \trasp \mbox{Month} + \eta_{1}(\mbox{Temp}) + \eta_{2}(\mbox{Wind}) + \eta_{3}(\mbox{Solar.R}) + u$$
where the errors $u=\sigma \;\eps$ are assumed to be independent, homoscedastic and with location parameter 0, Month$=(d_{6},d_{7},d_8,d_9)\trasp\in\real^4$ with $d_{\ell}=1$ if the month is the $\ell-$th one and 0 otherwise, for $\ell=6,\dots,9$, and $\bbe_0=(\beta_1,\beta_2,\beta_3,\beta_4)\trasp\in\real^4$. With this notation $\bZ=\mbox{Month}$ and $\bX=(X_1,X_2, X_3)\trasp=(\mbox{Temp}, \mbox{Wind}, \mbox{Solar.R})\trasp$.

To estimate the additive components, we use cubic $B-$splines. When estimating $\eta_j$, the   knots are taken as  the $\ell/(k+1)100\%$ quantiles, $\ell=1,\dots,k$, of the observed values of $X_j$. Taking into account that $n=111$, the basis dimension $k$ varies    between 4 and 13. Both the classical and the robust $BIC$ criteria introduced in \eqref{genBIC} selected $5$ terms to approximate  the additive functions. The loss function and tuning constants were selected as in the simulation study.

\begin{figure}[ht!]
\begin{center}
  $\weta_{1}$\hspace{4.7cm} $\weta_{2}$\hspace{4.7cm} $\weta_{3}$\\
\includegraphics[scale=0.28]{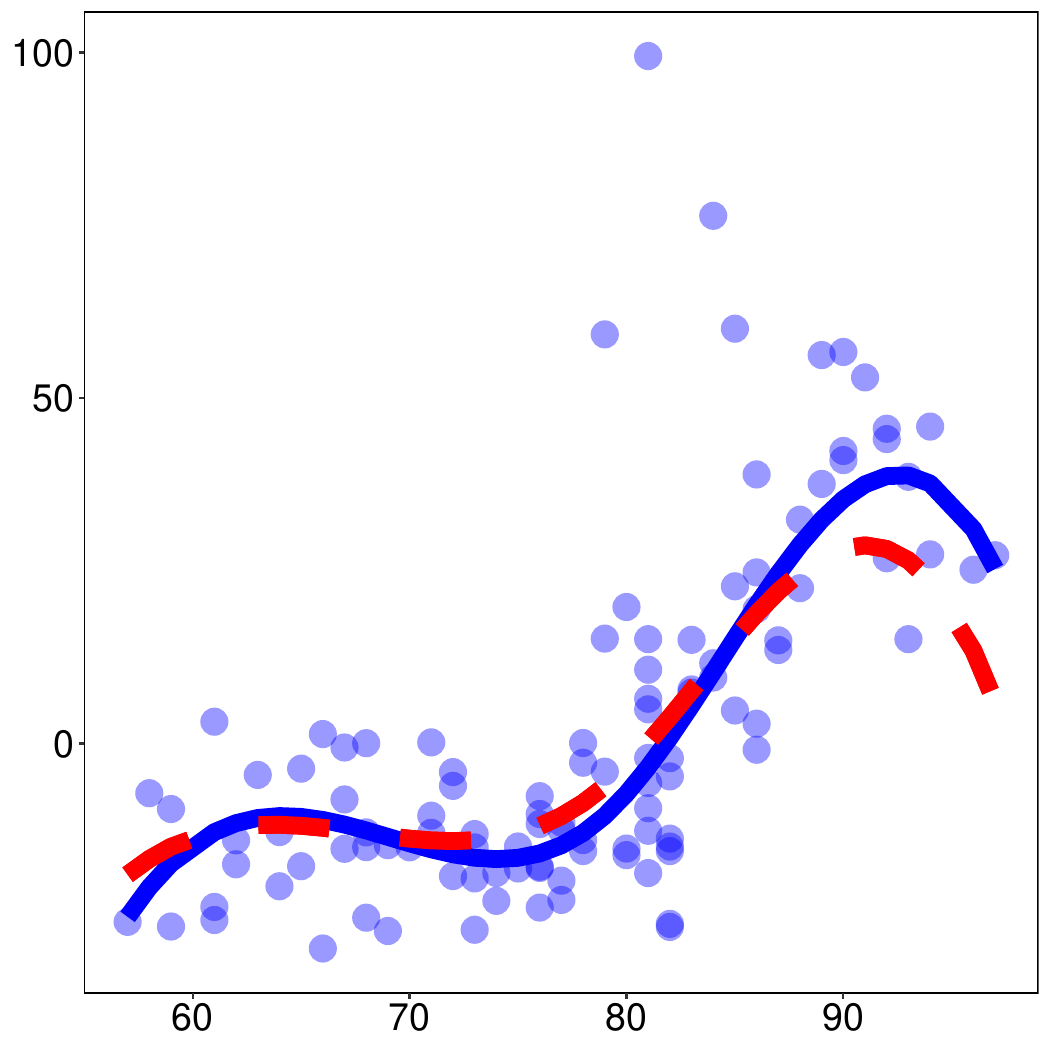} \hskip0.1in \includegraphics[scale=0.28]{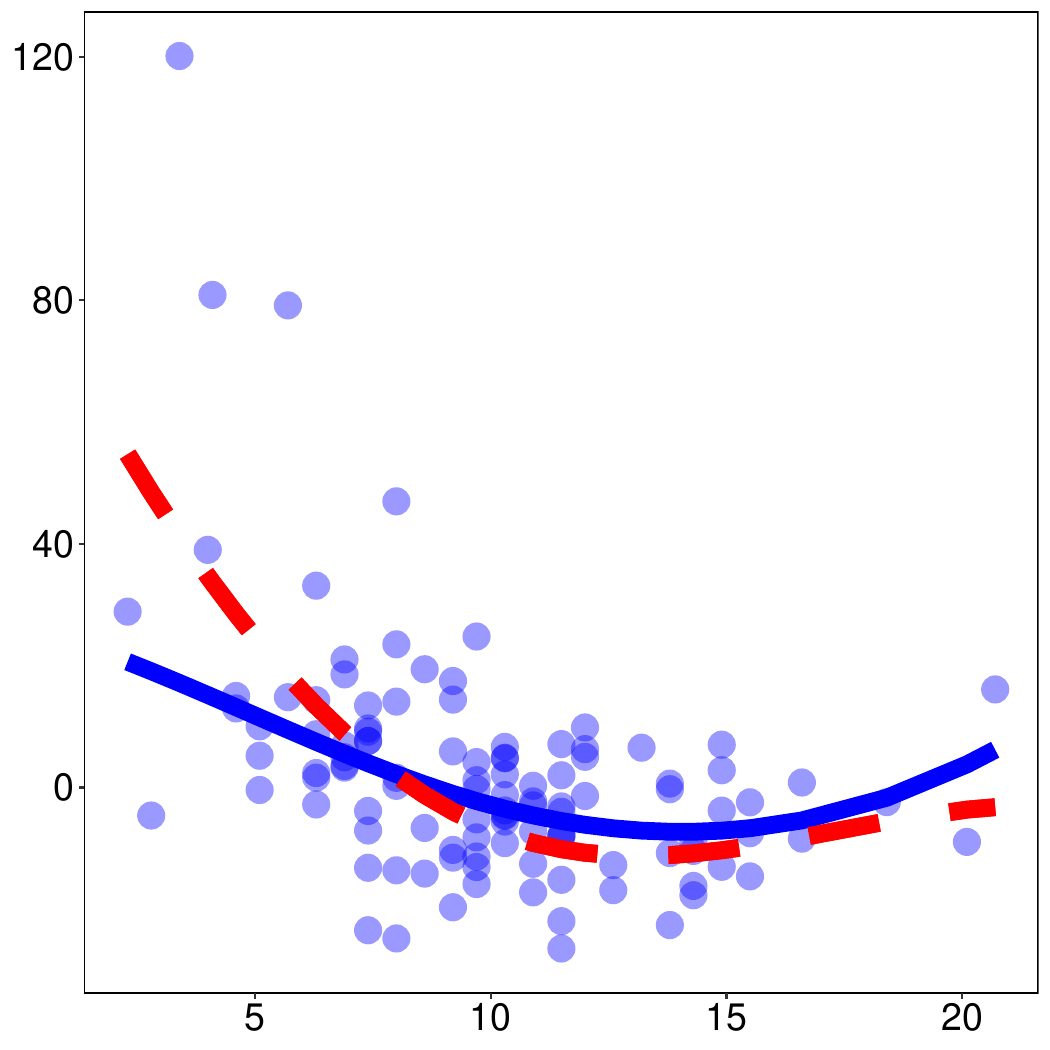}  \hskip0.1in
\includegraphics[scale=0.28]{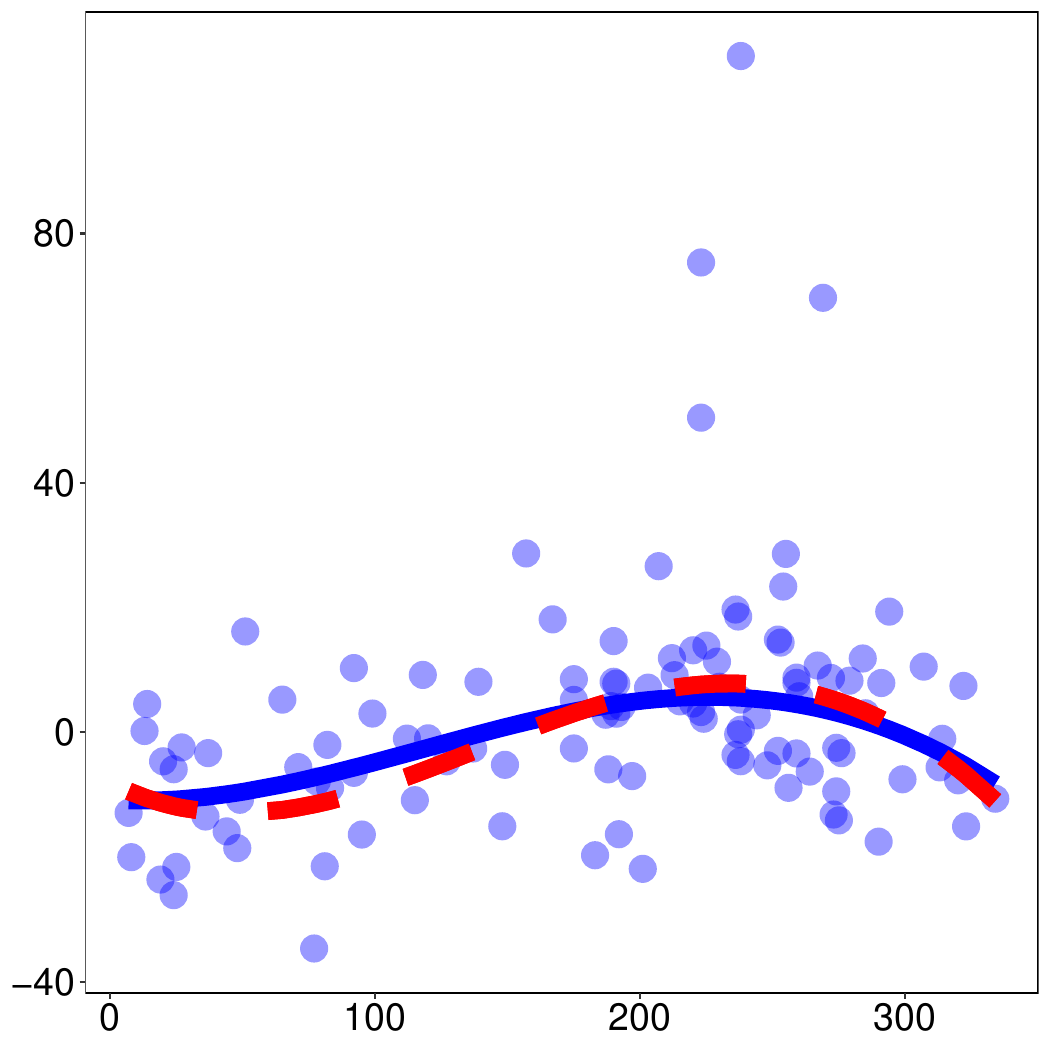}\\
\vskip-0.07in
  {\small{Temp}}\hspace{4.4cm} {\small{Wind}} \hspace{4.2cm} {\small{Solar.R}}\\
\end{center}
\vskip-0.1in 
\caption{\footnotesize Estimated curves for the classical (in red dashed lines) and robust  (in blue solid lines) estimators.}
\label{fig:ozone-original}
\end{figure}

From now on, we label with   the subscripts \textsc{cl} and \textsc{mm}  the estimators obtained through the classical and robust approach, respectively. 
The   obtained estimates of $\mu$ equal $\wmu_{\ls}=46.054$ and  $\wmu_{\eme\eme}= 40.651$, while   those of $\bbe $ are $ \wbbe_{\ls}=(-6.736, -4.614, 3.907, -12.008)\trasp$ and  $\wbbe_{\eme\eme}= (-5.641, -0.539, 5.167,-5.816)\trasp$. The presence of possible atypical data is suspected from the differences observed between these estimates. In particular, when considering the estimators of $\beta_2$, the classical procedure leads to an estimate which is almost 10 times larger than the robust one and the robust estimates of $\beta_4$ are a half than those obtained with the classical approach. 
 The estimators of $\eta_j$, $j=1,2,3$ are shown in  Figure \ref{fig:ozone-original} together with the partial residuals corresponding to the robust fit. Solid blue lines are used for the robust estimator and red dashed ones for the classical one.   Even though the shape of the estimates corresponding to the solar radiation is similar for both the robust and classical method, differences are observed in the estimation of $\eta_1$ and $\eta_2$. In particular, the classical estimate of the additive component related to temperature has a lower peak than the robust one for high values of temperature. Note that the obtained estimates of $\eta_j$ have a shape  quite similar to that obtained in  Boente \textsl{et al.} (2017) who considered an additive model and use a backfitting approach without including  the variable month.

\begin{figure}[ht!]
\begin{center}
\includegraphics[scale=0.35]{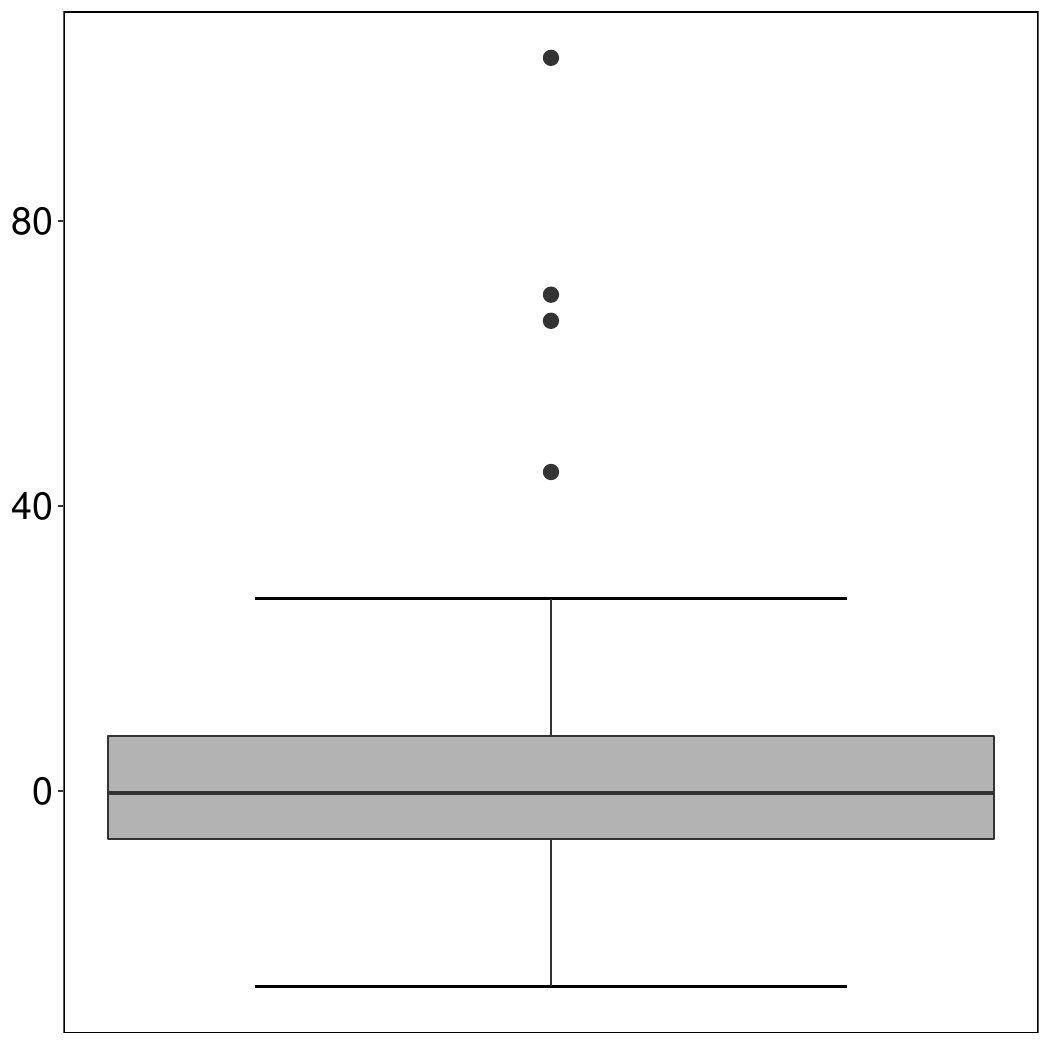}
\vskip-0.1in 
\caption{\footnotesize Boxplot of the residuals obtained using the robust fit.}
\label{fig:ozone-boxplots}
\end{center}
\end{figure}

\begin{figure}[ht!]
\begin{center}
  $\weta_{1}$\hspace{4.7cm} $\weta_{2}$\hspace{4.7cm} $\weta_{3}$\\
\includegraphics[scale=0.28]{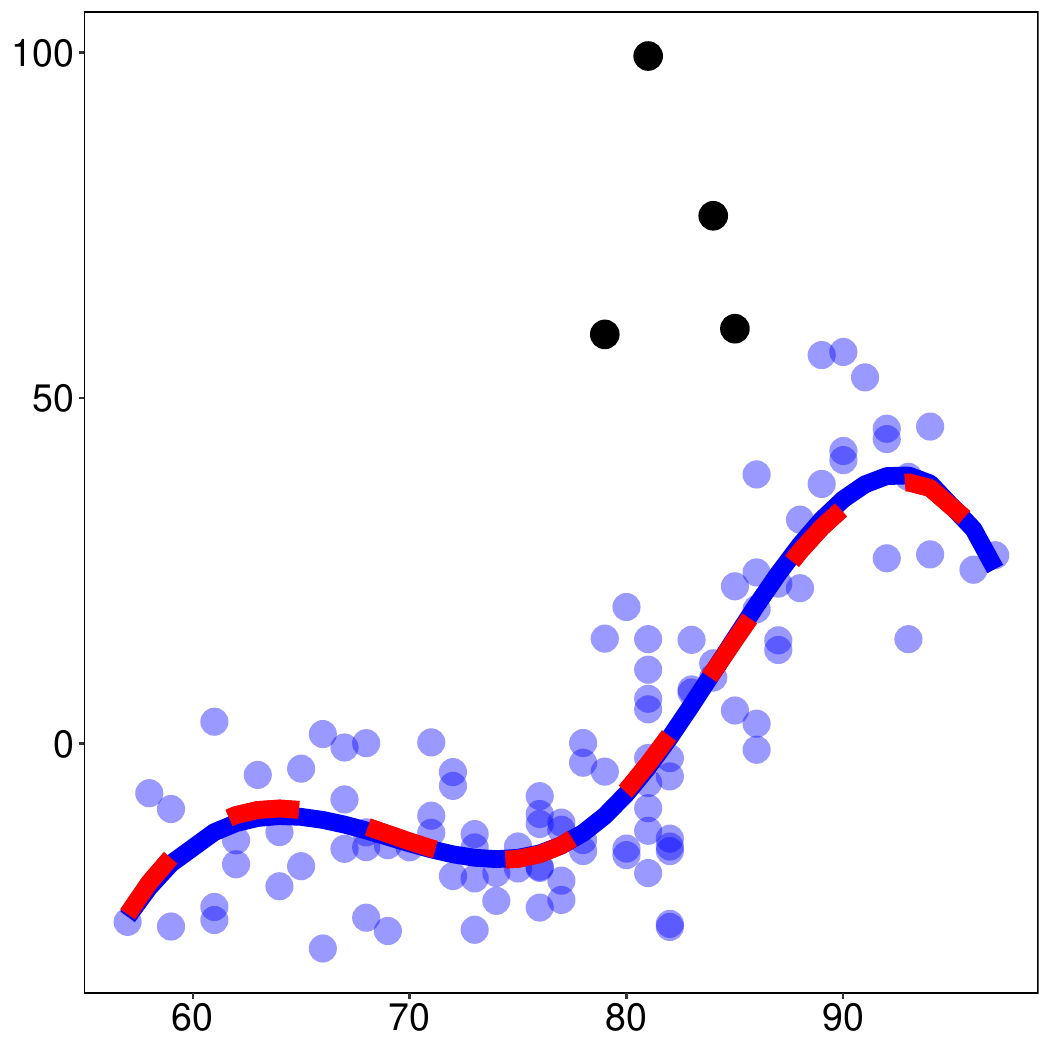} \hskip0.1in \includegraphics[scale=0.28]{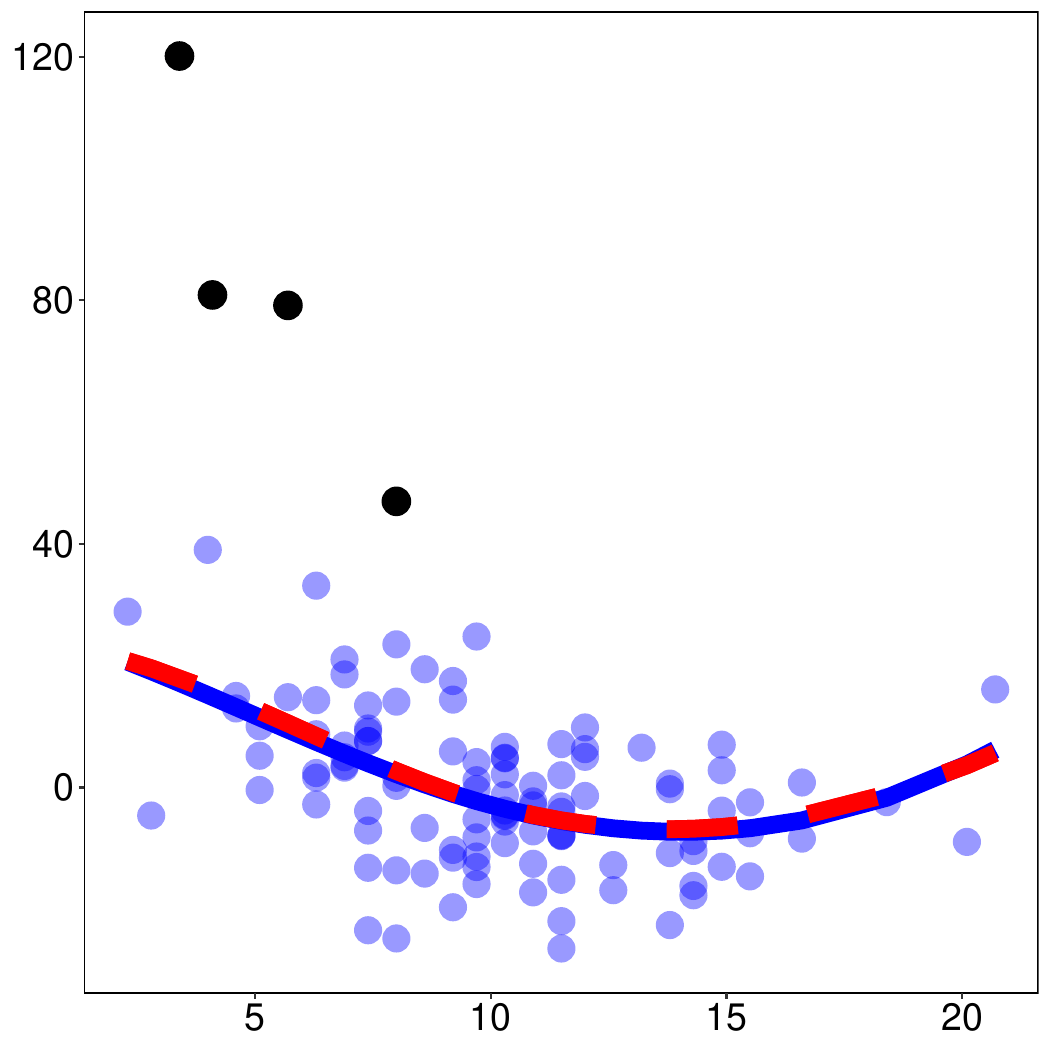}  \hskip0.1in
\includegraphics[scale=0.28]{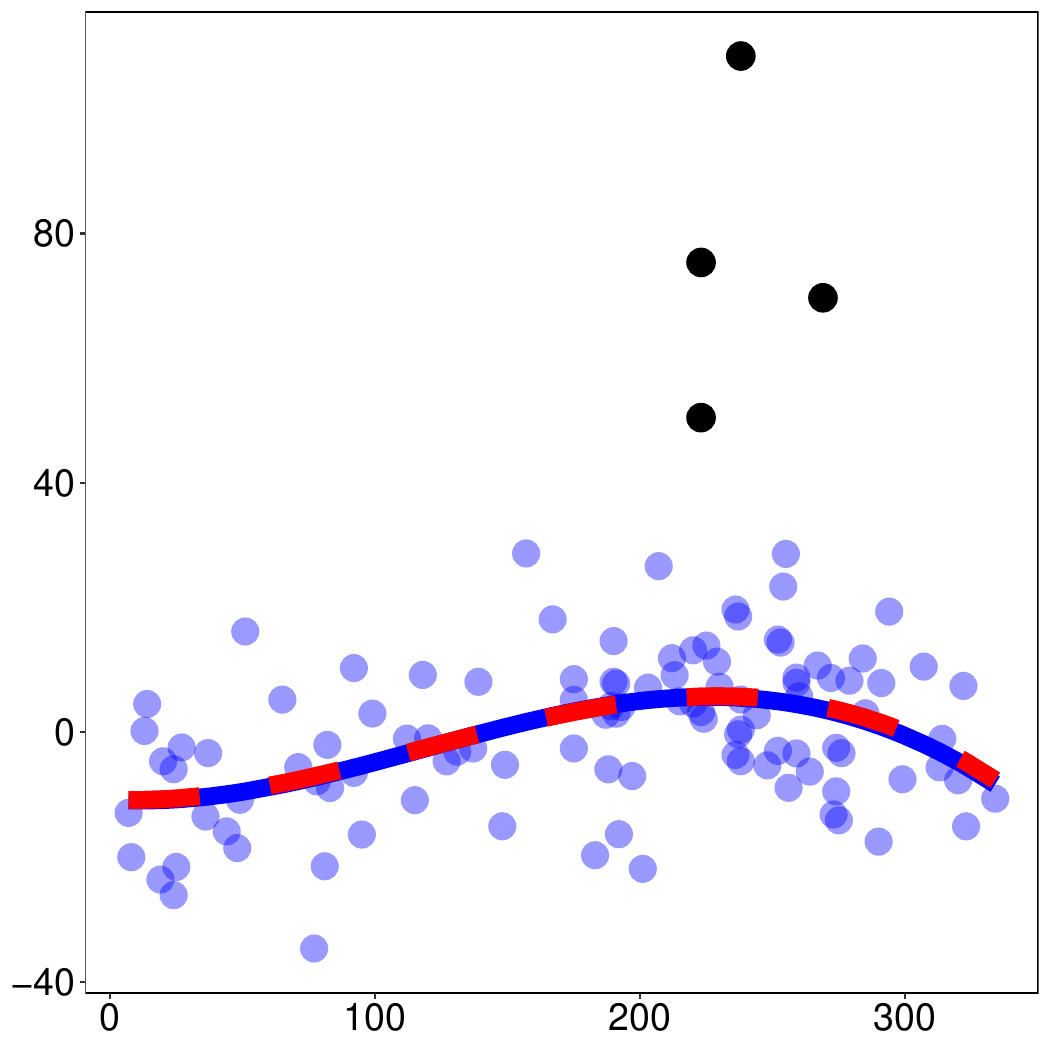}\\
\vskip-0.07in
  {\small{Temp}}\hspace{4.4cm} {\small{Wind}} \hspace{4.2cm} {\small{Solar.R}}\\
\end{center}
\vskip-0.1in 
\caption{\footnotesize Estimated curves for $\eta_j$, $j=1,2,3$. The red dashed line corresponds to the classical  estimators after the atypical data have been removed. The blue solid line indicated the  robust fit computed with the whole data set. 
}\label{fig:ozone-final}
\end{figure}

To identify potential outliers in the data, we use  a boxplot of the residuals obtained from the robust fit. This boxplot is displayed in   
Figure \ref{fig:ozone-boxplots} and identifies as atypical the  observations 23, 34, 53 and 77. These four observations were also detected in   Boente \textsl{et al.} (2017). 
We then compute the classical estimators after removing these potential atypical data. As above, the number of knots was the same for all components and a classical BIC was used to select the basis dimension $k$ resulting in $k=5$.  
The dashed red lines in Figure  \ref{fig:ozone-final} correspond to the classical fit computed without these possible atypical observations. The blue solid lines correspond to the robust estimators computed with the original data set.  Blue points correspond to the partial residuals obtained by the robust fit, while the black points identify the partial residuals corresponding to the four   observations detected as atypical. Note that the classical estimators computed without these potential 
outliers are very close to the
robust ones. In other words, the robust estimator behaves similarly to 
the classical one if one were able to manually remove suspected outliers. Besides, the classical estimates of $\mu$ and $\bbe$ after removing the four  outliers, denoted  $\wmu_{\ls}^{(-4)}$  and  $\wbbe_{\ls}^{(-4)} $, are   equal to  $\wmu_{\ls}^{(-4)} =39.374$ and $\wbbe_{\ls}^{(-4)} =(-4.920,   -0.126,  6.475,   -5.558)\trasp$. Even though there is a difference between the estimates of the second component of $\bbe$,   $\wmu_{\ls}^{(-4)}$  and  $\wbbe_{ \ls}^{(-4)}$ have values closer  to  $\wmu_{\eme\eme}$ and $\wbbe_{\eme\eme}$ than to $\wmu_{\ls} $  and  $\wbbe_{\ls}$.  

To conclude the robust procedure leads to more reliable results  automatically down--weighting potential outliers. It also allows to identify potential atypical observations and leads to estimated components $\eta_j$ that are almost identical to the classical ones when these detected outliers are removed.

\normalsize

\normalsize

\section{Final Comments}{\label{comentarios}}
In this paper, we presented a procedure to robustly estimate the regression parameter and the additive components under a partially linear additive model. The method combines $B-$splines to smoothly estimate the additive components and $MM-$estimators.  The use of $B-$splines avoids the higher complexity of kernels methods which require backfitting or marginal integration combined with a profile approach, see, for instance, Li (2000) or Ma and Yan (2011) for a discussion on the computational problems raising in partially linear additive model when using kernel methods. 
In contrast to Ma and Yan (2011), we do not consider an additional step in the procedure using kernel smoothers to obtain estimators of the nonparametric components for which the limiting distribution can be derived. However, our proposal can be extended to provide a robust spline-backfitted kernel smoother combining the ideas in Ma and Yang (2011) to construct pseudo--observations with the robust kernel smoothers defined in Boente and Fraiman (1989). This important issue is beyond the scope of the paper and will be object of future work. 

For the robust $B-$spline estimators  consistency results and rates of convergence are obtained. Besides,   the asymptotic distribution of the linear regression estimator is derived under mild conditions. 
The numerical results obtained illustrate the stability of the proposed methods under the considered contaminations and allow to conclude that our proposal has  good robustness and finite-sample statistical properties. We illustrate our method on the well known  air quality data set. The analysis  shows that the robust estimators automatically discard influential observations leading to reliable  estimates.

\noi\textbf{\small Acknowledgements.} {\small  This research was partially supported by  20020170100022\textsc{ba}    from the Universidad de Buenos Aires  and  \textsc{pict} 2018-00740 from \textsc{anpcyt} at  Argentina (Graciela Boente and Alejandra Mart\'{\i}nez), the Spanish Project {MTM2016-76969P} from the Ministry of Economy, Industry and Competitiveness, Spain (MINECO/AEI/FEDER, UE) (Graciela Boente) and  Proyectos Internos  CD-CBLUJ 301/19 and CD-CBLUJ 204/19  from the Departamento de Ciencias B\'asicas, Universidad Nacional de Luj\'an (Alejandra Mart\'{\i}nez) as well as  Proyecto de Investigadores en Formaci\'on  RESREC-LUJ 224/19, Universidad Nacional de Luj\'an at Argentina (Alejandra Mart\'{\i}nez).}

\setcounter{equation}{0}
\def\theequation{A.\arabic{equation}}
  \setcounter{section}{0}
\renewcommand{\thesection}{\Alph{section}}

\section{Appendix} 
From now on, for any measure $\qu$ and class of functions $\itF$, $N(\epsilon, \itF, L_s(\qu))$ and $N_{[\;]}(\epsilon, \itF, L_s(\qu))$ stand  for the covering  and bracketing  numbers  of the class $\itF $ with respect to the distance in $ L_s(\qu)$, as defined, for instance, in van der Vaart and Wellner (1996). 

\subsection{Proofs of Lemma \ref{lemma:FC}, Proposition \ref{prop:prop1} and Theorem \ref{teo:consist}}
The proof of  Lemma \ref{lemma:FC} follows the same steps as that of Lemma A.1.1 in Boente \textsl{et al.} (2021).

\noindent\textsc{Proof of Lemma \ref{lemma:FC}.}  Note that   $u=\sigma \eps/\varsigma$ also satisfies \ref{ass:densidad} since $\eps$ does. Then, from Lemma 3.1 of Yohai (1985) we obtain that  for all $c\neq 0$
\begin{equation}\label{A.1}
\esp\rho\left(\frac{\sigma\eps}{\varsigma}-c\right)=\esp\rho\left(u-c\right)>\esp\rho\left(u\right)=\esp\rho\left(\frac{\sigma\eps}{\varsigma}\right)\,.
\end{equation}
a) follows easily taking conditional expectation from the fact that $L(\mu,\bbe,\eta_{1},\dots,\eta_{p},\varsigma)=\esp\rho\left(u\right)$ while
$$ L(a,\bb,g_1,\dots,g_p,\varsigma)= \esp\rho\left(u+ \frac{(\mu-a)+ (\bbe-\bb)\trasp\bZ+\sum_{j=1}^p \left\{\eta_j(X_{j})-g_j(X_{j})\right\}}{\varsigma}\right)\,.$$ 

To derive b), fix $a\in \real$, $\bb\in\real^q$, $g_1\in\itG,\dots,g_p\in\itG$  and define $D(\bz,\bx)=a-\mu +(\bb-\bbe)\trasp\bz+\sum_{j=1}^p (g_j-\eta_{j})(x_j)$ and $H(\bz,\bx)=D(\bz,\bx)/\varsigma$. Denote $\itA_0=\{(\bz, \bx)\,:\,D(\bz,\bx)=0\}$, then \ref{ass:probacond} entails that $\prob(\itA_0)<1$ if $(a,\bb,g_1,\dots,g_p )\ne (\mu,\bbe,\eta_{1},\dots,\eta_{p})$.

The independence between the error  and the explanatory variables leads to
\begin{eqnarray*}
L(a,\bb,g_1,\dots,g_p,\varsigma)&=&  \esp\rho\left(u - H(\bZ,\bX)\right)\\
&=&\esp\left[\rho\left(u\right)\indica_{\itA_0}(\bZ,\bX)\right]+\esp\left[\rho\left(u-H(\bZ,\bX)\right)\indica_{\itA_0^c}(\bZ,\bX)\right]\\
&=&  \esp\left[\rho\left(u\right)\right]\prob(\itA_0)+\esp\left\{\indica_{\itA_0^c} (\bZ,\bX)\esp\left[\rho\left(u-H(\bZ,\bX)\right)\Big|(\bZ ,\bX)\right]\right\}
\,.
\end{eqnarray*}
Take $(\bz_0,\bx_0)\in\itA_0^c$, then   \eqref{A.1}  and   the independence between the error and the covariates entail that
$$
 \esp\left[\rho\left(u-H(\bZ,\bX)\right)\Big|(\bZ ,\bX)=(\bz_0,\bx_0)\right]=\esp\left[\rho\left(u-H(\bz_0,\bx_0)\right)\right] >\esp\rho\left(u\right)\,.$$
Then, using that $\prob(\itA_0^c)>0$ we obtain
\begin{eqnarray*}
L(a,\bb,g_1,\dots,g_p,\varsigma)&=&\esp\left[\rho\left(u\right)\right]\prob(\itA_0)+\esp\left\{\indica_{\itA_0^c}(\bZ,\bX)\esp\left[\rho\left(u-H(\bZ,\bX)\right)\Big|(\bZ ,\bX)\right]\right\}\\
&>&\esp\left[\rho\left(u\right)\right]\prob(\itA_0)+\esp\left\{\esp\rho\left(u\right)\indica_{\itA_0^c}(\bZ,\bX)\right\} =\esp\rho\left(u\right)\,,
\end{eqnarray*}
concluding the proof.\, \qed

The following Lemmas will be needed to prove Proposition \ref{prop:prop1} and Theorem \ref{teo:consist}.  We first state some notation that will be helpful in the sequel.
Given a loss function $\rho:\real\to\real$, we define the function 
\begin{align*}
L_n(a,\bb,g_1,\dots,g_p,\varsigma)&=\frac{1}{n}\sum_{i=1}^n \rho\left(\frac{Y_i-a-\bb\trasp\bZ_i-\sum_{j=1}^p g_j(X_{ji})}{\varsigma}\right)\,,
\end{align*}
which is the sample version of the function $L(a,\bb,g_1,\dots,g_p,\varsigma)$ defined in \eqref{eq:funcionL}. 

Recall that $\itS_{j}$, $1\le j\le p$, denote the linear spaces
spanned by the centered $B-$splines bases of order $\ell_j$ and size $k_j$ as defined in \eqref{eq:itSj}.
From now on, for $g_j (x)= \sum_{s=1}^{k_j-1} c_s^{(j)} \,   B_s^{(j)}(x)\in \itS_{j}$, $1\le j\le p$, and identifying the functions with their coefficients, we denote $s_n(a,\bb,g_1,\dots,g_p)= s_n(a,\bb,\bc^{(1)},\dots,\bc^{(p)})$ as defined in \eqref{eq:s-est} and 
$r_i(a,{\bb},  g_1,\dots, g_p)=r_i(a,{\bb}, \bc^{(1)}, \dots, \bc^{(p)})$
as defined in \eqref{eq:residuals} with $\bc^{(1)}=( c_1^{(j)}, \dots,  c_{k_j-1}^{(j)})\trasp$.

Recall that  $\bV^{(j)}(t)=(B_1^{(j)}(t),\dots,B_{k_j-1}^{(j)}(t))\trasp$. To derive uniform results, Lemma \ref{lemma:lema3} below provides a bound to the covering number of the class of functions 
\begin{equation}{\label{eq:claseFn}}
\itF_n=\left\{ f(y,\bz,\bx)=\rho\left( \frac{y-a-\bb\trasp\bz-\sum_{j=1}^p \bc^{(j)} \bV^{(j)}(x_j) }{\varsigma}\right)
\,,\, a\in \real,\bb\in\real^q,\bc^{(j)}\in\real^{k_j-1}, \varsigma>0\right\}
\end{equation}
 Lemma \ref{lemma:lema3} is a direct consequence of Lemma S.2.1 in Boente \textsl{et al.} (2020)  noting that the number of parameters involved is $q+K+1=q+\sum_{j=1}^p (k_j-1)+1$ and that the class  $\itF_n$ has envelope 1, for that reason, its proof is omitted.

\begin{lemma}{\label{lemma:lema3}}
Let  $\rho$ be a   function satisfying \ref{ass:rho_bounded_derivable}\textbf{(a)} and $\itF_n$ the class of functions given in \eqref{eq:claseFn}. Then,  for any $0<\epsilon<1$, there exists some constant $C>1$ independent of $n$ and $\epsilon$, such that
\begin{equation}\label{eq:cota}
N(2\,\epsilon, \itF_n,L_1(\qu))\leq \left[C q_n (16\,e)^{q_n}\left(\frac{1}{\epsilon}\right)^{q_n-1}\right]^2
\end{equation} 
where $q_n=2( q+\sum_{j=1}^p k_j-p+4)- 1$ and for any measure $\qu$, $N(\varepsilon, \itF_n,L_s(\qu))$   stands for the covering number of the class $\itF_n$ with respect to the distance $L_s(\qu)$, as defined,   in van der Vaart and Wellner (1996). 
\end{lemma}

\vskip0.1in
To derive consistency of the $MM-$estimators and the $S-$scale the following Lemma will be helpful. It shows that   $L_n(a,\bb,g_1,\dots,g_p,\varsigma)$ converges to $L(a,\bb,g_1,\dots,g_p,\varsigma)$ with probability one, uniformly over $a\in \real$, $\varsigma > 0$, $\bb\in \real^q$ and 
$\itS_{1}\times\dots \times \itS_{p}$ and its proof uses similar arguments to those considered in the proof of Lemma A.1.2 in Boente \textsl{et al.} (2021). We include it for the sake of completeness.

\vskip0.1in
\begin{lemma}{\label{lema:A1}}
Let $\rho$ be a  function satisfying \ref{ass:rho_bounded_derivable}   and assume that \ref{ass:kj}  holds.  Then,  
\begin{itemize}
\item[a)] $\sup_{\varsigma >0, a\in \real, \bb \in \real^q, g_1 \in \itS_{1}, \dots, g_p \in \itS_{p}} \left|L_n(a,\bb,g_1,\dots,g_p,\varsigma)- L(a,\bb,g_1,\dots,g_p,\varsigma)\right| \convpp 0$.
\item[b)] Furthermore, if we denote $K=\sum_{j=1}^p (k_j-1)$, we have that
$$\mathop{\sup_{ \varsigma >0, a\in \real, \bb \in \real^q}}_{g_1 \in \itS_{1}, \dots, g_p \in \itS_{p}}\left| \frac 1{n-q-K} \sum_{i=1}^n \left[\rho\left(\frac{Y_i - a- \bb\trasp \bZ_i-\sum_{j=1}^p g(X_{ij})}{\varsigma}\right)-L(a,\bb,g_1,\dots,g_p,\varsigma)\right]\right| \convpp 0\,.$$
\end{itemize}
\end{lemma}

\vskip0.2in
\noi \textsc{Proof.} b) follows immediately from a) noting that $n/(n-q-K)\to 1$, since $k_j=O(n^{\nu_j})$ with $\nu_j<1$. 

Let us show (a). First note that
$$\mathop{\sup_{ \varsigma >0, a\in \real, \bb \in \real^q}}_{g_1 \in \itS_{1}, \dots, g_p \in \itS_{p}}\left|L_n(a,\bb,g_1,\dots,g_p,\varsigma)- L(a,\bb,g_1,\dots,g_p,\varsigma)\right| =\sup_{f\in \itF_n} \left|P_n f - P f\right| \,,$$
where  the class $\itF_n$ is defined in \eqref{eq:claseFn} and we have used the empirical process notation as in van der Vaart and Wellner (1996) with $P_n$ the empirical distribution of $(Y_i,\bZ_i\trasp,\bX_i\trasp)$. Thus, to derive a) it will be enough to show that $(1/n) \log N(2\epsilon,\itF_n,L_1(P_n))\convpp 0$. Using \eqref{eq:cota} and that   $\log(q_n)/q_n<1$, where  $q_n=2( q+\sum_{j=1}^p k_j-p+4)- 1$, we get easily that
\begin{eqnarray*}
\log\left(N(2\,\epsilon, \itF_n, L_1(P_n)) \right) 
& \le & C_1 q_n \log\left(\frac 1\epsilon\right)\,,
\end{eqnarray*}
for $\epsilon<  \min((16e)^{-1}, e^{-C})$ and some constant $C_1>0$. Assumption \ref{ass:kj}  entails that $k_j=O(n^{\nu_j})$ with $\nu_j<1$, so $q_n/n\to 0$ leading to 
\begin{eqnarray*}
\frac 1n \log N(2\epsilon,\itF_n,L_1(P_n)) & \le &  C_1\, \frac{q_n}n  \,  \log\left(\frac 1\epsilon\right)\to 0\,,
\end{eqnarray*}
which concludes the proof. \qed
 
 \vskip0.1in

\noindent\textsc{Proof of Proposition \ref{prop:prop1}.}  
To avoid burden notation, we will use $\rho$ instead of $\rho_{0}$ and $\wmu$, $\wbbe$ and $\weta_{j}$, for $j=1,\dots,p$, instead of $\wmu_{\ini},\wbbe_{\ini}$ and $\weta_{j,\ini}$,  respectively.

We will show that for any $\delta>0$, with probability 1 there exists $n_0\geq 1$ such that for $n\geq n_0$, we have that $|\wsigma-\sigma|\leq \delta$. 
Lemma \ref{lema:A1} entails  there exists a null probability set $\itN_1$ such that, for any $\omega\notin\itN_1$, 
\begin{equation}\label{eq:A8}
\mathop{\sup_{ \varsigma >0, a\in \real, \bb \in \real^q}}_{g_1 \in \itS_{1}, \dots, g_p \in \itS_{p}}\left| \frac 1{n-q-K} \sum_{i=1}^n \left[\rho\left(\frac{Y_i - a- \bb\trasp \bZ_i-\sum_{j=1}^p g_j(X_{ij})}{\varsigma}\right)-L(a,\bb,g_1,\dots,g_p,\varsigma)\right]\right| \to 0
\end{equation}
holds. On the other hand, the boundedness of $\rho$, the strong law of large numbers together with the fact that $n/( n-q-K)\to 1$  and assumption \ref{ass:rho_bounded_derivable}\textbf{(a)} imply that 
$$\frac{1}{ n-q-K}\sum_{i=1}^n \rho\left(\frac{\sigma\eps_i}{\sigma+\delta}\right)\convpp\esp\rho\left(\frac{\sigma\eps}{\sigma+\delta}\right)<\esp\rho(\eps)=b\,.$$
Hence,  there exists a null probability set $\itN_2$ such that, for an $\omega\notin\itN_2$,
\begin{equation}\label{eq:A9}
A_n(\delta)=\frac{1}{ n-q-K}\sum_{i=1}^n \rho\left(\frac{\sigma\eps_i}{\sigma+\delta}\right)\to\esp\rho\left(\frac{\sigma\eps}{\sigma+\delta}\right)=b_1<\esp\rho(\eps)=b\,.
\end{equation}
Fix $\omega\notin \bigcup_{i=1}^2\itN_i$. For each $j=1,\dots,p$, using \ref{ass:etajCr} and Corollary 6.21 in Schumaker (1981), we obtain that there exists a spline of order $\ell_j$,  $\wteta_j(x)=\sum_{j=1}^{k_j} \lambda_s^{(j)}\wtB_s^{(j)}(x)$ such that 
$\|\wteta_j-\eta_{j}\|_{\infty}=O\left(n^{-\nu_j \,r_j }\right)$. First note that the fact that $\int_0^1 \eta_j(x) dx=0$, entails that $\int_0^1 \wteta_j(x) dx=O\left(n^{-\nu_j\, r_j}\right)$ and  denote as 
$$\widetilde{\wteta}_j(x)= \wteta_j(x) -\int_0^1 \wteta_j(x) dx= \sum_{j=1}^{k_j} \lambda_s^{(j)} B_s^{(j)}(x)$$
the centered spline. Then, we have that $\widetilde{\wteta}_j(x)= \sum_{j=1}^{k_j-1}c_s^{(j)} B_s^{(j)}(x)$ with $c_s^{(j)}=\lambda_s^{(j)}-\lambda_{k_j}^{(j)}$, so $\widetilde{\wteta}_j\in \itS_j$ and $\|\widetilde{\wteta}_j-\eta_{j}\|_{\infty}=O\left(n^{-\nu_j\, r_j}\right)$. 

A Taylor's expansion of order one leads to
\begin{eqnarray*}
\frac{1}{n-q-K }\sum_{i=1}^n \rho\left(\frac{Y_i-\mu-\bbe\trasp\bZ_i-\sum_{j=1}^p \widetilde{\wteta}_j(X_{ij})}{\sigma+\delta}\right)&=&\frac{1}{n-q-K }\sum_{i=1}^n \rho\left(\frac{\sigma\eps_i+\sum_{j=1}^p (\eta_{j}-\widetilde{\wteta}_j)(X_{ij})}{\sigma+\delta}\right)\\
&=&\frac{1}{n-q-K }\sum_{i=1}^n \rho\left(\frac{\sigma\eps_i}{\sigma+\delta}\right)+R_n\\
&=&A_n(\delta)+R_n\;,
\end{eqnarray*}
where $A_n(\delta) \convpp b_1$, 
$$R_n=\frac{1}{n-q-K}\sum_{i=1}^n{\psi(\xi_i)\frac{\sum_{j=1}^p (\eta_{j}-\widetilde{\wteta}_j)(X_{ij})}{\sigma+\delta}}$$ 
and $\xi_i$ is an intermediate point.  Using that
$$|R_n|\leq \frac{1}{n-q-K} \|\psi\|_{\infty}(\sigma+\delta)^{-1} \sum_{j=1}^p\|\widetilde{\wteta}_j-\eta_{j}\|_{\infty}= \frac{1}{n-q-K} \sum_{j=1}^p O\left(n^{-\nu_j\, r_j}\right)$$
and that $n/( n-q-K)\to 1$, we obtain that $|R_n|\to 0$. Therefore,  using \eqref{eq:A9} we conclude that
$$\frac{1}{n-q-K}\sum_{i=1}^n \rho\left(\frac{Y_i-\mu-\bbe\trasp\bZ_i-\sum_{j=1}^p \widetilde{\wteta}_j(X_{ij})}{\sigma+\delta}\right)\to b_1\,.$$
Choose $\delta_1>0$ such that $b_1+\delta_1<b$, then there exists $n_0\in\natu$ such that for $n\geq n_0$,
\begin{equation}\label{eq:A12}
\frac{1}{n-q-K}\sum_{i=1}^n \rho\left(\frac{Y_i-\mu-\bbe\trasp\bZ_i-\sum_{j=1}^p \widetilde{\wteta}_j(X_{ij})}{\sigma+\delta}\right)<b_1+\delta_1<b \,.
\end{equation}
Recall that
$$\frac{1}{n-q-K}\sum_{i=1}^n \rho\left(\frac{Y_i-\mu-\bbe\trasp\bZ_i-\sum_{j=1}^p \widetilde{\wteta}_j(X_{ij})}{s_n(\mu,\bbe,\widetilde{\wteta}_1,\dots,\widetilde{\wteta}_p)}\right)=b\,.$$
Thus, \eqref{eq:A12}  and the fact that $\rho$ is non-decreasing, imply that $s_n(\mu,\bbe,\widetilde{\wteta}_1,\dots,\widetilde{\wteta}_p)<\sigma+\delta$. 
 
Using that $\wsigma=\min_{a\in \real, \bb\in\real^q,g_1\in\itS_1,\dots,g_p\in\itS_p}s_n(a,\bb,g_1,\dots,g_p)$ and the fact that $\widetilde{\wteta}_1\in\itS_1,\dots,\widetilde{\wteta}_p\in\itS_p$, we conclude that for $n\geq n_0$, 
$$\wsigma \leq s_n(\mu,\bbe,\widetilde{\wteta}_1,\dots,\widetilde{\wteta}_p)<\sigma+\delta\,.$$

It remains to show that there exists $n_1\in\natu$ such that for any $n\geq n_1$, $\wsigma\geq \sigma-\delta$.
Lemma 3 in Salibi\'an-Barrera (2006) and assumptions \ref{ass:densidad} and \ref{ass:rho_bounded_derivable}\textbf{(a)} imply that 
$$L(\mu,\bbe,\eta_{1},\dots,\eta_{p},\sigma-\delta)>L(\mu,\bbe,\eta_{1},\dots,\eta_{p},\sigma)=b\,.$$
Let $\delta_2>0$ be such that $L(\mu,\bbe,\eta_{1},\dots,\eta_{p},\sigma-\delta)=b+\delta_2=b_2$. Using that \eqref{eq:A8} holds,   $n/(n-p-K)\to 1$  and  $\rho$ is bounded, we get that there exists $n_1\in\natu$ such that for any $n\geq n_1$, 
$$\mathop{\sup_{ \varsigma >0, a\in \real, \bb \in \real^q}}_{g_1 \in \itS_{1}, \dots, g_p \in \itS_{p}}\left|\frac{1}{n-q-K}\sum_{i=1}^n \rho\left(\frac{Y_i-a-\bb\trasp\bZ_i-\sum_{j=1}^p g_j(X_{ij})}{\varsigma}\right)-L(a,\bb,g_1,\dots,g_p,\varsigma)\right|<  \delta_2 \,.$$
Hence,  
$$\left|\frac{1}{n-q-K}\sum_{i=1}^n \rho\left(\frac{Y_i-\wmu-\wbbe\trasp\bZ_i-\sum_{j=1}^p \weta_j(X_{ij})}{\wsigma}\right)-L(\wmu,\wbbe,\weta_1,\dots,\weta_p,\wsigma)\right|< \delta_2 $$
leading to
\begin{equation}\label{eq:AA}
 L(\wmu,\wbbe,\weta_1,\dots,\weta_p,\wsigma)<\frac{1}{n-q-K}\sum_{i=1}^n \rho\left(\frac{Y_i-\wmu-\wbbe\trasp\bZ_i-\sum_{j=1}^p \weta_j(X_{ij})}{\wsigma}\right)+ \delta_2 =b+ \delta_2 \,.
\end{equation}
The Fisher-consistency derived in Lemma \ref{lemma:FC} entails that  $L(\mu,\bbe,\eta_{1},\dots,\eta_{p},\wsigma)\leq L(\wmu,\wbbe,\weta_1,\dots,\weta_p,\wsigma)$, which together with \eqref{eq:AA} leads to
$$L(\mu,\bbe,\eta_{1},\dots,\eta_{p},\wsigma)<b+\delta_2=L(\mu,\bbe,\eta_{1},\dots,\eta_{p},\sigma-\delta)\,,$$
 which entails that $\wsigma\geq \sigma-\delta$ for any $n\geq n_1$, concluding the proof.\, \qed

\vskip0.1in
\begin{lemma}{\label{lemma:lema2}}
Assume that $\rho$ satisfies \ref{ass:rho_bounded_derivable} and let $\itV=[\sigma_1,\sigma_2]$ with $0<\sigma_1<\sigma_2$ some neighbourhood of the errors scale $\sigma$.  Then, the function $L\left(a,\bb,g_1,\dots,g_p,\varsigma\right)$ satisfies the following equicontinuity condition: for any $\nu>0$ there exists $\delta>0$ such that for any $\varsigma_1,\varsigma_2\in \itV$, 
$$|\varsigma_1-\varsigma_2|<\delta\Rightarrow 
\mathop{\sup_{   a\in \real, \bb \in \real^q}}_{g_1 \in \itS_{1}, \dots, g_p \in \itS_{p}}|L(a,\bb,g_1,\dots,g_p,\varsigma_1)-L(a,\bb,g_1,\dots,g_p,\varsigma_2)|<\nu\,.$$
\end{lemma}

\noindent\textsc{Proof.} Let $\nu>0$ and $a\in\real$, $\bb\in\real^q$ and  $g_j(x)=\sum_{s=1}^{k_j-1} c_s^{(j)} B_s^{(j)}(x)\in \itS_j$, $1\le j\le p$. Using a Taylor expansion of order one and taking into account that $\rho^\prime=\psi$, we have that
\begin{eqnarray*}
&& L(a,\bb,g_1,\dots,g_p,\varsigma_1)-L(a,\bb,g_1,\dots,g_p,\varsigma_2) \\
&&= \esp\left[\psi\left(\frac{Y_1-a-\bb\trasp\bZ-\sum_{j=1}^p g_j(X_{j}) }{\xi}\right) \, \frac{Y_1-a-\bb\trasp\bZ-\sum_{j=1}^p g_j(X_{j})}{\xi^2}(\varsigma_2-\varsigma_1)\right] 
\end{eqnarray*}
where $\xi$ is an intermediate point between $\varsigma_1$ and $\varsigma_2$, so $ \xi \geq \sigma_1>0$. Using  that $\zeta(s)=s\psi(s)$ is bounded, we get the bound
$$|L(a,\bb,g_1,\dots,g_p,\varsigma_1)-L(a,\bb,g_1,\dots,g_p,\varsigma_2)|\leq \frac{\|\zeta\|_{\infty}}{\sigma_1}|\varsigma_2-\varsigma_1|\,.$$
Finally, by taking $\delta=\sigma_1\nu/(2\zeta)$ and noting that the bound does not depend on $a,\bb,g_1,\dots,g_p$, the result follows.\, \qed

\vskip0.1in 
In order to prove Theorem   \ref{teo:consist} and Proposition \ref{lemma:lema5}, we introduce some additional notation. 
The unit ball in $\itH_r$ will be denoted as $\itV_1^{(r)}=\left\{\eta\in\itH_r\,:\, \|\eta\|_{\itH_r}\leq 1\right\}$.
Besides, denote $\itB_q=\{\bb\in\real^q\,:\, \|\bb\|\leq 1\}$ the unit ball in $\real^q$.
The following result is needed to derive Theorem   \ref{teo:consist}.

\vskip0.1in

\begin{lemma}{\label{lemma:lema4}}
Assume that $\rho$ satisfies \ref{ass:rho_bounded_derivable} and that $L(\mu,\bbe,\eta_1,\dots,\eta_p,\sigma)=b_\rho<1$. Let $(\wmu, \wbbe,\weta_1,\dots,\weta_p)\in \real\times\real^q\times\itS_{1}\times\dots\times\itS_{p}$ be such that $L(\wmu,\wbbe,\weta_1,\dots,\weta_p,\sigma)\convpp L(\mu,\bbe,\eta_1,\dots,\eta_p,\sigma)$. Assume that $\esp\|\bZ\|^2<\infty$ and that \ref{ass:etajCr} and \ref{ass:proba} hold with $c<1-b_\rho$. Then, we have that there exists $M$ such that $$\prob\left(\bigcup_{m\in\natu}\bigcap_{n\geq m}\left\{|\wmu-\mu|+\|\wbbe-\bbe\|+\sum_{j=1}^p{\|\weta_j-\eta_j}\|_{\itH_j}\leq M\right\}\right)=1$$
\end{lemma}

\noindent\textsc{Proof.} 
Given $\delta>0$, let $C_\delta$ be such for for any $C\ge C_{\delta}$
\begin{equation}
\label{eq:ZC} 
\prob(\|\bZ\|\ge C)<\delta\,.
\end{equation}
Fix $\bt=(a,\bb,g_1,\dots,g_p)\in \itB=\{(a^{*}, \bb^{*}, g_1^{*},\dots, g_p^{*})\in\real\times\real^q\times C([0,1])\times\dots\times C([0,1])\,: |a^{*}|+\|\bb^{*}\|+ \sum_{j=1}^p \|g_j^{*}\|_{\infty}=1 \} [-1,1]\times \itB_q\times \itV_1^{(r_1)}\times\dots\times\itV_1^{(1)}$. Assumption \ref{ass:proba} allows to select a positive real number $\phi_\bt$ such that   $\phi_\bt$ is a continuity point of $\left|a+\bb\trasp\bZ+\sum_{j=1}^p g_j(X_j)\right|$ and
\begin{equation}
\label{eq:phibt}
\prob\left(\left|a+\bb\trasp\bZ+\sum_{j=1}^p g_j(X_j)\right|<\phi_\bt\right)<c\,.
\end{equation}
Denote $\vartheta_\bt= \phi_\bt /2$ and $v_\bt= \phi_\bt/(2(C+p+1))$. Then, if $a^{*}\in \real$, $\bb^{*}\in\real^q$ and $g^{*}_1,\dots,g_p^{*}\in C([0,1])$ are such that $|a-a^{*}|+\|\bb^{*}-\bb\|+\max_{1\leq j\leq p}\{\|g_j^{*}-g_j\|_{\infty}\}<v_\bt$, we have that 
\begin{align*}
\prob\left(\left|a^{*}+\bZ\trasp \bb^{*} +\sum_{j=1}^p g_j^{*}(X_j)\right| \geq
 \vartheta_\bt\right) & \geq \prob\left(\left|a+\bb\trasp\bZ+\sum_{j=1}^p g_j(X_j)\right|\geq  \phi_\bt \right)\\
 &-\prob\left(|a-a^{*}|+\|\bb^{*}-\bb\| \|\bZ\|+ \left|\sum_{j=1}^p (g_j^{*}-g_j)(X_j)\right| \geq \vartheta_\bt\right)
 \\ 
 & \ge  \prob\left(\left|a+\bb\trasp\bZ+\sum_{j=1}^p g_j(X_j)\right|\geq  \phi_\bt \right)-\prob\left(v_\bt (p+1+ \|\bZ\|)\geq \vartheta_\bt\right)=A(\bt)
\end{align*}
 Hence, noting that \eqref{eq:phibt} and \eqref{eq:ZC} imply that $A(\bt)>1-c-\delta$, we conclude that
 \begin{equation}\label{eq:infbound}
 \inf_{|a-a^{*}|+\|\bb^{*}-\bb\|+\max_{1\leq j\leq p}\{\|g_j^{*}-g_j\|_{\infty}\}<v_\bt} \prob\left(\left|a^{*}+ \bZ\trasp \bb^{*}+\sum_{j=1}^p g_j^{*}(X_j)\right|\geq
 \vartheta_\bt\right) \geq 1-c-\delta\,.
 \end{equation}
 Let us considering the covering of $\itB$ given by $\{B(\bt,v_\bt)\}_{\bt\in\itB}$, where $B(\bt,\nu)$ is defined as 
 $$B(\bt,\nu)=\left\{(a^{*}, \bb^{*}, g_1^{*},\dots, g_p^{*})\in\real\times\real^q\times C([0,1])\times\dots\times C([0,1])\,:\, |a-a^{*}|+\|\bb^{*}-\bb\|+\max_{1\leq j\leq p}\{\|g_j^{*}-g_j\|_{\infty}\} <\nu \right\}\,.$$
The fact that $\itV_1^{(1)}\times\dots\times\itV_1^{(1)}$ is a compact set in $C([0,1])\times\dots\times C([0,1])$ entails that there exists $\bt_\ell=(a_\ell,\bb_\ell,g_{1,\ell},\dots,g_{p,\ell})\in\itB$, for $1\leq \ell\leq s$, such that $\itB\subset\bigcup_{\ell=1}^{s}B(\bt_\ell,v_\ell)$ where we have denoted $v_\ell=v_{\bt_\ell}$. Therefore, from \ref{eq:infbound} we obtain that
$$\min_{1\leq \ell\leq s}\,\,\inf_{|a-a_{\ell}|+ \|\bb-\bb_{\ell}\|+ \max\{\|g_j-g_{j,\ell}\|_{\infty}\}<v_\ell} \prob\left(\left|a+\bb\trasp\bZ+\sum_{j=1}^p g_j(X_j)\right|>\vartheta_\ell\right)>1-c-\delta\,,$$
where $\vartheta_\ell= \vartheta_{\bt_\ell}$. Henceforth, for any $\bt=(a,\bb,g_1,\dots,g_p)\in\itB$, there exists $1\leq\ell\leq s$ such that
\begin{equation}\label{eq:eqell}
\prob\left(\left|a+\bb\trasp\bZ+\sum_{j=1}^p g_j(X_j)\right|>\vartheta_\ell\right)>1-c-\delta\,.
\end{equation}
Let $\itN$ be such that $\prob(\itN)=0$ and for each $\omega\notin\itN$, $L(\wmu,\wbbe,\weta_1,\dots,\weta_p,\sigma)\to L(\mu,\bbe,\eta_1,\dots,\eta_p,\sigma)=b_\rho$. Fix $\omega\notin\itN$ and let $\xi>0$ such that $b_\rho+\xi<1-c$. Then, there exists $n_0\in\natu$ such that for all $n\geq n_0$, $L(\wmu,\wbbe,\weta_1,\dots,\weta_p,\sigma)\leq b_\rho+\xi/2$.

We want to show that there exists $M>0$ such that, for $\omega\notin\itN$, $\limsup_{n\to \infty}\left\{|\wmu-\mu|+\|\wbbe-\bbe\|+\right. $ $\left.\sum_{j=1}^p\|\weta_j-\eta_j\|_{\itH_{1}}\right\}\leq M$. For that purpose, it will be enough to show that there exists $M>0$ such that 
$$\inf_{|a-\mu|+\|\bb-{\bbech}\|+\sum_{j=1}^p\|g_j-\eta_j\|_{\itH_{1}}>M}L(a,\bb,g_1,\dots,g_p)\geq b_\rho+\xi\,.$$  
Denote as $R(u)=\esp\rho\left(\eps-u/\sigma\right)$. First note that the independence between the errors and the covariates entails that
\begin{eqnarray*}
L(a,\bb,g_1,\dots,g_p,\sigma)&=&\esp\rho\left(\eps+\frac{\mu-a+(\bbe-\bb)\trasp\bZ+\sum_{j=1}^p (\eta_j-g_j)(X_j)}{\sigma}\right)\\
&=&\esp R\left( a-\mu +(\bb-\bbe)\trasp\bZ+\sum_{j=1}^p (g_j-\eta_j)(X_j)\right)\,.
\end{eqnarray*}
Using that $\lim_{|u|\to+\infty}R(u)=1$, we get that for any $\delta>0$, there exists $u_0$ such that $|u|\geq u_0$,
\begin{equation}\label{eq:Ru}
R(u)>1-\delta\,.
\end{equation}
Choose $M>u_0/(\min_{1\leq\ell\leq s}\vartheta_\ell)$ and let $(a_k,\bb_k,g_{1,k},\dots,g_{p,k})\in\real\times\real^q\times \itH_{1}\times\dots\times\itH_{1}$ such that $\nu_k=|a_k-\mu|+\|\bb_k-\bbe\| +\sum_{j=1}^p \|g_{j,k}-\eta_j\|_{\itH_1}>M$ and 
$$L(a_k,\bb_k,g_{1,k},\dots,g_{p,k},\sigma)\to\inf_{|a-\mu|+\|\bb-\bbech\| +\sum_{j=1}^p \|g_{j}-\eta_j\|_{\itH_1}>M} L(a,\bb,g_{1},\dots,g_{p},\sigma)\,.$$
Denote as $\wta_k=(a_k-\mu)/\nu_k$, $\wtbb_k=(\bb_k-\bbe)/\nu_k$ and $\wtg_{j,k}=(g_{j,k}-\eta_j)/\nu_k$, for $1\leq j\leq p$. Then $(\wta_k,\wtbb_k,\wtg_{1,k},\dots,\wtg_{p,k})\in\itB$. Thus, using \ref{eq:eqell} we obtain that there exists $1\leq \ell=\ell(k)\leq s$ such that 
\begin{equation}\label{eq:eqtilde}
\prob\left(\left|\wta_k+\wtbb_k\trasp\bZ+\sum_{j=1}^p \wtg_{j,k}(X_j)\right|>\vartheta_\ell\right)>1-c-\delta\,.
\end{equation}
Using that $\nu_k>M>u_0/\vartheta_\ell$ and denoting $u_k(\bZ,\bX)=\nu_k \left(\wta_k+\wtbb_k\trasp\bZ+\sum_{j=1}^p \wtg_{j,k}(X_j)\right)$, we obtain that $|u_k(\bZ,\bX)|>u_0$ whenever $\left|\wta_k+\wtbb_k\trasp\bZ+\sum_{j=1}^p\wtg_{j,k}(X_j)\right|>\vartheta_\ell$, which together with \ref{eq:Ru} leads to
\begin{eqnarray*}
L(a_k,\bb_k,g_{1,k},\dots,\wtg_{p,k},\sigma)&=&\esp R\left( a_k-\mu+(\bb_k-\bbe)\trasp\bZ+\sum_{j=1}^p (g_{j,k}-\eta_j)(X_j)\right)\\
&=&\esp R(u_k(\bZ,\bX))\\
&\geq & \esp R(u_k(\bZ,\bX))\indica_{|\wta_k+\wtbb_k\trasp\bZ+\sum_{j=1}^p \wtg_{j,k}(X_j)|>\vartheta_\ell}\\
&>&(1-\delta)\,\prob\left(\left|\wta_k+\wtbb_k\trasp\bZ+\sum_{j=1}^p \wtg_{j,k}(X_j)\right|>\vartheta_\ell\right)\\
&>& (1-\delta)(1-c-\delta)\,,
\end{eqnarray*}
where the last inequality follows from \ref{eq:eqtilde}. Therefore, 
$$\inf_{|a-\mu|+\|\bb-\bbech\| +\sum_{j=1}^p \|g_{j}-\eta_j\|_{\itH_1}>M} L(a,\bb,g_{1},\dots,g_{p},\sigma)\geq (1-\delta)(1-c-\delta)\,.$$
The proof follows now easily noting that $\lim_{\delta\to 0}(1-\delta)(1-c-\delta)=1-c>b_\rho+\xi$, so we can choose $\delta$ and consequently $M$ such that
$$
\inf_{|a-\mu|+\|\bb-\bbech\| +\sum_{j=1}^p \|g_{j}-\eta_j\|_{\itH_1}>M} L(a,\bb,g_{1},\dots,g_{p},\sigma)>b_\rho+\xi>L(\wmu,\wbbe,\weta_1,\dots,\weta_p,\sigma)$$
so $|\wmu-\mu|+\|\wbbe-\bbe\| +\sum_{j=1}^p \|\weta_j-\eta_j\|_{\itH_1}\leq M$, concluding the proof. \qed

\vskip0.1in

\noindent\textsc{Proof of Theorem \ref{teo:consist}.}  
Let $V_{a,\bb,g_1,\dots,g_p,\varsigma}=\rho\left((y-a-\bb\trasp\bz-\sum_{j=1}^pg_j(x_{j}))/\varsigma\right)$ and denote as $P$ the probability measure of $(Y,\bZ\trasp,\bX\trasp)\trasp$ and as $P_n$ its corresponding empirical measure. Then, $L_n(a,\bb,g_1,\dots,g_p,\varsigma)=P_n V_{a,\bb,g_1,\dots,g_p,\varsigma}$ and $L(a,\bb,g_1,\dots,g_p,\varsigma)=P V_{a,\bb,g_1,\dots,g_p,\varsigma}$.

The consistency of $\wsigma$ entails that given any neighbourhood  $\itV$ of $\sigma$, there exists a null set $\itN_{\itV}$ such that for $\omega\notin\itN_{\itV}$, there exists $n_0\in\mathbb{N}$, such that for all $n\geq n_0$ we have that $\wsigma\in\itV$.

Lemma \ref{lema:A1} implies that
\begin{equation}\label{eq:A1}
A_n=\mathop{\sup_{ \varsigma >0, a\in \real, \bb \in \real^q}}_{g_1 \in \itS_{1}, \dots, g_p \in \itS_{p}}|L_n(a,\bb,g_1,\dots,g_p,\varsigma)-L(a,\bb,g_1,\dots,g_p,\varsigma)|\convpp 0\,.
\end{equation}
On the other hand, from  Lemma \ref{lemma:FC} we have 
$$L(\mu, \bbe, \eta_1,\dots, \eta_p,\sigma)=\min_{a\in \real, \bb\in\real^q\,,g_1\in\itG,\dots,g_p\in\itG}L(a,\bb,g_1,\dots,g_p,\sigma)\,,$$ 
so, we have that
\begin{equation}\label{eq:A2}
0\leq L(\wbthe,\sigma )-L(\bthe,\sigma)=\sum_{s=1}^3 A_{n,s}
\end{equation}
with $A_{n,1}=L(\wbthe,\wsigma)-L_n(\wbthe,\wsigma)$, $A_{n,2}=L_n(\wbthe,\wsigma)-L(\bthe,\sigma)$ and $A_{n,3}=L(\wbthe,\sigma)-L(\wbthe,\wsigma)$.  Note that $|A_{n,1}|\leq A_n$, hence $A_{n,1}=o_{\as}(1)$. On the other hand,   Lemma \ref{lemma:lema2} and  \ref{ass:wsigma} imply that  $A_{n,3}=o_{\as}(1)$.

It remains to see that $A_{n,2}=o_{\as}(1)$. As in the proof of Proposition \ref{prop:prop1}, Corollary 6.21 in Schumaker (1981) entails that, for $1\le j\le p$,  there exists a centered spline $\widetilde{\wteta}_j$ such that $\widetilde{\wteta}_j\in \itS_j$ and $\|\widetilde{\wteta}_j-\eta_{j}\|_{\infty}=O\left(n^{-\nu_j\, r_j}\right)$.

Denote $\bthe_n=\left(\mu,\bbe,\widetilde{\wteta}_1,\dots,\widetilde{\wteta}_p\right)$  and let $S_{n,1}=(P_n-P)V_{\mu,\bbech,\widetilde{\wteta}_1,\dots,\widetilde{\wteta}_p,\wsigma}=L_n(\bthe_n,\wsigma)-L(\bthe_n,\wsigma)$ and $S_{n,2}=L(\bthe_n,\wsigma)-L(\bthe,\sigma)$. Note that $S_{n,1}\leq A_n$, so that from \eqref{eq:A1} we get that $S_{n,1}\convpp 0$. On the other hand, if we write $S_{n,2}=\sum_{s=1}^2 S_{n,2}^{(s)}$ where $S_{n,2}^{(1)}=L(\bthe_n,\wsigma)-L(\bthe_n,\sigma)$ and $S_{n,2}^{(2)}=L(\bthe_n,\sigma)-L(\bthe,\sigma)$, using that $\rho$ is a bounded continuous function, together with the fact that $\|\widetilde{\wteta}_j-\eta_{j}\|_{\infty}\to 0$ for all $j=1,\dots,p$ and the dominated convergence theorem we have that $S_{n,2}^{(2)}=o_{\as}(1)$. Besides, from Lemma \ref{lemma:lema2} and the strong consistency of $\wsigma$, we obtain that $S_{n,2}^{(1)}=o_{\as}(1)$. Then, $S_{n,2}=o_{\as}(1)$.

Using that $\wbthe$ minimizes $L_n$ over  $\real\times\real^q\times \itS_1\times\dots\times \itS_p$, we obtain that
\begin{equation}\label{eq:A3}
A_{n,2}=L_n(\wbthe,\wsigma)-L(\bthe,\sigma)\leq L_n(\bthe_n,\wsigma)-L(\bthe,\sigma)=S_{n,1}+S_{n,2} \,.
\end{equation}
Hence, from \eqref{eq:A2}  and  \eqref{eq:A3} and using that $A_{n,s}=o_{\as}(1)$ for $s=1,3$ and that $S_{n,s}=o_{\as}(1)$ for $s=1,2$, we obtain that
$
0\leq L(\wbthe,\sigma)-L(\bthe ,\sigma)=\sum_{j=1}^3 A_{n,j}\leq A_{n,1}+S_{n,1}+S_{n,2}+A_{n,3}= o_{\as}(1)
$,
so   $L(\wbthe,\sigma)\convpp L(\bthe ,\sigma)$. Thus, Lemma \ref{lemma:lema4} implies that  there exists $M$ such that $$\prob\left(\bigcup_{m\in\natu}\bigcap_{n\geq m}\left\{|\wmu-\mu|+\|\wbbe-\bbe\|+\sum_{j=1}^p{\|\weta_j-\eta_j}\|_{\itH_j}\leq M\right\}\right)=1\,,$$
and the proof follows now  from the fact that,  for any $\delta>0$,  if $\inf_{\bt\in\itA_{\delta}}L(a,\bb,g_1,\dots,g_p,\sigma)>L(\mu,\bbe,\eta_{1},\dots,\eta_{p},\sigma)$ then $\pi(\wbthe,\bthe)\convpp 0$.   \qed

\vskip0.1in

We now can proceed with the proof of Proposition \ref{lemma:lema5}.

\vskip0.1in
\noindent\textsc{Proof of  Proposition \ref{lemma:lema5}.}
 Recall that 
 $$\itA_{\delta}=\{\bt=(a,\bb,g_1,\dots,g_p)\,:\, a\in\real, \bb\in\real^q, g_j\in\itG\cap \itH_{r_j},\, |a-\mu|+\|\bb-\bbe\|+\sum_{j=1}^p{\|g_j-\eta_j}\|_{\itH_{r_j}}\leq M,  \pi(\bthe,\bt) \ge \delta\}$$
  where $\bthe=(\mu,\bbe,\eta_1,\dots,\eta_p)$. As in Lemma \ref{lemma:lema4}, let $\bt_k=(a_k,\bb_k,g_{1,k},\dots,g_{p,k})\in\itA_{\delta}$ be such that 
$$L(\bt_k,\sigma)\to \inf_{\bt\in\itA_{\delta}} L(\bt,\sigma)$$
and denote $\nu_k=|a_k-\mu|+\|\bb_k-\bbe\| +\sum_{j=1}^p \|g_{j,k}-\eta_j\|_{\itH_1}$. Using that $\bt_k\in\itA_{\delta}$, we get that the sequences $\{g_{j,k}-\eta_j\}_{k\ge 1}$ and their first derivatives   are uniformly bounded. Hence, the compactness of $\{(a,\bb)\in\real\times\real^q: |a-\mu|+\|\bb-\bbe\|\le M \}$ and the Arzela-Ascoli Theorem imply that there exists a subsequence $k_\ell$ such that $d_\ell=a_{k_\ell}-\mu\to d$, $\be_\ell=\bb_{k_\ell}-\bbe\to \be$ for some   $d\in \real$  and  $\be\in \real^q$, while   $f_{j,\ell}=g_{j,k_\ell}-\eta_j$, for $1\leq j\leq p$,  converge uniformly to some continuous functions $f_1,\dots,f_p$, respectively. Denote $\wta=d+\mu$, $\wtbb=\be+\bbe$, $\wtg_j=f_j+\eta_j$, $1 \le j\le p$,  the uniform limit of $a_{k_\ell}$, $\bb_{k_\ell}$ and $g_{j,k_\ell}$, $1 \le j\le p$, respectively. Denote $\wtbt=(\wta,\wtbb,\wtg_1,\dots,\wtg_p)$ and $\wtbt_{k_\ell}=(\wta_{k_\ell},\wtbb_{k_\ell},\wtg_{1,k_\ell},\dots,\wtg_{p,k_\ell})$. Then, we have that $\pi(\wtbt,\wtbt_{k_\ell})= |a_{k_\ell}-\wta|+\|\bb_{k_\ell}-\wtbb\| +\sum_{j=1}^p \| g_{j,k_\ell}-\wtg_j\|_\infty\to 0$. The fact that $\rho_1$ is a bounded continuous function  and the Bounded Convergence Theorem imply that $L(\wtbt_{k_\ell},\sigma)\to L(\wtbt,\sigma)$ which leads to $\inf_{\bt\in\itA_\delta} L(\bt,\sigma)=L(\wtbt,\sigma)$. Furthermore, $\pi(\wtbt,\bthe)\geq \delta$    since $\pi(\wtbt_{k,\ell},\bthe)\ge \delta$, $\pi(\wtbt,\wtbt_{k_\ell})\to 0$ and $\pi(\wtbt,\bthe)\geq  \pi(\wtbt_{k,\ell},\bthe)-\pi(\wtbt,\wtbt_{k_\ell})$, hence  from Lemma \ref{lemma:FC}  we get that $L(\wtbt,\sigma)>L(\bthe,\sigma)$   concluding the proof. \qed


\subsection{Proof of Theorem \ref{teo:ratesnew}}

Let denote as $\Theta=\real\times\real^q\times \itH_{r_1}\times\dots\times\itH_{r_p}$ and as $\Theta_n=\left\{\bt\in\Theta: g_j\in\itS_j, 1\leq j\leq p, \pi(\bt,\bthe)\leq \epsilon_0\right\}$ where $\epsilon_0$ is given in assumption \ref{ass:lowerbound}. Proposition  \ref{lemma:lema5} implies that $\pi(\wbthe, \bthe)\convpp 0$ where $\wbthe=(\wmu,\wbbe\trasp,\weta_1,\dots,\weta_p)\trasp$ are defined through \eqref{eq:estfinitos} and \eqref{eq:estimadoresgj}. Therefore, except for a null probability set, $\wbthe\in\Theta_n$, for $n$ large enough.

The following Lemma gives conditions under which \ref{ass:lowerbound} holds. Its proof follows the same arguments as those considered in Boente \textsl{et al.} (2021).

\begin{lemma}{\label{lemma:lemacond1}}
Let $\rho$ be a function satisfying \ref{ass:rho_bounded_derivable} and such that $\rho^\prime=\psi$ is continuously differentiable with bounded derivative $\psi^\prime$ and $\esp\psi^\prime(\eps)>0$. If there exists $C>0$ such that $\prob(\|\bZ\|\leq C)=1$, then \ref{ass:lowerbound} holds.
\end{lemma}

\noindent\textsc{Proof.} Using a Taylor expansion of order two, we have that 
\begin{eqnarray*}
L(\bt,\varsigma)-L(\bthe,\varsigma)&=&\esp\left[\rho\left(\frac{Y-a-\bb\trasp\bZ-\sum_{j=1}^p g_j(X_j)}{\varsigma}\right)-\rho\left(\frac{\sigma\eps }{\varsigma}\right)\right]\\
&=&\esp\left[\rho\left(\frac{\sigma\eps -(a-\mu)-(\bb-\bbe)\trasp\bX-\sum_{j=1}^p (g_j-\eta_j)(X_j)}{\varsigma}\right)-\rho\left(\frac{\sigma\eps }{\varsigma}\right)\right]\\
&=& \esp\left[\psi\left(\frac{\sigma\eps }{\varsigma}\right)\left((a-\mu)+(\bb-\bbe)\trasp\bZ+\sum_{j=1}^p(g_j-\eta_j)(X_j)\right)\right]\\
&&+\frac{1}{2}\esp\left[\psi\left(\frac{\sigma\eps +\xi}{\varsigma}\right)\left((a-\mu)+(\bb-\bbe)\trasp\bZ+\sum_{j=1}^p(g_j-\eta_j)(X_j)\right)^2\right]\\
&=&\frac{1}{2}\esp\left[\psi\left(\frac{\sigma\eps +\xi}{\varsigma}\right)\left((a-\mu)+(\bb-\bbe)\trasp\bZ+\sum_{j=1}^p(g_j-\eta_j)(X_j)\right)^2\right]
\end{eqnarray*}
where $\xi$ is an intermmediate point between $G(\bZ,\bX)=a-\mu+(\bb-\bbe)\trasp\bZ+\sum_{j=1}^p(g_j-\eta_j)(X_j)$ and $0$. Noting that $|(a-\mu)+(\bb-\bbe)\trasp\bZ+\sum_{j=1}^p(g_j-\eta_j)(X_j)|\leq |a-\mu|+\|\bb-\bbe\| \|\bZ\|+\sum_{j=1}^p\|g_j-\eta_j\|_\infty$, if $|a-\mu|+\|\bb-\bbe\| +\sum_{j=1}^p \|g_j-\eta_j\|_\infty< \epsilon_0$ and recalling that $P(\|\bZ\|\leq C)=1$, we get that $|\xi|< (C+2)\epsilon_0$ with probability $1$. 

The fact that $\varphi=\esp \psi^\prime(\eps )>0$ and the continuity of $\psi^\prime$ entail that for $\delta$ small enough
$$\inf_{\varsigma>0,\, |\varsigma-\sigma|<\delta,\, |d|<\delta}\esp\psi^\prime\left(\frac{\sigma\eps +d}{\varsigma}\right)>\frac{\varphi}{2}>0\,.$$
Hence, if $\itV=\{\varsigma>0\,:\, |\varsigma-\sigma|<\delta\}$ and $\epsilon_0=\delta/(C+2)$, we have that
\begin{eqnarray*}
L(\bt,\varsigma)-L(\bthe,\varsigma)&=&\frac{1}{2}\esp\left[\psi^{\prime}\left(\frac{\sigma\eps +\xi}{\varsigma}\right)\left((a-\mu)+(\bb-\bbe)\trasp\bZ+\sum_{j=1}^p(g_j-\eta_j)(X_j)\right)^2\right]\\
&=&\frac{1}{2}\esp\left[\left((a-\mu)+(\bb-\bbe)\trasp\bZ+\sum_{j=1}^p(g_j-\eta_j)(X_j)\right)^2 \esp\left\{\psi^{\prime}\left(\frac{\sigma\eps +\xi}{\varsigma}\right)\Big | (\bZ,\bX)\right\}\right]\\
&> & \frac{\varphi}{4}\esp\left[\left((a-\mu)+(\bb-\bbe)\trasp\bZ+\sum_{j=1}^p(g_j-\eta_j)(X_j)\right)^2\right]=\frac{\varphi}{4}\pi^2_\prob(\bt,\bbe)\,,
\end{eqnarray*}  
concluding the proof. \qed

For each $1\leq j\leq p$ and $\bc=(c_1,\dots, c_{k_j-1})\trasp\in\real^{k_j-1}$, to strength the dependence on the coefficients $c_s$, $1\le s\le k_j-1$, we denote $g_{j,\bc}(x)=\sum_{s=1}^{k_j-1}c_sB_s^{(j)}(x)$. In order to prove Theorem \ref{teo:ratesnew}, we will need the following Lemma whose proof follows similar arguments to those considered in Lemma S.2.5 in Boente \textsl{et al.} (2020).

\begin{lemma}{\label{lemma:lemabacketing}}
Let $\rho$ be a  function satisfying \ref{ass:rho_bounded_derivable}. Given fixed values $\bc_{0,j}\in\real^{k_j-1}$, $1\leq j\leq p$,   let $\bt_0=( \mu,\bbe\trasp,g_{0,1},\dots, g_{0,p})\trasp\in\real\times\real^q\times\itS_1\times\dots\times\itS_p$ be such that $ g_{0,j}=g_{j,\bc_{0,j}}$. Define the class of functions
$$\itG_{n,\kappa,\bt^{\star},\bt_0}=\left\{f_{\bt,\varsigma}=V_{\bt,\varsigma}-V_{\bt^{\star},\varsigma}\,:\, \pi(\bt, \bt_0) <\kappa,\, \bt\in\Theta_n,\, \varsigma\in \itV=[\sigma_1,\sigma_2]\right\}$$
with $\sigma_1=\sigma/2$, $\sigma_2=(3/2)\sigma$, $\bt^{\star}=(a^{\star}, \bb^{\star}, g_1^{\star}, \dots, g_p^{\star})\in \Theta$ is a fixed point and 
$$V_{\bt,\varsigma}=V_{\bt,\varsigma}(y,\bz,\bx)=\rho\left(\frac{y-a-\bb\trasp\bz-\sum_{j=1}^p g_j(x_j)}{\varsigma}\right)$$
for $\bt=(a,\bb\trasp,g_1,\dots,g_p)\trasp$. Assume that $\esp\|\bZ\|^2< \infty$. Then,  there exists some constants  $A>0$  and $A^{\star}$ independent of $n$, $\bc_{0,j}$ and $\epsilon$ such that
$$N_{[\,\,]}\left(\epsilon,\itG_{n,\kappa,\bt^{\star},\bt_0},L_2(P)\right)\leq A\left(\frac{A^{\star}\max(1,\kappa)}{\epsilon}\right)^{K+p+q+2}\,.$$
\end{lemma}
 
\noindent\textsc{Proof.}  First note that, for any $g\in \itS_j$, there exists $\bc=(c_1,\dots, c_{k_j-1})\trasp\in\real^{k_j-1}$, such that $g(x)=g_{\bc}(x)=\sum_{s=1}^{k_j-1}c_s B_s^{(j)}(x)$. Furthermore, taking into account that $ B_s^{(j)}(x)=\wtB_s^{(j)}(x)- \int_{\itI_j}\wtB_s^{(j)}(x) dx$  and that $\sum_{s=1}^{k_j} \wtB_s^{(j)}(x)=1$, we get that
\begin{align*}
g_{\bc}(x) &=\sum_{s=1}^{k_j-1}c_s B_s^{(j)}(x)= \sum_{s=1}^{k_j-1}c_s \wtB_s^{(j)}(x)-  \sum_{s=1}^{k_j-1}c_s \int_{\itI_j}\wtB_s^{(j)}(t) dt\\
&= \sum_{s=1}^{k_j-1}c_s \wtB_s^{(j)}(x)-  \left\{\sum_{\ell=1}^{k_j-1}c_\ell \int_{\itI_j}\wtB_\ell^{(j)}(t) dt\right\} \sum_{s=1}^{k_j} \wtB_s^{(j)}(x)
= \sum_{s=1}^{k_j} \lambda_{s, \bc} \wtB_s^{(j)}(x)\,,
\end{align*}
where $ \lambda_{s, \bc} = c_s -  \sum_{\ell=1}^{k_j-1}c_\ell \int_{\itI_j}\wtB_\ell^{(j)}(t) dt$, for $1\le s\le k_j-1$ and $  \lambda_{k_j, \bc} = \,-\, \sum_{\ell=1}^{k_j-1}c_\ell \int_{\itI_j}\wtB_\ell^{(j)}(t) dt$. Therefore,   there exists a constant $D_j$, $D_j\leq 1$, depending only on the degree $\ell_j$ of the considered splines such that 
\begin{equation}\label{eq:s13}
 D_j \|\bla_{\bc} \|_\infty\leq \|g_{ \bc }\|_\infty\leq \|\bla_{\bc} \|_\infty
\end{equation}  
where $\bla_{\bc}=( \lambda_{1, \bc}, \dots,  \lambda_{k_j, \bc})\trasp $  and for a vector $\ba\in\real^{k}$, $\|\ba\|_\infty=\max_{1\leq s\leq k}|a_s|$ (see de Boor, 1973, Section 3).

Thus if we denote as
\begin{align*}
\itH_{\kappa, g_{0,j}}^{(j)} & =\left\{g_{\bc}(x)=\sum_{s=1}^{k_j-1}c_s B_s^{(j)}(x)\,, \bc \in \real^{k_j-1}, \|g_{\bc}- g_{0,j}\|_{\infty} \le \kappa \right\}\,.
\end{align*}
 we have that $\itH_{\kappa, g_{0,j}}^{(j)}\subset \{\sum_{s=1}^{k_j} a_{s} \wtB_s^{(j)}(x), \ba \in \itB_{\blach_0, k_j}(\kappa_1)\}$ with  $\kappa_1=\kappa/  D  $, $D=\min_{1\le j\le p} D_j$, $\itB_{\blach_0, k_j}(\delta)  = \{\ba \in \real^{k_j}: \|\ba - \bla_0\|_{\infty} <\delta\}$, $\ba=  ( a_1,\dots,a_{k_j})\trasp$ and $\bla_0=\bla_{\bc_0}$.

Recall that the ball $\itB_{\blach_0,k}(\delta)$ can be covered by at most $\{(4\delta+\epsilon)/\epsilon\}^k$ balls of radius $\epsilon$, when $\epsilon<\delta$, while if $\epsilon>\delta$ the covering number equals $1$. Hence, using the upper bounds given in \eqref{eq:s13} and using that for any class of functions $\itH$, $N_{[\,\,]}(\epsilon,\itH, L_\infty)\leq N(\epsilon,\itH,L_\infty)$, we obtain that 
\begin{equation}\label{eq:s14}
\log N_{[\,\,]}(\epsilon,\itH_{\kappa, g_{0,j}}^{(j)},L_\infty)\leq k_j\log(5\, \kappa_1/\epsilon)
\end{equation}
for $0<\epsilon<\kappa_1$. Henceforth, using \eqref{eq:s14}, we get that, for any $0<\epsilon<\kappa_1$, $\itH_{\kappa, g_{0,j}}^{(j)}$ can be covered by a finite number $M_j(\epsilon)\leq (5(\kappa_1)/\epsilon)^{k_j}$ of $\epsilon-$brackets $\{[g_{\ell,L}^{(j)},g_{\ell,U}^{(j)}], 1\leq \ell\leq M_j(\epsilon)\}$, $1\leq j\leq p$,  that is,  for any $g\in \itH_{\kappa, g_{0,j}}^{(j)}$, we have that $g_{\ell,L}^{(j)}\leq g\leq g_{\ell,U}^{(j)}$ and $\|g_{\ell,U}^{(j)}-g_{\ell,L}^{(j)}\|_\infty<\epsilon $.

Analogously, the ball $\itB_{ \mu,1}(\delta)$ can be covered by at most $(4\delta+\epsilon)/\epsilon$ balls of radius $\epsilon$ and $\itB_{\bbech,q}^{(2)}(\delta)$  by at most $\{(4\delta+\epsilon)/\epsilon\}^q$, where  $\itB_{\bbech,q}^{(2)}(\delta)=\{\bb \in \real^q: \|\bb-\bbe\|<\delta\}$ is the usual euclidean ball. Note that for any  $f_{\bt,\varsigma}\in  \itG_{n,\kappa,\bt^{\star},\bt_0}$ we have that $a\in \itB_{ \mu,1}(\kappa)$, $\bb \in \itB_{\bbech,q}^{(2)}(\kappa)$. Thus,  for any $0<\epsilon<\kappa$, $\itB_{ \mu,1}(\kappa)$  can be covered by a finite number $M_{ a_0}(\epsilon)\leq 5\kappa/\epsilon$ of $\epsilon-$balls with centers  $a^{(r)}$, $1\leq r\leq M_{a_0}(\epsilon)$. Similarly,   $\itB_{\bbech,q}^{(2)}(\kappa)$ can be covered by a finite number $M_{\bbech}(\epsilon)\leq (5\kappa/\epsilon)^q$ of $\epsilon-$balls of centers   $\bb^{(m)}$.

On the other hand, the set $\itV=[\sigma_1,\sigma_2]=\{\varsigma\,:\, |\varsigma-\sigma|\leq \sigma/2\}$ can be covered by $M_\sigma(\epsilon)\leq C_\sigma(1/\epsilon)$ balls of radius $\epsilon$ (when $\epsilon<\sigma/2$) and centers $\sigma^{(s)}$, $1\leq s\leq M_\sigma(\epsilon)$, where $C_\sigma=3\sigma$.

Recall that $\psi$ is bounded, so that, for $\varsigma\in\itV$, 
$$\left|\frac{\partial}{\partial u}\rho\left(\frac{y-u}{\varsigma}\right)\right|\leq \frac{\|\psi\|_\infty}{\varsigma}\leq 2\frac{\|\psi\|_\infty}{\sigma}\,.$$
Define $\epsilon_1=\epsilon/A_1$ where 
$$A_1=\frac{4}{\sigma}\left[\|\psi\|_\infty(1+(\esp\|\bZ\|^2)^{1/2}+p)+2\|\zeta\|_\infty\right]\,.$$
Given $f_{\bt,\varsigma}\in\itG_{n,\kappa,\bt^{\star},\bt_0}$, let $r$, $m$, $\ell$ and $s$ be such that $1\leq r\leq M_{\mu}(\epsilon_1)$,  $1\leq m\leq M_{\bbech}(\epsilon_1)$, $1\leq \ell\leq M_j(\epsilon_1)$ and  $1\leq s\leq M_\sigma(\epsilon_1)$ and
$|a-a^{(r)}|<\epsilon_1$, $\|\bb-\bb^{(m)}\| <\epsilon_1$, for $1\leq j\leq p$,
$g_j\in [g_{\ell,L}^{(j)},g_{\ell,U}^{(j)}]$ with $\|g_{\ell,U}^{(j)}-g_{\ell,L}^{(j)}\|_\infty<\epsilon_1 $ and $|\varsigma-\sigma^{(s)}|<\epsilon_1$. Denote as
$$f_{r,m,\ell,\varsigma}(y,\bz,\bx)=\rho\left(\frac{y-a^{(r)}-\bb^{(m)\mbox{\footnotesize{\sc t}}}\bz-\sum_{j=1}^p g_{\ell,U}^{(j)}(x_j)}{\varsigma}\right)-\rho\left(\frac{y-a^{\star}- \bb^{\star\, \mbox{\footnotesize{\sc t}}} \bz-\sum_{j=1}^p  g_{j}^{\star}(x_j)}{\varsigma}\right)\,.$$
Using a Taylor's expansion of order one and the fact that $\zeta(u)=u\psi(u)$ is bounded, we get that
\begin{eqnarray*}
|f_{\bt,\varsigma}-f_{r,m,\ell,\sigma^{(s)}}|&\leq &|f_{\bt,\varsigma}-f_{r,m,\ell,\varsigma}|+|f_{r,m,\ell,\varsigma}-f_{r,m,\ell,\sigma^{(s)}}|\\\
&\leq & \frac{2}{\sigma}\|\psi\|_\infty\left(|a-a^{(r)}|+\|\bb-\bb^{(m)}\| \|\bZ\| + \sum_{j=1}^p \|g_j-g_{\ell,U}^{(j)}\|_\infty\right)+\\
&& 4\frac{\|\zeta\|_\infty}{\sigma}|\varsigma-\sigma^{(s)}|\\
&\leq & \epsilon_1\frac{2}{\sigma}\left\{\|\psi\|_\infty(1+\|\bZ\|+p)+2\|\zeta\|_\infty\right\}
\end{eqnarray*}
Define the functions
\begin{eqnarray*}
\phi_{r,m,\ell,s}^{(U)}(y,\bz,\bx)&=&f_{r,m,\ell,\sigma^{(s)}}(y,\bz,\bx)+\epsilon_1\frac{2}{\sigma}\left\{\|\psi\|_\infty(1+\|\bZ\|+p)+2\|\zeta\|_\infty\right\}\\
\phi_{r,m,\ell,s}^{(L)}(y,\bz,\bx)&=&f_{r,m,\ell,\sigma^{(s)}}(y,\bz,\bx)-\epsilon_1\frac{2}{\sigma}\left\{\|\psi\|_\infty(1+\|\bZ\|+p)+2\|\zeta\|_\infty\right\}\,.
\end{eqnarray*}
Then we have that $\phi_{r,m,\ell,s}^{(L)}\leq f_{\bt,\varsigma}\leq \phi_{r,m,\ell,s}^{(U)}$ and since $\esp\|\bZ\|^2<\infty$, we obtain 
$$\|\phi_{r,m,\ell,s}^{(U)}-\phi_{r,m,\ell,s}^{(L)}\|_{L^2(P)}\leq \epsilon_1\frac{4}{\sigma}\left[\|\psi\|_\infty(1+(\esp\|\bZ\|^2)^{1/2}+p)+2\|\zeta\|_\infty\right]=\epsilon\,.$$ 
Therefore, if $A=3\sigma$ and $A^{\star}=5A_1/ D $ the total number of brackets of size $\epsilon$ needed to cover $\itG_{n,\kappa,\bt^{\star},\bt_0}$ is bounded by
\begin{eqnarray*}
M_{ \mu,1}(\epsilon_1)M_{\bbech,q}(\epsilon_1)\prod_{j=1}^p M_j(\epsilon_1)\;M_{\sigma}(\epsilon_1)&\leq& \frac{5 \kappa  }{\epsilon_1}\left(\frac{5 \kappa  }{\epsilon_1}\right)^q\left(\frac{5 \kappa_1 }{\epsilon_1}\right)^{K+p} 3\sigma\left(\frac{1}{\epsilon_1}\right)\\
&\leq & A \left(\frac{ A^{\star} \max(1,\kappa)}{\epsilon}\right)^{K+p+q+2}\;,
\end{eqnarray*}
 where we have used that $   \kappa \le  \kappa_1= \kappa/ D$ since  $D\leq 1$, concluding the proof. \qed

\vskip0.1in

  In order to prove Theorem \ref{teo:ratesnew}, we need the following Lemma which is a direct consequence of Lemma A.2.3  in Boente \textsl{et al.} (2021). In the statement of Lemma \ref{lemma:lemapreviotasa}, we have in mind that $\Theta=\real\times\real^q\times \itH_{r_1}\times\dots\times\itH_{r_p}$, $\widetilde{\Theta}=\left\{\bt\in\Theta: g_j\in\itS_j, 1\leq j\leq p\right\}$ and  $\Theta_n= \left\{\bt\in\widetilde{\Theta}, \pi(\bt,\bthe)\leq \epsilon_0\right\}$ as defined above.

 \begin{lemma}{\label{lemma:lemapreviotasa}} Let $L_n$  be an stochastic process   indexed by $\widetilde{\Theta}\times (0,+\infty)\subset\Theta\times (0,+\infty)$. Furthermore, let $L:\Theta \times (0,+\infty)\to \real$ be a fixed function and  $\wsigma$ an estimator of $\sigma$ such that $\prob(\wsigma \in \itV)\to 1$  where $\itV\subset (0,+\infty)$  and $\bthe_n \in \Theta_n\subset \widetilde{\Theta}$. Let $\delta_n\ge 0 $ be a fixed sequence such that $\delta_n \to 0$ and fix $\upsilon>0$ with $0\le \delta_n< \upsilon$ for all $n$. Denote as $\wbthe\in  \widetilde{\Theta}$ the minimizer of  $L_n( \bt, \wsigma)$ over   $\widetilde{\Theta}$, that is,  $L_n( \wbthe, \wsigma)  \le  L_n( \bt, \wsigma)$, for any $\bt \in   \widetilde{\Theta}$. Assume that  $\prob(\wbthe \in \Theta_n)\to 1$,  
$\pi_{\prob}(\wbthe,  \bthe_n)   \convprob  0 $ and that there exists a function $\phi_n$ such that $\phi_n(\delta)/\delta$ is decreasing on $(\delta_n, \infty)$   and that  for any $\delta_n<\delta \le \upsilon$, we have
\begin{eqnarray}
\sup_{ \bt\in \Theta_{n,\delta}, \varsigma \in \itV} L(\bthe_n, \varsigma)- L( \bt,\varsigma) &\lesssim & -\delta^2 \,,
\label{eq:aprobar1}\\
 \esp^{*} \sup_{ \bt\in \Theta_{n,\delta}, \varsigma \in \itV}  \sqrt{n} \left |(L_n( \bt, \varsigma)- L( \bt,\varsigma))-(L_n(\bthe_n, \varsigma)- L( \bthe_n,\varsigma)) \right | & \lesssim &\phi_n(\delta) \,,
\label{eq:aprobar2}
\end{eqnarray}
where $\Theta_{n,\delta}=\{ \bt\in \Theta_n: \delta / 2  <  \pi_{\prob}( \bt,\bthe_n) \leq \delta\}$, the symbol $\lesssim$ means \textit{less or   equal up to a universal constant}
 and  $\esp^{*}$ stands for the outer expectation. Then, if $\gamma_n$ is such that $\delta_n\,\gamma_n=O(1)$ and 
 $ \gamma_n^2 \phi_n\left(\gamma_n^{-1}\right)\le \sqrt{n}$,
 for every $n$, we have that $\gamma_n \pi_{\prob}(\wbthe, \bthe_n) = O_{\prob}(1)$. 
 \end{lemma}

 \vskip0.2in
  
\textsc{Proof of Theorem \ref{teo:ratesnew}.} To simplify the notation from now on $\rho=\rho_1$. 
To derive the desired rates of convergence for the estimator $\wbthe=(\wmu,\wbbe\trasp,\weta_1,\dots,\weta_p)\trasp$ defined through \eqref{eq:estfinitos} and \eqref{eq:estimadoresgj}, Lemma \ref{lemma:lemapreviotasa} will be helpful.  

  As in the proof of Proposition \ref{prop:prop1}, let   $\widetilde{\wteta}_j\in \itS_j$ be the centered spline such that $\|\widetilde{\wteta}_j-\eta_{j}\|_{\infty}=O\left(n^{-\nu_j\, r_j}\right)$. Denote as $\bc^{(j)}\in\real^{k_j-1}$ the vectors such that $\widetilde{\wteta}_j(x)=\bc^{(j)}\bV^{(j)}(x)$  and define $\bthe_n=\left(\mu,\bbe\trasp,\widetilde{\wteta}_1,\dots,\widetilde{\wteta}_p\right)\trasp$.  Hence, for $n$ large enough we have that 
$\sum_{j=1}^p \|\widetilde{\wteta}_j-\eta_j\|_\infty <\min(\epsilon_0, 1/(2 A))$, 
with $\epsilon_0$ defined in \ref{ass:lowerbound} and    $A=4\, \sqrt{p\,(C_0+A_0)/C_0}$ with $A_0=\|\psi^{\prime}\|_{\infty}/2$ and $C_0$ the constant given in \ref{ass:lowerbound}. Hence, $\bthe_n\in \Theta_n$, as required in  Lemma \ref{lemma:lemapreviotasa}.


   Let $\delta_n= A\; \left\{\sum_{j=1}^p\|\eta_j- \widetilde{\wteta}_j\|_{\infty}^2\right\}^{1/2}$, then $\delta_n<1$. Note that we can assume without loss of generality that   the subset $\itV$ in \ref{ass:lowerbound}  is such that $\itV \subset [\sigma/2, 3\sigma/2]$. Using \ref{ass:wsigma}, we immediately obtain that $\prob(\wsigma \in \itV)\to 1$. Besides,  Proposition \ref{lemma:lema5} implies that  $\prob(\wtheta \in \Theta_n)\to 1$ as desired, while by definition  $\wbthe$ is the minimizer of  $L_n( \bt, \wsigma)$  over $\widetilde{\Theta}$. Hence, we only have to prove that \eqref{eq:aprobar1} and \eqref{eq:aprobar2} hold for any $\delta_n<\delta<1$ and a proper function $\phi_n$.

From \ref{ass:lowerbound} we have that  for any $\bt \in \Theta_n$ and $\varsigma\in\itV$, 
\begin{equation}
L(\bt, \varsigma)- L(\bthe, \varsigma) \ge C_0 \pi_\prob^2(\bt, \bthe)\,.
\label{eq:cotainfer}
\end{equation} 
Using that $\psi$ is an odd function and that the errors have a symmetric distribution, we get that $\esp \psi(\eps /\varsigma) =0$, for any $\varsigma>0$, which together with the independence between the errors and the covariates leads to
$$ \esp\left\{ \psi\left( \frac{y-\mu-\bbe\trasp \bZ- \sum_{j=1}^p \eta_j(X_j)} {\varsigma}\right) \sum_{j=1}^p \left(\widetilde{\wteta}_j(X_j)- \eta_j(X_j)\right)\right\} =0 \,.$$
Hence, using a Taylor's expansion of order two, we obtain
\begin{eqnarray*}
  L(\bthe_n, \varsigma)-L(\bthe, \varsigma) &=&    \esp\left\{ \psi\left(\frac{y - \mu- \bbe\trasp \bZ +\sum_{j=1}^p \eta_j(X_j )}{\varsigma}\right) \sum_{j=1}^p \left(\widetilde{\wteta}_j(X_j)- \eta_j(X_j)\right)\right\} \\
        & +& \frac 1{2}\;   \esp\left\{ \psi^{\prime}\left(\frac{y - \mu- \bbe\trasp \bZ +\sum_{j=1}^p \xi_j(X_j )}{\varsigma}\right) \left(\sum_{j=1}^p \left(\widetilde{\wteta}_j(X_j)- \eta_j(X_j)\right)\right)^2    \right\}  \\  
          &=& \frac 1{2}\;   \esp\left\{ \psi^{\prime}\left(\frac{y - \mu- \bbe\trasp \bZ +\sum_{j=1}^p \xi_j(X_j )}{\varsigma}\right) \left(\sum_{j=1}^p \left(\widetilde{\wteta}_j(X_j)- \eta_j(X_j)\right)\right)^2    \right\}  \\
  &\leq  & \frac 1{2}\;    \|\psi^{\prime}\|_{\infty}\esp \left(\sum_{j=1}^p \left(\widetilde{\wteta}_j(X_j)- \eta_j(X_j)\right)\right)^2  
  \\
  &\leq  &  A_0\,  \left(\sum_{j=1}^p \left\|\widetilde{\wteta}_j - \eta_j \right\|_{\infty}\right)^2  \le  p  A_0 \sum_{j=1}^p \left\|\widetilde{\wteta}_j - \eta_j \right\|_{\infty} ^2 =  \sum_{j=1}^p  O(n^{-2\,r_j\nu_j } )\,,
\end{eqnarray*} 
where $A_0=    \|\psi^{\prime}\|_{\infty}/2$ has been defined above and $\xi_j(X_j )$ is an intermediate value between $\eta_j(X_j)$ and $\widetilde{\wteta}_j(X_j)$.
Thus, using that 
$$\pi_{\prob}^2(\bt ,\bthe_{n})\le 2 \pi_{\prob}^2(\bt ,\bthe )+ 2 \pi_{\prob}^2(\bthe_{n},\bthe) \le  2 \pi_{\prob}^2(\bt,\bthe ) + 2\,  p \, \sum_{j=1}^p \left\|\widetilde{\wteta}_j - \eta_j \right\|_{\infty} ^2 $$ 
and that  $\delta / 2  <  \pi_{\prob}(\bt,\bthe_{n}) $ for $\bt \in \Theta_{n,\delta}$, together with \eqref{eq:cotainfer}, we obtain that
\begin{align*}
 L(\bt, \varsigma)-  L(\bthe_{n}, \varsigma) & \ge     L(\bt, \varsigma)-  L(\bthe, \varsigma) -  \left\{L(\bthe_{n}, \varsigma)-  L(\bthe, \varsigma)\right\} \\
&  \ge C_0\,\pi_{\prob}^2(\bt,\bthe) -  p  A_0 \sum_{j=1}^p \left\|\widetilde{\wteta}_j - \eta_j \right\|_{\infty} ^2 \\
&  \ge \frac{C_0}2 \pi_{\prob}^2(\bt,\bthe_n) -   
   p \left( C_0 +A_0\right)  \sum_{j=1}^p \left\|\widetilde{\wteta}_j - \eta_j \right\|_{\infty} ^2 \\
 &\ge   \frac{C_0}8 \delta^2 - \frac{1}{A^2}  p \left( C_0+A_0\right) \delta_n^2 =\frac{C_0}8 \delta^2- \frac{C_0}{16} \delta_n^2\ge \frac{C_0}{16} \delta^2\,,
 \end{align*}  
where the last inequality follows from the fact that $ \delta_n  <\delta$. Hence,  inequality \eqref{eq:aprobar1} in Lemma \ref{lemma:lemapreviotasa} holds.

We have to show that inequality \eqref{eq:aprobar2} in  in Lemma \ref{lemma:lemapreviotasa} holds for a proper function $\phi_n(\delta)$. Note that the left hand side of \eqref{eq:aprobar2} can be bounded as
$$ \esp^*\dst\sup_{  {\varsigma\in \itV\,,}{ \bt \in\Theta_{n, \delta}}} \sqrt{n} \left|L_n(\bt,\varsigma)- L(\bt,\varsigma) - \left(L_n(\bthe_n,\varsigma)- L(\bthe_n,\varsigma)\right)\right| \le \esp^{*} \dst \sup_{f\in \itF_{n,\delta}} \sqrt{n} |(P_n-P) f|\,, $$
where $\itF_{n,\delta} $ is the class of functions
$$\itF_{n,\delta} = \{V_{\bt, \varsigma}-V_{\bthech_n, \varsigma}:  \bt\in \Theta_{n,\delta}\,, \, \varsigma\in \itV\}\subset \{V_{\bt,  \varsigma}-V_{\bthech_n,  \varsigma}:  \bt\in \Theta_{n}\,, \, \varsigma\in \itV\}\,,$$   
and $V_{\bt, \varsigma}$ is defined in Lemma \ref{lemma:lemabacketing}, that is, 
$$V_{\bt,\varsigma}=V_{\bt,\varsigma}(y,\bz,\bx)=\rho\left(\frac{y-a-\bb\trasp\bz-\sum_{j=1}^p g_j(x_j)}{\varsigma}\right)\,.$$
Using that $\rho$ is a bounded function, we obtain that,  for any $f\in \itF_{n,\delta} $,  $\|f\|_{\infty} \le   2 \|\rho\|_{\infty}=2$.  
Besides, if $A_2= 2 \|\psi \|_{\infty}/ \sigma$, using that $\varsigma\in [\sigma/2, (3/2)\,\sigma]$, we get
\begin{align*}
|V_{\bt, \varsigma}-V_{\bthech_n, \varsigma}| &  = \left|\rho\left(\frac{Y-a-\bb\trasp \bZ- \sum_{j=1}^p g_j(X_j)}{\varsigma}\right) - \rho\left(\frac{Y-\mu-\bbe\trasp \bZ- \sum_{j=1}^p  \widetilde{\wteta}_j(X_j)}{\varsigma}\right)\right| \\
& \le A_2  \left|(\mu-a)+\bZ\trasp(\bbe-\bb) +  \sum_{j=1}^p  \left\{\widetilde{\wteta}_j(X_j)- g_j(X_j)\right\}\right|\,.
\end{align*}
Therefore, from the fact that $\pi_{\prob}(\bt,\bthe_n) \le \delta$,  we get that
$$P f^2\le A_2^2\, \esp \left[(\mu-a)+\bZ\trasp(\bbe-\bb) +  \sum_{j=1}^p  \left\{\widetilde{\wteta}_j(X_j)- g_j(X_j)\right\}\right]^2 = A_2^2 \,  \pi_{\prob}^2(\bt,\bthe_n)\le A_2^2 \, \delta^2\,.$$ 
Using  Lemma 3.4.2 in van der Vaart and Wellner (1996), we obtain that
$$\esp^{*} \sup_{f\in \itF_{n,\delta}} \sqrt{n} |(P_n-P) f|\lesssim  J_{[\;]}\left( A_2 \delta,\itF_{n,\delta}, L_2(P)\right) 
\left ( 1+ 2 \;\frac{J_{[\;]}(A_2 \,\delta,\itF_{n,\delta}, L_2(P))}{A_2^2 \delta^2 \; \sqrt{n}}   \right ) \,,$$ 
 where    $J_{[\;]}(\delta, \itF, L_2(P)) =\int_0^\delta \sqrt{1+ \log N_{[\;]}(\epsilon, \itF, L_2(P)) } d\epsilon$ is the bracketing integral. 
  
Note that $\sum_{j=1}^p \|\widetilde{\wteta}_j-\eta_j\|_\infty <\epsilon_0$, so given $\bt=(a,\bb\trasp, g_1, \dots, g_p)\trasp \in\Theta_{n}$, we have that  $\pi(\bt, \bthe_{n})\le \pi(\bt, \bthe)+\pi(\bthe , \bthe_{n})\le 2\epsilon_0$. Hence, taking $\kappa=2\epsilon_0$ and $\bt_0=\bt^{\star}=\bthe_n$,   we have that   
$$\itF_{n,\delta}\subset \{V_{\bt,  \varsigma}-V_{\bthech_n,  \varsigma}:  \bt\in \Theta_{n}\,, \, \varsigma\in \itV\}\subset \itG_{n,\kappa, \bt^{\star},\bt_{0}}\,$$ 
where  $\itG_{n,\kappa,\bt^{\star}, \bt_{0}}$ is defined in Lemma \ref{lemma:lemabacketing}, so 
$$N_{[\;]}\left( \epsilon,\itF_{n,\delta}, L_2(P)\right)\le A\left(\frac{A^{\star} }{\epsilon}\right)^{K+p+q+2}\,.$$
This implies that 
$$J_{[\;]}( A_3 \delta,\itF_{n,\delta}, L_2(P)) \lesssim \delta \sqrt{\log\left(\frac{1}{\delta}\right)} \sqrt{K+p+q+2}\,.$$
Let $\ell_n = K +p+ q+2$. Then, for some constant $A_3$ independent of $n$ and $\delta$,
$$\esp^{*} \sup_{\varsigma\in \itV, \bt\in \Theta_{n,\delta}} \sqrt{n}\left|(P_n-P)\left\{ V_{\bthech_n, \varsigma}- V_{\bt, \varsigma}\right\}\right| \leq A_3\,\,\phi_n(\delta)\,,  $$
where 
$$\phi_n(\delta)=\delta \, \ell_n^{1/2} \sqrt{\log\left(\frac{1}{\delta}\right)}  + \frac{ \ell_n  }{ \sqrt{n}} \log\left(\frac{1}{\delta}\right) \,.$$
Noting that $\phi_n(\delta)/\delta$ is decreasing in $\delta$, we conclude the proof of \eqref{eq:aprobar2}.

Note that, since  $\gamma_n= O( n^{\lambda})$  where $\lambda=\min_{1\le j\le p}(r_j\, \nu_j)$ and $\delta_n=A\; \left\{\sum_{j=1}^p\|\eta_j- \widetilde{\wteta}_j\|_{\infty}^2\right\}^{1/2}=O(\sum_{j=1}^p n^{-r_j\,\nu_j})= O(n^{-\lambda})$,  then   $\delta_n\, \gamma_n= O(1)$ as required in Lemma \ref{lemma:lemapreviotasa}. 
It remains  to prove that $\gamma_n^2\phi_n \left(1/{\gamma_n}\right)\lesssim  \sqrt{n}$, since  $\phi_n(c\delta)\le c \,\phi_n(\delta)$, for $c>1$. 
Note that 
$$
\gamma_n^2\phi_n \left(\frac{1}{\gamma_n}\right)=\gamma_n  \ell_n^{1/2} \, \sqrt{\log(\gamma_n)} + 
\gamma_n^2\, \log(\gamma_n)\; \frac{ \ell_n }{\sqrt{n}} =\sqrt{n}\; a_n(1+a_n) \, ,
$$ 
where $a_n=\gamma_n \, \sqrt{\log(\gamma_n)}\;   \ell_n^{1/2}/\sqrt{n}$.
Therefore, to prove that $\gamma_n^2\phi_n \left(1/{\gamma_n}\right)\lesssim \sqrt{n}$, we only have to obtain  that $a_n=O(1)$. Note that    $\ell_n=O(\sum_{j=1}^p n^{\nu_j})=O(n^\nu)$ where $\nu=\max_{1\le j\le p} \nu_j$ and $\gamma_n$ is such that $\gamma_n \, \sqrt{\log(\gamma_n)}= O(n^{(1-\nu)/2})$, then $a_n=O(1)$ as desired. Therefore,   Lemma \ref{lemma:lemapreviotasa} entails that  $\pi_{\prob} ( \bthe_n ,\wbthe) = O_{\prob}(\gamma_n)$. On the other hand, 
$$\pi_{\prob} ( \bthe_n ,\bthe) \le \sum_{j=1}^n \left\{\esp \left(\widetilde{\wteta}_j(X_j)-\eta_{j}(X_j)\right)^2\right\}^{1/2} \le \sum_{j=1}^n \|\widetilde{\wteta}_j-\eta_{j}\|_{\infty}$$  
and $\|\widetilde{\wteta}_j-\eta_{j}\|_{\infty}=O\left(n^{-\nu_j\, r_j}\right)$ which implies that
$\pi_{\prob} ( \bthe_n ,\bthe) \le O(n^{-\lambda})$,  or equivalently $\gamma_n  \pi_{\prob} ( \bthe_n ,\bthe) = O_{\prob}(1)$, which together with $\pi_{\prob} ( \bthe_n ,\wbthe) = O_{\prob}(\gamma_n)$ leads to the desired result. \qed



\subsection{Proof of Theorem \ref{teo:asymptnormal}}

Throughout this section,   to simplify the notation  we denote $\rho=\rho_1$ and $\psi=\psi_1=\rho_1^{\prime}$.
As in the proof of Theorem \ref{teo:ratesnew},  $P_n$   stands for the empirical probability measure of the observations $(Y_i, \bZ_i\trasp, \bX_i\trasp)\trasp$ and $P$ for the underlying probability measure. Furthermore, for any  $\bt= (a, \bb\trasp,g_1,\dots,g_p)\trasp\in \real^{q+1}\times \itG\times\dots \times \itG$  let us consider the function  $V_{\bt, \varsigma}$ already  defined in Lemma \ref{lemma:lemabacketing}, that is, 
$$ V_{\bt,\varsigma}(y,\bz,\bx)=\rho\left(\frac{y-a-\bb\trasp\bz-\sum_{j=1}^p g_j(x_j)}{\varsigma}\right)\,.$$
Then, $L_n( a,\bb,g_1,\dots,g_p,\varsigma)=P_n V_{ \bt,\varsigma}$ and $L(a,\bb,g_1,\dots,g_p,\varsigma)=P V_{\bt,\varsigma}$.

Moreover, denote as $V_{\bt,\varsigma}^{(\mu)}$ and $\bV_{\bt,\varsigma}^{(0)}=(V_{1,\bt,\varsigma}^{(0)},\dots,V_{q,\bt,\varsigma}^{(0)})\trasp$ the functions
\begin{align*}
V_{\bt,\varsigma}^{(\mu)} (y,\bz,\bx) &=\; -\;\frac{1}{\varsigma}\psi\left(\frac{y-a-\bb\trasp\bz-\sum_{j=1}^p g_j(x_j)}{\varsigma}\right)
\\
\bV_{\bt,\varsigma}^{(0)} (y,\bz,\bx) & =\; -\;\frac{1}{\varsigma}\psi\left(\frac{y-a-\bb\trasp\bz-\sum_{j=1}^p g_j(x_j)}{\varsigma}\right)\bz\,.
\end{align*} 
Note that $V_{\bt,\varsigma}^{(\mu)}$  and $\bV_{\bt,\varsigma}^{(0)}$ are the partial derivative of $V_{\bt,\varsigma}$ with respect to $a $ and $\bb$, respectively. Therefore,  using that $L_n(\wmu, \wbbe,\weta_1,\dots,\weta_p,\wsigma)\le  L_n(a, \bb,\weta_1,\dots,\weta_p,\wsigma)$ for any $(a,\bb\trasp)\trasp \in \real^{q+1}$ we obtain that
\begin{equation}
 P_n  V^{(\mu)}_{\wbthech,\wsigma} =0  \quad  \mbox{ and }\quad  P_n \bV^{(0)}_{\wbthech,\wsigma} =\bcero     \,.
 \label{eq:PnV0}
 \end{equation}
 Besides, using that $\esp \psi(a\eps)=0$ for any $a>0$  and the independence between the errors and covariates, we get that for any $\varsigma>0$
 \begin{equation}
  P  V^{(\mu)}_{\bthech,\varsigma} =0 \mbox{ and }\quad P \bV_{\bthech,\varsigma}^{(0)}=\bcero\,.
 \label{eq:PV0}
 \end{equation}  
Similarly, if $\itG_0$ stands for the class of measurable functions over $[0,1]$, we consider the operator  $V_{\bt,\varsigma}^{(j)}$ defined as
$$V_{\bt,\varsigma}^{(j)}[h](y,\bz,\bx) =\, -\, \frac{1}{\varsigma}\psi\left(\frac{y-a-\bb\trasp\bz-\sum_{\ell=1}^p g_\ell(x_\ell)}{\varsigma}\right)h(x_j)\qquad \mbox{for any $h \in \itG_0$}\,.$$
As above, $V_{\bt,\varsigma}^{(j)}[h] $ is the directional derivative of $V_{\bt,\varsigma}$, that is, 
$$V_{\bt,\varsigma}^{(j)}[h] =
 \left.\frac{\partial V_{a,\bb,g_1,\dots,g_{j-1},g_j+sh,g_{j+1},\dots,g_p}}{\partial s}\right|_{s=0}\,.$$
 Furthermore, for $\bh =(h_{1 },\dots,h_{q })\trasp$, \ we denote 
$$V_{\bt,\varsigma}^{(j)}[\bh ]=(V_{\bt,\varsigma}^{(j)}[h_{1 }],\dots,V_{\bt,\varsigma}^{(j)}[h_{q  }])\trasp\,.$$
Using   that,  for any $s\in \real $ and $g_j \in \itS_j$, 
$$L_n( \wmu, \wbbe,\weta_1,\dots,\weta_p,\wsigma)\le  L_n( \wmu, \wbbe,\weta_1,\dots,\weta_{j-1}, \weta_j+s\; g_j, \weta_{j+1},\dots\weta_p,\wsigma)\,,$$ 
 we obtain that 
 \begin{equation} 
 P_n V_{\wbthech,\wsigma}^{(j)}[g_j] =0\,,\qquad \mbox{ for any } \quad g_j \in \itS_j
 \label{eq:PnVj} 
 \end{equation}
  for $1\le j\le p$. On the other hand, the independence between the errors and covariates and the fact that  $\esp \psi(a\eps)=0$, for any $a>0$, guarantee  that 
  \begin{equation}
   P  V_{\bthech,\sigma}^{(j)}[h] =0\,, \mbox{for any $h\in \itG_0$}\,.
   \label{eq:PVj} 
   \end{equation}

Define for any $\bt=(a,\bb\trasp,g_1,\dots,g_p)\trasp$ and $\varsigma>0$ the function 
$$\bW_{\bt,\varsigma}(y,\bz,\bx) =\,-\, \frac{1}{\varsigma}\psi\left(\frac{y-a- \bb\trasp\bz-\sum_{j=1}^p g_j(x_j)}{\varsigma}\right)(\bz-\bh^*(\bx))\,,$$
where $\bh^*$ is defined in  \eqref{eq:11}.

Note that, under assumption \ref{ass:hstar}, we have that $(\bW_{\bt,\sigma})_m=(\bV_{\bt,\sigma}^{(0)})_m-\sum_{j=1}^p V^{(j)}_{\bt,\sigma}[h_{mj}^*]\,,$ 
 for $1\leq m\leq q$, meaning that
\begin{equation}
\label{eq:BW}
\bW_{\bt,\sigma}=\bV_{\bt,\sigma}^{(0)}-\sum_{j=1}^p V^{(j)}_{\bt,\sigma}[\bh_{j}^*]
\end{equation}
where $\bh_j^*=(h_{1j}^*,\dots,h_{q,j}^*)\trasp$.

From now on, $\itV$ will refer to a neighbourhood of $\sigma$, which we assume to be a subset of $[\sigma/2, 3\,\sigma/2]$.

For each $1\leq m\leq q$ and $1\leq j\leq p$, let $g_{j,\bc_{0,j}}\in \itS_j$ the centered spline with coefficients $\bc_{0,j}$ and $h$ a function  in $\itS_j$ such that $\| h_{mj}^* -h\|_\infty<\delta$. Let $\bthe_{\bc_{0,1},\dots,\bc_{0,p}}=(\mu, \bbe\trasp,g_{1,\bc_{0,1}},\dots,g_{p,\bc_{0,p}})\trasp$, $\bt_{\bc_1,\dots,\bc_p}=(a,\bb\trasp,g_{1,\bc_1},\dots,g_{p,\bc_p})\trasp$ for $\bc_s\in\real^{k_s-1}$, $1\le s\le p$, and $\bC_0 =\left(\bc_{0,1}\trasp, \dots, \bc_{0,p}\trasp\right)\trasp$. Let $\epsilon_0>0$ be a fixed value, for instance, $\epsilon_0=1$ or the value stated in assumption \ref{ass:lowerbound}, when it holds. The following classes of functions will be useful in the proof of Theorem \ref{teo:asymptnormal}
\begin{align}
\itE_{n,m,h,\delta,\bC_0}^{(j)}= & \left\{f =V^{(j)}_{\bt_{\bc_1,\dots,\bc_p}, \varsigma}[h_{mj}^* - h]\,:   \pi\left(\bt_{\bc_1,\dots,\bc_p},\bthe_{\bc_{0,1},\dots,\bc_{0,p}}\right)<\epsilon_0  , \varsigma\in\itV ,  \bc_\ell\in\real^{k_\ell-1},  1\leq \ell\leq p\right\}
\label{def:s1}\\   
\itF_{n,m,\delta,\bC_0}^{(j)}= & \left\{f =V^{(j)}_{\bt_{\bc_1,\dots,\bc_p},\varsigma}[h_{mj}^*]-V^{(j)}_{\bthech,\varsigma}[h_{mj}^*] \right.\,: \pi\left(\bt_{\bc_1,\dots,\bc_p},\bthe_{\bc_{0,1},\dots,\bc_{0,p}}\right)<\epsilon_0  \,,\, \varsigma\in\itV, \bc_\ell\in\real^{k_\ell-1}, 1\leq \ell\leq p, \nonumber \\
&  \left. 
 \pi_\prob(\bt_{\bc_1,\dots,\bc_p},\bthe_{\bc_{0,1},\dots,\bc_{0,p}})<\delta \, \right\}\label{def:s2}\\
 \itG_{n,j,\delta, \bC_0}  = & \{f=V_{j,\bt_{\bc_1,\dots,\bc_p},\varsigma}^{(0)} -V_{j,\bthech,\varsigma}^{(0)} \,:  \pi\left(\bt_{\bc_1,\dots,\bc_p},\bthe_{\bc_{0,1},\dots,\bc_{0,p}}\right)<\epsilon_0 \,, \varsigma\in  \itV, \bc_\ell\in\real^{k_\ell-1}, 1\leq \ell\leq p, \nonumber\\  
 & \pi_\prob (\bt_{\bc_1,\dots,\bc_p},\bthe_{\bc_{0,1},\dots,\bc_{0,p}})<\delta\,\}\,.
\label{eq:Gnjd}
\end{align}
Note that the family of functions $\itE_{n,m,h,\delta,\bC_0}^{(j)}$ depends on $\delta$ through the function $h\in \itS_j$ which is fixed and such that $\| h_{mj}^* -h\|_\infty<\delta$.

To simplify the notation, from now on we denote $\bd=(a,\bb\trasp)\trasp$, $\btau=(\mu, \bbe\trasp)\trasp$ and $\wtbz=(1,\bz\trasp)\trasp$. 

Note that
\begin{align*}
V^{(j)}_{\bt_{\bc_1,\dots,\bc_p}, \varsigma}[h_{mj}^* - h](y,\bz,\bx) & =   \frac{1}{\varsigma}\psi\left(\frac{y- \bd\trasp\wtbz-\sum_{j=1}^p g_{j,\bc_j}(x_j)}{\varsigma}\right)\left\{h(x_j)- h_{mj}^*(x_j)\right\}\\
\left(V^{(j)}_{\bt_{\bc_1,\dots,\bc_p},\varsigma}[h_{mj}^*]-V^{(j)}_{\bthech,\varsigma}[h_{mj}^*]\right)(y,\bz,\bx)= & \left\{\frac{1}{\varsigma} \psi\left(\frac{y-  \btau\trasp\wtbz- \sum_{\ell=1}^p \eta_{\ell}(x_\ell)}{\varsigma}\right)\right.\\
&\left. - \frac{1}{\varsigma} \psi\left(\frac{y-  \bd\trasp\wtbz- \sum_{\ell=1}^p g_{\ell,\bc_\ell}(x_\ell)}{\varsigma}\right)\right\}h_{mj}^*(x_j) \,,
\end{align*}
while
$$
\left(V_{j,\bt_{\bc_1,\dots,\bc_p},\varsigma}^{(0)} -V_{j,\bthech,\varsigma}^{(0)}\right)(y,\bz,\bx)=  \frac{1}{\varsigma} \left\{\psi\left(\frac{y-\btau\trasp\wtbz-\sum_{j=1}^p \eta_{j}( x_j)}{\varsigma}\right)-\psi\left(\frac{y-  \bd\trasp\wtbz-\sum_{j=1}^p g_{j,\bc_j} (x_j)}{\varsigma}\right)\right\} z_j\,.$$ 
Similar arguments to those considered in the proof of Lemma \ref{lemma:lemabacketing} and the fact that $\| h_{mj}^* -h\|_\infty<\delta$, allow to bound the bracketing number of the classes $\itE_{n,m,h,\delta,\bC_0}^{(j)}$ and $\itF_{n,m,\delta,\bC_0}^{(j)}$ and to obtain that 
for some generic constant $C$  independent of $n$ and $\delta$  
\begin{align}
\label{eq:entropiaE} 
J_{[\,\,]}( A_1\; \delta,\itE_{n,m,h,\delta,\bC_0}^{(j)},L_2(P)) & \leq C\delta \sqrt{K+p+q+2}    \\
J_{[\,\,]}(C_{mj}\; \delta,\itF_{n,m,\delta,\bC_0}^{(j)},L_2(P)) & \leq C\delta \sqrt{\log\left(\frac{1}{\delta}\right)}\sqrt{K+p+q+2}  \,,
\label{eq:entropiaF} 
\end{align}
where $K=\sum_{j=1}^p (k_j-1)$, $A_1=(2/\sigma)\|\psi\|_\infty$, $C_{mj}=8\|\psi^\prime\|_\infty\|h_{mj}^*\|_\infty/\sigma^2$. A similar bound holds for $ \itG_{n,j,\delta, \bC_0}$ since  $\esp \|\bZ\|^2<\infty$.

To derive Theorem \ref{teo:asymptnormal},   we will verify the conditions of the following lemma, which we give without proof since it is slight modification of Theorem  3 in Zhang \textsl{et al.} (2010). Note that \textbf{H3}(ii) corresponds to assumption (B3) in Zhang \textsl{et al.} (2010).

\begin{lemma}{\label{lemma:lemadistribucionasintotica}}
Let $\bthe=(\btau\trasp,\eta_1,\dots,\eta_p)\trasp$ and $\wbthe=(\wbtau\trasp,\weta_1,\dots,\weta_p)\trasp$ a consistent estimator of $\bthe$. Assume that \ref{eq:BW} holds and
\begin{enumerate}
\item[\textbf{H1}] $P_n \bV_{\wbthech,\wsigma}^{(0)}=o_\prob (n^{-1/2})$ and $P_nV_{\wbthech,\wsigma}^{(j)}[h_{mj}^{*}]=o_\prob(n^{-1/2})$, for $1\leq m\leq q$  and $1\leq j\leq d$,
\item[\textbf{H2}]
\begin{enumerate}
\item[(a)] $(P_n-P)\left\{\bV_{\wbthech,\wsigma}^{(0)}-\bV^{(0)}_{\bthech,\sigma}\right\}=o_\prob(n^{-1/2})$ and 
\item[(b)] $(P_n-P)\left\{V_{\wbthech,\wsigma}^{(j)}[h_{mj}^{*}]-V^{(j)}_{\bthech,\sigma}[h_{mj}^{*}]\right\}=o_\prob(n^{-1/2})$, for $1\leq m\leq q$ and $1\leq j\leq d$,  
\end{enumerate}
\item[\textbf{H3}]  $P\left\{\bW_{\wbthech,\wsigma}-\bW_{\bthech,\sigma}\right\}=\,-\, \bB_{\bthech,\wsigma}(\wbbe-\bbe)+o_\prob(n^{-1/2})$ 
 \end{enumerate}
hold. Then, if  \ref{ass:hstar} holds, $\bB_{\bthech,\wsigma}\convprob \bB_{\bthech,\sigma}$ and $\bB_{\bthech,\sigma}$ is non singular, we have that $n^{1/2}(\wbbe-\bbe)=n^{1/2}\bB_{\bthech,\sigma}^{-1}P_n\bW_{\bthech,\sigma}+o_\prob(1)$. Hence, if $\bD_{\bthech,\sigma}=\esp \bW_{\bthech,\sigma}\bW_{\bthech,\sigma}\trasp$, we have that 
$$n^{1/2}\left(\wbbe-\bbe\right)\convdist N\left(\textbf{0},\bB_{\bthech,\sigma}^{-1}\bD_{\bthech,\sigma}\bB_{\bthech,\sigma}^{-1 \mbox{\footnotesize{\sc t}}}\right)\,.$$
\end{lemma}
  
  \vskip0.1in
  
From now on, $\|\bbG_n\|_{\itF}$ stands for  $\|\bbG_n\|_{\itF}=\sup_{f\in \itF} \sqrt{n} |(P_n-P) f|$.

 \vskip0.1in
\noindent\textsc{Proof of Theorem \ref{teo:asymptnormal}.} In order to show that Lemma \ref{lemma:lemadistribucionasintotica} can be applied, the proof will be carried out in several steps. 

(i) We begin by deriving \textbf{H1}. Recall that according to \ref{eq:PnV0}  $P_n \bV^{(0)}_{\wbthech,\wsigma} =0$ so, we only need to verify 
\begin{equation}\label{eq:s4}
P_n V_{\wbthech,\wsigma}^{(j)}[h_{mj}^*]=o_\prob(n^{-1/2})\qquad \mbox{ for all }1\leq m \leq q \mbox{ and } 1\leq j\leq p.
\end{equation}
As in the proof of Proposition \ref{prop:prop1}, let first consider, for $1\leq s\leq p$, $\widetilde{\wteta}_j \in\itS_j$ such that $\|\widetilde{\wteta}_j-\eta_j\|_\infty=O(n^{-\nu_j r_j})=O(n^{-(1-\nu_j)/2})$. Let $\bc_{0,j}\in \real^{k_j-1}$ be such that  $\widetilde{\wteta}_j =g_{j, \bc_{0,j}}$ and $\bthe_n=\bthe_{\bc_{0,1},\dots,\bc_{0,p}}=(\btau\trasp,g_{1, \bc_{0,1}},\dots,g_{p, \bc_{0,p}})\trasp$.  Then, using that $\|\weta_j-\eta_j\|_\infty\convprob 0$, we obtain that $\|\weta_j-g_{j, \bc_{0,j}}\|_\infty\convprob 0$.

Let $1\leq m\leq q$ and $1\leq j\leq p$.   
Using that  \ref{ass:hstar} and   \ref{ass:kj} hold, as in Proposition \ref{prop:prop1}, from Schumaker (1981), we get that there exists $h_{n,m,j}\in\itS_j$ such that $\|h_{mj}^* - h_{n,m,j}\|_\infty=O(n^{-r_j/(1+2r_j)})=O(n^{-(1-\nu_j)/2})$. Hence, using  \eqref{eq:PnVj}, we conclude that to derive \eqref{eq:s4} it is enough to show that
\begin{equation}\label{eq:s6}
P_n V_{\wbthech,\wsigma}^{(j)}[h_{mj}^* - h_{n,m,j}]=o_\prob(n^{-1/2})\,.
\end{equation}
The term $P_n V_{\wbthech,\wsigma}^{(j)}[h_{mj}^* - h_{n,m,j}]$ can be written as $T_1+T_2$ where $T_1=(P_n-P)V_{\wbthech,\wsigma}^{(j)} [h_{mj}^* - h_{n,m,j}]$ and $T_2=P V_{\wbthech,\wsigma}^{(j)} [h_{mj}^* - h_{n,m,j}]$. 

Let us consider the family of functions   defined in  \eqref{def:s1}  with $h=h_{n,m,j}\in\itS_j$, $\bc_{0,j} $  such that  $\widetilde{\wteta}_j =g_{j, \bc_{0,j}}$ and $\delta=\delta_n = 2\max_{1\le j\le p}\|h_{mj}^*-h_{n,m,j}\|_\infty$. To avoid burden notation, let $\itE_n^{(j)}=\itE^{(j)}_{n,m,h_{n,m,j},\delta,\bC_0}$. For any $f\in\itE_n^{(j)}$, 
$$\|f\|_\infty=\|V_{a,\bb,g_{1, \bc_{0,1}},\dots,g_{p, \bc_{0,p}}}[h_{mj}^*-h]\|_\infty\leq \frac{2}{\sigma}\|\psi\|_\infty \|h_{mj}^*-h\|_\infty\leq M(\delta)\,,$$ 
where $M(\delta)=(2/\sigma)\|\psi\|_\infty\,\delta=A_1\,\delta$. Furthermore,
$$Pf^2=\esp\left[-\frac{1}{\varsigma}\psi\left(\frac{Y-a- \bb\trasp\bz-\sum_{s=1}^p g_{j,\bc_j}(X_j)}{\varsigma}\right)(h_{mj}^*(X_j)-h(X_j))\right]^2\leq \frac{4}{\sigma^2}\|\psi\|^2_\infty\|h_{mj}^*-h\|_\infty^2\leq M^2(\delta)\,.$$
  Hence, Lemma 3.4.2 in van der Vaart and Wellner (1996) entails that  
\begin{eqnarray*}
\esp^*\|\bbG_n\|_{\itE_n^{(j)}}&\lesssim &J_{[\,\,]}(M(\delta),\itE_n^{(j)},L_2(P))\left(1+\frac{J_{[\,\,]}(M(\delta),\itE_n^{(j)},L_2(P))}{M^2(\delta)\sqrt{n}}M(\delta)\right)\\
&&=J_{[\,\,]}(M(\delta),\itE_n^{(j)},L_2(P))\left(1+\frac{J_{[\,\,]}(M(\delta),\itE_n^{(j)},L_2(P))}{M(\delta)\sqrt{n}}\right)
\end{eqnarray*}
which together with \eqref{eq:entropiaE} leads to
$$\esp^*\|\bbG_n\|_{\itE_n^{(j)}}\leq C\delta \, (K+p+q+2)^{1/2}\left(1+\frac{C\, (K+p+q+2)^{1/2}}{A_1\sqrt{n}}\right)\,.$$
Using that $K =\sum_{j=1}^p O(n^{\nu_j})=O(n^{\nu})$ and $\delta = 2\max_{1\le j\le p}\|h_{mj}^*-h_{n,m,j}\|_\infty = O( n^{-(1-\nu)/2} ) $, we get that, for $n$ large enough  
\begin{align*}
\prob\left(\sqrt{n}|T_1|>\epsilon \cap \|\wbtau-\btau\|+\sum_{j=1}^p \|\weta_j-g_{j, \bc_{0,j}}\|_\infty<\epsilon_0\right)&\leq   \frac{1}{\epsilon}\esp^*\|\bbG_n\|_{\itE_n^{(j)}}\leq \frac{1}{\epsilon} 2\, C\delta\; K^{1/2}\left(1+\frac{2\,C\, K^{1/2}}{A_1\sqrt{n}}\right)\\
&\leq   C_1\; n^{-(1-\nu)/2}\;\, p^{1/2}\,  n^{\nu/2} \left(1+\frac{C_2}{A_1} \;  p^{1/2}\,  n^{\frac{\nu}2}  n^{-\frac 12}\right)\\
&\leq   \frac{1}{\epsilon}C_1 \, p^{1/2}\,  n^{-\frac{1-2\,\nu}2}\;\left(1+\frac{C_2}{A_1} p^{1/2}\,  n^{- \;\frac{1-  \nu}2}\right)\,,
\end{align*}
which converges to $0$ since  $r_j\geq 1$, i.e., $\nu< 1/2$. Hence, noting that $\|\wbtau-\btau\|+\sum_{s=1}^p \|\weta_s-g_{j, \bc_{0,j}}\|_\infty\convprob 0$, we obtain that $T_1=o_\prob(n^{-1/2})$.

To conclude the proof of  \eqref{eq:s6}  it remains to show that $T_2=o_\prob(n^{-1/2})$. Using the Fisher-consistency given in Lemma \ref{lemma:FC}, we have that $PV^{(j)}_{\bthech,\varsigma}[h_{mj}^*-h_{m,n,j}]=0$, for any $\varsigma>0$, thus,
$$T_2=P(V_{\wbthech,\wsigma}^{(j)}-V^{(j)}_{\bthech,\wsigma})[h_{mj}^*-h_{n,m,j}]$$
Denote as $\zeta_{\btauch}$ and $\zeta_j(x_j)$ intermediate values between $\btau$ and $\wbtau$ and $\eta_j(x_j)$ and $\weta_j(x_j)$, respectively. Then, using a first order Taylor's approximation and recalling that  $\wtbZ =(1,\bZ\trasp)\trasp$ we get that
\begin{eqnarray*}
|T_2|&=&\left|\esp\psi^\prime\left(\frac{Y-\zeta_{\btauch}\trasp\wtbZ-\sum_{s=1}^p \zeta_s(X_s)}{\wsigma}\right)\frac{1}{\wsigma^2}\left[(\btau-\wbtau)\trasp\wtbZ+\sum_{s=1}^p (\eta_s-\weta_s)(X_s)\right][h_{mj}^*-h_{n,m,j}]\right|\\
&\leq & \frac{4}{\sigma^2}\|\psi^\prime\|_\infty\|h_{mj}^*-h_{n,m,j}\|_\infty\esp\left|
(\btau-\wbtau)\trasp\wtbZ+\sum_{s=1}^p (\eta_s-\weta_s)(X_s)\right|\\
&\leq & \frac{4}{\sigma^2}\|\psi^\prime\|_\infty\|h_{mj}^*-h_{n,m,j}\|_\infty\left\{\esp\left(
(\btau-\wbtau)\trasp\wtbZ+\sum_{s=1}^p (\eta_s-\weta_s)(X_s)\right)^2\right\}^{1/2}\\\
&\leq &\frac{4}{\sigma^2}\|\psi^\prime\|_\infty\|h_{mj}^*-h_{n,m,j}\|_\infty\pi_\prob(\wbthe,\bthe)\,.
\end{eqnarray*}
Taking into account that  $  \pi_\prob(\wbthe,\bthe)=O_{\prob}( n^{-\,(1-\nu)/2+\omega})$, $\omega <(1-2\nu)/2$, see assumption \ref{ass:sobreZ},  and $\|h_{mj}^* - h_{n,m,j}\|_\infty=O(n^{-(1-\nu_j)/2})$,  we conclude that $|T_2|=O_\prob(n^{- (2-\nu-\nu_j)/2+\omega})=O_\prob(n^{-1/2})$ as desired.

(ii) We have to show that \textbf{H2} holds. We will only show that \textbf{H2}(b), since \textbf{H2}(a) follows in a similar way using the class of functions $ \itG_{n,j,\delta, \bC_0}$ which is bounded if for some $C>0$, $\prob(\|\bZ\|<C)=1$,   that is, $\bZ$ is bounded. If $\bZ$ is not bounded, one has to consider    the covering number of the family of functions $ \itG_{n,j,\delta, \bC_0}$ with respect to $L_2(P_n)$ and to use similar arguments to those described in van der Vaart and Wellner (1996) together with the strong law of large numbers and the fact that $\esp \|\bZ\|^2 <\infty$ to derive  \textbf{H2}(a). 

To prove that \textbf{H2}(b) holds, fix $1\leq m\leq q$ and $1\leq j\leq p$. Note that $(P_n-P)\left\{V^{(j)}_{\wbthech,\wsigma}[h_{mj}^*]-V^{(j)}_{\bthech,\sigma}[h_{mj}^*] \right\}=S_{1,n,m,j}+S_{2,n,m,j}$, where
\begin{eqnarray*}
S_{1,n,m,j}&=&(P_n-P)\left\{V^{(j)}_{\wbthech,\wsigma}[h_{mj}^*]-V^{(j)}_{\bthech,\wsigma}[h_{mj}^*]\right\}\,,\\
S_{2,n,m,j}&=&(P_n-P)\left\{V^{(j)}_{\bthech,\wsigma}[h_{mj}^*]-V^{(j)}_{\bthech,\sigma}[h_{mj}^*]\right\}\,.
\end{eqnarray*}
Recall that a class $\itF$ of functions is Donsker when $\int_0^{\infty} \sqrt{N_{[\;]}(\delta, \itF, L_2(P))}\,d\delta <\infty$, see van der Vaart and Wellner (1996). Using that $\psi$ has a bounded derivative, $h_{mj^*}$ is bounded (see assumption \ref{ass:hstar}), $\itV\subset[\sigma/2, 3\sigma/2]$ and Theorem 2.7.11 in van der Vaart and Wellner (1996),  we obtain easily that the family of functions  
$$\itF_{mj}=\left\{f(\eps,\bx)=V^{(j)}_{\bthech,\varsigma}[h_{mj}^*]=\,-\,\frac{1}{\varsigma}\psi\left(\frac{\sigma\eps}{\varsigma}\right)h_{mj}^*(x_j)\,;\, \varsigma\in\itV\right\}$$ 
is such that $\int_0^{\infty} \sqrt{N_{[\;]}(\delta, \itF_{mj}, L_2(P))}\,d\delta <\infty$. Thus, $\itF_{mj}$ is Donsker which together with the fact that $\wsigma\convprob 0$ leads to $\sqrt{n}S_{2,n,m,j}=o_\prob(n^{-1/2})$. Hence, to conclude the proof of \textbf{H2}(b), we have to show that $\sqrt{n}S_{1,n,m,j}=o_\prob(n^{-1/2})$.

For $1\leq j\leq p$, as in (i), let $\bc_{0,j}\in \real^{k_j-1}$ be such that   $g_{j, \bc_{0,j}}=\widetilde{\wteta}_j $  where $\widetilde{\wteta}_j \in\itS_j$  is the spline approximation to $\eta_j$, that is, $\|\widetilde{\wteta}_j-\eta_j\|_\infty=O(n^{-\nu_j\;r_j })=O(n^{-(1-\nu_j)/2})$. Then, for $n$ large enough $\sum_{j=1}^p \|\widetilde{\wteta}_j-\eta_j\|_\infty <\epsilon_0/2$.

Take $\delta=\delta_{n}=n^{- \alpha (1-\nu)/2}$ with $\alpha=3/4$. Then, $\pi_\prob(\bthe_{n},\bthe)<\delta/2$ for $n\geq n_0$ with $\bthe_n=\bthe_{\bc_{0,1},\dots,\bc_{0,p}}=(\btau\trasp, g_{1,\bc_{0,1}},\dots,g_{p,\bc_{0,p}})\trasp$.  Furthermore, using that $\gamma_n \pi_\prob(\wbthe,\bthe)=O_{\prob}(1)$, we conclude that $n^{  \alpha (1-\nu)/2} \pi_\prob(\wbthe,\bthe)\convprob 0$ since $\omega < (1-\nu)/8$. Hence, for $n$ large enough $\pi_\prob(\wbthe ,\bthe_n)<\delta$ with probability converging to 1.

Taking into account that   $\pi(\wbthe, \bthe)\convprob 0$, we have that with probability converging to $1$, $\pi(\wbthe, \bthe)<\epsilon_0/2$, which entails that $\pi(\wbthe, \bthe_n)<\epsilon_0$. Let us consider the probability set where $\pi(\wbthe, \bthe_n)<\epsilon_0$. Then,    for $n\ge n_0$,    $V^{(j)}_{\wbthech,\wsigma}[h_{mj}^*]-V^{(j)}_{\bthech,\wsigma}[h^*_{mj}]\in\itF_{n,m,\delta,\bC_0}^{(j)}$,  where $\itF_{n,m,\delta,\bC_0}^{(j)}$ is defined in \eqref{def:s2}.

For the sake of simplicity, denote $\itF^{\star}=\itF_{n,m,\delta_n,\bC_0}^{(j)}$. Let $f(y,\wtbz,\bx)$ be a function in $\itF^{\star}$, that is, 
$$f(y,\wtbz,\bx)=\,-\, \frac{1}{\varsigma}\left[\psi\left(\frac{y-\bd\trasp\wtbz-\sum_{\ell=1}^p g_{\ell,\bc_\ell}(x_\ell)}{\varsigma}\right)-\psi\left(\frac{y-\btau\trasp\wtbz-\sum_{\ell=1}^p \eta_\ell(x_\ell)}{\varsigma}\right)\right]h_{mj}^*(x_j)\,$$
for some $a\in \real$, $\bb \in \real^q$,  $\bd=(a,\bb\trasp)\trasp$,  $\varsigma\in\itV$ and $g_{\ell,\bc_\ell}\in\itS_\ell$, for $1\leq \ell\leq p$, such that $\pi(\bt_{\bc_1,\dots,\bc_p},\bthe_{n})<\epsilon_0$ and $\pi_\prob(\bt_{\bc_1,\dots,\bc_p},\bthe_{n})<\delta$. Then, $\|f\|_\infty\leq B_j=(4/\sigma)\|\psi\|_\infty\|h_{mj}^*\|_\infty$. On the other hand, using a Taylor's expansion of order one, we get that
$$f(y,\wtbz,\bx)=\frac{1}{\varsigma^2}\psi^\prime\left(\frac{y-\zeta_{\btauch}\trasp\wtbz-\sum_{\ell=1}^p \zeta_\ell(x_\ell)}{\varsigma}\right)\left[(a-\mu)+(\bb-\bbe)\trasp\bz+\sum_{\ell=1}^p (g_{\ell,\bc_{\ell}}-\eta_\ell)(x_\ell)\right]h_{mj}^*(x_j)$$
where $\wtbz=(1,\bz\trasp)\trasp$, $\zeta_{\btauch}$ and $\zeta_\ell(x_\ell)= \xi_\ell \, g_{\ell,\bc_{\ell}}(x_\ell)+ (1-\xi_\ell)\eta_\ell(x_\ell)$, $0<\xi_\ell<1$, are intermediate points between $(a,\bb\trasp)\trasp$ and $(\mu,\bbe\trasp)\trasp$ and $g_{\ell,\bc_{\ell}}(x_\ell)$ and $\eta_\ell(x_\ell)$, for $1\leq \ell\leq p$, respectively. Hence, from the bound 
$$|f(y,\bz,\bx)|\leq \frac{4}{\sigma^2}\;\|\psi^\prime\|_\infty\; \|h_{mj}^*\|_\infty \; \left|(\bd-\btau)\trasp\wtbz+\sum_{\ell=1}^p (g_{\ell,\bc_\ell}-\eta_\ell)(x_\ell)\right|$$ 
and the fact that $\pi_\prob(\bthe,\bt_{\bc_1,\dots,\bc_p})\leq \pi_\prob(\bthe_{n},\bt_{\bc_1,\dots,\bc_p})+\pi_\prob(\bthe,\bthe_n)\leq 2\delta$,  we conclude that
\begin{align*}
Pf^2 & \leq \frac{16}{\sigma^4}\|\psi^\prime\|^2_\infty\; \|h_{mj}^*\|^2_\infty\;\esp\left((a-\mu)+(\bb-\bbe)\trasp\bZ+\sum_{\ell=1}^p (g_{\ell,\bc_\ell}-\eta_\ell)(X_\ell)\right)^2=\frac{16}{\sigma^4}\|\psi^\prime\|^2_\infty\|h_{mj}^*\|^2_\infty\pi^2_\prob(\bthe,\bt_{\bc_{1},\dots,\bc_{p}})\\
& \leq C_{mj}^2\delta^2\,,
\end{align*}
with $C_{mj}^2=64\|\psi^\prime\|^2_\infty\|h_{mj}^*\|^2_\infty/\sigma^4$ as defined in \eqref{eq:entropiaF}. 
Using again Lemma 3.4.2 of var der Vaart and Wellner (1996) we get that
$$\esp^*\|\bbG_n\|_{\itF^{\star}} \lesssim J_{[\,\,]}(C_{mj}\; \delta,\itF^{\star},L_2(P))\left(1+\frac{J_{[\,\,]}(C_{mj}\;\delta,\itF^{\star},L_2(P))}{C_{mj}^2\,\delta^2\sqrt{n}}B_j\right)\,,$$
which together with \eqref{eq:entropiaF}  leads that for $n$ large enough
\begin{align*}
\esp^*\|\bbG_n\|_{\itF^{\star}} & \leq 2\, C\delta\; \sqrt{\log\left(\frac{1}{\delta}\right)}\; K^{1/2}\left(1+\frac{B_j}{C_{mj}^2}\frac{2\,C  \sqrt{\log\left(\frac{1}{\delta}\right)}\, K^{1/2}}{\delta\;\sqrt{n}}\right)\\
& \le 2\; C\; \delta\; \sqrt{\log\left(\frac{1}{\delta}\right)}\; K^{1/2}+\frac{4\; C^2\,B_j}{C_{mj}^2}   \log\left(\frac{1}{\delta}\right) \;K n^{-1/2}\,. 
\end{align*}
Denote as $\itB_n=\left\{\pi(\wbthe, \bthe_n)<\epsilon_0\,\cap\, \pi_\prob(\wbthe, \bthe_n)<\delta\right\}$.  Then, $\prob(\itB_n)\to 1$. Using that $\delta=n^{-\alpha (1-\nu)/2}$, $K=\sum_{\ell=1}^p O(n^{\nu_{\ell}})=O(n^\nu)$ and the Markov inequality, we obtain that
\begin{align*}
\prob\left(\sqrt{n}  |S_{1,n,m,j}|>\epsilon \cap \itB_n\right)&\leq   \frac{1}{\epsilon}\esp^* \|\bbG_n\|_{\itF^{\star}}
\leq \frac{1}{\epsilon}\left(2\; C\; \delta\; \sqrt{\log\left(\frac{1}{\delta}\right)}\; K^{1/2}+\frac{4\; C^2\,B_j}{C_{mj}^2}  \log\left(\frac{1}{\delta}\right) \;K \, n^{-1/2}\right)  \\
&\leq   4 C^{\star}  \frac{1-\nu}{8}\left(n^{-\, \frac 18 (3-7\,\nu)}+n^{-\frac{1-2\nu}{2}}\right) \log\left(n\right) 
\leq     \frac{C^{\star}}{2} \left(n^{-\, \frac 18 (3-7\,\nu)}+n^{-\frac{1-2\nu}{2}}\right) \log\left(n\right) \,,
\end{align*}
which converges to $0$ since the fact that $r_\ell\geq 1$ for all $\ell$ implies that $\nu< 3/7$. Hence, using that $\prob(\itB_n)\to 1$, we obtain that $S_{1,n,m,j}=o_\prob(n^{-1/2})$,   concluding the proof of \textbf{H2}(b).

(iii) To conclude the proof, we will now show that \textbf{H3} is fulfilled.   Using a Taylor expansion of order two around $\bthe=(\btau\trasp,\eta_1,\dots,\eta_p)\trasp$, we get
\begin{eqnarray*}
\bW_{\bt,\varsigma}
&=& \bW_{\bthech,\varsigma} + \psi^\prime\left(\frac{Y-\btau\trasp\wtbZ-\sum_{j=1}^p \eta_j(X_j)}{\varsigma}\right)\frac{1}{\varsigma^2}(\bZ-\bh^*(\bX))(\bZ-\bh^*(\bX))\trasp(\bb-\bbe)\\
&& + \psi^\prime\left(\frac{Y-\btau\trasp\wtbZ-\sum_{j=1}^p \eta_j(X_j)}{\varsigma}\right)\frac{1}{\varsigma^2}(\bZ-\bh^*(\bX))\left\{ (a-\mu)+\bh^*(\bX)\trasp(\bd-\btau)\right\}\\
&&+\psi^\prime\left(\frac{Y-\btau\trasp\wtbZ-\sum_{j=1}^p \eta_j(X_j)}{\varsigma}\right)\frac{1}{\varsigma^2}(\bZ-\bh^*(\bX))\sum_{j=1}^p (g_j-\eta_j)(X_j)\\
&&+\frac{1}{2}\psi^{\prime\prime}\left(\frac{Y-\zeta_{\btauch}\trasp\wtbZ-\sum_{j=1}^p \zeta_j(X_j)}{\varsigma}\right)\left(-\frac{1}{\varsigma^3}\right)(\bZ-\bh^*(\bX))\left\{(\bd-\btau)\trasp\wtbZ+\sum_{j=1}^p (g_j-\eta_j)(X_j)\right\}^2
\end{eqnarray*}
with $\zeta_{\btauch}$ is intermediate points between $\bd$ and $\btau$ and    $\zeta_j=\xi_j g_{j}+(1-\xi_j)\eta_j$, with $0<\xi_j<1$ and $1\leq j\leq p$, respectively. For any $\bt=(\bd\trasp,g_1,\dots,g_p)\trasp $ and $\varsigma\in\itV$, denote as
\begin{eqnarray*}
\bB_{\bt,\varsigma}&=&\,-\,\esp \psi^\prime\left(\frac{Y-a- \bb\trasp\bZ-\sum_{j=1}^p g_j(X_j)}{\varsigma}\right)\frac{1}{\varsigma^2}(\bZ-\bh^*(\bX))(\bZ-\bh^*(\bX))\trasp\\
\bF_{\bt,\varsigma}&=&\esp \psi^\prime\left(\frac{Y-a- \bb\trasp\bZ-\sum_{j=1}^p g_j(X_j)}{\varsigma}\right)\frac{1}{\varsigma^2}(\bZ-\bh^*(\bX))\bh^*(\bX)\trasp\\
\bg_{\bt,\varsigma}&=&\esp \psi^\prime\left(\frac{Y-a- \bb\trasp\bZ-\sum_{j=1}^p g_j(X_j)}{\varsigma}\right)\frac{1}{\varsigma^2}(\bZ-\bh^*(\bX))\\
\be_{\bt,\varsigma}(\wtg_1,\dots,\wtg_p)&=&\esp \psi^\prime\left(\frac{Y-a- \bb\trasp\bZ-\sum_{j=1}^p g_j(X_j)}{\varsigma}\right)\frac{1}{\varsigma^2}(\bZ-\bh^*(\bX))\sum_{j=1}^p (\wtg_j-\eta_j)(X_j)\,.
\end{eqnarray*}
Then, we have that for any $\varsigma\in\itV$, 
$$P\bW_{\wbthech,\varsigma}=P\bW_{\bthech,\varsigma}-\bB_{\bthech,\varsigma}(\wbbe-\bbe)+\bg_{\bthech,\varsigma}(\wmu-\mu)+ \bF_{\bthech,\varsigma}(\wbbe-\bbe)+\be_{\bthech,\varsigma}(\weta_1,\dots,\weta_p)+\frac{1}{2}\bR_n(\wbthe,\varsigma)$$
where $\bR_n(\bt,\varsigma)=(R_{n,1}(\bt,\varsigma),\dots,R_{n,q+1}(\bt,\varsigma))\trasp$ is defined as
$$\bR_n(\bt,\varsigma)=\esp \psi^{\prime\prime}\left(\frac{Y-\zeta_{\btauch}\trasp\wtbZ-\sum_{j=1}^p \zeta_j(X_j)}{\varsigma}\right)\left(-\frac{1}{\varsigma^3}\right)(\bZ-\bh^*(\bX))\left\{\wtbZ\trasp(\bd-\btau)+\sum_{j=1}^p (g_j-\eta_j)(X_j)\right\}^2\,.$$
The independence between the errors and the covariates and the definition of $\bh^*$ imply that, for any $\varsigma\in\itV$, 
\begin{eqnarray*}
\bF_{\bthech,\varsigma}&=&\frac{1}{\varsigma^2}\esp \psi^\prime\left(\frac{\sigma\eps }{\varsigma}\right)\esp\left\{(\bZ-\bh^*(\bX))\bh^*(\bX)\trasp\right\}=\textbf{0}\\
\bg_{\bthech,\varsigma}&=&\frac{1}{\varsigma^2}\esp \psi^\prime\left(\frac{\sigma\eps }{\varsigma}\right)
\esp\left\{(\bZ-\bh^*(\bX))\right\}=\textbf{0}\\
\be_{\bthech,\varsigma}(g_1,\dots,g_p)&=&\frac{1}{\varsigma^2}\esp \psi^\prime\left(\frac{\sigma\eps }{\varsigma}\right)\esp\left\{(\bZ-\bh^*(\bX))\sum_{j=1}^p(g_j-\eta_j)(X_j)\right\}=\textbf{0}\,.
\end{eqnarray*}
On the other hand, \eqref{eq:PV0} and \eqref{eq:PVj} together with \eqref{eq:BW} entail that $P\bW_{\bthech,\varsigma}=\textbf{0}$, hence, we obtain that 
$$P\left(\bW_{\wbthech,\wsigma}-\bW_{\bthech,\sigma}\right)=  P\left(\bW_{\wbthech,\wsigma}-\bW_{\bthech,\wsigma}\right) =\;-\;\bB_{\bthech,\wsigma}(\wbbe-\bbe)+\frac{1}{2}\bR_n(\wbthe,\wsigma)\,.$$  
From the consistency of $\wsigma$ and the fact that $\psi^{\prime}$ is a continuous  bounded function, it is easy to see that $\bB_{\bthech,\wsigma}\convprob \bB_{\bthech,\sigma}$.   Then, in order to show that \textbf{H3} holds, it only remains to prove that $\bR_n(\wbthe,\wsigma)=o_\prob(n^{-1/2})$.

Denote  $b_{\bt}(\bZ,\bX)=\wtbZ\trasp(\bd-\btau)+\sum_{j=1}^p (g_j-\eta_j)(X_j)$ and $R_{n,m}(\bt,\varsigma)$ the $m-$th component of $\bR_n(\bt,\varsigma)$, $1\le m\le q$. Using that the second derivative of $\rho$ is bounded   we get that
$$|R_{n,m}(\bt,\varsigma)| \le  \frac{8}{\sigma^3}\|\psi^{\prime\prime}\|_\infty\esp \left\{\left(| Z_{m}|+|h^*_{m}(\bX)|\right)b_{\bt}^2(\bZ,\bX)\right\}\,.$$
Hence, using that \ref{ass:hstar} entails that $h^*_{m}$ is bounded, we obtain that $| R_{n,m}(\bt,\varsigma)|\le  R_{n,m,1}(\bt,\varsigma)+ R_{n,m,2}(\bt,\varsigma)$  where
\begin{align*}
R_{n,m,1}(\bt,\varsigma)= & \frac{8}{\sigma^3}\|\psi^{\prime\prime}\|_\infty \esp \left\{  | Z_{m}|  b_{\bt}^2(\bZ,\bX)\right\} = \frac{8}{\sigma^3}\|\psi^{\prime\prime}\|_\infty R_{n,m,1}^{\star}(\bt,\varsigma)\\
 R_{n,m,2}(\bt,\varsigma)= & \frac{8}{\sigma^3}\|\psi^{\prime\prime}\|_\infty \|h^*_{m}\|_{\infty} \esp  b_{\bt}^2(\bZ,\bX) = \frac{8}{\sigma^3}\|\psi^{\prime\prime}\|_\infty \|h^*_{m}\|_{\infty} \pi_\prob^2(\bt,\bthe)\,.
\end{align*}
Note that the fact that $r_j\ge 1$ implies that $\nu\le 1/3$ so $(1-\nu)/8\le  (1-2\nu)/4$. Besides, $\pi_{\prob}(\wbthe,\bthe)=O_\prob(n^{-(1-\nu)/2+\omega})$ with $ \omega< (1-\nu)/8\le (1-2\nu)/4$ so $R_{n,m,2}(\wbthe,\varsigma)=o_\prob(n^{-1/2})$. Therefore, we only have to show that $R_{n,m,1}^{\star}(\wbthe,\varsigma)=o_\prob(n^{-1/2})$. 

Using the Cauchy-Schwartz inequality we get that 
\begin{align}
R_{n,m,1}^{\star}(\bt,\varsigma)  & =  \esp \left\{ \left|b_{\bt} (\bZ,\bX)\right| | Z_{m}|  \left|b_{\bt} (\bZ,\bX)  \right| \right\} \nonumber\\
  & \le  \left\{\esp  b_{\bt}^2 (\bZ,\bX)  Z_{m}^2\right\}^{\frac 12} \left\{\esp b_{\bt}^2  (\bZ,\bX)\right\}^{\frac 12} =\left\{\esp  b_{\bt}^2 (\bZ,\bX)  Z_{m}^2\right\}^{\frac 12} \pi_{\prob}(\bt,\bthe)  \,.
  \label{eq:acotoRnm1}
\end{align}
Using again the Cauchy--Schwartz inequality we obtain that 
$$ \esp  b_{\bt}^2 (\bZ,\bX)  Z_{m}^2 =\esp  b_{\bt}  (\bZ,\bX)  Z_{m}^2\, b_{\bt}  (\bZ,\bX)\le \pi_{\prob}(\bt,\bthe) \left\{\esp  b_{\bt}^2 (\bZ,\bX)  Z_{m}^4\right\}^{\frac 12} $$
which together with \eqref{eq:acotoRnm1} and the fact that $(a+b)^2 \le 2 (a^2+b^2)$ leads to 
\begin{align*}
R_{n,m,1}^{\star}(\bt,\varsigma)  & \le \left\{\esp  b_{\bt}^2 (\bZ,\bX)  Z_{m}^4\right\}^{\frac 14} \pi_{\prob}^{\frac 32}(\bt,\bthe) \\
 & \le 2^{\frac 14}  \left\{\esp  Z_{m}^4 \left((1+\|\bZ\|^2)\|\bd-\btau\|^2+ p^2 \max_{1\le j\le p} \|g_j-\eta_j\|_{\infty}^2\right) \right\}^{\frac 14} \,  \pi_{\prob}^{\frac 32}(\bt,\bthe)
\end{align*}
that is,
$$R_{n,m,1}^{\star}(\wbthe,\wsigma) \le  2^{\frac 14} \left\{\max(\esp  Z_{m}^4 \|\bZ\|^2, p^2 \esp  Z_{m}^4 )\right\}^{\frac{1}{4}} \left\{  \|\wbtau-\btau\|^2+ \max_{1\le j\le p} \|\weta_j-\eta_j\|_{\infty}^2  \right\}^{\frac 14}\;
\pi_{\prob}^{\frac 32}(\wbthe,\bthe)\,.$$
Therefore, if \ref{ass:sobreZ}(a) holds, $r_j>1$ for all $1\le j\le p$ which implies that $\nu< 1/3$, so using that   $\pi_{\prob}(\wbthe,\bthe)=O_\prob(n^{-(1-\nu)/2+\omega})$, we obtain that 
$$n^{\frac 12} \pi_{\prob}^{\frac 32}(\bt,\bthe)= O_{\prob}(1) n^{\frac{1}{2}-\frac{3(1-\nu)}4+\frac{3}{2}\,\omega} = O_{\prob}(1) n^{-\frac{1-3\nu}4+\frac{3}{2}\,\omega}=o_{\prob}(1)\,,$$
since $0<\omega< (1-3\nu)/6$ which allows to conclude that  $R_{n,m,1}^{\star}(\wbthe,\varsigma)=o_\prob(n^{-1/2})$.

Assume now that \ref{ass:sobreZ}(b) holds. Note that in this case $\nu=1/3$. Using again the Cauchy--Schwartz inequality, we get the  bound
$$\esp  b_{\bt}^2 (\bZ,\bX)  Z_{m}^4 \le \pi_{\prob}(\bt,\bthe) \, \left\{\esp  b_{\bt}^2 (\bZ,\bX)  Z_{m}^8\right\}^{\frac 12} $$
which leads to 
\begin{align*}
R_{n,m,1}^{\star}(\bt,\varsigma)  & \le \left\{\esp  b_{\bt}^2 (\bZ,\bX)  Z_{m}^4\right\}^{\frac 14} \pi_{\prob}^{\frac 32}(\bt,\bthe) \le \left\{\esp  b_{\bt}^2 (\bZ,\bX)  Z_{m}^8\right\}^{\frac 18} \pi_{\prob}^{\frac 74}(\bt,\bthe)\\
 & \le 2^{\frac 18}  \left\{\esp  Z_{m}^8 \left[(1+\|\bZ\|^2)\|\bd-\btau\|^2+ p^2 \|g_j-\eta_j\|_{\infty}^2\right] \right\}^{\frac 18} \,  \pi_{\prob}^{\frac 74}(\bt,\bthe)
 \\
 & \le 2^{\frac 18}  \left\{\max(\esp  Z_{m}^8 \|\bZ\|^2, p^2 \esp  Z_{m}^8 )\right\}^{\frac{1}{8}} \left\{\|\bd-\btau\|^2+  \|g_j-\eta_j\|_{\infty}^2  \right\}^{\frac 18} \,  \pi_{\prob}^{\frac 74}(\bt,\bthe)\,.
\end{align*}
Hence, using that   $\pi_{\prob}(\wbthe,\bthe)=O_\prob(n^{-1/3+\omega})$ implies that 
$$n^{\frac 12} \pi_{\prob}^{\frac 74}(\bt,\bthe)= O_{\prob}(1) n^{\frac{1}{2}-\frac{7}{12}+\frac{7}{4}\,\omega} = O_{\prob}(1) n^{-\frac{1}{12}+\frac{7}{4}\,\omega}=o_{\prob}(1)\,,$$
since $\omega<1/21$.

Finally to obtain the asymptotic variance of the estimators, it is enough to note that the independence between the errors and the covariates imply that $$
\bB_{\bthech,\sigma}=\, -\, \frac 1{\sigma^2}\esp \psi^\prime(\eps) \bA \qquad \mbox{ and }\qquad 
\bD_{\bthech,\sigma}=\esp \bW_{\bthech,\sigma}\bW_{\bthech,\sigma}\trasp = \frac 1{\sigma^2}\esp \psi^2(\eps) \bA\,.
$$
Therefore, the asymptotic covariance matrix is given by
$$\bSi =\bB_{\bthech,\sigma}^{-1}\bD_{\bthech,\sigma}\bB_{\bthech,\sigma}^{-1\mbox{\footnotesize{\sc t}}}=\sigma^2\frac{\esp \psi^2(\eps )}{\left\{\esp\psi^\prime(\eps )\right\}^2}\bA$$
concluding the proof.  \qed

\small


\end{document}